\documentclass[prb,aps,twocolumn,nofootinbib,floatfix,10pt,longbibliography,superscriptaddress,notitlepage]{revtex4-2}
\usepackage[english]{babel}
\usepackage{breakcites}
\usepackage{listings}
\usepackage[utf8]{inputenc}
\usepackage[T1]{fontenc}
\usepackage{textcomp}
\usepackage{gensymb,braket}

\usepackage{graphicx}
\usepackage{braket}
\usepackage{color}
\usepackage{epstopdf}
\usepackage{mathtools}
\usepackage{amsmath}
\usepackage{amssymb}
\usepackage{float}
\usepackage{comment}
\usepackage{amsmath}
\usepackage{multirow}

\usepackage{amsfonts}
\usepackage{bm}
\usepackage{bbold}

\usepackage{color}

\graphicspath{{./figs/}}
\usepackage{gensymb}
\usepackage[colorinlistoftodos,prependcaption]{todonotes}
\usepackage{xargs}  
\newcommandx{\unsure}[2][1=]{\todo[linecolor=red,backgroundcolor=red!25,bordercolor=red,#1]{#2}}
\newcommandx{\change}[2][1=]{\todo[linecolor=blue,backgroundcolor=blue!25,bordercolor=blue,#1]{#2}}
\newcommandx{\info}[2][1=]{\todo[linecolor=OliveGreen,backgroundcolor=OliveGreen!25,bordercolor=OliveGreen,#1]{#2}}
\newcommandx{\improvement}[2][1=]{\todo[linecolor=Plum,backgroundcolor=Plum!25,bordercolor=Plum,#1]{#2}}
\newcommandx{\thiswillnotshow}[2][1=]{\todo[disable,#1]{#2}}
\newcommandx{\greencom}[2][1=]
{\todo[inline, color=green!40,#1]{#2}}
\newcommandx{\bluecom}[2][1=]
{\todo[inline, color=blue!40,#1]{#2}}

\usepackage[colorlinks]{hyperref}
\definecolor{winered}{rgb}{0.5,0,0}
\hypersetup
{
colorlinks=true,
linkcolor=winered,
urlcolor={winered},
filecolor={winered},
citecolor={winered},
allcolors={winered}
}

\usepackage{letltxmacro}
\LetLtxMacro{\ORIGselectlanguage}{\selectlanguage}
\makeatletter
\DeclareRobustCommand{\selectlanguage}[1]{%
  \@ifundefined{alias@\string#1}
    {\ORIGselectlanguage{#1}}
    {\begingroup\edef\x{\endgroup
       \noexpand\ORIGselectlanguage{\@nameuse{alias@#1}}}\x}%
}
\newcommand{\definelanguagealias}[2]{%
  \@namedef{alias@#1}{#2}%
}
\makeatother

\definelanguagealias{en}{english}
\definelanguagealias{EN}{english}

\graphicspath{{figure/}}

\begin{document}
\title{
Quantized quasinormal mode theory of coupled lossy and amplifying 
resonators}
\author{
Sebastian Franke}
\email{sebastian.r.franke@gmail.com}
\affiliation{Technische Universit\"at Berlin, Institut f\"ur Theoretische Physik,
Nichtlineare Optik und Quantenelektronik, Hardenbergstra{\ss}e 36, 10623 Berlin, Germany}
\affiliation{\hspace{0pt}Department of Physics, Engineering Physics, and Astronomy, Queen's University, Kingston, Ontario K7L 3N6, Canada\hspace{0pt}}
\author{Juanjuan Ren}
\affiliation{\hspace{0pt}Department of Physics, Engineering Physics, and Astronomy, Queen's University, Kingston, Ontario K7L 3N6, Canada\hspace{0pt}}
 \author{Stephen Hughes}
\affiliation{\hspace{0pt}Department of Physics, Engineering Physics, and Astronomy, Queen's University, Kingston, Ontario K7L 3N6, Canada\hspace{0pt}}

\date{\today}

\begin{abstract}
In the presence of arbitrary three-dimensional linear media with 
material loss and amplification, we present an electromagnetic field quantization scheme for quasinormal modes (QNMs), extending previous work
for lossy media [Franke et al., Phys. Rev. Lett. 122, 213901 (2019)]. Applying a symmetrization transformation, we show two fundamentally different ways for constructing a QNM photon Fock space, including (i) where there is 
a separate operator basis for both gain and loss, and (ii) where the loss and gain degrees of freedom are combined into a single basis. These QNM operator bases are subsequently used to derive the associated QNM master equations, including the interaction with a quantum emitter, modelled as a quantized two-level system (TLS). We then compare the two different quantization approaches, and also show how commonly used phenomenological methods to quantize light in gain-loss resonators are corrected by several important aspects, such as a loss-induced and gain-induced intermode coupling, which appears through the rigorous treatment of loss and amplification on a dissipative mode level.  
For  specific resonator designs, 
modelled
in a  fully consistent way with the classical Maxwell equations with open boundary conditions, 
we then present numerical results for the quantum parameters and observables of a TLS weakly interacting with the medium-assisted field in a gain-loss microdisk resonator system, and discuss the validity of the different quantization approaches for several gain-loss parameter regimes.
\end{abstract}

\maketitle 
\section{Introduction}
Light-matter interactions in nano-optical structures, such as photonic crystal cavities~\cite{Yoshie_Nature_432_200_2004,sh2007,KamandarDezfouli2017} and semiconductor micropillars~\cite{micropillars,micropillars2,Reithmaier_Nature_432_197_2004}, is a major topic in quantum optics and laser science. These optical cavities provide an ideal platform for studying several phenomena in open system quantum optics, such as non-classical light generation~\cite{faraon2008coherent,brooks2012non}, and emerging applications such as quantum information processing~\cite{quantinfo,loss1998quantum}.
Recently, $\mathcal{PT}$-symmetrical systems and general gain-loss resonators have also gained much attraction as a basis to investigate exceptional point (EP) physics and non-Hermitian quantum physics, which comes with a large variety of applications, such as improved lasing and sensing~\cite{peng_loss-induced_2014,peng_paritytime-symmetric_2014,chang_paritytime_2014,chen_exceptional_2017,chen_parity-time-symmetric_2018,miri_exceptional_2019}.

While the quantum description of lossy resonator structures has been addressed from many different viewpoints, e.g., via coupling to a ``bath''~\cite{carmichael2009statistical,GirishBook1} or via dissipative modes~\cite{2ndquanho,PhysRevLett.122.213901}, the combination of lossy and amplifying environments on a quantized photon level is still an open and recently debated problem.
Moreover, although 
some progress has been made towards a description of gain through a Scully Lamb model and quantum-Langevin type equations with phenomenological gain and loss noise operators~\cite{Kepesidis_2016,PhysRevA.96.033806,PhysRevA.99.053806,PhysRevA.101.013812}, not much effort has been made in terms of a dissipative mode quantization for gain-loss structures.
Such an approach is especially important for 
 coupled resonators, where we expect the response to be dominated by a few dominant cavity modes~\cite{KamandarDezfouli2017,dezfouli2019molecular}.

From a macroscopic quantization viewpoint, nearly two decades ago a quantization approach for spatial-inhomogeneous and absorptive media was introduced~\cite{Dung,scheel1998qed,vogel2006}, where the familiar normal mode expansion of the electromagnetic fields in non-absorptive dielectric media was generalized in terms of a source-field expansion of the field operators using the photonic Green function together with a phenomenological noise operator basis. The Green function quantization method is based on earlier works from Huttner and Barnett~\cite{Huttner,barnett1992spontaneous} on quantization in spatial-homogeneous media, and was later justified by more rigorous quantization approaches~\cite{suttorp,philbin2010canonical}, based on a microscopic oscillator model for the medium degrees of freedom~\cite{hopfield1958theory} and a Fano diagonalization method~\cite{PhysRev.72.26}. Furthermore, the approach was extended to the case of amplifying media~\cite{raabe2008qed}, which is also justified by a complementary inverted microscopic oscillator model~\cite{PhysRevA.84.013806}. For the case of amplifying and lossy media, it was recently shown 
that the intrinsic nature of gain can lead to drastically different behavior of quantized light-matter interactions through the altered quantum vacuum fluctuations~\cite{franke2021fermi}.

The Green function quantization has already been successfully applied to a variety of phenomena~\cite{Scheel,dung2000spontaneous,philbin2011casimir,PhysRevB.92.205420} in quantum optics and quantum plasmonics. However, the presence of  a continuum of photon noise operators usually prevents the use of the theory for light-matter interaction beyond the weak coupling limit. For purely lossy media, this problem was recently tackled via a dissipative mode quantization scheme~\cite{PhysRevLett.122.213901,franke2020fluctuation,franke2020quantized} based on a quasinormal mode~\cite{Lai,LeungSP1D,Leung3,2ndquant2,Kristensen:20} (QNM) expansion, which allowed for a construction of a 
discrete photon Fock basis for a lossy resonator problem. 

The QNMs are solutions to the Helmholtz equation with open boundary conditions, leading to complex QNM eigenfrequencies. One of the main advantages of QNM theory is the fact that, for many resonator structures, only a few modes are necessary to accurately calculate different figures of merit for semiclassical light-matter applications, such as the Purcell factors and the radiative $\beta$-factors~\cite{muljarovPert,KristensenHughes,SauvanNorm,NormKristHughes,PhysRevA.98.043806,Lalanne_review,carlson2019dissipative}. 
In addition, it was recently shown, that the QNMs do not only provide a suitable basis to describe lossy cavity structures from the classical physics viewpoint, but also a combination of lossy and amplifying structures, as long as the QNM poles are located in the lower complex half plane~\cite{EPClassicalPaper}.
This motivates system-level mode quantization for such systems, where both loss and gain can be includes
in a self-consistent way.

In this work, we combine QNMs and the Green function quantization in amplifying and lossy media to formulate a new general theory of mode quantization in gain-loss systems, which can be regarded as a generalization of the quantization scheme from Ref.~\cite{PhysRevLett.122.213901} (restricted to lossy structures). We tackle this problem  in two different ways: (i) The degrees of freedom originating from gain and loss are combined into a single photon Hilbert space; since this is closely related to the more phenomenological quantum gain models, this first method will allow us to compare the usual normal mode approaches to the more rigorous QNM approaches on a quantum level, and also to validate the more phenomenological approaches in certain gain-loss parameter regimes. (ii)  Gain and loss are separated into two different statistically independent Hilbert spaces; we stress this is completely different to the usual approaches and can be used for a variety of gain-loss regimes, while preserving the definition of the vacuum state from the photon continuum operators. 

Our manuscript is organized  as follows: In Section~\ref{Sec: TheoBackground}, we 
introduce the theoretical background used for the QNM quantization scheme with gain and loss. Specifically, we present the macroscopic Green function quantization in the presence of lossy and amplifying media including a coupling to a quantum emitter as a two-level system (TLS), some aspects of classical QNM theory as well as a summary of the QNM quantization in purely lossy media from Ref.~\onlinecite{PhysRevLett.122.213901}.
In Section~\ref{Sec: QNMQuantization}, we then present the quantization scheme for QNMs in lossy and amplifying media for the two different photon Hilbert space treatments, as explained above. We introduce the respective QNM Fock space and symmetrized QNM operator basis, the quantum Langevin equation with the associated Hamiltonian,
as well as the respective QNM master equation.  
In Section~\ref{Sec: DiffQuantumModels}, we discuss the subtle differences between the two different approaches, and compare with the usual phenomenological models of gain-loss resonators, that are based on Scully-Lamb model in linear gain regime~\cite{PhysRevA.99.053806,PhysRevA.101.013812} as well as extended
models with a real coupled mode coefficient~\cite{PhysRevA.100.062131,PhysRevA.81.033843, doi:10.1080/09500340.2014.992992}.

In Section~\ref{Sec: Applications}, we then show quantitative results for a two-dimensional gain-loss ring resonator system based on the different quantization approaches. In a first step, we briefly present the numerical results of the QNM calculations for the hybrid resonator structure. In a second step, we introduce an adapted and improved phenomenological quantum gain model where the parameters are partly taken from rigorous QNM calculations, and compare the resulting coupling parameters with the results from the quantized QNM approach. In a third step, we then discuss the mode quantum properties and the validity of the quantum approaches with combined gain and loss operators for different configurations of the ring resonators.
Fourth, we  derive the bad cavity limit of the quantized QNM approaches as well as the improved phenomenological quantum gain approach
and investigate the steady state population of the quantum emitter within this limit. We compare the results to other more classical and semiclassical models that are typically used in the literature. 
We then summarize our findings and results in Section~\ref{Sec: Conclusion}. 

The main text is complemented by several appendices, including the derivation of the photon Hamiltonians as well as the light-matter Hamiltonian in the QNM picture, the derivation of quantum Langevin equation and the QNM master equations, and a detailed derivation of the bad cavity limit of the QNM master equation with gain and loss within a density matrix picture and a Bloch equation treatment, as well as an analogous derivation for the improved phenomenological quantum gain model.

\section{Theoretical background\label{Sec: TheoBackground}}
\subsection{Macroscopic quantization in amplifying and lossy  media\label{subsec: MacroQuantization}}
In this subsection, we briefly summarize the macroscopic Green function quantization approach in the presence of amplifying and lossy media~\cite{raabe2008qed}, which is an extension of the quantum theory for purely lossy media~\cite{Dung,vogel2006}. 

In the dipole and rotating-wave approximations, the Hamiltonian of a quantum emitter as a TLS interacting with the medium-assisted photon field is given via $H=H_{\rm em}+H_{\rm a}+ H_{\rm I}$, as~\cite{raabe2008qed}
\begin{subequations}\label{eq: Hamiltonian}
\begin{align}
H_{\rm em}&=\hbar\int_0^{\infty}{\rm d}\omega~\omega\int{\rm d}^3r~ {\rm sgn}[\epsilon_I]~\mathbf{b}^\dagger(\mathbf{r},\omega)\cdot\mathbf{b}(\mathbf{r},\omega)\label{eq:HBgain},\\
H_{\rm a}&=\hbar\omega_{\rm a}\sigma^+\sigma^-,\label{eq: H_atom}\\
H_{\rm I}&=-\left[\sigma^+\int_0^\infty{\rm d}\omega \mathbf{d}_{\rm a}\cdot\hat{\mathbf{E}}(\mathbf{r}_{\rm a},\omega)+{\rm H.a.}\right],\label{eq: HI}
\end{align}
\end{subequations}
where $H_{\rm em}$ is the energy of the combined system of the vacuum electric field and the amplifying and lossy medium, $\mathbf{b}^{(\dagger)}(\mathbf{r},\omega)$ are the corresponding vector-valued bosonic annihilation (creation) operators and the spatial integration is performed with respect to $\mathbb{R}^3$.

Importantly,
in contrast to the purely lossy case, a sign function ${\rm sgn}[\epsilon_I]$ appears in $H_{\rm em}$, where $\epsilon_I=\epsilon_I(\mathbf{r},\omega)$ is the imaginary part of the (in general) spatially inhomogeneous and $\omega$-dependent complex permittivity function $\epsilon(\mathbf{r},\omega)$, which describes passive (lossy) media through $\epsilon_I(\mathbf{r},\omega)\geq 0$ and active (amplifying) media through $\epsilon_I(\mathbf{r},\omega)<0$. 
The terms $H_{\rm a}$ and $H_{\rm I}$ represent the TLS
and emitter-field interactions, respectively; here, $\omega_{\rm a}$, $\mathbf{r}_{\rm a}$ and $\mathbf{d}_{\rm a}$ are the transition frequency, spatial position and dipole moment of the TLS, respectively. The raising and lowering (Pauli) operators of the TLS are given by
 $\sigma^{\pm}$.

The operator $\hat{\mathbf{E}}(\mathbf{r},\omega)$ is the medium-assisted field operator, which solves the quantized Helmholtz equation, 
\begin{equation}
\boldsymbol{\nabla}\times\boldsymbol{\nabla}\times\hat{\mathbf{E}}(\mathbf{r},\omega) -\frac{\omega^2}{c^2}\epsilon(\mathbf{r},\omega)\hat{\mathbf{E}}(\mathbf{r},\omega)=i\omega\mu_0\hat{\mathbf{j}}_{\mathrm N}(\mathbf{r},\omega),\label{eq: HelmholtzQuant2}
\end{equation}
where $c$ is the speed of light and $\mu_0$ is the vacuum permeability. 
Furthermore, $\hat{\mathbf{j}}_{\mathrm N}(\mathbf{r},\omega)$ is a 
quantum noise density, that is introduced to account for the interaction of the free electromagnetic field with the dielectric medium. In the presence of loss and amplifying media, it is given as the sum of loss and gain terms $\hat{\mathbf{j}}_{\mathrm N}(\mathbf{r},\omega)=\hat{\mathbf{j}}^{\rm L}_{\mathrm N}(\mathbf{r},\omega)+\hat{\mathbf{j}}^{\rm G}_{\mathrm N}(\mathbf{r},\omega)$, where~\cite{scheel1998qed,raabe2008qed}
\begin{subequations}\label{eq: NoiseCurrents}
\begin{align}
 \hat{\mathbf{j}}^{\rm L}_{\mathrm N}(\mathbf{r},\omega)&=\omega\sqrt{\frac{\hbar\epsilon_0}{\pi}\epsilon_I(\mathbf{r},\omega)}\Theta(\epsilon_I)\mathbf{b}(\mathbf{r},\omega),\\
 \hat{\mathbf{j}}^{\rm G}_{\mathrm N}(\mathbf{r},\omega)&=\omega\sqrt{\frac{\hbar\epsilon_0}{\pi}|\epsilon_I(\mathbf{r},\omega)|}\Theta(-\epsilon_I)\mathbf{b}^\dagger(\mathbf{r},\omega),
\end{align}
\end{subequations}
with the Heaviside function, $\Theta(\epsilon_I)$, defined through 
\begin{align}
    \Theta(\epsilon_I)=
    \begin{cases}
    1 & \epsilon_I(\mathbf{r},\omega) \geq 0 \\
    0 & \epsilon_I(\mathbf{r},\omega) < 0, 
    \end{cases}
\end{align}
which implicitly depends on $\mathbf{r}$ and $\omega$, and $\Theta(-\epsilon_I)=1-\Theta(\epsilon_I)$. Note that it can be shown that the specific choice of the noise operators, with gain (`G') and loss (`L') contributions from Eq.~\eqref{eq: NoiseCurrents}, preserves the fundamental commutation relations of the electromagnetic operators~\cite{raabe2008qed}.

A formal solution of Eq.~\eqref{eq: HelmholtzQuant2} is given as
\begin{equation}
\hat{\mathbf{E}}(\mathbf{r},\omega)=\frac{i}{\omega\epsilon_0}\int{\mathrm d}^3s\, \mathbf{G}(\mathbf{r},\mathbf{s},\omega)\cdot \hat{\mathbf{j}}_{\mathrm N}(\mathbf{s},\omega),\label{eq: SourceEField2}
\end{equation}
where 
$\mathbf{G}(\mathbf{r},\mathbf{r}',\omega)$ is the (classical) photon Green function, 
defined from
\begin{equation}
\boldsymbol{\nabla}\times\boldsymbol{\nabla}\times\mathbf{G}(\mathbf{r},\mathbf{r}',\omega) -\frac{\omega^2}{c^2}\epsilon(\mathbf{r},\omega)\mathbf{G}(\mathbf{r},\mathbf{r}',\omega)=\mathbb{1}\delta(\mathbf{r}-\mathbf{r}')\label{eq: GFHelmholtz2},
\end{equation}
with suitable radiation conditions. The total electric field operator is then obtained as the integration over all $\omega$,  through 
\begin{equation}
    \hat{\mathbf{E}}(\mathbf{r},t)=\int_0^\infty {\rm d}\omega  \hat{\mathbf{E}}(\mathbf{r},\omega,t)+{\rm H.a.} ,\label{eq: PositiveRotPart}
\end{equation}
where we emphasize that $\omega$ is not a Fourier variable of time, but rather a mode component of the medium-photon fields. The time dependence of $\hat{\mathbf{E}}(\mathbf{r},\omega,t)$ is governed by the Heisenberg equation of motion with respect to the Hamiltonian from Eq.~\eqref{eq: Hamiltonian}.

From Eqs.~\eqref{eq: NoiseCurrents} and \eqref{eq: SourceEField2}, one can readily see that, in contrast to purely lossy media, the electric field operator component $\hat{\mathbf{E}}(\mathbf{r},\omega,t)$ is a linear combination of photon-medium annihilation and creation operators, because the role of annihilation and creation of quanta is exchanged in the amplifying region.
Moreover, 
it should be noted that this reversed nature of annihilation and creation is accounted for as a (formal) negative photon energy part in $H_{\rm em}$, which preserves the consistency with the associated macroscopic Maxwell equations~\cite{raabe2008qed}. Consequently, $\int_0^\infty {\rm d}\omega \hat{\mathbf{E}}(\mathbf{r},\omega)$ constitutes the positive rotating part of the electric field as would be the case for purely lossy media, which is in turn consistent with the rotating-wave approximation applied to the light-matter interaction Hamiltonian $H_{\rm I}$ (Eq.~\eqref{eq: HI}).

We remark that the quantum theory is strictly only applicable in the case of {\it linear amplifying media}. 
One of the consequences of this  criterion, is  that the Green function  $\mathbf{G}(\mathbf{r},\mathbf{r}',\omega)$ is only allowed to have complex poles in the lower complex half plane, meaning that it must be analytic in the upper complex half plane\footnote{This is not a unique restriction for the quantum theory, as it also applies to classical linear Maxwell theory, which of course is also subject to causality conditions in the sense of linear response theory.}. Any poles in the upper half plane would lead to a breakdown of the fundamental field commutation relations~\cite{raabe2008qed}, which of course coincides with the breakdown of causality in the sense of linear response theory.

\subsection{Classical quasinormal mode theory}
\label{class}

Here, we summarize the key properties of the 
classical QNMs~\cite{2ndquant2,NormKristHughes,Lalanne_review,Kristensen:20} as well as 
 QNM expansion technique, which can be used to model open cavity systems.
The electric-field QNMs, $\tilde{\mathbf{f}}_{{\mu}}(\mathbf{r})$, are solutions to the 
source-free 
Helmholtz equation,
\begin{equation}\label{smallf}
\boldsymbol{\nabla}\times\boldsymbol{\nabla}\times\tilde{\mathbf{f}}_{{\mu}}\left(\mathbf{r}\right)-\left(\dfrac{\tilde{\omega}_{{\mu}}}{c}\right)^{2}
\epsilon(\mathbf{r},\tilde{\omega}_{\mu})\,\tilde{\mathbf{f}}_{{\mu}}\left(\mathbf{r}\right)=0,
\end{equation}
where 
$\tilde{\omega}_{{\mu}}= \omega_{{\mu}}-i\gamma_{{\mu}}$ is the complex eigenfrequency,
and $\epsilon(\mathbf{r},\tilde{\omega}_\mu)$ is the analytical continuation of the permittivity function (introduced in subsection~\ref{subsec: MacroQuantization}) into the complex frequency space.  

The open boundary conditions ensure the Silver-M\"uller radiation conditions~\cite{Kristensen2015}, and the quality factor of each resonance
is $Q_\mu = \omega_\mu/(2\gamma_\mu)$. 
 Normalization of the QNMs can be
done in different ways \cite{KristensenHughes,SauvanNorm,muljarovPert}, and additional care is needed for dispersive and absorptive cavity structures.
For the cavity structures investigated later, 
the  QNMs can be obtained
from an efficient dipole scattering approach 
in complex frequency~\cite{Bai}, 
using an inverse Green function approach.
 The total Green function
can also be obtained numerically
from the full dipole response (namely, without any modal approximation),
which can be used to check the accuracy of the QNM expansions and mode approximations.

For spatial positions near (or within) the scattering geometry, the Green function 
can be expanded in terms of the QNMs through~\cite{MDR1,GeNJP2014,Kristensen:20}
\begin{equation}
\mathbf{G}\left(\mathbf{r},\mathbf{r}_{0},\omega\right)= \sum_{\mu} A_{\mu}\left(\omega\right)\,\tilde{\mathbf{f}}_{\mu}\left({\bf r}\right)\tilde{\mathbf{f}}_{\mu}\left({\bf r}_{0}\right),\label{eq:GFwithSUM}
\end{equation}
with 
$A_{\mu}(\omega)={\omega}/[{2(\tilde{\omega}_{\mu}-\omega)}]$\footnote{Or one can also use
    $A\left(\omega\right)=\omega^{2}/[2\tilde{\omega}_{\mu}\left(\tilde{\omega}_{\mu}-\omega\right)]$
    since these are related through a sum relationship~\cite{Kristensen:20}.}.
The Green function enables us to connect to a wide range of physical
quantities such as the local density of states (LDOS). Thus, we can define a normalized 
(projected) LDOS
in terms of the QNM Green function,
\begin{equation}
\rho_{\rm a}^{\rm QNM} = \frac{{\bf n}_{\rm a} \cdot {\rm Im} {\bf G}({\bf r}_{\rm a},{\bf r}_{\rm a},\omega) \cdot{\bf n}_{\rm a}}{{\bf n}_{\rm a} \cdot {\rm Im} {\bf G}_{\rm B}({\bf r}_{\rm a},{\bf r}_{\rm a},\omega) \cdot{\bf n}_{\rm a}},    \label{eq: LDOS_normalized}
\end{equation}
 where ${\bf G}_{\rm B}$
   is the Green function for a homogeneous {\it background} medium
   and we recall that ${\bf r}_{\rm a}$ is the location of the dipole emitter, with dipole moment 
    $\mathbf{d}_{\rm a}$ (=\,d\,$\mathbf{n}_{\rm a}$).

It is important to note that the usual classical 
expressions for spontaneous emission and the Purcell factor do not work in the presence of a gain medium~\cite{franke2021fermi,EPClassicalPaper}. 
Nonetheless, the projected LDOS is still a useful quantity to define and for checking the  accuracy of a QNM expansion.

Finally, we mention that for fields far outside the resonator, a few  QNM expansion is  no longer convenient to use as they diverge due to the radiation condition in conjunction with the complex eigenfrequency.
In this case, one can use the Dyson equation to construct regularized
QNMs, derived as~\cite{GeNJP2014}
\begin{align}
    \tilde {\bf F}_\mu({\bf R},\omega)
    = \int_{\rm res} d {\bf r}\Delta \epsilon({\bf r},\omega)
    {\bf G}_{\rm B}({\bf R},{\bf r},\omega)
    \cdot\tilde{\bf f}_\mu({\bf r}).\label{eq: RegF}
\end{align}
Alternatively, these regularized fields can be constructed in the far field using 
using near-field to far-field transformations~\cite{ren_near-field_2020}.
For the resonators we use below,
however, the QNMs are already an excellent approximation at the spatial positions
we consider, especially for the
high $Q$ cavities that we will consider.  

\subsection{Quantization of lossy quasinormal modes\label{Subsec: LossyModeQuant}}
In this part, we summarize the main results from the lossy QNM quantization. For the case of purely lossy media (which includes the limit of a lossless dielectrics~\cite{franke2020fluctuation}), the noise current density $\hat{\mathbf{j}}_{\mathrm N}(\mathbf{r},\omega)$ is equal to the loss-induced noise $\hat{\mathbf{j}}_{\mathrm N}^{\rm L}(\mathbf{r},\omega)$, extending over all space, so that 
\begin{equation}
    \hat{\mathbf{E}}(\mathbf{r},\omega)=i\sqrt{\frac{\hbar}{\pi\epsilon_0}}\int{\mathrm d}^3s\,\sqrt{\epsilon_I(\mathbf{s},\omega)} \mathbf{G}(\mathbf{r},\mathbf{s},\omega)\cdot \mathbf{b}(\mathbf{s},\omega).
\end{equation}

By using the QNM Green function, Eq.~\eqref{eq:GFwithSUM}, the modal part of the quantized electric field operator near or inside the scattering structure can be regarded as a combined cavity system (with one set of QNMs), and  is  written as
\begin{equation}
\hat{\mathbf{E}}_{\rm QNM}(\mathbf{r})=i\sum_{\mu}\sqrt{\frac{\hbar\omega_\mu}{2\epsilon_0}}\tilde{\mathbf{f}}_\mu(\mathbf{r})\tilde{a}_\mu + {\rm H.a.}, 
\end{equation}
where $\tilde{a}_\mu$ are QNM operators as linear combinations of the noise source operators $\mathbf{b}(\mathbf{r},\omega)$:
\begin{equation}
    \tilde{a}_\mu = \lim_{\lambda\rightarrow\infty}\int_0^\infty {\rm d}\omega\int_{V(\lambda)}{\rm d}^3 r \tilde{\mathbf{L}}_{\mu}(\mathbf{r},\omega)\cdot\mathbf{b}(\mathbf{r},\omega),
\end{equation}
with 
\begin{equation}
\tilde{\mathbf{L}}_{\mu}(\mathbf{r},\omega)=\sqrt{\frac{2}{\pi\omega_\mu}}A_\mu(\omega)\sqrt{\epsilon_I^{(\alpha)}(\mathbf{r},\omega)}\tilde{\mathbf{f}}_\mu(\mathbf{r},\omega),
\end{equation}
where $\tilde{\mathbf{f}}_\mu(\mathbf{r},\omega)$ is equal to the QNM $\tilde{\mathbf{f}}_\mu(\mathbf{r})$ for positions near or inside the cavity region and equal to the regularized QNM $\tilde{\mathbf{F}}_\mu(\mathbf{r},\omega)$, Eq.~\eqref{eq: RegF}, for positions outside the cavity region. 
Note, that the integration over all space $\mathbb{R}^3$ is formally written as a limiting process over a sequence of volumes $V(\lambda)$, such that $V(\lambda)\rightarrow \mathbb{R}^3$ for $\lambda\rightarrow\infty$.

In order to rigorously account for radiative loss processes within the macroscopic Green function quantization, we introduced a sequence of permittivity functions $\epsilon^{(\alpha)}(\mathbf{r},\omega)=\epsilon(\mathbf{r},\omega)+\alpha\chi_{\rm L}(\omega)$, where $\alpha\chi_{\rm L}(\omega)$ is a spatially homogeneous Lorentz oscillator weighted by the parameters $\alpha\geq 0$. It is important to emphasize, that the limit $\alpha\rightarrow 0$ and $\lambda\rightarrow\infty$ are not interchangeable;  in fact the ordering of the limits with respect to $\lambda$ and $\alpha$ is not only a requirement to obtain meaningful radiation processes, but also to preserve the fundamental field commutation relations in the dielectric (non-absorptive) limit~(cf. Ref.~\onlinecite{franke2020fluctuation} for details).

The mode quantization scheme can be performed by first constructing proper annihilation and creation operators through a symmetrization transformation:
\begin{align}
    a_{\mu}=\sum_{\eta}\left[\mathbf{S}^{-1/2}\right]_{\mu\eta}\tilde{a}_\eta,
\end{align}
with $S_{\mu\eta}=[\tilde{a}_{\mu},\tilde{a}_{\eta}^\dagger]$, where 
\begin{equation}
    S_{\mu\eta}=\lim_{\lambda\rightarrow\infty}\int_0^\infty {\rm d}\omega\int_{V(\lambda)}{\rm d}^3 r \tilde{\mathbf{L}}_{\mu}(\mathbf{r},\omega)\cdot\tilde{\mathbf{L}}_{\eta}^*(\mathbf{r},\omega),
\end{equation}
is a dissipation-induced QNM overlap matrix, which yields a positive definite form~\cite{franke2020fluctuation,PhysRevLett.122.213901}. After applying the limit $\alpha\rightarrow 0$, $S_{\mu\eta}$ can be written as a sum $S_{\mu\eta}=S_{\mu\eta}^\mathrm{nrad}+S_{\mu\eta}^\mathrm{rad}$, where~\cite{franke2020fluctuation,PhysRevLett.122.213901}
\begin{align}
S_{\mu\eta}^\mathrm{nrad}&=\int_0^\infty{\mathrm d}\omega\frac{2A_\mu(\omega)A_\eta^*(\omega)}{\pi\sqrt{\omega_\mu\omega_\eta}}I^{\rm nrad}_{\mu\eta}(\omega),\label{eq:SnradGeneral}\\
S^{\mathrm{rad}}_{\mu\eta} &=\int_0^\infty{\mathrm d}\omega\frac{2A_\mu(\omega)A_\eta^*(\omega)}{\pi\sqrt{\omega_\mu\omega_\eta}}[I^{\rm rad}_{\mu\eta}(\omega)+I^{\rm rad*}_{\eta\mu}(\omega)] \label{eq:SradGeneral} ,
\end{align}
represent {\it non-radiative} and {\it radiation} contributions,
with 
\begin{align}
    I^{\rm nrad}_{\mu\eta}(\omega)&=\int_{V_{\rm S}} {\mathrm d}^3s\epsilon_I(\mathbf{s},\omega)\tilde{\mathbf{f}}_\mu(\mathbf{s})\cdot\tilde{\mathbf{f}}_\eta^*(\mathbf{s}),\\
    I^{\rm rad}_{\mu\eta}(\omega)&=\frac{1}{2\epsilon_0\omega}\oint_{\mathcal{S}} {\mathrm d}A_\mathbf{s}\left[\tilde{\mathbf{H}}_{\mu}(\mathbf{s},\omega)\times\mathbf{n}_{\mathbf{s}}\right]\cdot\tilde{\mathbf{F}}_{\eta}^*(\mathbf{s},\omega).
\end{align}

While $S_{\mu\eta}^\mathrm{nrad}$ describes the absorption inside the scattering volume $V_{\rm S}$, 
$S^{\mathrm{rad}}_{\mu\eta} $ reflects the power flow of the regularized QNM fields through an outer surface $\mathcal{S}$. For practical calculations, it is also useful to apply a pole approximation to the above frequency integrals to obtain the symmetrization matrices through~\cite{PhysRevB.101.205402}
\begin{align}
 S_{\mu\eta}^\mathrm{nrad}&=\frac{\sqrt{\omega_\mu\omega_\eta}}{i(\tilde{\omega}_\mu-\tilde{\omega}_\eta^*)}I^{\rm nrad,p}_{\mu\eta},\label{eq:SnradPole}\\
S^{\mathrm{rad}}_{\mu\eta} &=\frac{\sqrt{\omega_\mu\omega_\eta}}{i(\tilde{\omega}_\mu-\tilde{\omega}_\eta^*)}[I^{\rm rad,p}_{\mu\eta}+I^{\rm rad,p*}_{\eta\mu}] \label{eq:SradPole} ,
\end{align}
with 
\begin{align}
     I^{\rm nrad,p}_{\mu\eta}&=\int_{V_{\rm S}} {\mathrm d}^3s\sqrt{\epsilon_I(\mathbf{s},\omega_\eta)\epsilon_I(\mathbf{s},\omega_\eta)}\tilde{\mathbf{f}}_\mu(\mathbf{s})\cdot\tilde{\mathbf{f}}_\eta^*(\mathbf{s}),
\end{align}
and 
\begin{align}
    I^{\rm rad,p}_{\mu\eta}&=\frac{1}{2\epsilon_0\sqrt{\omega_\mu\omega_\eta}}\oint_{\mathcal{S}} {\mathrm d}A_\mathbf{s}\left[\tilde{\mathbf{H}}_{\mu}(\mathbf{s},\omega_\mu)\times\mathbf{n}_{\mathbf{s}}\right]\cdot\tilde{\mathbf{F}}_{\eta}^*(\mathbf{s},\omega_\eta).
\end{align}

By construction, the symmetrized operators fulfill the commutation relations $[a_{\mu},a_{\eta}]=[a_{\mu}^\dagger,a_{\eta}^\dagger]=0$ and $[a_{\mu},a_{\eta}^\dagger]=\delta_{\mu\eta}$, for all $\mu,\eta$ and can thus be regarded as proper annihilation and creation operators to construct QNM Fock states from the vacuum state $|{\rm vac}\rangle_{\rm em}$, which is defined via $\hat{b}_i(\mathbf{r},\omega)|{\rm vac}\rangle_{\rm em} = 0$.

The construction of the QNM Fock space together with the Hamiltonian, Eq.~\eqref{eq: Hamiltonian} (restricted to purely lossy media) can be utilized to derive an associated master equation for the density operator $\rho$ in the combined QNM-TLS space~\cite{PhysRevLett.122.213901,franke2020quantized}:
\begin{equation}
\partial_t \rho = -\frac{i}{\hbar}[H_{\rm S},\rho]+\mathcal{L}\rho,
\end{equation}
where $H_{\rm S}=H_{\rm QNM-a}+H_{\rm a}+H_{\rm QNM}$ is the system Hamiltonian consisting of three parts. First, 
\begin{equation}
    H_{\rm QNM-a}=\hbar\sum_{\mu}\tilde{g}_\mu^{\rm s}a_\mu^\dagger\sigma^- +{\rm H.a.},
\end{equation}
is the TLS-QNM interaction Hamiltonian, where $\tilde{g}_\mu^{\rm s}=-i\sum_\eta[\mathbf{S}^{1/2}]_{\eta\mu}\sqrt{\omega_\eta/(2\hbar\epsilon_0)}\mathbf{d}_{\rm a}\cdot \tilde{\mathbf{f}}_\eta(\mathbf{r}_{\rm a})$ is the TLS-QNM coupling constant.
Second,
\begin{equation}
    H_{\rm QNM}=\hbar\sum_{\mu\eta}\chi^{+}_{\mu\eta}a_\mu^\dagger a_\eta,
\end{equation}
is the QNM Hamiltonian, including off-diagonal photon coupling terms through $\chi^{+}_{\mu\eta}=[\chi_{\mu\eta}+\chi_{\eta\mu}^*]/2$ with $\chi_{\mu\eta}=\sum_{\nu}[\mathbf{S}^{-1/2}]_{\mu\nu}\tilde{\omega}_\nu[\mathbf{S}^{1/2}]_{\nu\eta}$. Third,  $H_{\rm a}$ represents the TLS energy [Eq.~\eqref{eq: H_atom}]. Furthermore, the dissipator term, $\mathcal{L}=\mathcal{L}_{\rm em}+\mathcal{L}_{\rm SE}$ is the rigorously derived Lindblad super-operator, which contains the QNM photon decay through
\begin{equation}
    \mathcal{L}_{\rm em}=\sum_{\mu\eta}\chi_{\mu\eta}^{-}\left[2a_\eta\rho a_\mu^\dagger - a_\mu^\dagger a_\eta \rho -a_\eta a_\mu^\dagger \rho\right],
\end{equation}
with $\chi^{-}_{\mu\eta}=i[\chi_{\mu\eta}-\chi_{\eta\mu}^*]/2$,
and the atomic spontaneous emission through $\mathcal{L}_{\rm SE}$. For more details, see Refs.~\onlinecite{PhysRevLett.122.213901,franke2020quantized}.

\section{QNM quantization in amplifying and lossy media\label{Sec: QNMQuantization}}
Next, we extend the theory of lossy mode quantization from Refs.~\onlinecite{PhysRevLett.122.213901,franke2020fluctuation,franke2020quantized} to the more general case of combined lossy and amplifying dielectric media. To do this, we first reformulate the lossy mode quantization scheme to separate the combined QNM operators into different spatial regions of the medium. Then, we discuss the geometry of interest for a general gain-loss resonator system. Subsequently, we apply the quantization procedure to two different formulations to the general case of amplifying and lossy media. 

\subsection{Extending lossy QNM quantization to include gain\label{subsec: ExtensionLossyQuantization}}
Instead of defining a single set of QNM operators (as done in subsection~\ref{Subsec: LossyModeQuant}), one can also construct multiple QNM operator sets $\{\dots,a_{k\mu},\dots\}$, each corresponding to a source volume $V_k$ and the same QNM basis $\{\mu\}$, so that 
\begin{equation}
    \tilde{a}_{k\mu} = \lim_{\lambda\rightarrow\infty}\sum_i\int_0^\infty {\rm d}\omega\int_{V(\lambda)}{\rm d}^3 r \tilde{L}_{k\mu,i}^{(\lambda)}(\mathbf{r},\omega)\hat{b}_i(\mathbf{r},\omega),
\end{equation}
with $\tilde{L}_{k\mu,i}^{(\lambda)}(\mathbf{r},\omega)=\chi_{V_{k}}^{(\lambda)}(\mathbf{r})\tilde{L}_{\mu,i}(\mathbf{r},\omega)$ and 
\begin{align}
\chi_{V_{k}}^{(\lambda)}(\mathbf{r})=\begin{cases}
    1 & \mathbf{r}\in V_k(\lambda) \\
    0 & \mathbf{r}\notin V_k(\lambda), 
    \end{cases},~
    \bigcup_k V_k(\lambda) = V(\lambda).
\end{align}

The QNM electric field operator can then simply be  rewritten as
\begin{equation}
\hat{\mathbf{E}}_{\rm QNM}(\mathbf{r})=i\sum_{k}\sum_{\mu}\sqrt{\frac{\hbar\omega_\mu}{2\epsilon_0}}\tilde{\mathbf{f}}_\mu(\mathbf{r})\tilde{a}_{k\mu} + {\rm H.a.}, 
\end{equation}
and the commutation relations of these operators is given by $[\tilde{a}_{k\mu},\tilde{a}_{k'\eta}^\dagger]=\delta_{kk'}S_{\mu\eta}^{(k)}$, with 
\begin{equation}
    S_{\mu\eta}^{(k)}=\lim_{\lambda\rightarrow\infty}\int_0^\infty {\rm d}\omega\int_{V(\lambda)}{\rm d}^3 r \tilde{\mathbf{L}}_{k\mu}(\mathbf{r},\omega)\cdot\tilde{\mathbf{L}}_{k\eta}^*(\mathbf{r},\omega).
\end{equation}
Thus, the operator sets can be symmetrized separately through 
\begin{equation}
    a_{k\mu}=\sum_{\eta}\left[\left(\mathbf{S}^{(k)}\right)^{-1/2}\right]_{\mu\eta} \tilde{a}_{k\eta}.
\end{equation}

One obvious disadvantage of such a separation would be the increasing dimension of the underlying photon Hilbert space, which may complicate the numerical calculations for, e.g., simulating multiphoton processes. 
However, as we discuss further below, in the case of amplifying media, the separation leads to a fundamentally different quantum description of the linear amplified system on a mode level, a fact that has not been considered in previous quantization schemes of gain-loss resonators (to the best of our knowledge).

\subsection{Coupled gain-loss resonator geometries of interest}
We now turn to the general case of gain and loss within the medium (i.e., before QNM construction and field quantization). It is instructive to apply the formal separation of the spatial integration over the bosonic noise source operators $\mathbf{b}(\mathbf{r},\omega)$ in two different regions, namely, the loss region $\mathbb{R}^3-V_{\rm G}$, and the amplifying region $V_{\rm G}$. We assume that $V_{\rm G}$ always remains finite. In addition, we choose the parameter $\alpha$ small enough, such that the medium in $V_{\rm G}$ is purely amplifying  ($\epsilon_I^{(\alpha)}(\mathbf{r},\omega)<0$ for all $\omega$ and $\mathbf{r}\in V_{\rm G}$) and purely absorptive in $\mathbb{R}^3- V_{\rm G}$ ($\epsilon_I^{(\alpha)}(\mathbf{r},\omega)<0$ for all $\omega$ and $\mathbf{r}\in \mathbb{R}^3-V_{\rm G}$). In this way, $\Theta[\epsilon_I]$ coincides with $\chi_{\mathbb{R}^3-V_{\rm G}}(\mathbf{r})$, while $\Theta[-\epsilon_I]$ coincides with $\chi_{V_{\rm G}}(\mathbf{r})$. 
\begin{figure}[h]
 \centering
 \includegraphics[width=1\columnwidth,angle=0]{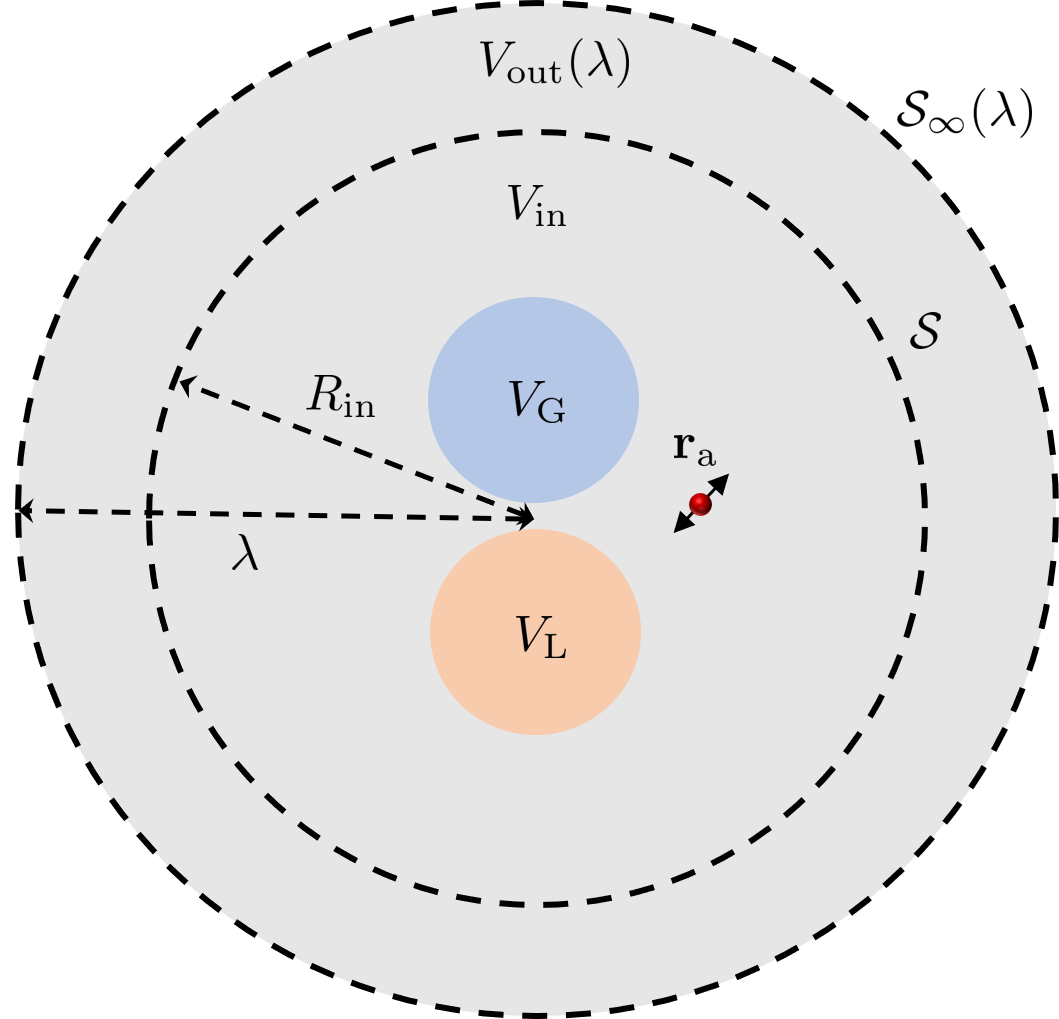}
 \vspace{-0.2cm}
 \caption[Visualization of the introduction of a sequence of permittivity functions]{{Schematic of the spatially inhomogeneous media with the physical permittivity $\epsilon(\mathbf{r},\omega)$, describing gain and absorption. The gain-loss system $V_{\rm G}+V_{\rm L}$ is contained in a fixed volume $V_{\rm in}$, which also includes an exemplary active quantum dipole at position $\mathbf{r}_{\rm a}$.  
 }}
\label{fig: SchematicGeometry}
\end{figure} 

As illustrated in Fig.~\ref{fig: SchematicGeometry}, we concentrate on a finite gain-loss scattering system (with regions $V_{\rm G}$ and $V_{\rm L}$), which is contained in a fixed ($\lambda$-independent) and finite spherical volume $V_{\rm in}$ with radius $R_{\rm in}$ and surface $\mathcal{S}$, else filled with background medium with permittivity $\epsilon_B^{(\alpha)} = 1+\alpha\chi_{\rm L}$. 
In addition, $V_{\rm in}$ is surrounded by a spherical shell $V_{\rm out}(\lambda)$ with radius $\lambda-R_{\rm in}$ and inner (outer) surface $\mathcal{S}$ ($\mathcal{S}_{\infty}(\lambda)$), so that the total volume is $V(\lambda)=V_{\rm in}+V_{\rm out}(\lambda)$.

\subsection{QNM quantization with separated gain and loss operators\label{Subsec: SEP_Gain_Loss}}

In this Section, we apply the QNM quantization scheme with the separation of the QNM basis sets into two different spatial domains, namely $V_{\rm G}$ and $\mathbb{R}^3-V_{\rm G}$. We formally term this method as {\it separated gain-loss operator} approach.

\subsubsection{QNM operator construction}
Similar to the lossy mode quantization, we start by inserting the QNM Green function (Eq.~\eqref{eq:GFwithSUM}) into the electric field operator (Eq.~\eqref{eq: SourceEField2}) at positions near the scattering geometry, to obtain the modal parts of the total electric field operator: 
\begin{equation}
\hat{\mathbf{E}}_{\rm QNM}(\mathbf{r})=i\sum_{\mu}\sqrt{\frac{\hbar\omega_\mu}{2\epsilon_0}}\tilde{\mathbf{f}}_\mu(\mathbf{r})[\tilde{a}_{{\rm L}\mu}+\tilde{a}_{{\rm G}\mu}^\dagger]+{\rm H.a.}\label{eq: SourceEFieldQNM_GL},
\end{equation}
where
\begin{align}
\tilde{a}_{{\rm L}\mu}=&\lim_{\lambda\rightarrow\infty}\sum_i\int_{V(\lambda)}{\rm d}^3s \int_0^\infty{\rm d}\omega\tilde{L}_{{\rm L}\mu,i}^{(\lambda)}(\mathbf{s},\omega)\hat{b}_i(\mathbf{s},\omega)
\label{eq: amuOmegaDefLoss},
\end{align}
is the lossy QNM operator part, and
\begin{align}
\tilde{a}_{{\rm G}\mu}^\dagger=\sum_i\int{\rm d}^3s \int_0^\infty{\rm d}\omega\tilde{L}_{{\rm G}\mu,i}(\mathbf{s},\omega)\hat{b}_i^\dagger(\mathbf{s},\omega)\label{eq: bmuOmegaDefGain},
\end{align}
is the amplifying QNM operator part. 

In contrast to the lossy mode quantization, the adjoint QNM gain operator $a_{{\rm G}\mu}^\dagger$ (rather then $a_{{\rm G}\mu}$) enters the positive rotating part of the modal electric field operator, as a consequence of the extension of the Green function quantization formalism to amplifying media. Since $V_{\rm G}$ is independent of $\lambda$,  the gain operator $\tilde{a}_{{\rm G}\mu}$ is also independent of $\lambda$, and $\tilde{L}_{{\rm G}\mu,i}(\mathbf{s},\omega)$ is simply given by
\begin{equation}
    \tilde{L}_{{\rm G}\mu,i}(\mathbf{s},\omega)=\chi_{V_{{\rm G}}}(\mathbf{s})\sqrt{\frac{2}{\pi\omega_\mu}}A_\mu(\omega)\sqrt{|\epsilon_I^{(\alpha)}(\mathbf{s},\omega)|}\,\tilde{\mathbf{f}}_\mu(\mathbf{s}).
\end{equation}

In contrast, $\tilde{L}_{{\rm L}\mu,i}^{(\lambda)}(\mathbf{s},\omega)$ can be written as 
\begin{align}
    \tilde{L}_{{\rm L}\mu,i}^{(\lambda)}&(\mathbf{s},\omega)=\chi_{V_{{\rm in}}-V_{\rm G}}(\mathbf{s})\sqrt{\frac{2}{\pi\omega_\mu}}A_\mu(\omega)\sqrt{\epsilon_I^{(\alpha)}(\mathbf{s},\omega)}\tilde{\mathbf{f}}_\mu(\mathbf{s})\nonumber\\
    +&\chi_{V_{{\rm out}}(\lambda)}(\mathbf{s})\sqrt{\frac{2}{\pi\omega_\mu}}A_\mu(\omega)\sqrt{\epsilon_I^{(\alpha)}(\mathbf{s},\omega)}\tilde{\mathbf{F}}_\mu(\mathbf{s},\omega).
\end{align}
Thus, the lossy QNM operator has a non-radiative contribution through the integration over the finite region $V_{\rm L}$ contained in $V_{\rm in}-V_{\rm G}$, where $\epsilon_I(\mathbf{r},\omega)>0$ and a radiative contribution through the integration over $V_{\rm out}(\lambda)$, where $\epsilon_I(\mathbf{r},\omega)=0$.

\subsubsection{Commutation relations and symmetrization}
From the spatial integration domains in both QNM operator parts, one can immediately deduce that the only non-vanishing commutation relations are
$
[\tilde{a}_{{\rm L}\mu},\tilde{a}_{{\rm L}\eta}^\dagger]\equiv S_{\mu\eta}^\mathrm{L}$ and $[\tilde{a}_{{\rm G}\mu},\tilde{a}_{{\rm G}\eta}^\dagger]\equiv S_{\mu\eta}^{\rm G}$. Here, the loss part is formally identical to the photon commutation relation in the lossy mode quantization, i.e., Eq.~\eqref{eq:SnradGeneral} and \eqref{eq:SradGeneral}, where $V_{\rm S}$ in $S_{\mu\eta}^\mathrm{nrad}$ is simply replaced by $V_{\rm L}$. 
Moreover, 
\begin{equation}
S_{\mu\eta}^\mathrm{G}=\int_0^\infty{\mathrm d}\omega\frac{2A_\mu^*(\omega)A_\eta(\omega)}{\pi\sqrt{\omega_\mu\omega_\eta}}I^{\rm G}_{\mu\eta}(\omega)\label{eq:SgainGeneral}, 
\end{equation}
with 
\begin{equation}
  I^{\rm G}_{\mu\eta}(\omega)=\int_{V_{\rm G}} {\mathrm d}^3s|\epsilon_I(\mathbf{s},\omega)|\,\tilde{\mathbf{f}}_\mu^*(\mathbf{s})\cdot\tilde{\mathbf{f}}_\eta(\mathbf{s}).
\end{equation}
Note, that compared to $S_{\mu\eta}^{\rm L}$, the conjugation with respect to complex QNM functions ($\mu$ and $\eta$) in $S_{\mu\eta}^{\rm G}$ is reversed. Applying a pole approximation similar to the non-radiative and radiative loss part leads to
\begin{equation}
S_{\mu\eta}^\mathrm{G}\approx\frac{\sqrt{\omega_\mu\omega_\eta}}{i(\tilde{\omega}_\mu-\tilde{\omega}_\eta^*)}I^{\rm G,p}_{\mu\eta}\label{eq:SgainPole} ,
\end{equation}
with 
\begin{equation}
  I^{\rm G,p}_{\mu\eta}=\int_{V_{\rm G}} {\mathrm d}^3s\sqrt{|\epsilon_I(\mathbf{s},\omega_\mu)\epsilon_I(\mathbf{s},\omega_\eta)|}\,\tilde{\mathbf{f}}_\mu^*(\mathbf{s})\cdot\tilde{\mathbf{f}}_\eta(\mathbf{s}).
\end{equation}

Next, we apply a symmetrization transformation to the gain- and loss-assisted QNM operators separately
to obtain proper Fock basis operators: 
\begin{equation}
a_{{\rm L(G)}\mu} = \sum_\eta \left[\left(\mathbf{S}^\mathrm{L(G)}\right)^{1/2}\right]_{\mu\eta}\tilde{a}_{{\rm L(G)}\eta}.
\end{equation}
In this way, $a_{{\rm L}\mu}^{(\dagger)}$ are annihilation (creation) operators for the ``loss'' Fock basis and $a_{{\rm G}\mu}^{(\dagger)}$ are annihilation (creation) operators for the ``gain'' Fock basis. They share the same vacuum state $|{\rm vac}\rangle_{\rm em}$, defined through 
\begin{equation}
b_i(\mathbf{r},\omega)|{\rm vac}\rangle_{\rm em}=0.
\end{equation}

The respective modal electric field operator then reads $\hat{\mathbf{E}}(\mathbf{r})=\hat{\mathbf{E}}^{\rm L}_{\rm QNM}(\mathbf{r})+\hat{\mathbf{E}}^{\rm G}_{\rm QNM}(\mathbf{r})$, with
\begin{align}
\hat{\mathbf{E}}^{\mathrm{L}}_{\rm QNM}(\mathbf{r})&=i\sum_\mu\sqrt{\frac{\hbar\omega_\mu}{2\epsilon_0}}\tilde{\mathbf{f}}_\mu^{\mathrm{s,L}}(\mathbf{r})a_{{\rm L}\mu}+\mathrm{H.a.}\label{eq: ESymLoss}, \\
\hat{\mathbf{E}}^{\mathrm{G}}_{\rm QNM}(\mathbf{r})&=i\sum_\mu\sqrt{\frac{\hbar\omega_\mu}{2\epsilon_0}}\tilde{\mathbf{f}}_\mu^{\mathrm{s,G}}(\mathbf{r})a_{{\rm G}\mu}^\dagger+\mathrm{H.a.}\label{eq: ESymGain}, 
\end{align}
with symmetrized QNM functions 
\begin{align}
\tilde{\mathbf{f}}_\mu^{\mathrm{s,L}}(\mathbf{r})&=\sum_\eta \left[\left(\mathbf{S}^\mathrm{L}\right)^{1/2}\right]_{\eta\mu}\tilde{\mathbf{f}}_\eta(\mathbf{r})\sqrt{\frac{\omega_\eta}{\omega_\mu}},\\
\tilde{\mathbf{f}}_\mu^{\mathrm{s,G}}(\mathbf{r})&=\sum_\eta \left[\left(\mathbf{S}^\mathrm{G}\right)^{1/2}\right]_{\mu\eta}\tilde{\mathbf{f}}_\eta(\mathbf{r}) \sqrt{\frac{\omega_\eta}{\omega_\mu}}.
\end{align}

Since two different Fock bases appear for the same number of QNMs, the total photonic degrees of freedom is $2N$ ($N$ is the number of QNMs). We also note the reversed index pair of the symmetrization factor ($\mu \rightarrow \eta$) for the loss and gain part, because of the exchange of {\it annihilation} and {\it creation} in these regimes. 

\subsubsection{Quantum Langevin equations}
We next inspect the Heisenberg equation of motion for the TLS emitter and QNM operators with respect to the initial Hamiltonian, Eq.~\eqref{eq: Hamiltonian}. Using an approximate separation of the photon continuum (represented by $b_{i}(\mathbf{r},\omega)$) into the QNM part and a bosonic non-QNM part, and applying a Markov approximation similar to Ref.~\onlinecite{franke2020quantized}, yields the Heisenberg equations of motion
\begin{subequations}\label{eq: HOM_Mode+TLS}
\begin{align}
\dot{a}_{k\mu} =& -\frac{i}{\hbar}[a_{k\mu},H_{\rm S}]-\sum_\eta\chi_{\mu\eta}^{(k)-}a_{k\eta}+F_\mu^{(k)}\label{eq: a_prime_HOM}, \\
\dot{\sigma}^- =& -\frac{i}{\hbar}[\sigma^-,H_{\rm S}]+\frac{\Gamma^{\rm B}}{2}\sigma_z\sigma^- -\sigma_z F_{\rm a},
\end{align}
\end{subequations}
with $k={\rm L,G}$, and the quantum noise terms that are discussed below.

Here, $H_{\rm S}=H_{\rm QNM}+H_{\rm QNM-a}+H_{\rm a}$ is the combined photon-emitter system Hamiltonian, with an electromagnetic QNM part:
\begin{equation}
    H_{\rm QNM}=\hbar\sum_{\mu\eta}\chi_{\mu\eta}^{\rm L+}a_{{\rm L}\mu}^\dagger a_{{\rm L}\eta}-\hbar\sum_{\mu\eta}\chi_{\mu\eta}^{\rm G+}a_{{\rm G}\mu}^\dagger a_{{\rm G}\eta},
\end{equation}
an emitter-QNM interaction part:
\begin{equation}
    H_{\rm QNM-a}=\hbar\sum_{\mu} [\tilde{g}_\mu^{\rm s,L} a_{{\rm L}\mu}\sigma^+ + \tilde{g}_\mu^{\rm s,G} a_{{\rm G}\mu}^\dagger\sigma^+] + {\rm H.a.},
\end{equation}
and the TLS energy $H_{\rm a}$, Eq.~\eqref{eq: H_atom}. 

Similar to the lossy mode case, an inherent {\it inter-mode} coupling appears between the QNMs, determined by the coupling matrices $\chi_{\mu\eta}^{\rm L(G)+}$, 
which is the Hermitian part 
of the QNM frequency matrix, 
\begin{align}
    \chi^{\rm L}_{\mu\eta}=\sum_\nu\left[\mathbf{S}^{\rm L-1/2}\right]_{\mu\nu}\tilde{\omega}_\nu \left[\mathbf{S}^{\rm L 1/2}\right]_{\nu\eta},\\
    \chi^{\rm G}_{\mu\eta}=\sum_\nu\left[\mathbf{S}^{\rm G-1/2}\right]_{\mu\nu}\tilde{\omega}_\nu^* \left[\mathbf{S}^{\rm G 1/2}\right]_{\nu\eta},
\end{align}
for the loss and gain parts, respectively. Note, that this expression of the coupling constant is an approximated form consistent with the pole approximations applied to the symmetrization matrices $S_{\mu\eta}^{\rm L(G)}$ (cf. App.~\ref{app:Sec_QLE}).

In addition to the positive lossy mode energy, a negative energy contribution appears connected to $a_{{\rm G}\mu}$ as a consequence of the appearance of the sign function in Eq.~\eqref{eq:HBgain}. Furthermore, two different emitter-photon interaction processes appear, one through the gain region and one through the loss region with coupling constant, 
\begin{equation}
    \tilde{g}_\mu^{\rm s,L(G)} = -i\sqrt{\frac{\omega_\mu}{2\hbar\epsilon_0}}\mathbf{d}_{\rm a}\cdot\tilde{\mathbf{f}}_\mu^{\mathrm{s,L(G)}}(\mathbf{r}_{\rm a}).
\end{equation}
We highlight that there appear non-preserving terms with respect to the excitation number, namely the terms $\sigma^+a_{{\rm G}\mu}^\dagger$, induced by the reversed nature of annihilation and creation in the gain region. However, because $a_{{\rm G}\mu}^\dagger$ is rotating with (approximately) $e^{-i\omega_\mu t}$, these terms are fully consistent with the rotating wave approximation.

Apart from the Hamiltonian part, different decay and noise terms appear in the Heisenberg equations of motion: The damping of the QNM gain and loss modes enters the model via the decay matrices $\chi^{\rm L-}_{\mu\eta}$ and $\chi^{\rm G-}_{\mu\eta}$, respectively. These matrices are closely related to the anti-Hermitian part of $\chi^{\rm L(G)}_{\mu\eta}$ through $\chi^{\rm L(G)}_{\mu\eta}=\chi_{\mu\eta}^{\rm L(G)+}\mp i\chi_{\mu\eta}^{\rm L(G)-}$. 

On the other hand, the TLS is damped via the 
{\rm background spontaneous emission} into the non-QNM space (originating from $H_{\rm I}$, Eq.~\eqref{eq: HI}), with the rate:
\begin{equation}
   \Gamma^{\rm B}= \frac{2}{\hbar\epsilon_0}\mathbf{d}_{\rm a}\cdot{\rm Im}[\mathbf{G}_{\rm B}(\mathbf{r}_{\rm a},\mathbf{r}_{\rm a},\omega_{\rm a})]\cdot\mathbf{d}_{\rm a},
\end{equation}
where $\mathbf{G}_{\rm B}$ is the background Green function, which is obtained from Eq.~\eqref{eq: GFHelmholtz2} with $\epsilon(\mathbf{r},\omega)=\epsilon_{\rm B}$. 

In addition to the decay terms, noise operators appear in Eqs.~\eqref{eq: HOM_Mode+TLS}, that counteract the respective damping processes to preserve the commutation relations of the system operators. Specifically $F_{\eta}^{\rm L}, F_{\eta}^{\rm G}$ represent statistically independent noise operators for the loss and gain contributions, while $F_{\rm a}$ counteracts the spontaneous emission decay. The precise definition of the noise forces and their derivation is shown in App.~\ref{app:Sec_QLE}.  

We remark that apart from the system Hamiltonian, the quantum Langevin equations in the separated gain-loss operator approach, Eqs.~\eqref{eq: HOM_Mode+TLS}, are formally identical to results obtained for the purely lossy case from Section~\ref{subsec: ExtensionLossyQuantization} (with the separation into two spatial regions).

\subsubsection{QNM master equation}
Applying the Markov approximation for the quantum Langevin equation for arbitary system operators,  and following the steps in Refs.~\onlinecite{PhysRevLett.122.213901,franke2020quantized}, the relevant master equation can be formulated as
\begin{equation}
    \partial_t \rho = -\frac{i}{\hbar}[H_{\rm S},\rho]+\mathcal{L}_{\rm L}[\mathbf{a}_{\rm L}]\rho + \mathcal{L}_{\rm G}[\mathbf{a}_{\rm G}]\rho + \mathcal{L}_{\rm SE}[\sigma^-]\rho,\label{eq: ME_separated}
\end{equation}
with the Lindblad dissipators,
\begin{align}
    \mathcal{L}_{(k)}&[\mathbf{a}_{k}]\rho\nonumber\\
    &=\sum_{\mu,\eta}\chi^{(k)-}_{\mu\eta}\left[2a_{k\eta}\rho a_{k\mu}^\dagger - a_{k\mu}^{\dagger}a_{k\eta}\rho -\rho a_{k\mu}^{\dagger}a_{k\eta} \right],
\end{align}
with $k={\rm L,G}$,  as well as
\begin{align}
    \mathcal{L}_{\rm SE}[\sigma^-]\rho&=\frac{\Gamma^{\rm B}}{2}\left[2\sigma^-\rho\sigma^+ - \sigma^+\sigma^-\rho - \rho \sigma^+\sigma^-\right].
\end{align}

We have considered the same inverted Lorentzian model (for material gain) as in the general case without mode quantization from Ref.~\onlinecite{franke2021fermi}, based on the idea in Ref.~\onlinecite{PhysRevA.55.1623}, to justify the vacuum state as the input state for $c_{{\rm G}\mu}^{\rm in}$, connected to the amplifying part;
thus
\begin{align}
    \langle c_{k\mu}^{\rm in}(t)c_{k'\eta}^{\rm in\dagger}(t')\rangle &\approx \delta_{kk'}\delta_{\mu\eta}\delta(t-t')\label{eq: Input_corr_separated},\\
    \langle c_{\rm a}^{\rm in}(t)c_{\rm a}^{\rm in\dagger}(t')\rangle &\approx \delta(t-t'),
\end{align}
for $k,k'={\rm L,G}$ within the applied Markov approximation (cf. App.~\ref{app: Bath_assumptions}). Here, the input operators $c_{k\mu}^{\rm in}$ are related to $F_\mu^{(k)}$ via 
\begin{equation}
    F_\mu^{(k)}=-\sqrt{2}\sum_\eta\left[\left(\boldsymbol{\chi}^{(k)-}\right)^{1/2}\right]_{\mu\eta}c_{k\eta}^{\rm in},
\end{equation}
and $F_{\rm a}=-\sqrt{\Gamma^{\rm B}}c_{\rm a}^{\rm in}$.

Subsequently, using the 
noise forces $F_\mu^{(k)}$ and $F_{\rm a}$ as a basis,  one can also write the vacuum input assumption as
\begin{equation}
    \langle F_{\mu}^{(k)}(t)F_{\eta}^{(k')\dagger}(t')\rangle\approx 2\delta_{kk'}\chi_{\mu\eta}^{(k)-}\delta(t-t'),
\end{equation}
and $\langle F_{\rm a}(t)F_{\rm a}^{\dagger}(t')\rangle\approx \Gamma^{\rm B}\delta(t-t')$.
In contrast to Ref.~\onlinecite{franke2021fermi}, the main coupling part of emitter to the photonic environment, i.e., the QNM part, is still treated on the {\it system level} without any bath approximations. It can thus be used for multi-photon quantum simulations in the strong light-matter coupling regime, and to model entangled photons and matter.

\subsection{QNM quantization with  unified gain-loss operators\label{Subsec: UNI_Gain_Loss}}
Next, we present a formulation to combine the gain and loss contributions into a single QNM operator basis, which is more closely related to the original lossy QNM quantization scheme~\cite{PhysRevLett.122.213901,franke2020fluctuation,franke2020quantized}. This method is termed in the following as {\it unified gain-loss operator} approach.

\subsubsection{QNM  operator construction}
First, we simply write the QNM expanded electric field, Eq.~\eqref{eq: SourceEFieldQNM_GL}, as
\begin{equation}
    \hat{\mathbf{E}}_{\rm QNM}(\mathbf{r})=i\sum_{\mu}\sqrt{\frac{\hbar\omega_\mu}{2\epsilon_0}}\tilde{\mathbf{f}}_\mu(\mathbf{r})\tilde{a}_\mu'+ \mathrm{H.a.},\label{eq: SourceEFieldQNMTotal_Unified}
\end{equation}
where 
\begin{equation}
    \tilde{a}_\mu'=\tilde{a}_{{\rm L}\mu}+\tilde{a}_{{\rm G}\mu}^\dagger,
\end{equation}
is identified as a {\it combined QNM operator}, containing both loss and gain degrees of freedom. Note that we have added a prime superscript to this operator basis to distinguish it with the purely lossy mode case.

The operators, $\tilde{a}_\mu',\tilde{a}_\eta^{\prime\dagger}$, fulfill the commutation rules:
\begin{align}
    [\tilde{a}_\mu',\tilde{a}_\eta']=&[\tilde{a}_\mu^{\prime\dagger},\tilde{a}_\eta^{\prime\dagger}]=0,\\
    [\tilde{a}_\mu',\tilde{a}_\eta^{\prime\dagger}]=&S_{\mu\eta}^\mathrm{L}-S_{\mu\eta}^{\rm G*}.
\end{align}

\subsubsection{Symmetrization and Fock space construction}
While the combined operator approach can lead to a reduction of the degrees of freedom compared to the description with $\tilde{a}_{{\rm L}\mu}$ and $\tilde{a}_{{\rm G}\mu}$, one has to be more careful here, since the commutation matrix, $\mathbf{S}'\equiv \mathbf{S}^{\rm L}-\mathbf{S}^{\rm G*}$, now appears as the difference between two positive definite matrices, which is in general no longer of positive definite form. 
Yet, this (positive definiteness) property of $\mathbf{S}'$ is a requirement to apply the symmetrization transformation for the construction of photon Fock states.

To investigate the positive definiteness in a more detailed way, we rewrite the symmetrization factor as $\mathbf{S}'=\mathbf{S}^{\rm rad}+\Delta\mathbf{S}'$, where 
\begin{align}
    [\Delta&\mathbf{S}']_{\mu\eta}\nonumber\\
    =&\int_0^\infty{\mathrm d}\omega\frac{2A_\mu(\omega)A_\eta^*(\omega)}{\pi\sqrt{\omega_\mu\omega_\eta}}\int_{V_{\rm G}+V_{\rm L}} {\mathrm d}^3s\epsilon_I(\mathbf{s},\omega)\tilde{\mathbf{f}}_\mu(\mathbf{s})\cdot\tilde{\mathbf{f}}_\eta^*(\mathbf{s}),
\end{align}
and thus it would be sufficient to prove that $\Delta\mathbf{S}'$ is positive definite, since the sum of positive definite forms is again positive definite (and $\mathbf{S}^{\rm rad}$ is always positive definite).
Although one cannot in general prove that $\Delta\mathbf{S}'$ is indeed positive definite, for cases in which $S_{\mu\mu}^{\rm nrad}\gg S_{\mu\mu}^{\rm G}$ (dominant loss contribution) and the off-diagonal elements are small, $\Delta\mathbf{S}'$ can be assumed to be positive definite. 
We note that these are stricter conditions than the concept of linear amplification in the sense that $\gamma_{\mu}>0$, where the fundamental field commutation relation are preserved and the method in Section ~\ref{Subsec: SEP_Gain_Loss} is applicable.

In cases where the positive definiteness is fulfilled, we can then meaningfully apply the symmetrization transformation to define new photon operators,
\begin{equation}
    a_\mu' = \sum_{\eta}\left[\mathbf{S}^{\prime-1/2}\right]_{\mu\eta}\tilde{a}_\eta', 
\end{equation}
where the symmetrized operators $a_\mu^{\prime(\dagger)}$ can be written as
\begin{align}
    a_\mu' =& \sum_i\int{\rm d}^3 r\int_0^\infty{\rm d}\omega L_{{\rm L}\mu,i}^{\prime}(\mathbf{r},\omega)b_i(\mathbf{r},\omega)\nonumber\\
    &+\sum_i\int{\rm d}^3 r\int_0^\infty{\rm d}\omega L_{{\rm G}\mu,i}^{\prime}(\mathbf{r},\omega)b_i^\dagger(\mathbf{r},\omega) ,\label{eq: c_muFormalDef}
\end{align}
with 
\begin{equation}
    L_{{\rm L(G)}\mu,i}^{\prime}(\mathbf{r},\omega)=\sum_{\eta}\left[\mathbf{S}^{\prime-1/2}\right]_{\mu\eta}\tilde{L}_{{\rm L(G)}\eta,i}(\mathbf{r},\omega).
\end{equation}

These new {\it symmetrized} operators  fulfill bosonic commutation relations, i.e., $[a_\mu', a_\eta^{\prime\dagger}]=\delta_{\mu\eta}$, and thus fulfill the algebraic requirements to be regarded as photon annihilation and creation operators in the QNM subspace. However, in contrast to the two separated operator basis ($a_{{\rm L}\mu},a_{{\rm G}\mu}$), they do not share the same vacuum state as the continuum operators $b_i(\mathbf{r},\omega)$. Indeed, applying the corresponding number operator $a_\mu^{\prime\dagger} a_\mu'$ on the vacuum state $|{\rm vac}\rangle_{\rm em}$ of the $b_i(\mathbf{r},\omega)$ basis would yield
\begin{equation}
    a_\mu^{\prime\dagger} a_\mu' |{\rm vac}\rangle_{\rm em} =  n^{\rm vac}_\mu|{\rm vac}\rangle_{\rm em},
\end{equation}
where 
\begin{equation}
    n^{\rm vac}_\mu=\sum_i\int{\rm d}^3r\int_0^\infty{\rm d}\omega |L_{{\rm G}\mu,i}^{\prime}(\mathbf{r},\omega)|^2\neq 0.
\end{equation}

The question is then: how much does $n^{\rm vac}_{\mu}$ deviate from zero? To investigate this, we rewrite $n^{\rm vac}_{\mu}$ as
\begin{equation}
    n^{\rm vac}_\mu=\sum_{\eta,\eta'}\left[\mathbf{S}^{\prime-1/2}\right]_{\mu\eta}S_{\eta\eta'}^{\rm G}\left[\mathbf{S}^{\prime-1/2}\right]_{\eta'\mu}.
\end{equation}
Summing over all $\mu$ would yield $n^{\rm vac}=\sum_\mu n^{\rm vac}_\mu$, so that
\begin{equation}
    n^{\rm vac}=\sum_{\eta,\eta'}S_{\eta\eta'}^{\rm G}\left[\mathbf{S}^{\prime-1}\right]_{\eta'\eta}={\rm tr}[\mathbf{S}^{\rm G}\cdot\mathbf{S}^{\prime -1}]\label{eq: n_vac}.
\end{equation}
Within the condition that $S^{\rm L}_{\mu\mu}\gg S^{\rm G}_{\mu\mu}$ (and small off-diagonal elements), $n^{\rm vac}$ is a very small number, i.e. $n^{\rm vac}\ll 1$, so that one can approximately identify $|{\rm vac}\rangle_{\rm em}$ with the vacuum state of the QNM photon space within the unified gain-loss operator description. We note that this does not necessarily coincide with the conditions for positive definiteness, as we discuss in more details in Section~\ref{Sec: Applications}.

We also note that it may be possible to define a supplementary basis of states with reversed symmetry of annihilating and creating a photon within the gain geometry, defined through
\begin{align}
    b_i^\dagger(\mathbf{r},\omega)|{\rm vac}\rangle_{\widetilde{{\rm em}}}&=\Theta[\epsilon_I]|1_{i,\mathbf{r},\omega}\rangle ,\\
    b_i(\mathbf{r},\omega)|{\rm vac}\rangle_{\widetilde{{\rm em}}}&=\Theta[-\epsilon_I]|\tilde{1}_{i,\mathbf{r},\omega}\rangle, 
\end{align}
where $|\tilde{1}_{i,\mathbf{r},\omega}\rangle$ can be regarded as a ``negative'' photon number (one photon is missing). In this way, we obtain
\begin{equation}
     a_\mu^{\prime\dagger} a_\mu' |{\rm vac}\rangle_{\widetilde{{\rm em}}} = 0,
\end{equation}
as the well-defined QNM photon vacuum state without any quanta. For the remaining parts of this subsection, we assume that there indeed exists a well-defined vacuum state, and will investigate the respective criteria (positive definiteness and dominant loss contribution) for a particular gain-loss structure in more detail in Section~\ref{Sec: Applications}.

In the new symmetrized basis (for combined operators), the electric field operator takes the form
\begin{equation}
    \hat{\mathbf{E}}_{\rm QNM}(\mathbf{r})=i\sum_{\mu}\sqrt{\frac{\hbar\omega_\mu}{2\epsilon_0}}\tilde{\mathbf{f}}_\mu^{\prime\rm s}(\mathbf{r})a_\mu'+ \mathrm{H.a.},\label{eq: SourceEFieldQNMTotal_Unified_Sym}
\end{equation}
with symmerized QNM functions,
\begin{equation}
    \tilde{\mathbf{f}}_\mu^{\prime\rm s}(\mathbf{r})=\sum_\eta \left[\mathbf{S}^{\prime1/2}\right]_{\eta\mu}\tilde{\mathbf{f}}_\eta(\mathbf{r})\sqrt{\frac{\omega_\eta}{\omega_\mu}}.\label{eq: SymmQNM_unified}
\end{equation}

\subsubsection{Quantum Langevin equations}
For the unified gain-loss operator approach, we obtain the Heisenberg equation of motion for the emitter and QNM operators as (cf.~App.~\ref{app:Sec_QLE})
\begin{subequations}\label{eq: HOM_Mode+TLS_Unified}
\begin{align}
\dot{a}_{\mu}' =& -\frac{i}{\hbar}[a_{\mu}',H_{\rm S}']-\sum_\eta[\chi_{\mu\eta}^{\prime \rm L-}-\chi_{\mu\eta}^{\prime \rm G-}]a_{\eta}'+F_{\rm L\mu}^{\prime}+F_{\rm G\mu}^{\prime\dagger}, \\
\dot{\sigma}^- =& -\frac{i}{\hbar}[\sigma^-,H_{\rm S}']+\frac{\Gamma^{\rm B}}{2}\sigma_z\sigma^--\sigma_z F_{\rm a},
\end{align}
\end{subequations}
where the system Hamiltonian takes the form $H_{\rm S}'=H_{\rm QNM}'+H_{\rm QNM-a}'+H_{\rm a}$. The first term is the QNM Hamiltonian is given by
\begin{equation}
    H_{\rm QNM}'=\hbar\sum_{\mu,\eta}\chi_{\mu\eta}^{\prime\rm L+}a_\mu^{\prime\dagger}a_\eta' - \hbar\sum_{\mu,\eta}\chi_{\mu\eta}^{\prime\rm G+}a_\mu' a_\eta^{\prime\dagger},
\end{equation}
which consists of a normal ordered and an anti-normal ordered operator part, where the the latter reflects the gain contribution and the former the loss contribution. 

Using the bosonic commutation relations of $a_{\mu}',a_\mu^{\prime\dagger}$, these terms can be recast into
the form
\begin{equation}
    H_{\rm QNM}'=\hbar \sum_{\mu,\eta}[\chi^{\prime\rm L+}_{\mu\eta}-\chi^{\prime\rm G+}_{\eta\mu}]a_\mu^{\prime\dagger} a_\eta' - C^{\rm gain}\mathbb{1},\label{eq: H_QNM_uni}
\end{equation}
with a constant energy $C^{\rm gain}=\sum_{\mu}\chi^{\prime\rm G+}_{\mu\mu}$. In addition, one can combine the sum of the coupling matrices as $\chi^{\prime +}_{\mu\eta}=\chi^{\prime\rm L+}_{\mu\eta}-\chi^{\prime\rm G+}_{\eta\mu}$, which is the Hermitian part of 
\begin{equation}
    \chi^{\prime}_{\mu\eta}=\sum_{\nu,\nu'}\left[\mathbf{S}^{\prime-1/2}\right]_{\mu\nu}\tilde{\omega}_\nu \left[\mathbf{S}^{\prime1/2}\right]_{\nu\eta}, 
\end{equation}
yielding a positive definite form $\chi^{\prime +}_{\mu\eta}$ as long as $S_{\nu\nu'}'$ is positive definite. This is a very interesting reformulation of $H_{\rm QNM}$, as the negative energy part is now fully encoded in the constant term $-C^{\rm gain}\mathbb{1}$. 

The QNM-emitter interaction is now given by 
\begin{equation}
    H_{\rm QNM-a}'=\hbar\sum_{\mu} \tilde{g}_\mu^{\prime\rm s} a_{\mu}'\sigma^+  + {\rm H.a.},
\end{equation}
which is formally identical to the purely lossy case using a combined QNM operator basis; the TLS-QNM coupling constants are given by
\begin{equation}
    \tilde{g}_\mu^{\prime\rm s} = -i\sqrt{\frac{\omega_\mu}{2\hbar\epsilon_0}}\mathbf{d}_{\rm a}\cdot\tilde{\mathbf{f}}_\mu^{\prime\mathrm{s}}(\mathbf{r}_{\rm a}),
\end{equation}
where the gain part is reflected through the bosonic operators $a_\mu'$ and the reduction of the coupling constant $\tilde{g}_\mu$ due to symmetrization from the gain contribution of the QNM overlap integrals. 

In this formalism, the QNM operators are driven by two (quantum noise) forces, the loss-induced forces $F_{{\rm L}\eta}^{\prime}$ and the adjoint gain-induced forces $F_{{\rm G}\eta}^{\prime\dagger}$. In addition, there is a damping and a pumping term, connected to $\chi_{\mu\eta}^{\prime \rm L-}$ and $-\chi_{\mu\eta}^{\prime \rm G-}$, respectively. Thus, apart from the differences in system Hamiltonian, the coupling to the reservoir drastically changes from the purely lossy case. 

\subsubsection{QNM master equation with a unified gain-loss operator approach}
To obtain a QNM master equation in the unified gain-loss operator approach, the steps in Refs.~\onlinecite{PhysRevLett.122.213901,franke2020quantized} can be extended to allow for two different noise sources for a single QNM operator basis. This can be done in a straightforward way, once the corresponding Markovian quantum Langevin equation is known. The associated master equation can be formulated as 
\begin{equation}
    \partial_t \rho = -\frac{i}{\hbar}[H_{\rm S}',\rho]+\mathcal{L}_{\rm L}'[\mathbf{a}']\rho + \mathcal{L}_{\rm G}'[\mathbf{a}^{\prime\dagger}]\rho + \mathcal{L}_{\rm SE}[\sigma^-]\rho,\label{eq: ME_unified}
\end{equation}
with the Lindblad dissipators:
\begin{align}
    \mathcal{L}_{\rm L}'[\mathbf{a}']\rho&=\sum_{\mu,\eta}\chi^{\rm \prime L-}_{\mu\eta}\left[2a_{\eta}'\rho a_{\mu}^{\prime\dagger} - a_{\mu}^{\prime\dagger}a_{\eta}'\rho -\rho a_{\mu}^{\prime\dagger}a_{\eta}' \right],\\
    \mathcal{L}_{\rm G}'[\mathbf{a}^{\prime\dagger}]\rho&=\sum_{\mu,\eta}\chi^{\rm \prime G-}_{\mu\eta}\left[2a_{\mu}^{\prime\dagger}\rho a_{\eta}^{\prime} - a_{\eta}^{\prime}a_{\mu}^{\prime\dagger}\rho -\rho a_{\eta}^{\prime}a_{\mu}^{\prime\dagger} \right].
\end{align}

Here again we have considered the same inverted Lorentzian model as in the general case without mode quantization from Ref.~\onlinecite{franke2021fermi,PhysRevA.55.1623}, to justify the vacuum state as the input state for $c_{{\rm G}\mu}^{\prime\rm in}$, connected to the amplifying part, so that 
\begin{equation}
    \langle c_{k\mu}^{\prime\rm in}(t)c_{k'\eta}^{\prime\rm in\dagger}(t')\rangle\approx \delta_{kk'}\delta_{\mu\eta}\delta(t-t'),\label{eq: Input_corr_unified}
\end{equation}
for $k,k'={\rm L,G}$ within the applied Markov approximation (cf. App.~\ref{app: Bath_assumptions}). Note, that similar to the separated gain-loss operator approach, these input operators are linear combinations of their respective noise forces $F^{\prime}_{k\mu}$. The assumptions regarding the TLS decay are identical to the separated gain-loss operator approach. 
In contrast to the purely lossy case, an incoherent and intrinsic mode pumping term $\mathcal{L}_{\rm G}'[\mathbf{a}^{\prime\dagger}]$ appears, characterized by the reversed ordering of photon annihilation and creation operators.

\section{Discussion on the different quantum models with gain\label{Sec: DiffQuantumModels}}

\subsection{Comparison of the two QNM representations}
We now summarize the main differences between the two QNM quantization approaches for the combined loss and amplifying media.

First, the vacuum state is generally different in both cases; while in the separated gain-loss operator description, the vacuum state is (exactly) identical to the vacuum state $|\rm vac\rangle_B$ of $b_i(\mathbf{r},\omega)$, this is no longer (exactly) the case for the unified gain-loss operator approach. In the latter case, the action of the number operator on $|\rm vac\rangle_B$ yields a finite number $n^{\rm vac}\neq 0$, which depends on the gain contribution to the total symmetrization. 

Second, the positive definiteness of the symmetrization matrices for the unified gain-loss operator approach does not obviously coincide with the causality condition of the Green function or equivalently with the condition $\gamma_\mu >0$ for all QNMs $\mu$, as would be the case for the separated gain-loss operator description. However,
if the symmetrization matrix $\mathbf{S}'$ is in a positive definite form, then the photon Hamiltonian in the unified gain-loss operator approach can be recast into a sum of a positive form and a negative constant, while in the separated gain-loss operator approach, there is a strict negative (non-constant) part with respect to the photon subspace related to  $a_{{\rm G}\mu}$. 

Third, in the separated gain-loss operator approach, the master equation only contains strict decay processes in the Lindblad dissipators of the form $\mathcal{D}[\mathbf{a}_{{\rm G(L)}}]$, similar to the purely lossy case, while in the unified gain-loss operator approach, decay as well as pumping terms of the form $\mathcal{D}[\mathbf{a}^{\prime\dagger}]$ appear, which depend on the amplifying media and cross coupling between the QNMs. This is induced by the combination of loss and gain on the system level into one operator basis. Thus the strict negativity of the photon Hamiltonian from $a_{{\rm G}\mu}$ degrees of freedom is resolved in a vacuum state reformulation, a pumping term from the gain region as well as a constant negative energy part in the $a_\mu'$ picture. 

Fourth, we can also compare the form of the TLS-QNM interaction; while in the separated gain-loss operator description there are two-particle creation processes through terms like $\sigma^+a_{{\rm G}\mu}^\dagger$ (atomic raising operator and negative photon creation), in the unified gain-loss operator approach, only quanta-preserving operator terms such as $a_\mu'\sigma^+$ appear on the formal level, because of the redefinition of annihilation and creation. However, one should be very careful here with respect to a well-defined ground state, since the unified gain-loss operator approach still must met the criteria of linear amplification, so that there could be no unrealistic excitation of the added emitter system without violating the positive definiteness of $\mathbf{S}'$. 

\subsection{Comparison to phenomenological results in the current literature on gain-loss quantum systems}
Interestingly, the results from the unified gain-loss operator description are similar to those found in the current literature on quantum theories for gain-loss systems~\cite{PhysRevA.101.013812,PhysRevA.96.033806,Kepesidis_2016}.  A suitable example is the theory developed in Ref.~\onlinecite{PhysRevA.101.013812},  where a similar quantum Langevin equation as Eq.~\eqref{eq: a_prime_HOM} (for two modes)  was proposed (and where one mode is pumped) for linear amplification, having the form~\cite{PhysRevA.101.013812}
\begin{subequations}\label{eq: Phenomenological_QLE}
\begin{align}
    \dot{b}_1 &= -\frac{i}{\hbar}\left[b_1,H_{\rm ph}\right] +\frac{A-\Gamma_1}{2}b_1  + \sqrt{A}f_1^\dagger + \sqrt{\Gamma_1}l_1,\\
       \dot{b}_2 &=-\frac{i}{\hbar}\left[b_1,H_{\rm ph}\right]  -\frac{\Gamma_2}{2}b_2 + \sqrt{\Gamma_2}l_2.
\end{align}
\end{subequations}

Here, $\Gamma_{1(2)}$ is the phenomenological decay rate of the normal mode $1(2)$ with operators $b_1 (b_2)$, $A$ is the pumping rate of mode $1$, and $l_i(f_i)$ ($i=1,2$) are the noise operators for dissipation (gain), assumed to behave as white noise. Furthermore, 
\begin{equation}
    H_{\rm ph}=\hbar\sum_{i=1,2}\omega_ib_i^\dagger b_i + i\hbar\kappa[b_1^\dagger b_2 - b_2^\dagger b_1]\label{eq: H_ph_ScullyLamb},
\end{equation}
is the corresponding photon Hamiltonian, where $\kappa$ is the mode coupling constant.

Despite having a form similar to the unified gain-loss operator approach, there are some fundamental caveats with these approaches, since the respective starting point is not a rigorous QNM model with inherent dissipation. These differences will be discussed in  detail below.

First, the mode coupling rate, $\kappa$, is included phenomenologically as a real-valued coupling constant, which is in contrast to the method presented here, where (despite the complex classical coupling between the loss and gain bare resonator modes) the photon coupling matrix is complex and naturally appears through the symmetrization transformation, necessary to construct photon Fock states for QNMs. This is of course connected to the assumption, that the photon operators $b_1,b_2$ fulfil bosonic commutation relations, which is for general open systems not valid without introducing a symmetrization transformation (but can be a good approximation for high $Q$ factors).  

Second, the noise operators appearing in Eq.~\eqref{eq: Phenomenological_QLE} are phenomenologically introduced to counteract the dissipation, which is not the case in the presented theory, where the gain and loss noise come from the same macroscopic Green function quantization approach. As a consequence of the inter-QNM coupling and the non-bosonic nature of the initial photon operators, all modes are indirectly pumped in the quantized QNM picture, and assuming the pumping of just one mode from the bare resonator picture as done in the above approach is clearly a vague approximation in the case of general $Q$ factors. The solution must respect the dissipation-induced cross-coupling as well as the known classical modes for coupled loss and gain resonators, and we know that the coupled
system forms hybrid QNMs, that belong to both the gain and loss resonators~\cite{EPClassicalPaper}.

Third, and most importantly, the range of validity of the phenomenological approaches is not obvious at all, while we show for our unified gain-loss operator approach (which can be regarded as a rigorous version of the former) that there are indeed criteria that must be met for using such a quantum Langevin equation. These are connected to the symmetrization transformation and thus the dissipation-induced mode coupling itself, which is of course missing in the phenomenological approaches.
Indeed, there are possibly situations where $\mathbf{S}'$ is not positive definite, in which case one has to use the separated gain-loss operator approach, which doubles the dimension of the underlying photon Hilbert space.

\section{Applications for a coupled gain-loss microdisk resonators \label{Sec: Applications}}

\begin{figure}[t]
    \centering
    \includegraphics[width=0.99\columnwidth]{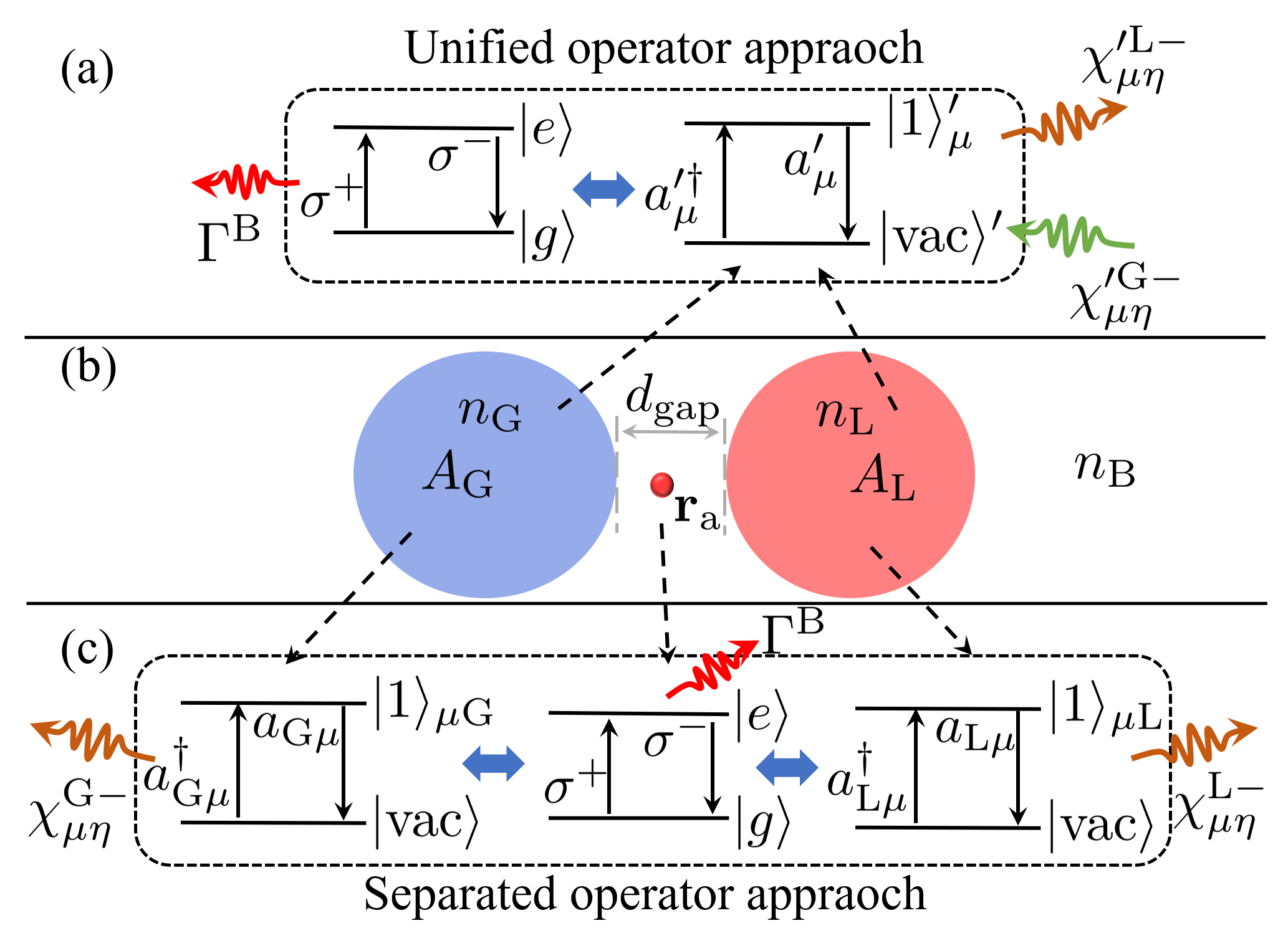} 
    \caption{(a) Quantum picture from the unified gain-loss operator approach. The dipole emitter with exited state $|e\rangle$ and ground state $|g\rangle$ is coupled to a single set of QNM photon states $|n_\mu\rangle$, and decays via spontaneous emission with rate $\Gamma^{\rm B}$. The corresponding QNM photon states are pumped incoherently via the gain matrix $\chi_{\mu\eta}^{\prime\rm G-}$ and decay via the loss matrix  $\chi_{\mu\eta}^{\prime\rm L-}$. (b) Schematic of a $z$-polarized dipole emitter at position $\mathbf{r}_{\rm a}$ placed in an effective two-dimensional gain-loss ring resonator system with respective areas $A_{\rm G}$ and $A_{\rm L}$ with minimal distance $d_{\rm gap}$. The complex refractive index of the lossy (amplifying) ring is given through $n_{\rm L(G)}$, while $n_{\rm B}$ describes the refractive index of the surrounding free space.  (c) Quantum picture of the separated gain-loss operator approach. The TLS dipole emitter is coupled to two sets of QNM photon states, which decay separately via the dissipation matrices  $\chi_{\mu\eta}^{\rm L-}$ and $\chi_{\mu\eta}^{\rm G-}$. The pumping mechanism is encoded in the TLS-QNM interaction. 
}\label{fig:scheme_applications}
\end{figure}

We next present some concrete numerical calculations, 
where we will consider a coupled gain-loss resonator system, which consists of a pair of two-dimensional microdisks separated by a distance $d_{\rm gap}$. The lossy (amplifying) microdisk with area $A_{\rm L}$ ($A_{\rm G}$) is described by the complex refractive index $n_{\rm L}={\rm Re}[n_{\rm L}]+i{\rm Im}[n_{\rm L}]$ ($n_{\rm G}={\rm Re}[n_{\rm G}]+i{\rm Im}[n_{\rm G}]$) with ${\rm Im}[n_{\rm L}]>0$ (${\rm Im}[n_{\rm G}]<0$). The resonators are surrounded by homogeneous free space with refractive index $n_{\rm B}$, cf. Fig.~\ref{fig:scheme_applications}(a). 
For the following investigations, the refractive index of the lossy resonator is set to $n_{\rm L}=2+10^{-5}i$, while the refractive index of the amplifying resonator is varied from $n_{\rm G}=2-2\cdot 10^{-6}i$ to $n_{\rm G}=2- 10^{-6}i$ and $n_{\rm G}=2- 10^{-7}i$  
to cover a wide range of amplification regimes, ranging from large to small gain contributions. Note, that in the following, we define and use $\alpha_{\rm G}=|{\rm Im}[n_{\rm G}]|$ as the gain coefficient.

We first briefly discuss the numerical calculation and accuracy of  the classical two-QNM expansion for the hybrid structure by comparing to a direct and fully numerical simulation of the Maxwell's equations. Subsequently, we introduce an adapted and improved form of the phenomenological quantum gain approaches for the specific geometry and derive crucial quantum parameters for the former as well as the unified gain-loss operator approach to discuss the appearing differences and the validity of these methods. Finally, we show and discuss results for  quantum metrics of a quantum emitter, that is placed between the ring resonators, modelled through the different quantum treatments of gain in the bad cavity limit.

\begin{figure*}[t]
    \centering
    \includegraphics[width=1.99\columnwidth]{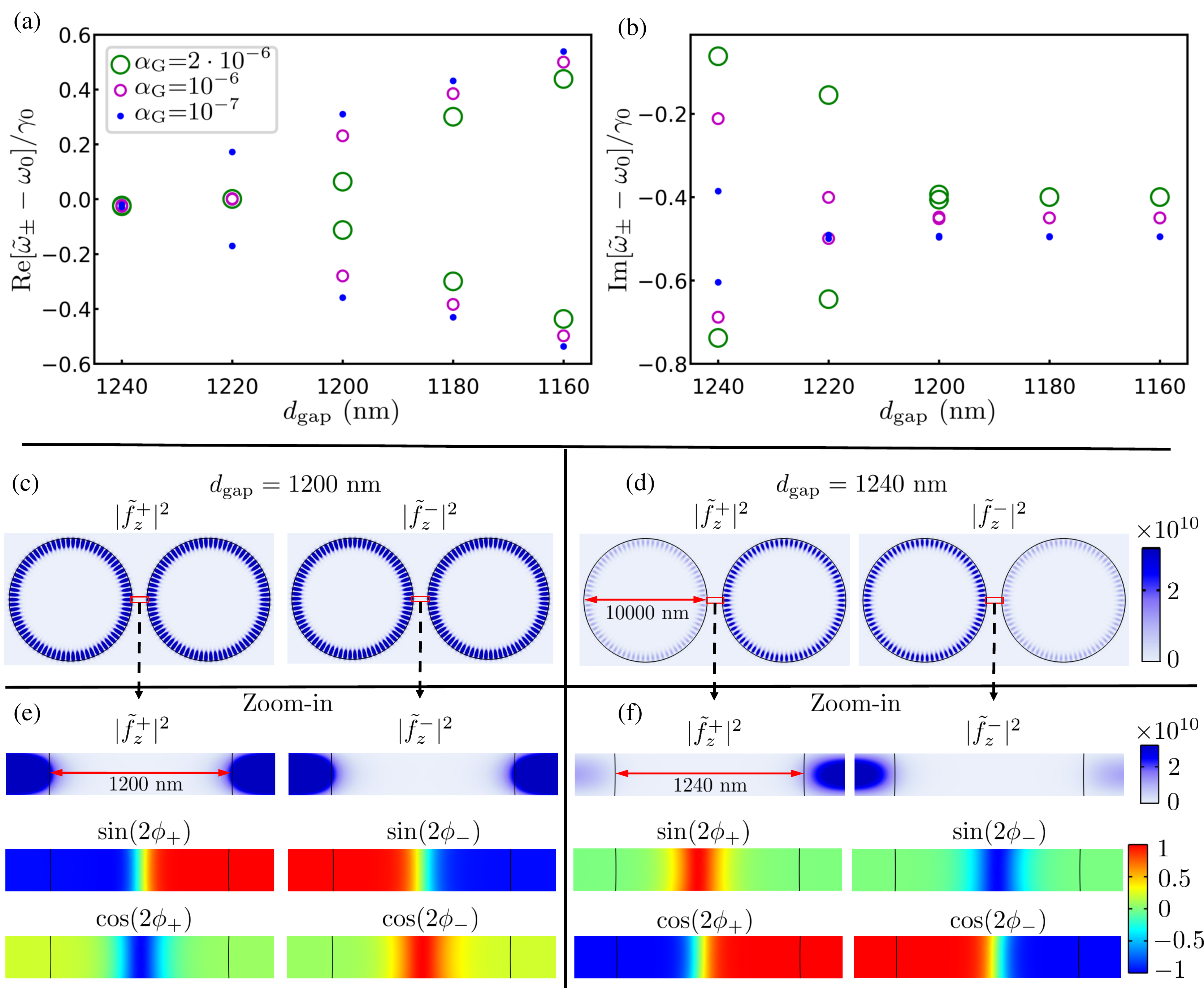} 
    \caption{Top: Real part (a) and imaginary part (b) of the eigenfrequencies $\tilde{\omega}_{\pm}$ of the microdisk hybrid structure depicted in Fig.~\ref{fig:scheme_applications}(b) obtained from CMT approach as function of gap size $d_{\rm gap}$ for the different absorption coefficients $\alpha_{\rm G}=|{\rm Im}[n_{\rm G}]|$ and $n_{\rm L}=2+10^{-5}i$. Note, that 
    $\tilde{\omega}_0=0.833717-4.120496\cdot 10^{-6}i~({\rm eV})$
    is the reference frequency of the isolated lossy ring resonator with  $n_{\rm L}=2+10^{-5}i$. Bottom: Absolute square of $z$-components the QNM eigenfunctions $\tilde{f}_z^{\pm}$  obtained from CMT with $\alpha_{\rm G}=2\cdot10^{-6}$ for the gap size $d_{\rm gap} = 1200$~nm (c), close to the lossy EP region, and $d_{\rm gap} = 1240$~nm (d), further away from the lossy EP region. In (e) and (f) a zoom-in of the gap region for the absolute square values of the QNM eigenfunctions as well as the corresponding cos and sin of QNM phases $2\phi_{\pm}$ are shown. 
}\label{fig:Eigfreq_Eigfunctions}
\end{figure*}

\subsection{Numerical calculations of the QNM parameters and classical Green functions}

The QNMs of the single resonators and coupled resonator can be obtained numerically
using a dipole normalization technique
implemented in COMSOL~\cite{Bai}. This is basically an inverse Green function approach and allows one to obtain the normalized QNMs using a dipole source.
For non-dispersive media, the normalized QNMs satisfy:
\begin{equation}
\braket{\braket{\tilde {\bf f}_1|\hat \epsilon_1|\tilde {\bf f}_1}}
\rightarrow 
\int d{\bf r} \epsilon_1({\bf r}) \tilde {\bf f}_1({\bf r}) \tilde {\bf f}_1({\bf r}) =1,
\end{equation}
where some coordinate transform has been applied to regularize the outgoing surface fields~\cite{kristensen2012generalized, SauvanNorm,muljarovPert}. The generalization to dispersive materials is straightforward,
but is not needed for our numerical examples
presented later.

To understand the mode coupling between the cavities from 
an underlying physics perspective, we have also utilized a generalized QNM coupled mode theory (CMT) for intrinsically dissipative modes~\cite{EPClassicalPaper}, where the two fundamental complex hybrid frequencies $\tilde{\omega}_\pm$ can be obtained from a eigenvalue problem through
\begin{equation}
\label{eq: HybridFreqCMT}
\tilde \omega_{\pm}=
\frac{\tilde\omega_{\rm L}+\tilde\omega_{\rm G}}{2}
\pm \frac{\sqrt{4\tilde \kappa_{\rm LG}\tilde \kappa_{\rm GL} + (\tilde\omega_{\rm L}-\tilde\omega_{\rm G})^2}}{2}.
\end{equation}

Here, $\tilde\omega_{\rm L}$ ($\tilde\omega_{\rm G}$) is the dominant complex QNM eigenfrequency of the isolated lossy (amplifying) disk, which are obtained from the respective single resonator solutions. 
Furthermore, $\tilde{\kappa}_{\rm LG}$ and $\tilde{\kappa}_{\rm GL}$ are the (complex) CMT coupling parameters, given by the overlap integrals,
\begin{equation}\label{eq: QNMCMT_coup}
\tilde \kappa_{\rm LG}=
\frac{\tilde \omega_{\rm G}}{2} \braket{\braket{\tilde{\bf f}_{\rm L}
|\Delta\hat\epsilon_{\rm L}|\tilde{\bf f}_{\rm G}}},~ \tilde \kappa_{\rm GL}=
\frac{\tilde \omega_{\rm L}}{2} \braket{\braket{\tilde{\bf f}_{\rm G}
|\Delta\hat\epsilon_{\rm G}|\tilde{\bf f}_{\rm L}}},
\end{equation}
where $\Delta\hat\epsilon_{\rm L(G)}=\hat{\epsilon}_{\rm L(G)}-\hat{\epsilon}_{\rm B}$ is the permittivity difference bound to the lossy (amplifying) cavity region, and $\tilde{\bf f}_{\rm L(G)}$ are the eigenfunctions of the single loss (gain) resonator problem. It is important to stress that the coupling parameters $\tilde \kappa_{\rm LG}$ are in general complex and that $\tilde \kappa_{\rm LG}\neq \tilde \kappa_{\rm GL}$ and $\tilde \kappa_{\rm LG}\neq \tilde \kappa_{\rm GL}^*$. This is in contrast to the more usual assumptions of real coupling parameters from classical normal mode theories and more phenomenological quantum gain models associated to Eq.~\eqref{eq: Phenomenological_QLE}; this assumption  can lead to dramatic changes for general $Q$ factors, even on the classical level~\cite{EPClassicalPaper}. 
An alternative QNM coupled mode theory for lossy coupled resonators is also discussed in Ref.~\cite{tao_coupling_2020}.

The eigenfrequencies $\tilde\omega_\pm$ are shown in Fig.~\ref{fig:Eigfreq_Eigfunctions}(a-b) as a function of gap distance $d_{\rm gap}$, for different gain contributions. In the following, we concentrate on three particularly interesting gap separations: (i) $d_{\rm gap}=1240~{\rm nm}, 1220~{\rm nm}$, where $\omega_+\sim\omega_-$, (ii) $d_{\rm gap}=1160~{\rm nm}, 1180~{\rm nm}$, where $\gamma_+\sim\gamma_-$ and (iii) $d_{\rm gap}=1200~{\rm nm}$, where $\tilde\omega_+\sim\tilde\omega_-$, very close to a lossy exceptional point, at which the eigenfrequencies become nearly degenerate. Note, that these regions slightly change for the different gain coefficients. It should be further noted here, that the imaginary part of the QNM eigenfrequencies of this specific hybrid structure are dominantly originating from non-radiative loss/gain mechanism, i.e., the radiation to the far field is negligible.

Similar to the analytic CMT eigenfrequencies, the hybrid QNM eigenfunctions can also be obtained analytically from the CMT coupling constant and single resonator QNM functions:
\begin{align}\label{eq: QNMs_pm}
\ket{\tilde{\bf f}^\pm}&=
\frac{\tilde\omega_\pm-\tilde\omega_{\rm G}}{\sqrt{(\tilde\omega_\pm-\tilde\omega_{\rm G})^2+ \tilde \kappa_{\rm GL}^2}}
\ket{\tilde{\bf f}_{\rm L}}
\nonumber \\
&
+ \frac{-\tilde\kappa_{\rm GL}}{\sqrt{(\tilde\omega_\pm-\tilde\omega_{\rm G})^2+ \tilde \kappa_{{\rm GL}}^2}} 
\ket{\tilde{\bf f}_{\rm G}}.
\end{align}

For the convenience of the analysis, we could define these eigenfunctions with QNM phases as $\mathbf{\tilde{f}}^{\pm}(\mathbf{r})=|\mathbf{\tilde{f}}^{\pm}(\mathbf{r})|e^{i\phi_{\pm}(\mathbf{r})}$. In the projected LDOS, Eq.~\eqref{eq: LDOS_normalized}, 
it is that the square of the eigenfunction matters, so the amplitude $|\mathbf{\tilde{f}}^{\pm}(\mathbf{r})|^2$ as well as the cos and sin of the complex phase $2\phi_{\pm}(\mathbf{r})$ of the hybrid functions for $d_{\rm gap}=1200~{\rm nm}$ and $d_{\rm gap}=1240~{\rm nm}$, with gain coefficient $\alpha_{\rm G}=2\cdot 10^{-6}$, are shown in Fig.~\ref{fig:Eigfreq_Eigfunctions}(c-f). We see that although the bare resonator functions are close to real, the hybrid eigenfunctions have a significant complex phase due to the intercavity coupling. For more details on the QNM CMT and the numerical calculation of the single resonator problem, see Ref.~\onlinecite{EPClassicalPaper}.  

The underlying Green function of the total hybrid system can then be reconstructed, analytically,  as a sum of the hybrid QNM functions through the diagonal form
\begin{equation}
    \mathbf{G}(\mathbf{r},\mathbf{r}_0,\omega)=\omega\frac{\tilde{\bf f}^+(\mathbf{r})\tilde{\bf f}^+(\mathbf{r}_0)}{2(\tilde\omega_+ - \omega)}+\omega\frac{\tilde{\bf f}^-(\mathbf{r})\tilde{\bf f}^-(\mathbf{r}_0)}{2(\tilde\omega_- - \omega)}.
\end{equation}
Alternatively, it can be expressed entirely in terms of the bare QNMs, which can have certain advantages if studying the response at the
divergent exceptional point, or if applying a normal mode approximation. The corresponding Green function has a non-diagonal form and reads~\cite{EPClassicalPaper}
 \begin{align}
{\bf G}&(\mathbf{r},\mathbf{r}_0,\omega)\nonumber\\
=& \frac{\omega(\tilde \omega_{\rm G}-\omega) \tilde{\mathbf{f}}_{\rm L}(\mathbf{r})\tilde{\mathbf{f}}_{\rm L}(\mathbf{r}_0)}{2(\tilde \omega_{+}-\omega)(\tilde \omega_{-}-\omega)} 
+ \frac{\omega\tilde \kappa_{\rm LG} \tilde{\mathbf{f}}_{\rm L}(\mathbf{r})\tilde{\mathbf{f}}_{\rm G}(\mathbf{r}_0)}{2(\tilde \omega_{+}-\omega)(\tilde \omega_{-}-\omega)}
\nonumber \\
&+\frac{\omega\tilde \kappa_{\rm GL} \tilde{\mathbf{f}}_{\rm G}(\mathbf{r})\tilde{\mathbf{f}}_{\rm L}(\mathbf{r}_0)}{2(\tilde \omega_{+}-\omega)(\tilde \omega_{-}-\omega)}
+
\frac{\omega(\tilde \omega_{\rm L}-\omega) \tilde{\mathbf{f}}_{\rm G}(\mathbf{r})\tilde{\mathbf{f}}_{\rm G}(\mathbf{r}_0)}{2(\tilde \omega_{+}-\omega)(\tilde \omega_{-}-\omega)}.\label{eq: GFNonDiag}
\end{align}

\begin{figure}[h]
    \centering
    \includegraphics[width=0.99\columnwidth]{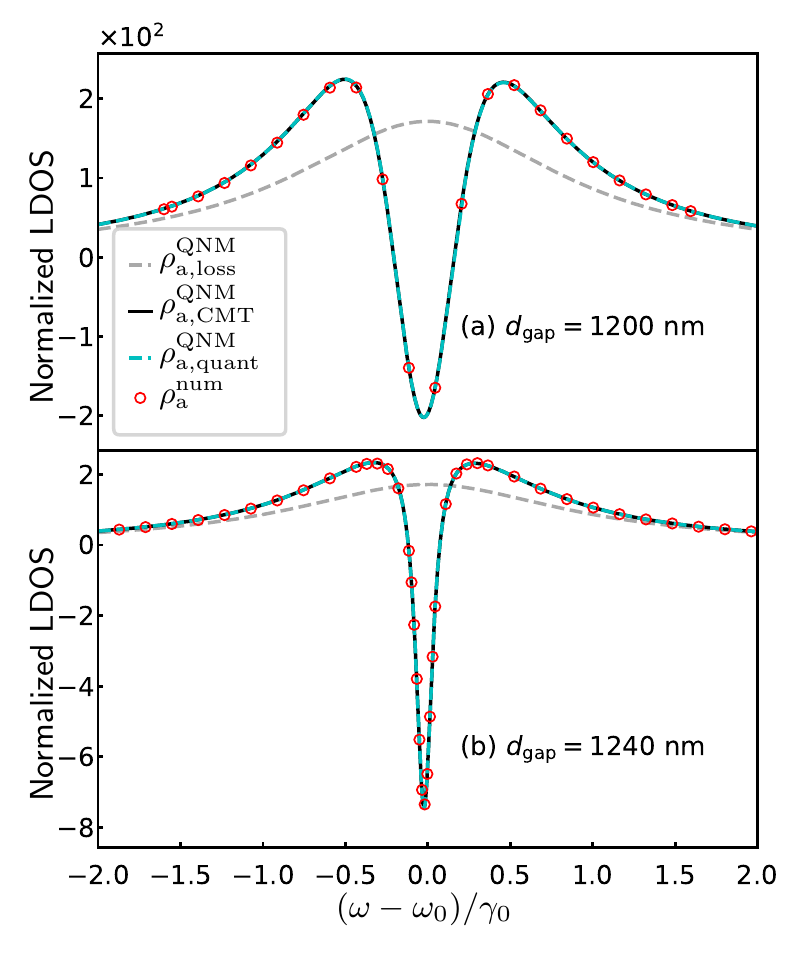}
    \caption{Dipole-projected LDOS 
    (Eq.~\eqref{eq: LDOS_normalized}) using QNM CMT expansion of the Green function over frequency $\omega$ for $\mathbf{r}_{\rm a}=\mathbf{r}_1$, $\alpha_{\rm G}=2\cdot 10^{-6}$ and (a) $d_{\rm gap}=1200~{\rm nm}$, (b) $d_{\rm gap}=1240~{\rm nm}$. The grey dashed curve reflects the solution of the single QNM expansion of the isolated lossy disk resonator. Furthermore, the cyan dashed curve reflects the result obtained from the bad cavity limit of the separated gain-loss operator approach, Eq.~\eqref{eq: LDOS_quant}. 
}\label{fig:LDOS_vs_fullDipole}
\end{figure} 

To verify the accuracy of above QNM expansion of the Green function from CMT, it is instructive to compare the LDOS (as introduced in Eq.~\eqref{eq: LDOS_normalized}) to the classical power flow from a fully numerical simulation of the classical Maxwell-dipole equations. Both quantities can be related to each other via the time-averaged Poynting theorem 
\begin{equation}
    \oint_S \mathbf{n}\cdot\mathbf{S}_{\rm poynting}(\mathbf{s}){\rm d}A_{\mathbf{s}}=-{\rm Re}\left[\int_V \mathbf{j}^*(\mathbf{r})\cdot\mathbf{E}_{\rm S}(\mathbf{r}){\rm d}\mathbf{r}\right]\label{eq: Poynting},
\end{equation}
where $V$ is a small volume around the classical dipole emitter (not intersecting with the scattering structure) with surface $S$ and normal vector $\mathbf{n}$.
In the fully numerical approach, the scattered fields $\mathbf{E}_{\rm S},\mathbf{H}_{\rm S}$ are calculated via COMSOL~\cite{comsol} and are used as input for deriving the Poynting vector $\mathbf{S}_{\rm poynting}={\rm Re}[\mathbf{E}_{\rm S}\times\mathbf{H}_{\rm S}^*]/2$. Using $\mathbf{E}_{\rm S}(\mathbf{s})=i\int_V {\rm d}\mathbf{r} \mathbf{G}(\mathbf{s},\mathbf{r})\cdot \mathbf{j}(\mathbf{r})/\epsilon_0\omega$ and $\mathbf{j}(\mathbf{r})=-i\omega\mathbf{d}\delta(\mathbf{r}-\mathbf{r}_{\rm a})$, the RHS of Eq.~\eqref{eq: Poynting} is indeed proportional to the (projected) LDOS. In particular, dividing Eq.~\eqref{eq: Poynting} by a reference power flow in the absence of scattering structures yields
\begin{equation}
    \rho_{\rm a}^{\rm num}= \frac{\oint_S \mathbf{n}\cdot\mathbf{S}_{\rm poynting}(\mathbf{s}){\rm d}A_\mathbf{s}}{\oint_S \mathbf{n}\cdot\mathbf{S}_{\rm poynting}^{(0)}(\mathbf{s}){\rm d}A_\mathbf{s}},
\end{equation}
which is an alternative expression for the normalized LDOS, compared to Eq.~\eqref{eq: LDOS_normalized}. As demonstrated in Fig.~\ref{fig:LDOS_vs_fullDipole}, the QNM CMT expansion of the LDOS is in excellent agreement with the fully numerically simulation of Maxwell's equations.

We emphasize that for the case of also including an amplifying medium, the LDOS is not directly connected to the (usual classical) Purcell factor, as would be the case for purely lossy media. In fact, as demonstrated in Ref.~\onlinecite{franke2021fermi}, Fermi's golden rule for amplifying media leads to additional terms for the spontaneous emission rates, that do not only depend on the photonic LDOS. Thus, a negative LDOS is not unphysical in the case of gain-loss resonators. This is fully taken into account for the presented quantization methods here, which are based on the same Green function quantization approach used in Ref.~\onlinecite{franke2021fermi}.

\subsection{Improved phenomenological quantum gain model and relation to the unified operator approach}
As a first application, we investigate the phenomenological quantum gain models for the specific resonator structure and compare the quantum mode parameters and formal appearance with the more rigorous quantized QNM approaches in more detail. Here we concentrate on the comparison to the unified gain-loss operator approach, since it is more closely related to the usual quantum gain models. 

To do so, we first adapt and improve Eqs.~\eqref{eq: Phenomenological_QLE} and Eq.~\eqref{eq: H_ph_ScullyLamb} to the investigated example of the gain-loss resonator system, interacting with a TLS by also taking into account our knowledge of the rigorous QNM CMT results. In the inspected case, $1=\rm{L}$, $2={\rm G}$, and $\Gamma_{\rm G}=0$ (since the intrinsic loss vanishes for the purely amplifying media without radiative loss). Furthermore, we identify the decay rate of the lossy disk as $\Gamma_{\rm L}\equiv 2\gamma_{\rm L}$ and the gain rate of the amplifying ring as $A=2\gamma_{\rm G}$. Lastly, we  replace the photon coupling Hamiltonian, i.e., $i\hbar\kappa[b_{\rm L}^\dagger b_{\rm G} - b_{\rm G}^\dagger b_{\rm L}]\rightarrow -\hbar\kappa[b_{\rm L}^\dagger b_{\rm G} + b_{\rm G}^\dagger b_{\rm L}]$
and assume that the photon-photon coupling, $\kappa$, is the mean value of the real parts of the QNM CMT coupling constants $\kappa = {\rm Re}(\kappa_{\rm GL}+\kappa_{\rm LG})/2$. Lastly, the TLS-mode coupling constant is assumed as $g_{\rm L(G)}=\sqrt{\omega_{0}/(2\hbar\epsilon_0)}d|\hat{\mathbf{z}}\cdot\bf f_{\rm L(G)}(\mathbf{r}_{\rm a})|$, where $\bf f_{\rm L(G)}(\mathbf{r}_{\rm a})$ is the real part of the complex QNM function, representing a usually assumed lossless mode. 

The resulting {\it improved phenomenological quantum gain model} is deeply connected to a normal mode approximation of the non-diagonal QNM Green function, Eq.~\eqref{eq: GFNonDiag}, as explained in detail in Ref.~\cite{2108.10194} for the purely lossy case. For the more general gain-loss case, the normal mode approximation would lead to
\begin{align}
{\bf G}_{\rm phen}&(\mathbf{r},\mathbf{r}_0,\omega)\nonumber\\
=& \frac{\omega(\tilde \omega_{\rm G}-\omega) \mathbf{f}_{\rm L}(\mathbf{r})\mathbf{f}_{\rm L}(\mathbf{r}_0)}{2(\tilde \omega_{+}'-\omega)(\tilde \omega_{-}'-\omega)} 
+ \frac{\omega\kappa \mathbf{f}_{\rm L}(\mathbf{r})\mathbf{f}_{\rm G}(\mathbf{r}_0)}{2(\tilde \omega_{+}'-\omega)(\tilde \omega_{-}'-\omega)}
\nonumber \\
&+\frac{\omega \kappa\mathbf{f}_{\rm G}(\mathbf{r})\mathbf{f}_{\rm L}(\mathbf{r}_0)}{2(\tilde \omega_{+}'-\omega)(\tilde \omega_{-}'-\omega)}
+
\frac{\omega(\tilde \omega_{\rm L}-\omega) \mathbf{f}_{\rm G}(\mathbf{r})\mathbf{f}_{\rm G}(\mathbf{r}_0)}{2(\tilde \omega_{+}'-\omega)(\tilde \omega_{-}'-\omega)},\label{eq: GFNonDiagNM}
\end{align}
where 
\begin{equation}
    \tilde\omega_\pm'=\frac{\tilde\omega_{\rm L}+\tilde\omega_{\rm G}}{2}+\sqrt{\kappa^2 +\frac{(\tilde\omega_{\rm L}-\tilde\omega_{\rm G})^2}{4}}.
\end{equation}
We emphasize that the choice of the sign or phase for the photon coupling constant in the Hamiltonian is very important to recapture the correct behavior of the emitter-photon interaction, and is not an obvious improvement at all. Indeed choosing a form as in Eq.~\eqref{eq: H_ph_ScullyLamb} can lead to wrong predictions of the TLS properties ~\cite{2108.10194}. 

The relevant master equation within the above adaption and improvement of the usual models then reads
\begin{align}
    \mathcal{L}^{\rm phen}\rho =& -\frac{i}{\hbar}[H_{\rm S}^{\rm phen},\rho]+\gamma_{\rm L}[2b_{\rm L} \rho b_{\rm L}^\dagger-b_{\rm L}^\dagger b_{\rm L}\rho-\rho b_{\rm L}^\dagger b_{\rm L}]\nonumber\\
    &+\gamma_{\rm G}[2b_{\rm G}^\dagger \rho b_{\rm G}-b_{\rm G}b^\dagger_{\rm G}\rho-\rho b_{\rm G}b_{\rm G}^\dagger]\label{eq: Pheno_ME},
\end{align}
with the system Hamiltonian 
\begin{align}
    H_{\rm S}^{\rm phen}=&\hbar\sum_{i=\rm G,L}\omega_{i}b_{i}^\dagger b_{i}-\hbar\kappa[b_{\rm L}^\dagger b_{\rm G}+{\rm H.a.}]\nonumber\\
    &+i\hbar \sum_{i=\rm L,G}g_i(b_i\sigma^++{\rm H.a.})-\hbar\omega_{\rm a}\sigma^+\sigma^-.
\end{align}

To compare to a unified gain-loss operator approach, we diagonalize the photon Hamiltonian which leads to a transformed photon Liouvillian 
\begin{align}
    \mathcal{L}^{\rm phen}_{\rm em}\rho =& -i\sum_i \Omega_i^{\rm eig}[B_i^\dagger B_i,\rho]\nonumber\\
    &+\sum_{i,j}\gamma_{\rm L}^{ij}[2B_j \rho B_i^\dagger- B_i^\dagger B_j\rho-\rho B_{i}^\dagger B_{j}]\nonumber\\
    &+\sum_{i,j}\gamma_{\rm G}^{ij}[2B_{i}^\dagger \rho B_{j}-B_{j}B^\dagger_{i}\rho-\rho B_{j}B_{i}^\dagger],
\end{align}
where $i,j=1,2$; $\gamma_{\rm L(G)}^{ij}$ are the mode loss (pump) matrices in the transformed picture; and  $\Omega_i^{\rm eig}$ are the eigenvalues of the Hermitian part of the complex photon matrix,
\begin{equation}
    \tilde{\boldsymbol{\Omega}}=\begin{pmatrix}
    \tilde{\omega}_{\rm L} & -\kappa \\
    -\kappa & \tilde{\omega}_{\rm G}\label{eq: TildeOmegaNM}
    \end{pmatrix}.
\end{equation}
Analogously, one can unitarily diagonalize the photon Hamiltonian in the unified gain-loss operator approach, so that the effect of mode coupling is fully captured in the transformed dissipation matrices. In doing so, the photon Hamiltonian, Eq.~\eqref{eq: H_QNM_uni}, takes a diagonal form, and the electromagnetic Liouvillian acting on the density operator is
\begin{align}
    \mathcal{L}_{\rm em}\rho=&-i\sum_\mu\nu_\mu [A_\mu^{\prime\dagger}A_\mu',\rho]\nonumber\\
    &+\sum_{\mu,\eta}\Gamma^{\rm \prime L}_{\mu\eta}\left[2A_{\eta}'\rho A_{\mu}^{\prime\dagger} - A_{\mu}^{\prime\dagger}A_{\eta}'\rho -\rho A_{\mu}^{\prime\dagger}A_{\eta}' \right]\nonumber\\
 &+\sum_{\mu,\eta}\Gamma^{\rm \prime G}_{\mu\eta}\left[2A_{\mu}^{\prime\dagger}\rho A_{\eta}^{\prime} - A_{\eta}^{\prime}A_{\mu}^{\prime\dagger}\rho -\rho A_{\eta}^{\prime}A_{\mu}^{\prime\dagger} \right],
\end{align}
where $\nu_\mu$ are the eigenvalues of the photon-photon coupling matrix $\boldsymbol\chi^{\prime +}$, and $\Gamma^{\rm \prime (L)G}_{\mu\eta}$ are the QNM loss and pump matrices in the transformed picture.

Thus, the master equation of the improved phenomenological quantum gain model and the unified gain-loss operator approach are brought into a identical form. 
Comparing the different quantum parameters for the specific resonator structure leads to the result, that they are nearly identical in both approaches. 
Thus, we see that the partial inclusion of rigorous QNM theory can in fact improve the usual adopted quantum approaches for coupled gain and loss resonators that are based on a normal mode coupling.
However, a crucial difference is the fact, that in the QNM models the TLS-mode coupling constant as well as the off-diagonal pump/loss matrix elements are complex-valued, while in the improved phenomenological quantum gain model, these are still assumed as real values independent on the amount of dissipation. Since the quality factors $Q_\mu$ of the QNMs are very large for the inspected example, the imaginary parts of the pump and loss matrices are very small and one could expect a similar behaviour in such cases. We emphasize, that for smaller $Q$ factors this can drastically change, similar to the purely lossy case, where, e.g., certain Fano interference effects in plasmonic-dielectric resonators can only be properly described by taking into account the intrinsic lossy nature of the QNMs~\cite{KamandarDezfouli2017,PhysRevLett.122.213901}.

\subsection{Vacuum state occupation and positive definiteness within the unified gain-loss operator framework}
Next, we inspect the positive definiteness of the symmetrization matrix $\mathbf{S}'$ and the vacuum state occupation $n_{\rm vac}$ within the framework of the unified gain-loss operator basis. The former will reflect its validity in different gain-loss regimes while the latter will pinpoint to the difference of the vacuum state to the separated gain-loss operator approach. As shown in the last subsection, the improved phenomenological quantum gain approach is closely related to the unified gain-loss operator approach. Thus, the latter property will also characterize the Fock space, that is simply constructed ad-hoc in the improved phenomenological quantum gain approach. 
\begin{figure}[h]
    \centering
    \includegraphics[width=0.99\columnwidth]{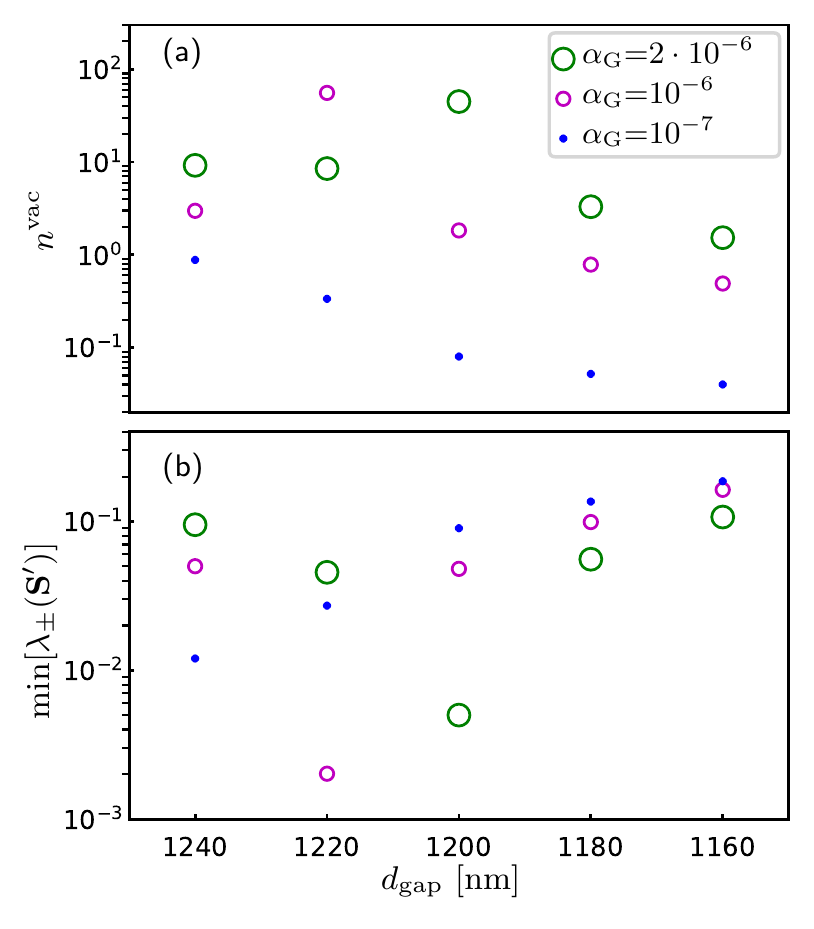}
    \caption{Vacuum state occupation $n^{\rm vac}$ (a) and smallest eigenvalue ${\rm min}[\lambda(\mathbf{S}')_\pm]$ of the symmetrization matrix $\mathbf{S}'$ (b) from the unified gain-loss operator approach as function of gap distances for three different gain cases with gain coefficients $\alpha_{\rm G}$. Note, that the eigenvalues are divided by ${\rm max}[\lambda(\mathbf{S}')_\pm]$, and that the $y$-axis is scaled logarithmic.  
}\label{fig:nvac_lambda_pm}
\end{figure}
First, to check the positive definiteness, we calculated the eigenvalues $\lambda[\mathbf{S}']_\pm$ of the $2\times2$-matrix $\mathbf{S}'$ as function of gap distance and gain contribution. As demonstrated in Fig.~\ref{fig:nvac_lambda_pm} (bottom), for all inspected gap distances and gain contributions, $\lambda[\mathbf{S}']_\pm >0$, and thus $\mathbf{S}'$ is positive definite. Therefore, the criteria for the validity of the unified gain-loss operator approach is fulfilled. Furthermore it is interesting to note that there appears a minimum of the smaller eigenvalues for $\alpha_{\rm G}=2\cdot 10^{-6}$ (around $d_{\rm gap}=1200~{\rm nm}$) and $\alpha_{\rm G}=10^{-6}$ (around  $d_{\rm gap}=1220~{\rm nm}$), which is located close to the respective EP region.

Second, we inspect the vacuum state occupation. As demonstrated in Fig.~\ref{fig:nvac_lambda_pm}, for small values $\alpha_{\rm L}/\alpha_{\rm G}$, the vacuum state occupation differs significantly from $0$, and for $d_{\rm gap}=1220~{\rm nm}$ even goes up to the value of $n^{\rm vac}\sim 50$. Even at small gain regions, such as $\alpha_{\rm G}/\alpha_{\rm L}=100$, $n^{\rm vac}$ is still noticeably different from $0$. Therefore, the Fock states construction of the unified gain-loss operator approach and thus of the improved phenomenological quantum gain approach significantly differs from the separated gain-loss operator approach. 

We note that similar to the eigenvalues of the symmetrization matrix, there appears a maximum of $n^{\rm vac}$ around the EP region for $\alpha_{\rm G}=2\cdot 10^{-6}$ and $\alpha_{\rm G}=10^{-6}$.
Thus, we can conclude that these quantum parameter are very sensitive around the classical EP region.

\subsection{Bad cavity limit of the quantum gain models\label{subsec: BadCav}}

After discussing the differences and similarities of the unified gain-loss operator approach and an improved phenomenological quantum gain approach as well as the fundamental quantum properties, we now focus on dynamics and behavior of the TLS interacting with the lossy and amplifying medium. 
In this work, we investigate the {\it weak light-matter coupling} limit, where a comparison to other classical or semiclassical models is possible, which helps to validate the quantized QNM models and in the presence of gain and loss. 
\begin{figure}[h]
   \centering
    \includegraphics[width=0.99\columnwidth]{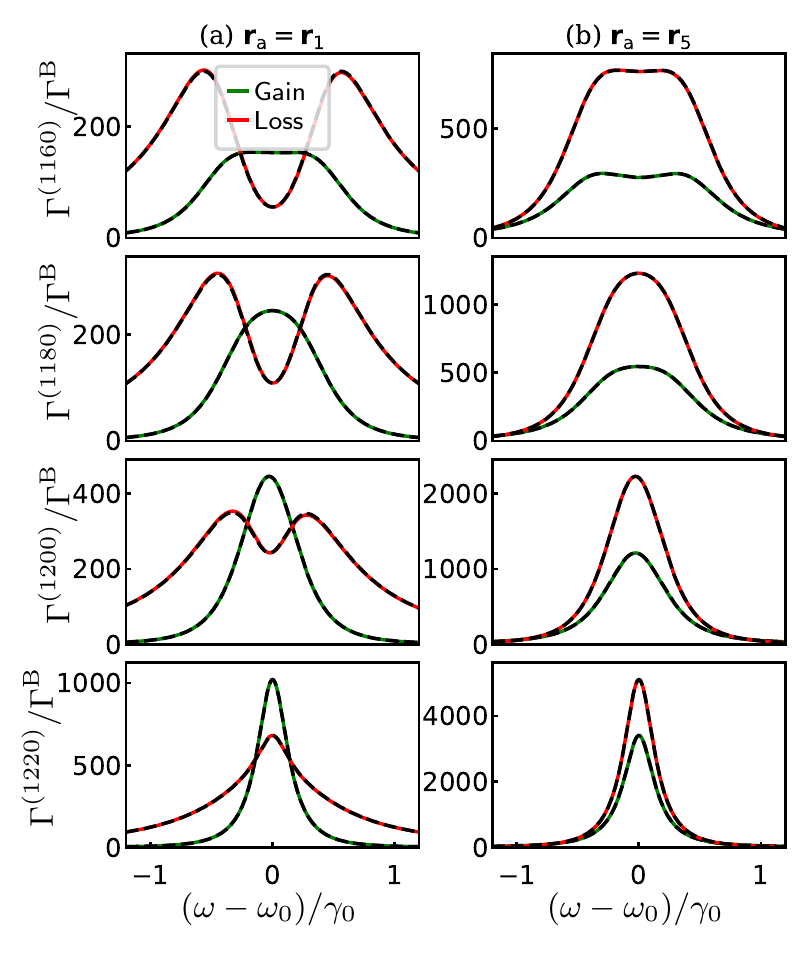}
    \caption{Gain and loss related rates (normalized by the free space rate $\Gamma^{\rm B}$) of the TLS at the position $\mathbf{r}_1$ near the lossy cavity (a) and position $\mathbf{r}_5$ near the amplifying cavity (b) as function of TLS frequency obtained from the bad cavity limit of the quantized QNM operator approaches (solid) and the improved phenomenological quantum gain model (black) for different gap distances $d_{\rm gap}$, ranging from $1160~{\rm nm}$ (top) to $1220~{\rm nm}$ (bottom). The gain coefficient is chosen as $\alpha_{\rm G}=2\cdot 10^{-6}$.
}\label{fig:Gamma_loss_gain}
\end{figure}

To obtain this limit, we assume a relatively small dipole moment, $d=d_0$, to apply a procedure similar to the purely lossy case Ref.~\onlinecite{PhysRevLett.122.213901} (following the approach from Cirac~\cite{Cirac}). However, we note here that there are several differences compared to the bad cavity limit of the purely lossy case. In the separated gain-loss operator approach, the presence of interaction terms $\sigma^+ a_{\rm G\mu}^\dagger$ induces a different ordering of the atomic operators with respect to the gain contributions (cf. App.~\ref{app: MarkovBad}), while in the unified gain-loss operator approach, the incoherent Lindblad dissipator $\mathcal{L}_{\rm G}'[\mathbf{a}^{\prime\dagger}]$ is responsible for such a change. We emphasize, that the origin of the correct gain and loss TLS rates are completely different in the latter approach, where the TLS-QNM Hamiltonian is in a mixed photon operator ordering (with respect to gain and loss contributions) by construction of the underlying Fock operator basis. However, both approaches lead to the same results in the weak coupling limit as proved for the demonstrative case of a single QNM in App.~\ref{app: BlochEquations}, by employing a Bloch equations treatment instead of a density matrix approach. 

Taking the changes over the purely lossy case into account and carefully applying the weak coupling limit, we obtain the TLS master equation for the atomic density operator $\rho_{\rm a}={\rm tr}_{\rm em}\rho$ within the quantized QNM models:
\begin{align}
    \partial_t\rho_{\rm a}&=-\frac{i}{\hbar}[H_{\rm a},\rho_{\rm a}]+ \frac{\Gamma^{\rm B}}{2}\mathcal{D}[\sigma^-]\rho_{\rm a}\nonumber\\
    &+\frac{\Gamma^{\rm loss}}{2}\mathcal{D}[\sigma^-]\rho_{\rm a}+\frac{\Gamma^{\rm gain}}{2}\mathcal{D}[\sigma^+]\rho_{\rm a}, \label{eq: BadCavMaster}
\end{align}
where
\begin{equation}
    \Gamma^{\rm loss}=\sum_{\eta,\eta '}\tilde{g}_\eta S_{\eta\eta '}^{\rm L}\tilde{g}_{\eta '}^*\frac{i(\omega_\eta-\omega_{\eta '})+(\gamma_{\eta }+\gamma_{\eta '})}{(\Delta_{\eta a}-i\gamma_{\eta })(\Delta_{\eta ' a}+i\gamma_{\eta '})}\label{eq: GammaEnh},
\end{equation}
is the modified SE rate and
\begin{equation}
    \Gamma^{\rm gain}=\sum_{\eta,\eta '}\tilde{g}_\eta S_{\eta'\eta }^{\rm G}\tilde{g}_{\eta '}^*\frac{i(\omega_\eta-\omega_{\eta '})+(\gamma_{\eta }+\gamma_{\eta '})}{(\Delta_{\eta a}-i\gamma_{\eta })(\Delta_{\eta ' a}+i\gamma_{\eta '})}\label{eq: GammaPump},
\end{equation}
is the gain-induced pump rate.

Here we have neglected the photonic Lamb shifts.
Apparently, the main difference over the purely lossy case is the presence of an incoherent pumping term with a rate $\Gamma^{\rm gain}$, which is in line with rigorous treatment of Fermi's golden rule in amplifying media~\cite{franke2021fermi}. In fact, when calculating the difference $\Gamma^{\rm loss}-\Gamma^{\rm gain}$, we observe that it {\it precisely matches the projected LDOS} through 
\begin{equation}
    \rho^{\rm QNM}_{\rm a,quant}=1+\frac{\Gamma^{\rm loss}-\Gamma^{\rm gain}}{\Gamma^{\rm B}},\label{eq: LDOS_quant}
\end{equation}
as demonstrated in Fig.~\ref{fig:LDOS_vs_fullDipole} by the blue dashed lines, fully consistent with the results obtained in Ref.~\cite{franke2021fermi}.

To explain the origin of the pronounced Fano dip in the LDOS in the quantum theory and to demonstrate the impact of the EP region, we show the concrete elements of the symmetrization matrix for the example of $\alpha=2\cdot10^{-6}$ in Table~\ref{table: Sparameters}. First of all, we recognize that the off-diagonal elements of the loss and gain symmetrization matrices are in the range of the diagonal elements, which indicates a large overlap of the hybridized QNMs. Second, we see a drastic change of the parameters by a factor of $\sim 5$, if we get close to the (classical) EP region. Depending on the position of the quantum emitter, this can lead to a significant change of the symmetrized TLS-QNM coupling constant (proportional to $\mathbf{S}^{\rm L(G)1/2}$) compared to the initial coupling, which would be not at all captured by a model, that neglects the off-diagonal elements of the $\mathbf{S}$ matrices, i.e., that neglects the effects of dissipation and amplification on the hybridized mode level. It is further noteworthy, that for larger gap distances $d_{\rm gap}$, the deviation between the diagonal elements $S_{++}$ and $S_{--}$ increases.

\begin{table*}
\caption {A selection of symmetrization matrices $S_{\mu\eta}^{\rm L}$ (equal to $S^{\rm nrad}_{\mu\eta}$ from Eq.~\eqref{eq:SnradPole}) and $S_{\mu\eta}^{\rm G}$ (Eq.~\eqref{eq:SgainPole}) for the coupled gain-loss microdisk resonators with $\alpha=2\cdot10^{-6}$ close to the EP region ($\sim 1200~{\rm nm}$).}

\label{table: Sparameters} 
\begin{center}
\begin{tabular}{ |c|c|c|c| } 
 \hline
$d_{\rm gap}=1160~$nm &$d_{\rm gap}=1180~$nm & $d_{\rm gap}=1200~$nm & $d_{\rm gap}=1220~$nm\\ 
\hline
$S_{++}^{\rm L}=2.1221$ &
 $S_{++}^{\rm L}=2.7945$ & $S_{++}^{\rm L}=8.5390$ & $S_{++}^{\rm L}=4.6738$ \\ 
 $S_{+-}^{\rm L}=-0.2847 - 1.4036i$ &
$S_{+-}^{\rm L}=0.4000 - 2.1998i$ & $S_{+-}^{\rm L}=-0.5947 - 8.3613i$ & $S_{+-}^{\rm L}=-0.0007 + 2.7960i$   \\ 
  $S_{-+}^{\rm L}=-0.2847 + 1.4036i$ & $S_{-+}^{\rm L}=0.4000 + 2.1998i$ & $S_{-+}^{\rm L}=-0.5947 - 8.3613i$ & $S_{-+}^{\rm L}=-0.0007 - 2.7960i$ \\ 
  $S_{--}^{\rm L}=2.1221$ &
 $S_{--}^{\rm L}=2.7953$ & $S_{--}^{\rm L}=8.6300$ & $S_{--}^{\rm L}=2.6751$  \\ 
  $S_{++}^{\rm G}=0.4246$ &
  $S_{++}^{\rm G}=0.5588$ & $S_{++}^{\rm G}=1.6732$ & $S_{++}^{\rm G}=2.2237$ \\ 
  $S_{-+}^{\rm G}=-0.2847 - 0.0317i$ &
 $S_{-+}^{\rm G}=-0.4000 - 0.2000i$ & $S_{-+}^{\rm G}=-0.5946 - 1.5675i$ & $S_{-+}^{\rm G}=-0.0006 + 0.5592i$ \\ 
 $S_{+-}^{\rm G}=-0.2847 + 0.0317i$ &
 $S_{+-}^{\rm G}=-0.4000 - 0.2000i$ & $S_{+-}^{\rm G}=-0.5946 + 1.5675i$ & $S_{+-}^{\rm G}=-0.0006 - 0.5592i$\\ 
 $S_{--}^{\rm G}=0.4243$ &
$S_{--}^{\rm G}=0.5592$ &$S_{--}^{\rm G}=1.7616$ & $S_{--}^{\rm G}=0.2249$
\\ 
 \hline
\end{tabular}
\end{center}
\end{table*}

On the other hand, applying the bad cavity limit to the improved phenomenological quantum gain model (based on Eq.~\eqref{eq: Pheno_ME}) would lead to a formally identical master equation as Eq.~\eqref{eq: BadCavMaster} (again neglecting the Lamb shift terms),  with the quantum loss rate (cf. App.~\ref{app: BlochEquationsPheno}) $\Gamma^{\rm loss}_{\rm phen}=\Gamma_{\rm phen}^{\rm LDOS}+\Gamma^{\rm gain}_{\rm phen}$, where 
\begin{equation}
    \Gamma_{\rm phen}^{\rm LDOS}=2\sum_{i,j,k}g_ig_k{\rm Re}\left(T_{ikj}\frac{i}{\omega_{\rm a}-\tilde{\Omega}^{\rm eig}_j}\right),
\end{equation}
and 
\begin{equation}
    \Gamma^{\rm gain}_{\rm phen}=2\sum_{i,k,j}g_ig_k{\rm Re}\left(T_{ikj}'\frac{i}{\omega_{\rm a}-\tilde{\Omega}^{\rm eig}_j}\right),
\end{equation}
is the gain rate. Here, 
\begin{align}
    T_{ikj}&=[\mathbf{V}]_{ij}[\mathbf{V}^{-1}]_{jk},\nonumber\\
    T_{ikj}'&=\sum_n[\mathbf{V}]_{ij}[\mathbf{V}]_{kn}^* \frac{2\gamma_{\rm G}}{i(\tilde{\Omega}^{\rm eig}_j-\tilde{\Omega}^{\rm eig*}_n)}[\mathbf{V}^{-1}]_{j\rm G}[\mathbf{V}^{-1}]_{n\rm G}^*,
\end{align}
and $\mathbf{V}$ ($\tilde{\Omega}_j^{\rm eig}$) is the eigenmatrix ($j$-th eigenvalue) of $\tilde{\boldsymbol{\Omega}}$ (Eq.~\eqref{eq: TildeOmegaNM}). 
We note that $\Gamma_{\rm phen}^{\rm LDOS}$ can be reformulated in a form proportional to the imaginary part of $\mathbf{G}_{\rm phen}(\mathbf{r}_{\rm a},\mathbf{r}_{\rm a},\omega_{\rm a})$ (Eq.~\eqref{eq: GFNonDiagNM}), i.e., 
\begin{equation}
     \Gamma_{\rm phen}^{\rm LDOS}=\frac{2}{\hbar\epsilon_0}\mathbf{d}\cdot{\rm Im}\left[\mathbf{G}_{\rm phen}(\mathbf{r}_{\rm a},\mathbf{r}_{\rm a},\omega_{\rm a})\right]\cdot\mathbf{d},
\end{equation}
only because of the specific choice of the sign in the photon coupling Hamiltonian, i.e., $-\hbar\kappa$. Otherwise the elements in $\Gamma_{\rm phen}$ proportional to $g_{\rm L}g_{\rm G}$ would deviate from the corresponding terms in $\mathbf{G}_{\rm phen}(\mathbf{r}_{\rm a},\mathbf{r}_{\rm a})$ by a minus sign~\cite{2108.10194}. However, in contrast to the purely lossy case (where $\Gamma^{\rm LDOS}_{\rm phen}$ would be the only appearing decay rate), the gain rate $\Gamma^{\rm gain}$ and its dependence on $\kappa$ could have an even bigger impact on the TLS behavior with respect the sign choice.

We recognize that, within the bad cavity limit of Eq.~\eqref{eq: Pheno_ME}, the corresponding quantum LDOS becomes 
\begin{equation}
    \rho^{\rm phen}_{\rm a,quant}=1+\frac{\Gamma_{\rm phen}^{\rm LDOS}}{\Gamma^{\rm B}},\label{eq: LDOS_quantNM}
\end{equation}
which also fully captures the pronounced interference effect, because the normal mode approximation was applied on the bare resonator picture (not shown).

After discussing the vadility of the different models in the bad cavity limit, we next concentrate on two figures of merit for the TLS. 

We initially study the gain and loss related rates of the TLS, $\Gamma^{\rm gain}$ and $\Gamma^{\rm loss}$, respectively, and the corresponding quantities in the improved phenomenological quantum gain approach. These rates are shown as function of TLS frequency in Fig.~\ref{fig:Gamma_loss_gain} for the specific gain coefficient $\alpha_{\rm G}=2\cdot10^{-6}$. We first recognize a nearly perfect agreement of $\Gamma^{\rm gain}$ and $\Gamma^{\rm loss}$ between the quantized QNM approach (solid lines) and the improved phenomenological quantum gain model (black dashed) for all inspected gap distances and TLS positions. This is again a consequence of the very high $Q$ factor of the bare resonators, and underlines the validity of the improved phenomenological quantum gain model in such regimes.

Second, we see a crucial dependence of interference effect on the gap distance as well as TLS position: For $\mathbf{r}_{\rm a}=\mathbf{r}_1$ (near the lossy resonator), the dip of the loss related rate becomes less pronounced when increasing the gap distance, because the overlap of the modes decreases. Although less pronounced, this trend holds also true for the gain related rate, and also for the TLS position $\mathbf{r}_{\rm a}=\mathbf{r}_5$ (close to the amplifying resonator).

Third, we mention, that for $\mathbf{r}_{\rm a}=\mathbf{r}_1$ as well as $\mathbf{r}_{\rm a}=\mathbf{r}_5$, the peak of the rates can drastically change when increasing the gap distance. Indeed, $\Gamma^{\rm gain}$ at $d_{\rm gap}=1220~{\rm nm}$ is nearly one order of magnitude larger compared to its value at $d_{\rm gap}=1160~{\rm nm}$. Lastly, we recognize that the interference behaviour is not necessarily identical for the gain and loss related rates. While one can have a Lorentzian-like form, the other can take a non-Lorentzian form, as can be seen in the exemplary case of $\mathbf{r}_{\rm a}=\mathbf{r}_1$ and $d_{\rm gap}=1180~{\rm nm}$.

Next, we study the steady-state occupation of the TLS excited state, defined through 
\begin{equation}
    n_{e,\rm ss} \equiv \langle \sigma^+\sigma^-\rangle (t\rightarrow\infty).
\end{equation}

\begin{figure}[h]
    \centering
    \includegraphics[width=0.99\columnwidth]{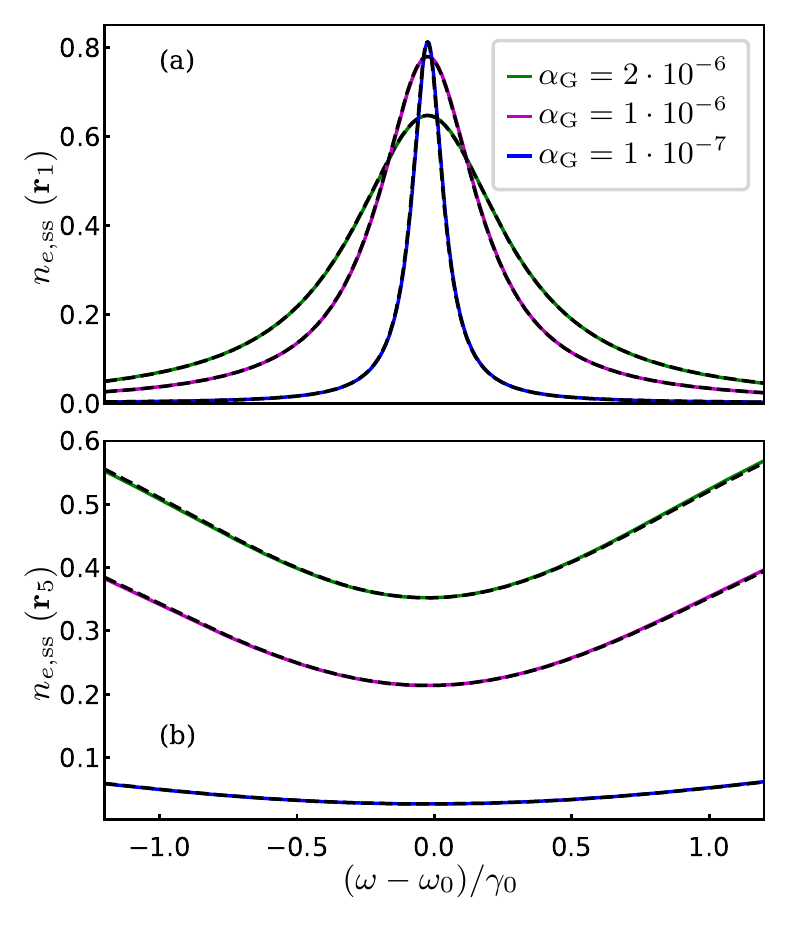}
    \caption{Steady state occupation of the excited state of the TLS at the position $\mathbf{r}_1$ near the lossy cavity (a) and position $\mathbf{r}_5$ near the amplifying cavity (b) as function of TLS frequency for different gain coefficients $\alpha_{\rm G}$ and $d_{\rm gap}=1200~{\rm nm}$. The solid (dashed) lines reflect the solutions from the quantized QNM (improved phenomenological quantum gain) model. 
}\label{fig:SteadyState_ne_ss_over_w_rlossAlt}
\end{figure}

It is easy to  show that the excited state population evolves as~\cite{franke2021fermi}
\begin{equation}\label{eq: ne_Bad_Temp}
    \dot{n}_{e}=-(\Gamma^{\rm loss}+\Gamma^{\rm B})n_e+\Gamma^{\rm gain}n_{\rm g},
\end{equation}
and, analogously, with the corresponding TLS rates in the improved phenomenological quantum gain model.
Using $n_{g}(t)+n_{e}(t)=1$,
Eq.~\eqref{eq: ne_Bad_Temp} can be solved analytically to obtain
 \begin{align}
     n_{e}(t)=&e^{-(\Gamma^{\rm loss}+\Gamma^{\rm gain}+\Gamma^{\rm B})t}n_{e}(0)\nonumber\\
     &+\frac{\Gamma^{\rm gain}}{\Gamma^{\rm loss}+\Gamma^{\rm gain}+\Gamma^{\rm B}}\left[1-e^{-(\Gamma^{\rm loss}+\Gamma^{\rm gain}+\Gamma^{\rm B})t}\right].
 \end{align}
Inspecting the limit $t\rightarrow\infty$, then leads to
\begin{equation}
    n_{e,\rm ss}=\frac{\Gamma^{\rm gain}}{\Gamma^{\rm gain}+\Gamma^{\rm loss}+\Gamma^{\rm B}}.\label{eq: BadCavity_steadystate}
\end{equation}
In Figs.~\ref{fig:SteadyState_ne_ss_over_w_rlossAlt}, we show results for $n_{e,\rm ss}$ for different gain contributions, utilizing the quantized QNM approaches (solid lines) and the improved phenomenological quantum gain approach (dashed lines).

To begin, we inspect the emitter position close to the lossy ring, $\mathbf{r}_{\rm d}=\mathbf{r}_{1}$. As a first important observation, even for very small gain cases, steady state occupations up to $n_{e,\rm ss}\sim 0.8$ can be achieved. The reason for this is that $n_{e,\rm ss}$ does not only depend on the amount of $\Gamma^{\rm gain}$, but rather on the ratio between gain and loss. To see this more clearly, we can rewrite the above expression as  
\begin{equation}
    n_{e,\rm ss}=\frac{\delta}{1+\delta},\label{eq: BadCavity_steadystate2}
\end{equation}
where
\begin{equation}
    \delta=\frac{\Gamma^{\rm gain}}{\Gamma^{\rm loss}+\Gamma^{\rm B}}.
\end{equation}
Thus, for very large ratios $\delta$, $n_{e,\rm ss}$ tends to 1.
However, we note that for smaller gain contributions, $\Gamma^{\rm gain}$ becomes even smaller than the free space loss rate $\Gamma^{\rm B}$, at which point $n_{e,\rm ss}$ also decreases, independent on the amount of $\Gamma^{\rm loss}$. 

In addition, we observe a broadening of the $n_{e,\rm ss}$ values as function of $\omega_{\rm a}$ for larger $\alpha_{\rm G}$, and thus the smaller gain cases are much more sensitive with respect to the TLS frequency. This can be attributed to the very small TLS rates, which narrow the response of the emitter to the photonic environment.

In contrast, when choosing a TLS position close to the amplifying resonator, the behavior is completely opposite: near the maximum of the absolute value of the LDOS, $n_{e,\rm ss}$ has a minimum, and these values scale proportionally to $\alpha_{\rm G}$. This is a very interesting observation, and once again shows the pronounced role of the spatially-dependent interference,
an effect that is captured by the complex QNM interaction in the quantum picture. {\it This striking difference is completely missing in the more phenomenological normal mode approaches}, if the lossless mode approximation is applied at the hybridized mode level. For high-$Q$ bare resonators (where $\tilde{\mathbf{f}}_{\rm L(G)}\approx {\rm Re}[\tilde{\mathbf{f}}_{\rm L(G)}]$), these phase effects are predominantly induced by the complex bare resonator eigenfrequencies, which can be seen from the definition of the hybridized eigenfunctions in terms of the bare eigenfunctions, Eq.~\eqref{eq: QNMs_pm}. This is captured by the earlier introduced improved phenomenological quantum gain approach.  However, we emphasize again, that this will 
likely significantly change for larger dissipation and lower $Q$-factors, where the imaginary part of the CMT coupling constants and the bare eigenfunctions become more important and where even the improved phenomenological quantum gain model will break down.

\section{Conclusions\label{Sec: Conclusion}}

We have presented a quantization scheme for
QNMs, in the presence of arbitrary three-dimensional media with linear material loss and amplification, and exemplified the theory with detailed calculations
for coupled ring resonators containing loss or gain. Two different quantized QNM approaches were developed, a separated and a unified gain-loss operator approach. In the former approach, amplification is captured through a negative Hamiltonian part and non-quanta preserving terms in the TLS-QNM interaction, while in the latter approach, the gain is captured by an incoherent mode pumping term. 

On a formal level, the separated gain-loss operator approach differs significantly from more phenomenological quantum gain models, that are usually based on a normal mode coupling in conjunction with a Scully-Lamb model in a linear gain regime. On the other hand, while the unified gain-loss operator approach is formally similar to the phenomenological models, the latter ones clearly neglect certain effects of dissipation on the quantum level, since the mode decay is included ad-hoc, so that one can expect drastic differences in the general case. Thus, the former approach can be regarded as a generalization of the usually used phenomenological quantum gain models. Moreover, a fundamental difference between the separated and the unified gain-loss operator approach (and thus also the phenomenological quantum gain approaches) was found for larger gain contributions due to the different photon vacuum state definition. 
This was demonstrated quantitatively for the inspected resonator structure, where occupation $n^{\rm vac}>1$ of the total photon number in the unified approach with respect to the initial vacuum state from the macroscopic Green function quantization approach were calculated from first principles. 

It was also shown how the results for the quantum rates of the TLS from the quantized QNM approaches in the bad cavity limit are in nearly perfect agreement (within numerical error) with the Green function quantization in the weak light-matter coupling regime, which rigorously justifies the symmetrization procedure and the obtained quantum parameters in this limit. Moreover, although the unified and separated gain-loss operator approaches were shown to yield identical results, the origin of the TLS decay and pump rate is fundamentally different due to the fact, that the photon operators couple to the loss reservoir operator and the adjoint gain reservoir operator in the former approach. As a consequence, the TLS-QNM interaction Hamiltonian cannot be written in normal photon operator ordering. This is similar for the improved phenomenological quantum gain model. In contrast, in the latter approach, the QNM-TLS interaction Hamiltonian can indeed be written in normal photon operator ordering, as the gain and loss contribution are separated on the system level.

Finally, we observed  interesting behavior of the steady-state populations of the TLS in the bad cavity limit, where these values can be above $0.5$, even for very small gain contributions, although we stress that the inspected gain-loss systems are located in a regime below the lasing threshold, due to the intrinsic condition of positive mode decay rates. 
While an improved phenomenological quantum gain model recovers these results of the quantized QNM approaches for the high-$Q$ situation, we caution that it will ultimately fail for smaller $Q$ values, where the effect of the dissipation on the photon quantization becomes more significant. Regardless, the partial inclusion of rigorous QNM theory in the phenomenological quantum gain models can lead to significant improvements of these theories.

Overall, we conclude that the quantized QNM approaches are rigorous methods for general gain-loss regimes, while the improved phenomenological quantum gain approach can still yield a valid representation for larger $Q$ values. 
Both QNM approaches reproduce the reference calculations in the weak coupling regime and are based on a rigorous macroscropic Green function quantization, and can thus can be regarded as a solid and rigorous basis for describing multi-photon effects in cavity QED setups with gain and loss.

The presented theory has a range of potential future applications in quantum optics, including the study of higher-order photon correlation functions in the stronger light-matter coupling regimes (i.e., beyond the bad cavity limit), such as the Fano factor, which gives a measure of the dispersion of a probability distribution~\cite{PhysRev.72.26}.
Also, it would be interesting to revisit the Petermann factor (which gives a measure of noise) from a quantum perspective, which was previously investigated with theories based on a normal mode quantization with phenomenological mode coupling terms in the corresponding master equations~\cite{PhysRevA.60.2529,PhysRevA.61.023806}.
In particular, these quantities could be investigated for arbitrary resonator structures,
beyond the simpler examples shown here, such as  
three-dimensional cavities with smaller $Q$ factors (including gain compensated metals), where the effects of dissipation on the mode quantization are expected to be much more drastic.

\section*{Acknowledgements}
We  acknowledge funding from Queen's University,
the Canadian Foundation for Innovation, 
the Natural Sciences and Engineering Research Council of Canada, and CMC Microsystems for the provision of COMSOL Multiphysics.
We also acknowledge support from the 
Alexander von Humboldt Foundation through a Humboldt Research Award.
We thank Marten Richter and Andreas Knorr for useful comments and support.

\clearpage
\onecolumngrid
\appendix

\section{Derivation of the quantum Langevin equations\label{app:Sec_QLE}}
In this first appendix, we give details on the derivation of the quantum Langevin equations within the unified gain-loss operator approach, Eqs.~\eqref{eq: HOM_Mode+TLS_Unified}, and discuss the complementary derivation for the separated gain-loss operator approach,  Eqs.~\eqref{eq: HOM_Mode+TLS}.  

\subsection{Photon Hamiltonian in the unified gain-loss operator approach}
First, we separate the full medium photon-space operators $\{b_i^{(\dagger)}(\mathbf{r},\omega)\}$ into a QNM part $\{a_\mu^{\prime(\dagger)}\}$ and a non-QNM part---in the following denoted as $\{c_i^{(\dagger)}(\mathbf{r},\omega)\}$---, similar to Ref.~\onlinecite{franke2020quantized}, so that:  
\begin{equation}
    b_i(\mathbf{r},\omega) = \sum_\mu [L_{{\rm L}\mu,i}^{\prime*}(\mathbf{r},\omega)a_\mu' -  L_{{\rm G}\mu,i}^{\prime}(\mathbf{r},\omega)a_\mu^{\prime\dagger}] + c_i(\mathbf{r},\omega).\label{eq: Separation}
\end{equation}
One can easily check that this separation is consistent with the definition of $a_\mu'$ from Eq.~\eqref{eq: c_muFormalDef}, with the sum rule
\begin{align}
    \sum_i&\int{\rm d}^3 r\int_0^\infty{\rm d}\omega L_{{\rm L}\mu,i}^{\prime}(\mathbf{r},\omega)c_i(\mathbf{r},\omega)+\sum_i\int{\rm d}^3 r\int_0^\infty{\rm d}\omega L_{{\rm G}\mu,i}^{\prime}(\mathbf{r},\omega)c_i^\dagger(\mathbf{r},\omega)=0\label{eq: Sumrule_dOps},
\end{align}
which also implies 
\begin{equation}
    [c_i(\mathbf{r},\omega),a_\mu^{\prime\dagger}]=[c_i(\mathbf{r},\omega),a_\mu']=0,
\end{equation}
and 
\begin{align}
    [c_i(\mathbf{r},\omega),c_j^\dagger(\mathbf{r}',\omega')]=&\delta_{ij}\delta(\mathbf{r}-\mathbf{r}')\delta(\omega-\omega')-\sum_\mu L_{{\rm L}\mu,i}^{\prime *}(\mathbf{r},\omega) L_{{\rm L}\mu,j}^{\prime}(\mathbf{r}',\omega')-\sum_\mu L_{{\rm G}\mu,i}^{\prime}(\mathbf{r},\omega) L_{{\rm G}\mu,j}^{\prime*}(\mathbf{r}',\omega').
\end{align}

Next, by using the separation, Eq.~\eqref{eq: Separation} and the corresponding adjoint equation, we rewrite the photon Hamiltonian, Eq.~\eqref{eq:HBgain}, as $H_{\rm em}=H_{\rm QNM}' + H_{\rm QNM-R}'+H_{\rm R}'$, where
\begin{align}
    H_{\rm QNM}'&=\hbar \sum_{\mu,\eta}\chi^{\prime\rm L+}_{\mu\eta}a_\mu^{\prime\dagger} a_\eta' - \hbar \sum_{\mu,\eta}\chi^{\prime\rm G+}_{\mu\eta}a_\mu' a_\eta^{\prime\dagger}, 
\end{align}
is the QNM Hamiltonian,
\begin{align}
  H_{\rm QNM-R}'&=\hbar\sum_i\sum_\mu\int{\rm d}^3 r\int_0^\infty{\rm d}\omega ~\omega L_{{\rm L}\mu,i}^{\prime *}(\mathbf{r},\omega) c_i^\dagger(\mathbf{r},\omega)a_\mu'+{\rm H.a.},\nonumber\\
    &-\hbar\sum_i\sum_\mu\int{\rm d}^3 r\int_0^\infty{\rm d}\omega ~\omega L_{{\rm G}\mu,i}^{\prime}(\mathbf{r},\omega) c_i^\dagger(\mathbf{r},\omega)a_\mu^{\prime\dagger} +{\rm H.a.}, 
\end{align}
reflects the interaction between the QNMs and the non-QNM continuum, and 
\begin{equation}
    H_{\rm R}'=\hbar\sum_i\int{\rm d}^3 r\int_0^\infty{\rm d}\omega ~\omega~ {\rm sgn}[\epsilon_I]  c_i^\dagger(\mathbf{r},\omega)c_i(\mathbf{r},\omega),
\end{equation}
is the non-QNM continuum energy.
Here, the QNM photon coupling matrices are given by
\begin{align}
    \chi^{\prime\rm L+}_{\mu\eta}&=\sum_i\int{\rm d}^3 r\int_0^\infty{\rm d}\omega ~\omega L_{{\rm L}\mu,i}^{\prime}(\mathbf{r},\omega)L_{{\rm L}\eta,i}^{\prime *}(\mathbf{r},\omega),\\
    \chi^{\prime\rm G+}_{\mu\eta}&=\sum_i\int{\rm d}^3 r\int_0^\infty{\rm d}\omega ~\omega L_{{\rm G}\mu,i}^{\prime* }(\mathbf{r},\omega)L_{{\rm G}\eta,i}^{\prime }(\mathbf{r},\omega),
\end{align}
where we have exploited the fact that $ L_{{\rm L}\mu,i}^{\prime}(\mathbf{r},\omega)$ and $L_{{\rm G}\mu,i}^{\prime }(\mathbf{r},\omega)$ are defined on distinct spatial regions through $\chi_{\mathbb{R}^3-V_{\rm G}}(\mathbf{r})$ and $\chi_{V_{\rm G}}(\mathbf{r})$ (or analogously through Heaviside functions $\Theta[\epsilon_I]$ and $\Theta[-\epsilon_I]$).

Next, we rewrite the interaction part $H_{\rm QNM-R}'$; it consists of two terms, one related to the gain contribution and one related to the loss contribution. Similar to Ref.~\onlinecite{franke2020quantized}, we rewrite the appearing integral kernels, e.g., $\omega L_{{\rm L}\mu,i}^{\prime}(\mathbf{r},\omega)$,  as
\begin{align}
    \omega L_{{\rm L}\mu,i}^{\prime}(\mathbf{r},\omega)=\left[\mathbf{S}^{\prime-1/2}\right]_{\mu\eta}[\omega-\tilde{\omega}_\eta] \tilde{L}_{{\rm L}\eta,i}(\mathbf{r},\omega)+\left[\mathbf{S}^{\prime-1/2}\right]_{\mu\eta}\tilde{\omega}_\eta \tilde{L}_{{\rm L}\eta,i}(\mathbf{r},\omega).
\end{align}
While the second term vanishes within the full interaction Hamiltonian due to the sum rule Eq.~\eqref{eq: Sumrule_dOps}, the first term remains as a non-resonant term (the simple QNM pole is cancelled out), yielding $H_{\rm QNM-R}'=H_{\rm QNM-R}^{\prime \rm L}-H_{\rm QNM-R}^{\prime\rm \rm G}$, with
\begin{align}
    H_{\rm QNM-R}^{\prime \rm L}=\hbar\sum_{i,\mu}\int{\rm d}^3 r\int_0^\infty{\rm d}\omega  g_{{\rm L}\mu,i}^{\prime *}(\mathbf{r},\omega) c_i^\dagger(\mathbf{r},\omega)a_\mu' +{\rm H.a.},
\end{align}
and 
\begin{align}
    H_{\rm QNM-R}^{\prime \rm G}=\hbar\sum_{i,\mu}\int{\rm d}^3 r\int_0^\infty{\rm d}\omega  g_{{\rm G}\mu,i}^{\prime *}(\mathbf{r},\omega) c_i(\mathbf{r},\omega)a_\mu' +{\rm H.a.},
\end{align}
with coupling constants 
\begin{equation}
    g_{{\rm L(G)}\mu,i}^{\prime}(\mathbf{r},\omega)=\left[\mathbf{S}^{\prime-1/2}\right]_{\mu\eta}[\omega-\tilde{\omega}_\eta] \tilde{L}_{{\rm L(G)}\eta,i}(\mathbf{r},\omega).
\end{equation}

\subsection{Atom-field Hamiltonian in the unified gain-loss operator approach\label{App_Subsec: Atom-fieldHam}}
Next, we formulate the atom-field Hamiltonian in terms of a scattering part and a background part. For this, we use the terms in Eq.~\eqref{eq: Separation} to rewrite the electric field operator contribution $\mathbf{E}(\mathbf{r}_{\rm a},\omega)$, at the emitter position, as
\begin{equation}
    \mathbf{E}(\mathbf{r}_{\rm a},\omega)=\mathbf{E}^{\rm a}(\mathbf{r}_{\rm a},\omega)+\mathbf{E}^{\rm c}(\mathbf{r}_{\rm a},\omega),
\end{equation}
where 
\begin{equation}
    \mathbf{E}^{\rm a}(\mathbf{r}_{\rm a},\omega)=i\sum_\mu\sqrt{\frac{\hbar}{\pi\epsilon_0}}\mathbf{K}_\mu(\mathbf{r}_{\rm a},\omega) a_\mu',
\end{equation}
is the QNM related part, with 
\begin{align}
   \mathbf{K}_\mu&(\mathbf{r}_{\rm a},\omega)= \int{\rm d}^3r\Theta[\epsilon_I]\sqrt{\epsilon_I(\mathbf{r},\omega)}\mathbf{G}(\mathbf{r}_{\rm a},\mathbf{r},\omega)\cdot \mathbf{L}^{\prime*}_{{\rm L}\mu}(\mathbf{r},\omega)-\int{\rm d}^3r\Theta[-\epsilon_I]\sqrt{|\epsilon_I(\mathbf{r},\omega)|}\mathbf{G}(\mathbf{r}_{\rm a},\mathbf{r},\omega)\cdot \mathbf{L}^{\prime*}_{{\rm G}\mu}(\mathbf{r},\omega),
\end{align}
and 
\begin{align}
    \mathbf{E}^{\rm c}(\mathbf{r}_{\rm a},\omega)=i\sqrt{\frac{\hbar}{\pi\epsilon_0}}\int{\rm d}^3r\left[\Theta[\epsilon_I]\sqrt{\epsilon_I(\mathbf{r},\omega)}\mathbf{G}(\mathbf{r}_{\rm a},\mathbf{r},\omega)\cdot \mathbf{c}(\mathbf{r},\omega)+\Theta[-\epsilon_I]\sqrt{|\epsilon_I(\mathbf{r},\omega)|}\mathbf{G}(\mathbf{r}_{\rm a},\mathbf{r},\omega)\cdot \mathbf{c}^\dagger(\mathbf{r},\omega)\right].
\end{align}

In the following, we assume that the emitter is located in the background region, where $\epsilon = \epsilon_B$. 
Then, we use the fact that the Green function with both spatial positions in a common spatial region (with the same permittivity) can be written as a sum of a scattering part and a background part~\cite{buhmann2013dispersionII}, namely,
\begin{equation}
   \mathbf{G}(\mathbf{r}_{\rm a},\mathbf{r},\omega)=\mathbf{G}_{\rm S}(\mathbf{r}_{\rm a},\mathbf{r},\omega)+\mathbf{G}_{\rm B}(\mathbf{r}_{\rm a},\mathbf{r},\omega). 
\end{equation}
In the cases where $\mathbf{r}_{\rm a}$ and $\mathbf{r}$ are not in a common spatial region, then the full Green function is given by the scattering part $\mathbf{G}_{\rm S}(\mathbf{r}_{\rm a},\mathbf{r},\omega)$. We also assume that $\mathbf{G}_{\rm S}(\mathbf{r}_{\rm a},\mathbf{r},\omega)$ is fully determined by the QNM Green function together with a regularization, which is of course consistent with the mode operator construction. However, one should note, that one can also allow for additional scattering contributions, not allocated to the QNM expansion, which will result in other non-modal scattering parts~\cite{GeNJP2014}. 

Within this Green function formulation, we first investigate $\mathbf{E}^{\rm c}(\mathbf{r}_{\rm a},\omega)$. To do so, the contribution associated to $\Theta[\epsilon_I]$ must again be split into two parts, where we have to explicitly reintroduce the permittivity sequences:
\begin{equation}
    \Theta[\epsilon_I]=\Theta[\alpha\chi_I-\epsilon_I]+\Theta[\epsilon_I-\alpha\chi_I].
\end{equation}
Here, the first part accounts for the artificial background contribution, while the second part accounts for the lossy scattering structures. In the part, where $\Theta[\alpha\chi_I-\epsilon_I]$ (or analogously $V_{\rm out}(\lambda)$) appears, the sum of $\mathbf{G}_{\rm S}$ and $\mathbf{G}_{\rm B}$ must be taken into account, while in all other parts, $\mathbf{G}=\mathbf{G}_{\rm S}$ (under the limit $\alpha\rightarrow 0$ on the finite regime $V_{\rm in}-V_{\rm G}$, cf. Fig.~\ref{fig: SchematicGeometry}). Thus, we can rewrite $\mathbf{E}^{\rm c}(\mathbf{r}_{\rm a},\omega)$ as $\mathbf{E}^{\rm c}(\mathbf{r}_{\rm a},\omega)=\mathbf{E}^{\rm c,S}(\mathbf{r}_{\rm a},\omega)+\mathbf{E}^{\rm c,B}(\mathbf{r}_{\rm a},\omega)$, with
\begin{align}
    \mathbf{E}^{\rm c,S}(\mathbf{r}_{\rm a},\omega)=i\sqrt{\frac{\hbar}{\pi\epsilon_0}}\int{\rm d}^3r\left[\Theta[\epsilon_I]\sqrt{\epsilon_I(\mathbf{r},\omega)}\mathbf{G}_{\rm S}(\mathbf{r}_{\rm a},\mathbf{r},\omega)\cdot \mathbf{c}(\mathbf{r},\omega)+\Theta[-\epsilon_I]\sqrt{|\epsilon_I(\mathbf{r},\omega)|}\mathbf{G}_{\rm S}(\mathbf{r}_{\rm a},\mathbf{r},\omega)\cdot \mathbf{c}^\dagger(\mathbf{r},\omega)\right],
\end{align}
and 
\begin{align}
    \mathbf{E}^{\rm c,B}(\mathbf{r}_{\rm a},\omega)=i\sqrt{\frac{\hbar}{\pi\epsilon_0}}\int{\rm d}^3r&\Theta[\alpha\chi_I-\epsilon_I]\sqrt{\epsilon_I(\mathbf{r},\omega)}\mathbf{G}_{\rm B}(\mathbf{r}_{\rm a},\mathbf{r},\omega)\cdot \mathbf{c}(\mathbf{r},\omega).
\end{align}

Inserting the QNM expansion for $\mathbf{G}_{\rm S}(\mathbf{r}_{\rm a},\mathbf{r},\omega)$, and integrating $\mathbf{E}^{\rm c,S}(\mathbf{r}_{\rm a},\omega)$ over all $\omega$, yields $\int_0^\infty{\rm d}\omega\mathbf{E}^{\rm c,S}(\mathbf{r}_{\rm a},\omega)=i\sum_\mu\sqrt{\hbar\omega_\mu}\tilde{\mathbf{f}}_\mu^{\rm s}(\mathbf{r}_{\rm a})c_\mu/\sqrt{2\epsilon_0}$, with
\begin{align}
    c_\mu=&\sum_i\sum_{\mu}\int{\rm d}^3 r\int_0^\infty{\rm d}\omega L_{{\rm L}\mu,i}^{\prime}(\mathbf{r},\omega)c_i(\mathbf{r},\omega)
    +\sum_i\int{\rm d}^3 r\int_0^\infty{\rm d}\omega L_{{\rm G}\mu,i}^{\prime}(\mathbf{r},\omega)c_i^\dagger(\mathbf{r},\omega).
\end{align}
However, this is precisely the LHS of the sum rule, Eq.~\eqref{eq: Sumrule_dOps}, which implies $\int_0^\infty{\rm d}\omega\mathbf{E}^{\rm c,S}(\mathbf{r}_{\rm a},\omega)=0$. The remaining total electric field operator of the non-QNM part is thus identical to $\int_0^\infty{\rm d}\omega\mathbf{E}^{\rm c,B}(\mathbf{r}_{\rm a},\omega)$ and connected to the background contribution only.

Second, we inspect the part $\mathbf{E}^{\rm a}$: Applying the same spatial separation as above leads to $\mathbf{K}_\mu(\mathbf{r}_{\rm a},\omega)=\mathbf{K}_\mu^{\rm B}(\mathbf{r}_{\rm a},\omega)+\mathbf{K}_\mu^{\rm S}(\mathbf{r}_{\rm a},\omega)$. Inserting the QNM expansion for $\mathbf{G}_{\rm S}$, we then arrive at 
\begin{align}
    \mathbf{K}_\mu^{\rm S}(\mathbf{r}_{\rm a},\omega)
    =\sum_{\nu\nu'\eta\eta'}\sqrt{\frac{\pi\omega_\nu}{2}}\tilde{\mathbf{f}}_\nu(\mathbf{r}_{\rm a},) [\mathbf{S}^{1/2}]_{\nu\nu'}[\mathbf{S}^{-1/2}]_{\nu'\eta}S_{\eta\eta'}(\omega)[\mathbf{S}^{-1/2}]_{\eta'\mu},
\end{align}
which implies that the total scattered electric field operator is
\begin{equation}
    \int_0^\infty{\rm d}\omega\mathbf{E}^{\rm a,S}(\mathbf{r}_{\rm a},\omega)=i\sum_\mu\sqrt{\frac{\hbar\omega_\mu}{2\epsilon_0}}\tilde{\mathbf{f}}^{\prime\rm s}_\mu(\mathbf{r}_{\rm a}) a_\mu',
\end{equation}
with the exact symmetrized QNM function as in Eq.~\eqref{eq: SymmQNM_unified}.

Assuming that the background contribution of the QNM part is small, i.e., $|\mathbf{d}\cdot\mathbf{E}^{\rm a,S}(\mathbf{r}_{\rm a},\omega)|\gg |\mathbf{d}\cdot\mathbf{E}^{\rm a,B}(\mathbf{r}_{\rm a},\omega)|$, the term associated to $\mathbf{K}_\mu^{\rm B}(\mathbf{r}_{\rm a},\omega)$ can be neglected, and the atom-field interaction Hamiltonian can be written as $H_{\rm I} = H_{\rm QNM-a}' + H_{\rm a-R}$
with
\begin{equation}
    H_{\rm QNM-a}'=\hbar\sum_\mu[\tilde{g}_\mu^{\prime\rm s} \sigma^+ a_\mu' + {\rm H.a.}],~
    H_{\rm a-R}=\hbar\sum_i\int{\rm d}^3r\int_0^\infty{\rm d}\omega g_{{\rm a},i}(\mathbf{r},\omega)\sigma^+ c_i(\mathbf{r},\omega),
\end{equation}
where
\begin{equation}
    \mathbf{g}_{\rm a}(\mathbf{r},\omega)=-i\sqrt{\frac{\epsilon_I(\mathbf{r},\omega)}{\pi\hbar\epsilon_0}}\Theta[\alpha\chi_I-\epsilon_I]\mathbf{d}_{\rm a}\cdot\mathbf{G}_{\rm B}(\mathbf{r}_{\rm a},\mathbf{r},\omega),
\end{equation}
is the atom-reservoir coupling constant.

\subsection{Heisenberg equation of motion and Markov approximation in the unified gain-loss operator approach}
Having derived the full Hamiltonian after separating into QNM and non-QNM contributions, we now derive the Heisenberg equation of motion of an arbitrary system operator $A$ (in the QNM-atom space):
\begin{align}
    \dot{A}=&-\frac{i}{\hbar}[A,H_{\rm S}']-i\sum_{i,\mu}\int{\rm d}^3 r\int_0^\infty{\rm d}\omega  g_{{\rm L}\mu,i}^{\prime*}(\mathbf{r},\omega) c_i^\dagger(\mathbf{r},\omega)[A,a_\mu']
    -i\sum_{i,\mu}\int{\rm d}^3 r\int_0^\infty{\rm d}\omega  g_{{\rm L}\mu,i}^{\prime}(\mathbf{r},\omega) [A,a_\mu^{\prime\dagger}]c_i(\mathbf{r},\omega)\nonumber\\&+i\sum_{i,\mu}\int{\rm d}^3 r\int_0^\infty{\rm d}\omega  g_{{\rm G}\mu,i}^{\prime*}(\mathbf{r},\omega) [A,a_\mu'] c_i(\mathbf{r},\omega)+i\sum_{i,\mu}\int{\rm d}^3 r\int_0^\infty{\rm d}\omega  g_{{\rm G}\mu,i}^{\prime}(\mathbf{r},\omega) c_i^\dagger(\mathbf{r},\omega)[A,a_\mu^{\prime\dagger}]\nonumber\\
    &-i\sum_{i}\int{\rm d}^3 r\int_0^\infty{\rm d}\omega  g_{{\rm a},i}^{*}(\mathbf{r},\omega) c_i^\dagger(\mathbf{r},\omega)[A,\sigma^-]-i\sum_{i}\int{\rm d}^3 r\int_0^\infty{\rm d}\omega  g_{{\rm a},i}(\mathbf{r},\omega) [A,\sigma^+]c_i(\mathbf{r},\omega),\label{eq: HOM_SysOp}
\end{align}
where $H_{\rm S}'=H_{\rm a}+H_{\rm QNM-a}'+H_{\rm QNM}'$.
In order to obtain the Markovian quantum Langevin equation, Eq.~\eqref{eq: HOM_Mode+TLS_Unified}, we apply three approximations to the temporal evolution and coupling constants.

We initially approximate $c_i^{(\dagger)}(\mathbf{r},\omega)$ by bosonic operators, similar to Ref.~\onlinecite{franke2020quantized}, i.e.,
\begin{equation}
    [c_i(\mathbf{r},\omega),c_j^\dagger(\mathbf{r}',\omega')]\approx  \delta_{ij}\delta(\mathbf{r}-\mathbf{r}')\delta(\omega-\omega').
\end{equation}
In this way, we obtain the Heisenberg equation of motion of $c_i(\mathbf{r},\omega)$ as
\begin{align}
\dot{c}_i(\mathbf{r},\omega)=&-i\omega\, {\rm sgn}[\epsilon_I]c_i(\mathbf{r},\omega)-i\sum_{\mu}  g_{{\rm L}\mu,i}^{\prime*}(\mathbf{r},\omega) a_\mu'+i\sum_{\mu}g_{{\rm G}\mu,i}^{\prime}(\mathbf{r},\omega) a_\mu^{\prime\dagger}-i g_{{\rm a},i}^{*}(\mathbf{r},\omega) \sigma^-.\label{eq: HOM_dOps}
\end{align}
Formally solving Eq.~\eqref{eq: HOM_dOps}, leads to
\begin{align}
c_i(\mathbf{r},\omega,t)=&e^{-i\omega {\rm sgn}[\epsilon_I](t-t_0)}c_i(\mathbf{r},\omega,t_0)-i\sum_{\mu}  g_{{\rm L}\mu,i}^{\prime*}(\mathbf{r},\omega) \int_{t_0}^t {\rm d\tau}e^{-i\omega {\rm sgn}[\epsilon_I](t-\tau)}a_\mu'(\tau)\nonumber\\
&+i\sum_{\mu}g_{{\rm G}\mu,i}^{\prime}(\mathbf{r},\omega) \int_{t_0}^t {\rm d\tau}e^{-i\omega {\rm sgn}[\epsilon_I](t-\tau)}a_\mu^{\prime\dagger}(\tau)-i g_{{\rm a},i}^{*}(\mathbf{r},\omega) \int_{t_0}^t {\rm d\tau}e^{-i\omega {\rm sgn}[\epsilon_I](t-\tau)}\sigma^-(\tau),\label{eq: Sol_dOps}
\end{align}
with initial time $t_0<t$ (time-retarded solution), and inserting the solution back into Eq.~\eqref{eq: HOM_SysOp}, yields
\begin{align}
    \dot{A}=&-\frac{i}{\hbar}[A,H_{\rm S}]-\sum_\mu F_{{\rm L}\mu}^{\prime\dagger}[A,a_\mu']
    +\sum_\mu[A,a_\mu^{\prime\dagger}]F_{{\rm L}\mu}^{\prime}-\sum_\mu [A,a_\mu']F_{{\rm G}\mu}^{\prime}
    +\sum_\mu F_{{\rm L}\mu}^{\prime\dagger}[A,a_\mu^{\prime\dagger}]- F_{\rm a}^{\dagger}[A,\sigma^-]
    +[A,\sigma^+]F_{\rm a}\nonumber\\
    &-\frac{1}{\pi}\sum_{\mu,\eta}[A,a_\mu']\int_0^\infty{\rm d}\omega  g_{\mu\eta}^{\prime\rm G*}(\omega) \int_{t_0}^te^{i\omega(t-\tau)} a_\eta^{\prime\dagger}(\tau)+\frac{1}{\pi}\sum_{\mu,\eta}\int_0^\infty{\rm d}\omega  g_{\mu\eta}^{\prime\rm G}(\omega) \int_{t_0}^te^{-i\omega(t-\tau)} a_\eta'(\tau)[A,a_\mu^{\prime\dagger}]\nonumber\\
     &+\frac{1}{\pi}\sum_{\mu,\eta}\int_0^\infty{\rm d}\omega  g_{\mu\eta}^{\prime\rm L*}(\omega) \int_{t_0}^te^{i\omega(t-\tau)} a_\eta^{\prime\dagger}(\tau)[A,a_\mu']-\frac{1}{\pi}\sum_{\mu,\eta}[A,a_\mu^{\prime\dagger}]\int_0^\infty{\rm d}\omega  g_{\mu\eta}^{\prime\rm L}(\omega) \int_{t_0}^te^{-i\omega(t-\tau)} a_\eta'(\tau)\nonumber\\
    &+\frac{1}{\pi}[A,\sigma^-]\int_0^\infty{\rm d}\omega  g_{\rm a}^{*}(\omega) \int_{t_0}^te^{i\omega(t-\tau)} \sigma^+(\tau)-\frac{1}{\pi}\int_0^\infty{\rm d}\omega  g_{\rm a}(\omega) \int_{t_0}^te^{-i\omega(t-\tau)} \sigma^-(\tau)[A,\sigma^+],\label{eq: HOM_SysOp2}
\end{align}
with noise forces 
\begin{align}
    F_{{\rm L}\mu}^{\prime}(t)&=-i\sum_{i}\int{\rm d}^3 r\int_0^\infty{\rm d}\omega  g_{{\rm L}\mu,i}^{\prime}(\mathbf{r},\omega) c_i(\mathbf{r},\omega,t_0)e^{- i\omega(t-t_0)},\nonumber\\
     F_{{\rm G}\mu}^{\prime}(t)&=-i\sum_{i}\int{\rm d}^3 r\int_0^\infty{\rm d}\omega  g_{{\rm G}\mu,i}^{\prime *}(\mathbf{r},\omega) c_i(\mathbf{r},\omega,t_0)e^{ i\omega(t-t_0)}
\end{align}
and
\begin{align}
    F_{\rm a}(t)=-i\sum_{i}\int{\rm d}^3 r&\int_0^\infty{\rm d}\omega  g_{{\rm a},i}(\mathbf{r},\omega) c_i(\mathbf{r},\omega,t_0)e^{-i\omega(t-t_0)},
\end{align}
and with the $\omega$-dependent coupling constants
\begin{align}
    g_{\mu\eta}^{\prime\rm L}(\omega)&=\pi\sum_i\int{\rm d}^3rg_{{\rm L}\mu,i}^{\prime}(\mathbf{r},\omega)g_{{\rm L}\eta,i}^{\prime*}(\mathbf{r},\omega),\\
    g_{\mu\eta}^{\prime\rm G}(\omega)&=\pi\sum_i\int{\rm d}^3r g_{{\rm G}\mu,i}^{\prime}(\mathbf{r},\omega)g_{{\rm G}\eta,i}^{\prime*}(\mathbf{r},\omega),\\
     g_{\rm a}(\omega)&=\pi\sum_i\int{\rm d}^3rg_{{\rm a},i}(\mathbf{r},\omega)g_{{\rm a},i}^{*}(\mathbf{r},\omega).
\end{align}

Note, we have used the fact that any spatial integral involving the products with mixed gain and loss coupling constants, e.g., $\sum_i \int{\rm d}^3 r g_{{\rm L}\mu,i}^{\prime}(\mathbf{r},\omega)g_{{\rm G}\eta,i}^{\prime*}(\mathbf{r},\omega)$, vanish. Also note, we 
have 
neglected cross-terms associated with both the atom-reservoir and photon-reservoir coupling, and they are assumed in the following as independent bath contributions, i.e., $[F_{{\rm L(G)}\mu}^{\prime}(t),F_{\rm a}^\dagger(t')]=0$.

To obtain a time-local quantum Langevin equation, we apply a resonance approximation to the coupling constants and extend the lower $\omega$ integral bound to $-\infty$:
\begin{align}
    \dot{A}=&-\frac{i}{\hbar}[A,H_{\rm S}]-\sum_\mu F_{{\rm L}\mu}^{\prime\dagger}[A,a_\mu']
    +\sum_\mu[A,a_\mu^{\prime\dagger}]F_{{\rm L}\mu}^{\prime}-\sum_\mu [A,a_\mu']F_{{\rm G}\mu}^{\prime}
    +F_{{\rm G}\mu}^{\prime\dagger}\sum_\mu[A,a_\mu^{\prime\dagger}]- F_{\rm a}^{\dagger}[A,\sigma^-]
    +[A,\sigma^+]F_{\rm a}\nonumber\\
    &-\sum_{\mu,\eta}\chi_{\mu\eta}^{\prime\rm G-}\left([A,a_\eta']a_\mu^{\prime\dagger} + a_\eta'[A,a_\mu^{\prime\dagger}]\right)+\sum_{\mu,\eta}\chi_{\mu\eta}^{\prime\rm L-}\left(a_\mu^{\prime\dagger}[A,a_\eta']-[A,a_\mu^{\prime\dagger}]a_\eta'\right)+\frac{\Gamma^{\rm B}}{2}\left(\sigma^+[A,\sigma^-] -   [A,\sigma^+]\sigma^-\right).\label{eq: HOM_SysOp3}
\end{align}
In Eq.~\eqref{eq: HOM_SysOp3}, the 
averaged photon decay matrices are explicitly given by 
\begin{align}
    \chi_{\mu\eta}^{\prime\rm L-}&\approx \frac{i}{2}\sum_{\nu\nu'}\left[\mathbf{S}^{\prime-1/2}\right]_{\mu\nu}(\tilde{\omega}_\nu-\tilde{\omega}_{\nu'}^*)S_{\nu\nu'}^{\prime\rm L}\left[\mathbf{S}^{\prime-1/2}\right]_{\nu'\eta},\\
     \chi_{\mu\eta}^{\prime\rm G-}&\approx \frac{i}{2}\sum_{\nu\nu'}\left[\mathbf{S}^{\prime-1/2}\right]_{\mu\nu}(\tilde{\omega}_\nu-\tilde{\omega}_{\nu'}^*)S_{\nu'\nu}^{\prime\rm G}\left[\mathbf{S}^{\prime-1/2}\right]_{\nu'\eta},
\end{align}
where similar approximations as in Ref.~\onlinecite{franke2020quantized} were used. Similarly, the photon coupling matrix $\chi_{\mu\eta}^{\prime +}$ is approximated as
\begin{equation}
    \chi_{\mu\eta}^{\prime+}\approx \frac{1}{2}\sum_{\nu\nu'}\left[\mathbf{S}^{\prime-1/2}\right]_{\mu\nu}(\tilde{\omega}_\nu+\tilde{\omega}_{\nu'}^*)S_{\nu\nu'}^{\prime}\left[\mathbf{S}^{\prime-1/2}\right]_{\nu'\eta},
\end{equation}
which is in line with the resonance approximation from above.
Furthermore, the averaged TLS decay rate is defined as
\begin{align}
     \Gamma^{\rm B}&\approx \pi g_{\rm a}(\omega_{\rm a})=\frac{2}{\hbar\epsilon_0}\mathbf{d}_{\rm a}\cdot\mathbf{K}_{\rm B}(\mathbf{r}_{\rm a},\mathbf{r}_{\rm a},\omega_{\rm a})\cdot\mathbf{d}_{\rm a},
\end{align}
with
\begin{align}
     \mathbf{K}_{\rm B}(\mathbf{r}_{\rm a},\mathbf{r}_{\rm a},\omega_{\rm a})&=\int{\rm d}^3r\epsilon_I(\mathbf{r},\omega_{\rm a})\Theta[\alpha\chi_I-\epsilon_I]\mathbf{G}_{\rm B}(\mathbf{r}_{\rm a},\mathbf{r},\omega_{\rm a})\cdot\mathbf{G}_{\rm B}^*(\mathbf{r},\mathbf{r}_{\rm a},\omega_{\rm a})\nonumber\\
     &=\lim_{\lambda\rightarrow\infty}\int_{V_{\rm out}(\lambda)}{\rm d}^3r\epsilon_{{\rm B},I}^{(\alpha)}(\mathbf{r},\omega_{\rm a})\mathbf{G}_{\rm B}(\mathbf{r}_{\rm a},\mathbf{r},\omega_{\rm a})\cdot\mathbf{G}_{\rm B}^*(\mathbf{r},\mathbf{r}_{\rm a},\omega_{\rm a}),
\end{align}
where we used the fact that the Heaviside function is defined on $V_{\rm out}(\lambda)$ and $\epsilon_{\rm B}^{(\alpha)}=1+\alpha\chi_{\rm L}$. In the limit of $\alpha\rightarrow 0$, one can add the trivial term $\int_{B}{\rm d}^3r\epsilon_{{\rm B},I}^{(\alpha)}\mathbf{G}_{\rm B}(\mathbf{r}_{\rm a},\mathbf{r})\cdot\mathbf{G}_{\rm B}(\mathbf{r},\mathbf{r}_{\rm a})$ on a bound volume $B$. Choosing $B=V_{\rm in}$ simply leads to 
\begin{equation}
    \lim_{\alpha\rightarrow 0}\mathbf{K}_{\rm B}(\mathbf{r}_{\rm a},\mathbf{r}_{\rm a},\omega_{\rm a})={\rm Im}[\mathbf{G}_{\rm B}(\mathbf{r}_{\rm a},\mathbf{r}_{\rm a},\omega_{\rm a})],
\end{equation}
so that 
\begin{equation}
    \Gamma^{\rm B}=\frac{2}{\hbar\epsilon_0}\mathbf{d}_{\rm a}\cdot{\rm Im}[\mathbf{G}_{\rm B}(\mathbf{r}_{\rm a},\mathbf{r}_{\rm a},\omega_{\rm a})]\cdot\mathbf{d}_{\rm a},
\end{equation}
which readily coincides with the usual formula for free space (or a homogeneous medium) decay of a quantum dipole.

To complete the Markovian description of the quantum Langevin equation, we  assume the following approximate relations for the non-equal time commutation relations of the noise forces,
\begin{align}
    [F_{{\rm L}\mu}^{\prime}(t),F_{{\rm L}\mu'}^{\prime\dagger}(t')]&\approx 2\chi_{\mu\mu'}^{\prime\rm L-}\delta(t-t'),\\
    [F_{{\rm G}\mu}^{\prime}(t),F_{{\rm G}\mu'}^{\prime\dagger}(t')]&\approx 2\chi_{\mu'\mu}^{\rm \prime G-}\delta(t-t'),\\
    [F_{\rm a}(t),F_{\rm a}^{\dagger}(t')]&\approx\Gamma^{\rm B}\delta(t-t'),
\end{align}
which is consistent with the resonance approximations applied to the decay rates. 

Inserting $A=a_\mu'$ into Eq.~\eqref{eq: HOM_SysOp3} 
coincides with Eq.~\eqref{eq: a_prime_HOM}, which completes this derivation.

\subsection{Comments on the derivation for the separated gain-loss operator approach}
As mentioned at the beginning of this appendix, the behavior of the noise operators acting on the QNM system in the separated gain-loss operator approach is formally identical to the purely lossy case described in Subsection~\ref{subsec: ExtensionLossyQuantization}, except that the number of photon operators per QNMs is doubled. Thus, the derivation of the photon Hamiltonian as well as the Markovian limit would carry over from the derivation in Ref.~\onlinecite{franke2020quantized}, leading to the quantum Langevin equations~\eqref{eq: HOM_Mode+TLS}. 
The crucial difference over the purely lossy case is rather captured by the system Hamiltonian which has a significantly different structure due to the altered medium-assisted electric field operator and the sign function in $H_{\rm em}$. 

This is in contrast  to the unified gain-loss operator approach, where the structure of the noise operators in the Heisenberg equation of motion is completely different, but where the formal appearance of the system Hamiltonian is nearly identical to the purely lossy case (except a negative constant in the photon Hamiltonian). 

Furthermore, we note that the separation of QNM and non-QNM contributions in the atom-field Hamiltonian, as demonstrated in Subsection~\ref{App_Subsec: Atom-fieldHam}, can be easily adapted to the separated gain-loss operator approach, yielding the same free spontaneous emission rate for the background part, induced by the same artificial noise operator in the background medium.

\section{Bath state assumptions and dielectric permittivity model for the derivation of the master equations\label{app: Bath_assumptions}}

In this appendix, we give further details on the assumed vacuum correlation functions of the QNM noise forces, from which the quantum Langevin equation can be connected to the associated master equations (Eq.~\eqref{eq: ME_unified} within the unified gain-loss operator approach,  and Eq.~\eqref{eq: ME_separated} within the separated gain-loss operator approach) through quantum Ito-Stratonovich calculus (cf. Ref.~\onlinecite{PhysRevLett.122.213901} for details with respect to the purely lossy case, which is based on work from Gardiner~\cite{Gardiner1}).

To determine a well-defined bath state for the case with amplifying media, we assume a simple Lorentz oscillator model for dielectric permittivity in the non-vacuum media similar to Ref.~\onlinecite{franke2021fermi} (following the idea in Ref.~\onlinecite{PhysRevA.55.1623}):
\begin{equation}
    \epsilon(\mathbf{r},\omega)=1-\frac{N_l(\mathbf{r})-N_u(\mathbf{r})}{N_l(\mathbf{r})+N_u(\mathbf{r})}\,\frac{\omega_{\rm p}^2}{\omega_0^2-\omega^2-i\gamma_{0}\omega},\label{eq: OscillatorModel1}
\end{equation}
where $N_l(\mathbf{r})$ is the occupation of the lower state and $N_u(\mathbf{r})$ is the occupation of the upper state of the oscillator.
This model is also
consistent with how one would typically
describe gain in Maxwell's equations through the gain medium material properties.

For the purely lossy region, where $N_l>N_u$ for all $\omega$, one can determine the bath correlation function (assuming a thermal state) as 
\begin{align}
    {\rm tr}_{\rm R,loss}&[c_i^{\dagger}(\mathbf{r},\omega)c_j(\mathbf{r}',\omega')\rho_{\rm R}^{\rm loss}]=n(\omega,T)\delta(\mathbf{r}-\mathbf{r}')\delta_{ij}\delta(\omega-\omega'), \label{eq: LossCorr}
\end{align}
where $\mathbf{r},\mathbf{r}'\in V_{\rm L}$, $ {\rm tr}_{\rm R,loss}$ is the trace over the lossy medium degrees of freedom, and one can formulate $n(\omega,T)$ in equivalent forms~\cite{PhysRevA.55.1623,PhysRevLett.110.153602}
\begin{equation}
    n(\omega,T)=\frac{1}{\exp(\hbar\omega/k_B T)-1}=\frac{N_u}{N_l-N_u}.\label{eq: numberThermal}
\end{equation}

For the purely amplifying region, where $N_l<N_u$ for all $\omega$, we assume an associated thermal state with effective negative temperatures~\cite{PhysRevA.55.1623,PhysRevLett.110.153602}, and replace $n(\omega,T)\rightarrow n(\omega,|T|)$ (as in more detail justified in Ref.~\onlinecite{franke2021fermi}), to obtain
\begin{align}
    {\rm tr}_{\rm R,amp}&[c_i^{\dagger}(\mathbf{r},\omega)c_j(\mathbf{r}',\omega')\rho_{\rm R}^{\rm amp}]=n(\omega, |T|)\delta(\mathbf{r}-\mathbf{r}')\delta_{ij}\delta(\omega-\omega')\label{eq: AmpCorr},
\end{align}
for $\mathbf{r},\mathbf{r}'\in V_{\rm G}$. 
Using  $n(\omega,|T|)=-n(\omega,T)-1$ and Eq.~\eqref{eq: numberThermal} yields
\begin{equation}
    n(\omega,|T|)=\frac{N_l}{N_u-N_l}.
\end{equation}
Physically, this corresponds to an inverted oscillator, where the lower and upper states are exchanged.
The construction of thermal states for negative temperatures was also discussed in Refs.~\onlinecite{PhysRevE.95.012125,PhysRevE.90.062116} using complementary Gibbs states. 
Note, that $\rho_{\rm R} = \rho_{\rm R}^{\rm amp}\rho_{\rm R}^{\rm loss}$, since both subspaces are independent from each other without any interactions. Needless to say, that one also obtains the trivial expectation values 
\begin{equation}
    \langle c_i(\mathbf{r},\omega)\rangle = \langle c_i^\dagger(\mathbf{r},\omega)\rangle =0,
\end{equation}
in lossy as well as amplifying regions. Therefore, one can summarize Eq.~\eqref{eq: LossCorr}
 and Eq.~\eqref{eq: AmpCorr} as follows:
\begin{align}
    {\rm tr}_{\rm R}[c_i^{\dagger}(\mathbf{r},\omega)c_j(\mathbf{r}',\omega')\rho_{\rm R}]=&\{\Theta[\epsilon_I]n(\omega, T)+\Theta[-\epsilon_I]n(\omega, |T|)\}\delta(\mathbf{r}-\mathbf{r}')\delta_{ij}\delta(\omega-\omega')\label{eq: TotalCorr},
\end{align}
 because of the linearity of the trace and property of the partial trace.
 If we now assume $N_l(\mathbf{r})=0$ (corresponding to a perfectly inverted oscillator) for $\mathbf{r}\in V_{\rm G}$ and $N_u(\mathbf{r})=0$ (corresponding to a ground-state oscillator) for $\mathbf{r}\in V_{\rm L}$ then yields $n(\omega,T)=0$, and $n(\omega,|T|)=0$, which leads to effective vacuum correlation functions, 
 \begin{subequations}\label{eq: VacCorr}
 \begin{align}
    {\rm tr}_{\rm R}[c_i^{\dagger}(\mathbf{r},\omega)c_j(\mathbf{r}',\omega')\rho_{\rm R}]&=0,~{\rm tr}_{\rm R}[c_i(\mathbf{r},\omega)c_j^{\dagger}(\mathbf{r}',\omega')\rho_{\rm R}]=\delta(\mathbf{r}-\mathbf{r}')\delta_{ij}\delta(\omega-\omega').
\end{align}
\end{subequations}
Since the QNM noise operators are linear combinations of $c_i^{(\dagger)}(\mathbf{r},\omega)$, the above result carries over to the specific QNM input sources, and equally justifies the correlation functions in Eq.~\eqref{eq: Input_corr_unified} and Eq.~\eqref{eq: Input_corr_separated}, which are the starting point to apply the procedure from Ref.~\onlinecite{Gardiner1} for deriving the master equations, Eq.~\eqref{eq: ME_unified}  and Eq.~\eqref{eq: ME_separated}, respectively.  

Strictly speaking, we only have to use the oscillator model, Eq.~\eqref{eq: OscillatorModel1}, for the amplifying part. In this way, one is not limited to the same Lorentzian in the lossy as well as the amplifying part, namely 
\begin{equation}
    \epsilon(\mathbf{r},\omega)=1-\chi_{V_{\rm L}}(\mathbf{r})\frac{N_l^{\rm loss}-N_u^{\rm loss}}{N_l^{\rm loss}+N_u^{\rm loss}}\,\frac{\omega_{\rm p,L}^2}{\omega_{0,\rm L}^2-\omega^2-i\gamma_{0,\rm L}\omega}-\chi_{V_{\rm G}}(\mathbf{r})\frac{N_l^{\rm gain}-N_u^{\rm gain}}{N_l^{\rm gain}+N_u^{\rm gain}}\,\frac{\omega_{\rm p,G}^2}{\omega_{0,\rm G}^2-\omega^2-i\gamma_{0,\rm G}\omega},
\end{equation}
with two different Lorentz oscillators and $N^{\rm loss}_l>N^{\rm loss}_u$ ($N^{\rm gain}_l<N^{\rm gain}_u$) for the lossy and amplifying regions ($\chi_{V_{\rm L(G)}}(\mathbf{r})$ was defined earlier). In this case,
\begin{equation}
    n(\omega,T)=\frac{1}{\exp(\hbar\omega/k_B T)-1}=\frac{N_u^{\rm loss}}{N_l^{\rm loss}-N_u^{\rm loss}}.
\end{equation}
for the lossy region and 
\begin{equation}
    n(\omega,|T|)=\frac{N_l^{\rm gain}}{N_u^{\rm gain}-N_l^{\rm gain}}.
\end{equation}
for the amplifying region.

\section{Bad cavity limit derivation using the separated gain-loss operator approach\label{app: MarkovBad}}
In this appendix, we present a detailed derivation of the bad cavity limit master equation, Eq.~\eqref{eq: BadCavMaster} within the separated gain-loss operator approach (the application to the unified gain-loss operator approach is also possible, as discussed at the end of this Appendix). The starting point for deriving the bad cavity limit is the full quantum master equation from the separated operator approach (Eq.~\eqref{eq: ME_separated}), slightly rewritten as
\begin{equation}
    \partial_t\rho = [\mathcal{L}_{\rm a}+\mathcal{L}_{\rm em}+\mathcal{L}_{I}]\rho,
\end{equation}
with superoperators 
\begin{align}
   \mathcal{L}_{\rm a}\rho &= -\frac{i}{\hbar}[H_{\rm a},\rho]+\mathcal{L}_{\rm SE}\rho, \\
   \mathcal{L}_{\rm em}\rho &= -\frac{i}{\hbar}[H_{\rm em},\rho]+\mathcal{L}[\mathbf{a}_{\rm L}]\rho+\mathcal{L}[\mathbf{a}_{\rm G}]\rho,\\
   \mathcal{L}_{\rm I}\rho &= -\frac{i}{\hbar}[H_{\rm I},\rho],
\end{align}
and where $H_{\rm em}\equiv H_{\rm QNM}$ and $H_{\rm I}\equiv H_{\rm QNM-a}$. Following Ref.~\onlinecite{Cirac}, we transform to the electromagnetic picture via 
\begin{equation}
    \tilde{\rho}\equiv e^{-\mathcal{L}_{\rm em}t}\rho,
\end{equation}
such that 
\begin{equation}
    \partial_t\tilde{\rho}=\mathcal{L}_a\tilde{\rho} + e^{-\mathcal{L}_{\rm em}t}\mathcal{L}_{I}\rho.
\end{equation}

Next, we split the electromagnetic operators into a gain and a loss part,
\begin{align}
    \mathcal{L}_{I}&=\mathcal{L}_{I}^{\rm gain}+\mathcal{L}_{I}^{\rm loss},\\
    \mathcal{L}_{I}^{\rm gain}\rho&=-i\sum_\mu\left\{ \tilde{g}_\mu^{\mathrm{s,G}*}(a_{\rm G\mu}[\sigma^-,\rho] + [a_{\rm G\mu},\rho]\sigma^-)+\tilde{g}_\mu^{\mathrm{s,G}}([\sigma^+,\rho]a_{\rm G\mu}^\dagger + \sigma^+[a_{\rm G\mu}^\dagger,\rho]) \right\},\label{eq:LIgain}\\
    \mathcal{L}_{I}^{\rm loss}\rho&=-i\sum_\mu\left\{ \tilde{g}_\mu^{\mathrm{s,L}*}([\sigma^-,\rho]a_{\rm L\mu}^\dagger + \sigma^-[a_{\rm L\mu}^\dagger,\rho])+\tilde{g}_\mu^{\mathrm{s,L}}(a_{\rm L\mu}[\sigma^+,\rho] + [a_{\rm L\mu},\rho]\sigma^+) \right\} ,
\end{align}
and 
\begin{align}
    \mathcal{L}_{\rm em}&=\mathcal{L}_{\rm em}^{\rm gain}+\mathcal{L}_{\rm em}^{\rm loss},\\
    \mathcal{L}_{\rm em}^{\rm gain}\rho &=-\frac{i}{\hbar}[H_{\rm em}^{\rm gain},\rho]+\mathcal{L}[\mathbf{a}_{\rm G}]\rho,\\
    \mathcal{L}_{\rm em}^{\rm loss}\rho &=-\frac{i}{\hbar}[H_{\rm em}^{\rm loss},\rho]+\mathcal{L}[\mathbf{a}_{\rm L}]\rho.
\end{align}

In this way, we obtain 
\begin{equation}
    \partial_t\tilde{\rho}=\mathcal{L}_a\tilde{\rho} + e^{-\mathcal{L}_{\rm em}^{\rm gain}t}\mathcal{L}_{I}^{\rm gain}\rho^{\rm loss}+e^{-\mathcal{L}_{\rm em}^{\rm loss}t}\mathcal{L}_{I}^{\rm loss}\rho^{\rm gain},
\end{equation}
where we used $\mathcal{L}_{\rm em }^{\rm loss}\mathcal{L}_{\rm em }^{\rm gain}\rho=\mathcal{L}_{\rm em }^{\rm gain}\mathcal{L}_{\rm em }^{\rm loss}\rho$ and where 
\begin{equation}
    \rho^{\rm gain/loss}=e^{-i\mathcal{L}_{\rm em}^{\rm gain/loss}t}\rho.
\end{equation}

To obtain a closed form master equation for $\tilde{\rho}$ we use properties of the operator exponential together with the commutation relations of $a_{\rm G\mu}^{(\dagger)}$, $a_{\rm L\mu}^{(\dagger)}$, yielding
\begin{equation}
    \partial_t\tilde{\rho}=\mathcal{L}_a\tilde{\rho} +\tilde{\mathcal{L}}_{I}^{\rm gain}\tilde{\rho}+\tilde{\mathcal{L}}_{I}^{\rm loss}\tilde{\rho},\label{eq: masterrhotilde}
\end{equation}
where 
\begin{align}
\tilde{\mathcal{L}}_{I}^{\rm loss}\tilde{\rho}=&-i\sum_{\eta}e^{-\gamma_\eta t}\Bigg[\sum_{\mu,\nu}\Big( \tilde{g}_\mu^{\rm s,L*}[\sigma^-,\tilde{\rho}]a_{\rm L\nu}^\dagger\left(\mathbf{S}^{\rm L\frac{1}{2}}\right)_{\nu\eta} e^{i\omega_\eta t}\left(\mathbf{S}^{\rm L-\frac{1}{2}}\right)_{\eta\mu} \nonumber\\
&\hspace{95pt}+\tilde{g}_\mu^{\rm s,L}\left(\mathbf{S}^{\rm L-\frac{1}{2}}\right)_{\mu\eta} e^{-i\omega_\eta t}\left(\mathbf{S}^{\rm L\frac{1}{2}}\right)_{\eta\nu}a_{\rm L\nu}[\sigma^+,\tilde{\rho}]\Big)\Bigg]\nonumber\\
&-i\sum_{\eta} e^{\gamma_\eta t}\Bigg[\sum_{\mu,\nu}\Big(\tilde{g}_\mu^{\rm s,L}\left(\mathbf{S}^{\rm L\frac{1}{2}}\right)_{\mu\eta} e^{-i\omega_\eta t}\left(\mathbf{S}^{\rm L-\frac{1}{2}}\right)_{\eta\nu}\left[a_{\rm L\nu},\tilde{\rho}\right]\sigma^+ \nonumber\\
&\hspace{95pt}+\tilde{g}_\mu^{\rm s,L*}\sigma^-\left[a_{\rm L\nu}^\dagger,\tilde{\rho}\right]\left(\mathbf{S}^{\rm L-\frac{1}{2}}\right)_{\nu\eta} e^{i\omega_\eta t}\left(\mathbf{S}^{\rm L\frac{1}{2}}\right)_{\eta\mu}\Big)\Bigg]\nonumber\\
\equiv&\sum_\eta e^{-\gamma_\eta t}\left(L_1^{\rm loss,\eta}(t)\tilde{\rho}(t)\right)+\sum_\eta e^{\gamma_\eta t}\left(L_2^{\rm loss,\eta}(t)\tilde{\rho}(t)\right).\label{eq: LItildRhotildeLOSS}
\end{align}

For the gain part, we instead get
\begin{align}
\tilde{\mathcal{L}}_{I}^{\rm gain}\tilde{\rho}=&-i\sum_{\eta}e^{-\gamma_\eta t}\Bigg[\sum_{\mu,\nu}\Big( \tilde{g}_\mu^{\rm s,G}[\sigma^+,\tilde{\rho}]a_{\rm G\nu}^\dagger\left(\mathbf{S}^{\rm G\frac{1}{2}}\right)_{\nu\eta} e^{-i\omega_\eta t}\left(\mathbf{S}^{\rm G-\frac{1}{2}}\right)_{\eta\mu} \nonumber\\
&\hspace{95pt}+\tilde{g}_\mu^{\rm s,G*}\left(\mathbf{S}^{\rm G-\frac{1}{2}}\right)_{\mu\eta} e^{i\omega_\eta t}\left(\mathbf{S}^{\rm G\frac{1}{2}}\right)_{\eta\nu}a_{\rm G\nu}[\sigma^-,\tilde{\rho}]\Big)\Bigg]\nonumber\\
&-i\sum_{\eta} e^{\gamma_\eta t}\Bigg[\sum_{\mu,\nu}\Big(\tilde{g}_\mu^{\rm s,G*}\left(\mathbf{S}^{\rm G\frac{1}{2}}\right)_{\mu\eta} e^{i\omega_\eta t}\left(\mathbf{S}^{\rm G-\frac{1}{2}}\right)_{\eta\nu}\left[a_{\rm G\nu},\tilde{\rho}\right]\sigma^- \nonumber\\
&\hspace{95pt}+\tilde{g}_\mu^{\rm s,G}\sigma^+\left[a_{\rm G\nu}^\dagger,\tilde{\rho}\right]\left(\mathbf{S}^{\rm G-\frac{1}{2}}\right)_{\nu\eta} e^{-i\omega_\eta t}\left(\mathbf{S}^{\rm G\frac{1}{2}}\right)_{\eta\mu}\Big)\Bigg]\nonumber\\
\equiv&\sum_\eta e^{-\gamma_\eta t}\left(L_1^{\rm gain,\eta}(t)\tilde{\rho}(t)\right)+\sum_\eta e^{\gamma_\eta t}\left(L_2^{\rm gain,\eta}(t)\tilde{\rho}(t)\right).\label{eq: LItildRhotildeGAIN}
\end{align}

Next, the density matrix $\rho_\mathrm{a}\equiv \text{tr}_{\text{em}}(\tilde{\rho})$ of the TLS degrees of freedom only is defined, where $\text{tr}_{\text{em}}=\text{tr}_{\text{em}}^{\rm loss}\text{tr}_{\text{em}}^{\rm gain}$ is the trace over the loss and gain QNM degrees of freedom. 
Applying the trace $\text{tr}_{\text{em}}$ on Eq.~\eqref{eq: masterrhotilde} then yields
\begin{equation}
\partial_t\rho_\mathrm{a}(t)=-\frac{i}{\hbar}[H_{\mathrm{a}},\rho_\mathrm{a}(t)]+\mathcal{L}_\mathrm{SE}\rho_\mathrm{a}+\sum_\eta\text{tr}_{\text{em}}\left(L_1^{\rm loss,\eta}(t)e^{-\gamma_\eta t}\tilde{\rho}(t)\right)+\sum_\eta\text{tr}_{\text{em}}\left(L_1^{\rm gain,\eta}(t)e^{-\gamma_\eta t}\tilde{\rho}(t)\right),\label{masterrhoAtom}
\end{equation}
where the relation $\text{tr}_{\text{em}}\left(L_2^{\rm loss/gain,\nu}(t)\tilde{\rho}(t)\right)=0$ was used, which follows from 
\begin{equation}
\text{tr}_{\text{em}}\left([a_{\rm L\eta}^\dagger,\tilde{\rho}]\right)=\text{tr}_{\text{em}}\left(a_{\rm L\eta}^\dagger\rho-\rho a_{\rm L\eta}^\dagger\right)=\text{tr}_{\text{em}}\left(a_{\rm L\eta}^\dagger\rho-a_{\rm L\eta}^\dagger\rho \right)=0,
\end{equation}
with the help of properties of the trace. Obviously, this holds also true for  $\text{tr}_{\text{em}}[a_{\rm L\eta}^\dagger,\tilde{\rho}]$, $\text{tr}_{\text{em}}[a_{\rm G\eta}^\dagger,\tilde{\rho}]$ and $\text{tr}_{\text{em}}[a_{\rm G\eta},\tilde{\rho}]$. 
To obtain a closed form master equation for the TLS density matrix $\rho_a$, Eq.~\eqref{eq: masterrhotilde} is formally solved for $\tilde{\rho}(t)$: 
\begin{align}
e^{-\gamma_\eta t}\tilde{\rho}(t)&= e^{-\gamma_\eta t}e^{-\frac{i}{\hbar}H_{\mathrm{a}} (t-t_0)}\tilde{\rho}(t_0)e^{\frac{i}{\hbar}H_{\mathrm{a}} (t-t_0)}+e^{-\gamma_\eta t}\mathcal{L}_\mathrm{SE}\tilde{\rho}(t)\nonumber\\
&+e^{-\gamma_\eta t}\int_{t_0}^{t}\mathrm{d}\tau\sum_{\eta '}e^{-\gamma_{\eta '}\tau} e^{-\frac{i}{\hbar}H_{\mathrm{a}} (t-\tau)}     \left(L_1^{\rm loss,\eta '}(\tau)\tilde{\rho}(\tau)\right)        e^{\frac{i}{\hbar}H_\mathrm{a} (t-\tau)}\nonumber\\
&+e^{-\gamma_\eta t}\int_{t_0}^{t}\mathrm{d}\tau\sum_{\eta '}e^{\gamma_{\eta '}\tau} e^{-\frac{i}{\hbar}H_{\mathrm{a}} (t-\tau)}     \left(L_2^{\rm loss,\eta '}(\tau)\tilde{\rho}(\tau)\right)        e^{\frac{i}{\hbar}H_{\mathrm{a}} (t-\tau)}\nonumber\\
&+e^{-\gamma_\eta t}\int_{t_0}^{t}\mathrm{d}\tau\sum_{\eta '}e^{-\gamma_{\eta '}\tau} e^{-\frac{i}{\hbar}H_{\mathrm{a}} (t-\tau)}     \left(L_1^{\rm gain,\eta '}(\tau)\tilde{\rho}(\tau)\right)        e^{\frac{i}{\hbar}H_\mathrm{a} (t-\tau)}\nonumber\\
&+e^{-\gamma_\eta t}\int_{t_0}^{t}\mathrm{d}\tau\sum_{\eta '}e^{\gamma_{\eta '}\tau} e^{-\frac{i}{\hbar}H_{\mathrm{a}} (t-\tau)}     \left(L_2^{\rm gain,\eta '}(\tau)\tilde{\rho}(\tau)\right)        e^{\frac{i}{\hbar}H_{\mathrm{a}} (t-\tau)}.\label{eq: FormSolrhoa}
\end{align}
Following the argumentation presented in Ref.~\cite{Cirac}, the terms on the first, second as well as fourth line on the RHS of Eq.~\eqref{eq: FormSolrhoa} are neglected as a consequence of the bad-cavity limit parameter regime, $\gamma_\eta\gg |\tilde{g}_\eta^{\rm L/G, s}|$ and $\gamma_\eta\gg \Gamma^{\rm B}$, which implies that the terms in the third and fifth line on the RHS of Eq.~\eqref{eq: FormSolrhoa} are the dominant contributions for times $t\geq \gamma_\eta^{-1}$.

Without loss of generality, in the following, the initial time $t_0$ is set to $t_0=0$. Inserting then the dominant term of Eq.~\eqref{eq: FormSolrhoa} back into Eq.~\eqref{masterrhoAtom} yields
\begin{align}
\partial_t\rho_{\rm a}(t)=&-\frac{i}{\hbar}[H_\mathrm{a},\rho_\mathrm{a}(t)]+\mathcal{L}_\mathrm{SE}\rho_\mathrm{a}(t)\nonumber\\
&+\sum_\nu\text{tr}_{\text{em}}\left(e^{-\gamma_\eta t}L_1^{\rm loss,\eta}(t)\int_{0}^{t}{\mathrm d}\tau\sum_{\eta '}e^{\gamma_{\eta '}\tau}    \left(\bar{L}_2^{\rm loss,\eta '}(\tau)\tilde{\rho}'(\tau)\right)     \right)\nonumber\\
&+\sum_\nu\text{tr}_{\text{em}}\left(e^{-\gamma_\eta t}L_1^{\rm loss,\eta}(t)\int_{0}^{t}{\mathrm d}\tau\sum_{\eta '}e^{\gamma_{\eta '}\tau}    \left(\bar{L}_2^{\rm gain,\eta '}(\tau)\tilde{\rho}'(\tau)\right)     \right)\nonumber\\
&+\sum_\nu\text{tr}_{\text{em}}\left(e^{-\gamma_\eta t}L_1^{\rm gain,\eta}(t)\int_{0}^{t}{\mathrm d}\tau\sum_{\eta '}e^{\gamma_{\eta '}\tau}    \left(\bar{L}_2^{\rm gain,\eta '}(\tau)\tilde{\rho}'(\tau)\right)     \right)\nonumber\\
&+\sum_\nu\text{tr}_{\text{em}}\left(e^{-\gamma_\eta t}L_1^{\rm gain,\eta}(t)\int_{0}^{t}{\mathrm d}\tau\sum_{\eta '}e^{\gamma_{\eta '}\tau}    \left(\bar{L}_2^{\rm loss,\eta '}(\tau)\tilde{\rho}'(\tau)\right)     \right)
,\label{masterrhoAtom2}
\end{align}
where 
\begin{align}
\tilde{\rho}'(\tau)=e^{-\frac{i}{\hbar}H_\mathrm{a} (t-\tau)} \tilde{\rho}(\tau)  e^{\frac{i}{\hbar}H_\mathrm{a} (t-\tau) },~  \bar{L}_2^{\rm loss/gain,\eta '}(\tau)=e^{-\frac{i}{\hbar}H_\mathrm{a} (t-\tau)} L_2^{\rm loss/gain,\eta '}(\tau) e^{\frac{i}{\hbar}H_\mathrm{a} (t-\tau) }.
\end{align}

Using Eqs.~\eqref{eq: LItildRhotildeLOSS}, \eqref{eq: LItildRhotildeGAIN} together with the Baker-Campell-Hausdorff formula~\cite{hall2015lie}, $\bar{L}_2^{\rm loss/gain,\eta '}(\tau)\tilde{\rho}'(\tau)$ is rewritten as
\begin{align}
\bar{L}_2^{\rm loss,\eta '}(\tau)\tilde{\rho}'(\tau)=-i\sum_{\mu,\nu}\Big(&\tilde{g}_\mu^{\rm L,s}\left(\mathbf{S}^{\rm L\frac{1}{2}}\right)_{\mu\eta'} e^{-i\omega_{\eta'} \tau}\left(\mathbf{S}^{\rm L-\frac{1}{2}}\right)_{\eta'\nu}\left[a_{\rm L\nu},\tilde{\rho}'\right]\sigma^+ e^{-i\omega_\mathrm{a}(t-\tau)}\nonumber\\
&+\tilde{g}_\mu^{\rm L, s*}\sigma^-e^{i\omega_\mathrm{a}(t-\tau)}\left[a_{\rm L\nu}^\dagger,\tilde{\rho}'\right]\left(\mathbf{S}^{\rm L-\frac{1}{2}}\right)_{\nu\eta'} e^{i\omega_{\eta'} \tau}\left(\mathbf{S}^{\rm L\frac{1}{2}}\right)_{\eta'\mu}\Big),
\end{align}
and 
\begin{align}
\bar{L}_2^{\rm gain,\eta '}(\tau)\tilde{\rho}'(\tau)=-i\sum_{\mu,\nu}\Big(&\tilde{g}_\mu^{\rm G,s*}\left(\mathbf{S}^{\rm G\frac{1}{2}}\right)_{\mu\eta'} e^{i\omega_{\eta'} \tau}\left(\mathbf{S}^{\rm G-\frac{1}{2}}\right)_{\eta'\nu}\left[a_{\rm G\nu},\tilde{\rho}'\right]\sigma^- e^{i\omega_\mathrm{a}(t-\tau)}\nonumber\\
&+\tilde{g}_\mu^{\rm G,s}\sigma^+e^{-i\omega_\mathrm{a}(t-\tau)}\left[a_{\rm G\nu}^\dagger,\tilde{\rho}'\right]\left(\mathbf{S}^{\rm G-\frac{1}{2}}\right)_{\nu\eta'} e^{-i\omega_{\eta'} \tau}\left(\mathbf{S}^{\rm G\frac{1}{2}}\right)_{\eta'\mu}\Big).
\end{align}
Next, a new time variable $\tau\rightarrow t-\tau$ of the time integral in Eq.~\eqref{masterrhoAtom2} is introduced, which leads to 
\begin{align}
\partial_t\rho_\mathrm{a}(t)=&-\frac{i}{\hbar}[H_\mathrm{a},\rho_\mathrm{a}(t)]+\mathcal{L}_\mathrm{SE}\rho_\mathrm{a}\nonumber\\
&+\sum_\eta\text{tr}_{\text{em}}\left(e^{-\gamma_\eta t}L_1^{\rm loss,\eta}(t)\int_{0}^{t}{\mathrm d}\tau\sum_{\eta '}e^{\gamma_{\eta '}(t-\tau)}    \left(\bar{L}_2^{\rm loss,\eta '}(t-\tau)\tilde{\rho}'(t-\tau)\right)     \right)\nonumber\\
&+\sum_\eta\text{tr}_{\text{em}}\left(e^{-\gamma_\eta t}L_1^{\rm loss,\eta}(t)\int_{0}^{t}{\mathrm d}\tau\sum_{\eta '}e^{\gamma_{\eta '}(t-\tau)}    \left(\bar{L}_2^{\rm gain,\eta '}(t-\tau)\tilde{\rho}'(t-\tau)\right)     \right)\nonumber\\
&+\sum_\eta\text{tr}_{\text{em}}\left(e^{-\gamma_\eta t}L_1^{\rm gain,\eta}(t)\int_{0}^{t}{\mathrm d}\tau\sum_{\eta '}e^{\gamma_{\eta '}(t-\tau)}    \left(\bar{L}_2^{\rm loss,\eta '}(t-\tau)\tilde{\rho}'(t-\tau)\right)     \right)\nonumber\\
&+\sum_\eta\text{tr}_{\text{em}}\left(e^{-\gamma_\eta t}L_1^{\rm gain,\eta}(t)\int_{0}^{t}{\mathrm d}\tau\sum_{\eta '}e^{\gamma_{\eta '}(t-\tau)}    \left(\bar{L}_2^{\rm gain,\eta '}(t-\tau)\tilde{\rho}'(t-\tau)\right)     \right)
.\label{masterrhoAtom3}
\end{align}
During the characteristic time $t\geq \gamma_\eta^{-1}$ of the QNM decay, only the fast evolution from the bare TLS Hamiltonian $H_\mathrm{a}$ is siginificant, and therefore the approximation $\tilde{\rho}(t-\tau)\approx e^{\frac{i}{\hbar}H_\mathrm{a} (\tau)}\tilde{\rho}(t)e^{-\frac{i}{\hbar}H_\mathrm{a} (\tau)} $ can be applied~\cite{Cirac}. 
Therefore, it follows that 
\begin{equation}
\tilde{\rho}'(t-\tau)=e^{-\frac{i}{\hbar}H_a \tau} \tilde{\rho}(t-\tau)  e^{\frac{i}{\hbar}H_a \tau }\approx\tilde{\rho}(t).
\end{equation}
The remaining $\tau$-dependent terms are classical numbers and the master equation can be rewritten as 
\begin{align}
\partial_t\rho_\mathrm{a}(t)=&-\frac{i}{\hbar}[H_\mathrm{a},\rho_\mathrm{a}(t)]+\mathcal{L}_\mathrm{SE}\rho_\mathrm{a}\nonumber\\
&+\sum_\eta\sum_{\eta '}e^{-\gamma_\eta t}e^{\gamma_{\eta '}t}\int_0^{t}{\mathrm d}\tau e^{-\gamma_{\eta '}
\tau}    \text{tr}_{\text{em}}\left(L_1^{\rm loss,\eta}(t) \left(\bar{L}_2^{\rm loss,\eta '}(t-\tau)\tilde{\rho}(t)\right)\right)\nonumber\\
&+\sum_\eta\sum_{\eta '}e^{-\gamma_\eta t}e^{\gamma_{\eta '}t}\int_0^{t}{\mathrm d}\tau e^{-\gamma_{\eta '}
\tau}    \text{tr}_{\text{em}}\left(L_1^{\rm loss,\eta}(t) \left(\bar{L}_2^{\rm gain,\eta '}(t-\tau)\tilde{\rho}(t)\right)\right)\nonumber\\
&+\sum_\eta\sum_{\eta '}e^{-\gamma_\eta t}e^{\gamma_{\eta '}t}\int_0^{t}{\mathrm d}\tau e^{-\gamma_{\eta '}
\tau}    \text{tr}_{\text{em}}\left(L_1^{\rm gain,\eta}(t) \left(\bar{L}_2^{\rm loss,\eta '}(t-\tau)\tilde{\rho}(t)\right)\right)\nonumber\\
&+\sum_\eta\sum_{\eta '}e^{-\gamma_\eta t}e^{\gamma_{\eta '}t}\int_0^{t}{\mathrm d}\tau e^{-\gamma_{\eta '}
\tau}    \text{tr}_{\text{em}}\left(L_1^{\rm gain,\eta}(t) \left(\bar{L}_2^{\rm gain,\eta '}(t-\tau)\tilde{\rho}(t)\right)\right)\nonumber\\
\equiv&-\frac{i}{\hbar}[H_\mathrm{a},\rho_\mathrm{a}(t)]+\mathcal{L}_\mathrm{SE}\rho_\mathrm{a}+\sum_{\eta,\eta'}\text{tr}_{\text{em}}\left\{\left[\mathcal{L}^{\eta\eta'}_{\rm ll}(t)+\mathcal{L}^{\eta\eta'}_{\rm lg}(t)+\mathcal{L}^{\eta\eta'}_{\rm gl}(t)+\mathcal{L}^{\eta\eta'}_{\rm gg}(t)\right]\tilde{\rho}(t)\right\},\label{masterrhoAtom4}
\end{align}
with 
\begin{align}
\bar{L}_2^{\rm loss,\eta '}(t-\tau)\tilde{\rho}(t)=-i\sum_{\mu,\nu}\Big(&\tilde{g}_\mu^{\rm L,s}\left(\mathbf{S}^{\rm L\frac{1}{2}}\right)_{\mu\eta'} e^{-i\omega_{\eta'} (t-\tau)}\left(\mathbf{S}^{\rm L-\frac{1}{2}}\right)_{\eta'\nu}\left[a_{\rm L\nu},\tilde{\rho}(t)\right]\sigma^+ e^{-i\omega_\mathrm{a}\tau}\nonumber\\
&+\tilde{g}_\mu^{\rm L,s*}\sigma^-e^{i\omega_\mathrm{a}\tau}\left[a_{\rm L\nu}^\dagger,\tilde{\rho}(t)\right]\left(\mathbf{S}^{\rm L-\frac{1}{2}}\right)_{\nu\eta'} e^{i\omega_{\eta'} (t-\tau)}\left(\mathbf{S}^{\rm L\frac{1}{2}}\right)_{\eta'\mu}\Big),
\end{align}
and 
\begin{align}
\bar{L}_2^{\rm gain,\eta '}(t-\tau)\tilde{\rho}(t)=-i\sum_{\mu,\nu}\Big(&\tilde{g}_\mu^{\rm G,s*}\left(\mathbf{S}^{\rm G\frac{1}{2}}\right)_{\mu\eta'} e^{i\omega_{\eta'} (t-\tau)}\left(\mathbf{S}^{\rm G-\frac{1}{2}}\right)_{\eta'\nu}\left[a_{\rm G\nu},\tilde{\rho}(t)\right]\sigma^- e^{i\omega_\mathrm{a}\tau}\nonumber\\
&+\tilde{g}_\mu^{\rm G,s}\sigma^+e^{-i\omega_\mathrm{a}\tau}\left[a_{\rm G\nu}^\dagger,\tilde{\rho}(t)\right]\left(\mathbf{S}^{\rm G-\frac{1}{2}}\right)_{\nu\eta'} e^{-i\omega_{\eta'} (t-\tau)}\left(\mathbf{S}^{\rm G\frac{1}{2}}\right)_{\eta'\mu}\Big).
\end{align}

Furthermore, $\mathcal{L}^{\eta\eta'}(t)\tilde{\rho}(t)$ is defined as 
\begin{align}
\mathcal{L}^{\eta\eta'}_{\rm ll}(t)\tilde{\rho}(t)&=\sum_{i=1}^4 C_{{\rm ll},i}^{\eta\eta'}(t) \mathcal{M}_{{\rm ll},i}^{\eta\eta'}\tilde{\rho},\\
\mathcal{L}^{\eta\eta'}_{\rm gg}(t)\tilde{\rho}(t)&=\sum_{i=1}^4 C_{{\rm gg},i}^{\eta\eta'}(t) \mathcal{M}_{{\rm gg},i}^{\eta\eta'}\tilde{\rho},\\
\mathcal{L}^{\eta\eta'}_{\rm lg}(t)\tilde{\rho}(t)&=\sum_{i=1}^4 C_{{\rm lg},i}^{\eta\eta'}(t) \mathcal{M}_{{\rm lg},i}^{\eta\eta'}\tilde{\rho},\\
\mathcal{L}^{\eta\eta'}_{\rm gl}(t)\tilde{\rho}(t)&=\sum_{i=1}^4 C_{{\rm gl},i}^{\eta\eta'}(t) \mathcal{M}_{{\rm gl},i}^{\eta\eta'}\tilde{\rho}.
\end{align}
The explicit form of the time-independent superoperators, e.g., $\mathcal{M}_{{\rm ll},1}^{\eta,\eta '}$ acting on $\tilde{\rho}(t)$ is given via
\begin{equation}
\mathcal{M}_{{\rm ll},1}^{\eta\eta'}\tilde{\rho}=-\sum_{\mu,\nu}\sum_{\mu',\nu'}\left(\tilde{g}_\mu^{\rm L,s*}\tilde{g}_{\mu '}^{\rm L,s*}\left(\mathbf{S}^{\rm L\frac{1}{2}}\right)_{\nu\eta}\left(\mathbf{S}^{\rm L-\frac{1}{2}}\right)_{\eta\mu}\left(\mathbf{S}^{\rm L-\frac{1}{2}}\right)_{\nu'\eta'}\left(\mathbf{S}^{\rm L\frac{1}{2}}\right)_{\eta'\mu'}\left[\sigma^- ,\sigma^-\left[a_{\rm L\nu '}^{\dagger},\tilde{\rho}\right]\right]a_{\rm L\nu}^{\dagger}\right).
\end{equation}

Next, the trace $\mathrm{tr}_\mathrm{em}$ for every $\mathcal{M}_j^{\eta\eta '}$ is performed. Since the trace and the commutator are linear operations, one can perform the trace internally to the electromagnetic degrees of freedom. Applying the trace to, e.g., $\mathcal{M}_{\rm ll,1}^{\eta\eta '}$ yields
\begin{equation}
\text{tr}_{\text{em}}\left[\sigma^- ,\sigma^-\left[a_{\rm L\nu'}^{\dagger},\tilde{\rho}\right]\right]a_{\rm L\nu}^{\dagger}=\left[\sigma^- ,\sigma^-\text{tr}_{\text{em}}\left(\left[a_{\rm L\nu'}^{\dagger},\tilde{\rho}\right]a_{\rm L\nu}^{\dagger}\right)\right]=0,
\end{equation} 
where the cyclic property of the trace and the fact, that $a_{\rm L\nu'}^{\dagger}$ and $a_{\rm L\nu}^{\dagger}$ commute were used, i.e.,
\begin{align}
\text{tr}_{\text{em}}\left(\left[a_{\rm L\nu'}^{\dagger},\tilde{\rho}\right]a_{\rm L\nu}^{\dagger}\right)=&\text{tr}_{\text{em}}\left(a_{\rm L\nu'}^{\dagger}\tilde{\rho}a_{\rm L\nu}^{\dagger}-\tilde{\rho}a_{\rm L\nu'}^{\dagger}a_{\rm L\nu}^{\dagger}\right)\nonumber\\
=&\text{tr}_{\text{em}}\left(a_{\rm L\nu'}^{\dagger}\tilde{\rho}a_{\rm L\nu}^{\dagger}-\tilde{\rho}a_{\rm L\nu}^{\dagger}a_{\rm L\nu'}^{\dagger}\right)\nonumber\\=&\text{tr}_{\text{em}}\left(a_{\rm L\nu}^{\dagger}a_{\rm L\nu'}^{\dagger}\tilde{\rho}-a_{\rm L\nu}^{\dagger}a_{\rm L\nu'}^{\dagger}\tilde{\rho}\right)=0.
\end{align}

The same result holds true for the trace over $\mathcal{M}_{{\rm ll},4}^{\eta\eta '},\mathcal{M}_{{\rm gg},1}^{\eta\eta '}, \mathcal{M}_{{\rm gg},4}^{\eta\eta '}$ as well as for all gain-loss cross terms $\mathcal{M}_{{\rm gl},i}^{\eta\eta '} (\mathcal{M}_{{\rm lg},i}^{\eta\eta '})$.  The trace for $\mathcal{M}_{{\rm gg},2}^{\eta\eta '},\mathcal{M}_{{\rm gg},3}^{\eta\eta '}$ and $\mathcal{M}_{{\rm ll},2}^{\eta\eta '},\mathcal{M}_{{\rm ll},3}^{\eta\eta '}$ are non-trivial, since the QNM gain/loss annihilation and creation operators do not commute for identical QNM indices. The explicit forms of these remaining contributions are 
\begin{subequations}
\label{operatorsM}
\begin{align}
\mathcal{M}_{{\rm gg},2}^{\eta\eta'}\tilde{\rho}&=-\sum_{\mu\nu}\sum_{\mu ',\nu'}\left(\tilde{g}_\mu^{\rm G,s*}\tilde{g}_{\mu '}^{\rm G,s}\left(\mathbf{S}^{\rm G-\frac{1}{2}}\right)_{\mu\eta}\left(\mathbf{S}^{\rm G\frac{1}{2}}\right)_{\eta\nu}\left(\mathbf{S}^{\rm G-\frac{1}{2}}\right)_{\nu'\eta'}\left(\mathbf{S}^{\rm G\frac{1}{2}}\right)_{\eta'\mu'}a_{\rm G\nu}\left[\sigma^- ,\sigma^+\left[a_{\rm G\nu'}^{\dagger},\tilde{\rho}\right]\right]\right),\\
\mathcal{M}_{{\rm ll},2}^{\eta\eta'}\tilde{\rho}&=-\sum_{\mu\nu}\sum_{\mu ',\nu'}\left(\tilde{g}_\mu^{\rm L,s}\tilde{g}_{\mu '}^{\rm L,s*}\left(\mathbf{S}^{\rm L-\frac{1}{2}}\right)_{\mu\eta}\left(\mathbf{S}^{\rm L\frac{1}{2}}\right)_{\eta\nu}\left(\mathbf{S}^{\rm L-\frac{1}{2}}\right)_{\nu'\eta'}\left(\mathbf{S}^{\rm L\frac{1}{2}}\right)_{\eta'\mu'}a_{\rm L\nu}\left[\sigma^+ ,\sigma^-\left[a_{\rm L\nu'}^{\dagger},\tilde{\rho}\right]\right]\right),\\
\mathcal{M}_{{\rm ll},3}^{\eta\eta '}\tilde{\rho}&=-\sum_{\mu,\nu}\sum_{\mu',\nu'}\left(\tilde{g}_\mu^{\rm L,s*}\tilde{g}_{\mu '}^{\rm L,s}\left(\mathbf{S}^{\rm L\frac{1}{2}}\right)_{\mu '\eta'}\left(\mathbf{S}^{\rm L-\frac{1}{2}}\right)_{\eta'\nu'}\left(\mathbf{S}^{\rm L\frac{1}{2}}\right)_{\nu\eta}\left(\mathbf{S}^{\rm L-\frac{1}{2}}\right)_{\eta\mu}\left[\sigma^- ,\left[a_{\rm L\nu'},\tilde{\rho}\right]\sigma^+\right]a_{\rm L\nu}^{\dagger}\right),\\  
\mathcal{M}_{{\rm gg},3}^{\eta\eta '}\tilde{\rho}&=-\sum_{\mu,\nu}\sum_{\mu',\nu'}\left(\tilde{g}_\mu^{\rm G,s}\tilde{g}_{\mu '}^{\rm G,s*}\left(\mathbf{S}^{\rm G\frac{1}{2}}\right)_{\mu '\eta'}\left(\mathbf{S}^{\rm G-\frac{1}{2}}\right)_{\eta'\nu'}\left(\mathbf{S}^{\rm G\frac{1}{2}}\right)_{\nu\eta}\left(\mathbf{S}^{\rm G-\frac{1}{2}}\right)_{\eta\mu}\left[\sigma^+ ,\left[a_{\rm G\nu'},\tilde{\rho}\right]\sigma^-\right]a_{\rm G\nu}^{\dagger}\right).
\end{align}
\end{subequations}
The corresponding $t$-dependent classical numbers $C_{\nu,\nu '}^{(j)}$ are given as
\begin{subequations}
\label{timeconstants}
\begin{align}
C^{\eta\eta '}_{{\rm gg},2}&=e^{(\gamma_{\eta '}-\gamma_\eta)t}e^{i(\omega_{\eta }-\omega_{\eta'})t}\int_0^{t}{\mathrm d}\tau e^{-\gamma_{\eta '}\tau}e^{-i(\omega_a-\omega_{\eta '}) \tau},\\
 C^{\eta\eta '}_{{\rm ll},2}&=e^{(\gamma_{\eta '}-\gamma_\eta)t}e^{i(\omega_{\eta '}-\omega_\eta)t}\int_0^{t}{\mathrm d}\tau e^{-\gamma_{\eta '}\tau}e^{i(\omega_a-\omega_{\eta '}) \tau},\\
C^{\eta\eta '}_{{\rm ll},3}&=e^{(\gamma_{\eta '}-\gamma_\eta)t}e^{-i(\omega_{\eta '}-\omega_\eta)t}\int_0^{t}{\mathrm d}\tau e^{-\gamma_{\eta '}\tau}e^{-i(\omega_a-\omega_{\eta '}) \tau},\\
C^{\eta\eta '}_{{\rm gg},3}&=e^{(\gamma_{\eta '}-\gamma_\eta)t}e^{-i(\omega_{\eta }-\omega_{\eta'})t}\int_0^{t}{\mathrm d}\tau e^{-\gamma_{\eta '}\tau}e^{i(\omega_a-\omega_{\eta '}) \tau}.
\end{align}
\end{subequations}
 
For deriving the traces over the remaining superoperators, we obtained the following relations:
\begin{align}
\left[\sigma^+ ,\sigma^-\text{tr}_{\text{em}}\left(a_{\rm L\nu}\left[a_{\rm L\nu'}^{\dagger},\tilde{\rho}\right]\right)\right]&=\left[\sigma^+ ,\sigma^-\text{tr}_{\text{em}}\left(a_{\rm L\nu}a_{\rm L\nu'}^{\dagger}\tilde{\rho}-a_{\rm L\nu}\tilde{\rho}a_{\rm L\nu'}^{\dagger}\right)\right]\nonumber\\
&=\left[\sigma^+ ,\sigma^-\text{tr}_{\text{em}}\left(\delta_{\nu\nu '}\tilde{\rho}\right)\right]=\delta_{\nu\nu '}\left(\sigma^+\sigma^-\rho_a - \sigma^-\rho_a\sigma^+ \right),
\end{align}
for the operators terms in $\mathcal{M}_{{\rm ll},2}^{\eta\eta '}$, 
\begin{align}
\left[\sigma^- ,\text{tr}_{\text{em}}\left(\left[a_{\rm L\nu'},\tilde{\rho}\right]a_{\rm L\nu}^\dagger\right)\sigma^+\right]&=\left[\sigma^- ,\text{tr}_{\text{em}}\left(a_{\rm L\nu'}\tilde{\rho}a_{\rm L\nu}^\dagger-\tilde{\rho}a_{\rm L\nu'}a_{\rm L\nu}^\dagger\right)\sigma^+\right]\nonumber\\
&=\left[\sigma^- ,\text{tr}_{\text{em}}\left(-\delta_{\nu\nu '}\tilde{\rho}\right)\sigma^-\right]=\delta_{\nu\nu '}\left(\rho_a\sigma^+\sigma^- - \sigma^-\rho_a\sigma^+ \right),
\end{align}
for the operators terms in $\mathcal{M}_{{\rm ll},3}^{\eta,\eta '}$,
\begin{align}
\left[\sigma^- ,\sigma^+\text{tr}_{\text{em}}\left(a_{\rm G\nu}\left[a_{\rm G\nu'}^{\dagger},\tilde{\rho}\right]\right)\right]&=\left[\sigma^- ,\sigma^+\text{tr}_{\text{em}}\left(a_{\rm G\nu}a_{\rm G\nu'}^{\dagger}\tilde{\rho}-a_{\rm G\nu}\tilde{\rho}a_{\rm G\nu'}^{\dagger}\right)\right]\nonumber\\
&=\left[\sigma^- ,\sigma^+\text{tr}_{\text{em}}\left(\delta_{\nu\nu '}\tilde{\rho}\right)\right]=\delta_{\nu\nu '}\left(\sigma^-\sigma^+\rho_a - \sigma^+\rho_a\sigma^- \right),
\end{align}
for the operators terms in $\mathcal{M}_{{\rm gg},2}^{\eta,\eta '}$, and
\begin{align}
\left[\sigma^+ ,\text{tr}_{\text{em}}\left(\left[a_{\rm G\nu'},\tilde{\rho}\right]a_{\rm G\nu}^\dagger\right)\sigma^-\right]&=\left[\sigma^+ ,\text{tr}_{\text{em}}\left(a_{\rm G\nu'}\tilde{\rho}a_{\rm G\nu}^\dagger-\tilde{\rho}a_{\rm G\nu'}a_{\rm G\nu}^\dagger\right)\sigma^-\right]\nonumber\\
&=\left[\sigma^+ ,\text{tr}_{\text{em}}\left(-\delta_{\nu\nu '}\tilde{\rho}\right)\sigma^-\right]=\delta_{\nu\nu '}\left(\rho_a\sigma^-\sigma^+ - \sigma^+\rho_a\sigma^- \right),
\end{align}
for the operators terms in $\mathcal{M}_{{\rm gg},3}^{\eta,\eta '}$.

Inserting these back into Eqs.~\eqref{operatorsM} yields the four remaining terms:
\begin{align}
\mathcal{M}_{{\rm gg},2}^{\eta\eta '}\tilde{\rho}&=\sum_{\mu,\mu'}\left(\tilde{g}_\mu^{\rm G,s*}\tilde{g}_{\mu '}^{\rm G,s}\left(\mathbf{S}^{\rm G-\frac{1}{2}}\right)_{\mu\eta}\delta_{\eta\eta '}\left(\mathbf{S}^{\rm G\frac{1}{2}}\right)_{\eta'\mu '}\right)\left(\sigma^+\rho_a\sigma^- - \sigma^-\sigma^+\rho_a \right),\\
\mathcal{M}_{{\rm ll},2}^{\eta\eta '}\tilde{\rho}&=\sum_{\mu,\mu'}\left(\tilde{g}_\mu^{\rm L,s}\tilde{g}_{\mu '}^{\rm L,s *}\left(\mathbf{S}^{\rm L-\frac{1}{2}}\right)_{\mu\eta}\delta_{\eta\eta '}\left(\mathbf{S}^{\rm L\frac{1}{2}}\right)_{\eta'\mu '}\right)\left(\sigma^-\rho_a\sigma^+ - \sigma^+\sigma^-\rho_a \right),\\
\mathcal{M}_{{\rm ll},3}^{\eta\eta '}\tilde{\rho}&=\sum_{\mu,\mu'}\left(\tilde{g}_\mu^{\rm L,s *}\tilde{g}_{\mu '}^{\rm L,s}\left(\mathbf{S}^{\rm L\frac{1}{2}}\right)_{\mu'\eta'}\delta_{\eta\eta '}\left(\mathbf{S}^{\rm L-\frac{1}{2}}\right)_{\eta\mu}\right)\left( \sigma^-\rho_a\sigma^+ - \rho_a\sigma^+\sigma^-\right),\\
\mathcal{M}_{{\rm gg},3}^{\eta\eta '}\tilde{\rho}&=\sum_{\mu,\mu'}\left(\tilde{g}_\mu^{\rm G,s}\tilde{g}_{\mu '}^{\rm G,s*}\left(\mathbf{S}^{\rm G\frac{1}{2}}\right)_{\mu'\eta'}\delta_{\eta\eta '}\left(\mathbf{S}^{\rm G-\frac{1}{2}}\right)_{\eta\mu}\right)\left( \sigma^+\rho_a\sigma^- - \rho_a\sigma^-\sigma^+\right), 
\end{align}
where the the Kronecker delta $\delta_{\nu\nu '}$ was applied and the multiplication of the matrix $\mathbf{S}^{\rm L/G\frac{1}{2}}$ with its inverse $\mathbf{S}^{\rm L/G-\frac{1}{2}}$ was also performed.

Next, the time integrations in $C^{\eta\eta '}_{{\rm ll,i}}$ and $C^{\eta\eta '}_{{\rm gg,i}}$ (corresponding to the non-vanishing operator contributions) from Eqs.~\eqref{timeconstants} are derived in the spirit of the Markov approximation, i.e, the upper integral limit is set to $t\rightarrow\infty$~\cite{carmichael2009statistical,Cirac}. In this way, the integrals can be analytically solved 
as 
\begin{align}
 C^{\eta\eta '}_{{\rm gg,2}}&=e^{(\gamma_{\eta '}-\gamma_\eta)t}e^{i(\omega_{\eta}-\omega_{\eta'})t}\frac{-i}{\omega_a-\omega_{\eta '}-i\gamma_{\eta '}},\\
 C^{\eta\eta '}_{{\rm ll,2}}&=e^{(\gamma_{\eta '}-\gamma_\eta)t}e^{i(\omega_{\eta '}-\omega_\eta)t}\frac{i}{\omega_a-\omega_{\eta '}+i\gamma_{\eta '}},\\
C^{\eta\eta '}_{{\rm ll,3}}&=e^{(\gamma_{\eta '}-\gamma_\eta)t}e^{-i(\omega_{\eta '}-\omega_\eta)t}\frac{-i}{\omega_a-\omega_{\eta '}-i\gamma_{\eta '}},\\
C^{\eta\eta '}_{{\rm gg,3}}&=e^{(\gamma_{\eta '}-\gamma_\eta)t}e^{-i(\omega_{\eta}-\omega_{\eta'})t}\frac{i}{\omega_a-\omega_{\eta '}+i\gamma_{\eta '}}.
\end{align}

Finally, the four terms are summarized to obtain
\begin{align}
\sum_{\eta,\eta'}&\text{tr}_{\text{em}}\left\{\left[\mathcal{L}^{\eta\eta'}_{\rm ll}(t)+\mathcal{L}^{\eta\eta'}_{\rm lg}(t)+\mathcal{L}^{\eta\eta'}_{\rm gl}(t)+\mathcal{L}^{\eta\eta'}_{\rm gg}(t)\right]\tilde{\rho}(t)\right\}\nonumber\\
&=\sum_{\mu,\mu '}\tilde{g}_\mu^{\rm L,s}\tilde{g}_{\mu '}^{\rm L,s *}\left\{\sum_\eta\left(\mathbf{S}^{\rm L-\frac{1}{2}}\right)_{\mu\eta}\frac{i}{\omega_a-\omega_{\eta }+i\gamma_{\eta }}\left(\mathbf{S}^{\rm L\frac{1}{2}}\right)_{\eta\mu '}\right\}\left(\sigma^-\rho_a\sigma^+ - \sigma^+\sigma^-\rho_a \right)\nonumber\\
&+\sum_{\mu,\mu '}\tilde{g}_\mu^{\rm L,s *}\tilde{g}_{\mu '}^{\rm L,s}\left\{\sum_{\eta}\left(\mathbf{S}^{\rm L\frac{1}{2}}\right)_{\mu' \eta}\frac{-i}{\omega_a-\omega_{\eta}-i\gamma_{\eta}}\left(\mathbf{S}^{\rm L-\frac{1}{2}}\right)_{\eta\mu}\right\}\left( \sigma^-\rho_a\sigma^+ - \rho_a\sigma^+\sigma^-\right)\nonumber\\
&+\sum_{\mu,\mu '}\tilde{g}_\mu^{\rm G,s}\tilde{g}_{\mu '}^{\rm G,s*}\left\{\sum_{\eta}\left(\mathbf{S}^{\rm G-\frac{1}{2}}\right)_{\eta\mu}\frac{i}{\omega_a-\omega_{\eta}+i\gamma_{\eta}}\left(\mathbf{S}^{\rm G\frac{1}{2}}\right)_{\mu' \eta}\right\}\left( \sigma^+\rho_a\sigma^- - \rho_a\sigma^-\sigma^+\right)\nonumber\\
&+\sum_{\mu,\mu '}\tilde{g}_\mu^{\rm G,s*}\tilde{g}_{\mu '}^{\rm G,s}\left\{\sum_\eta\left(\mathbf{S}^{\rm G\frac{1}{2}}\right)_{\eta\mu '}\frac{-i}{\omega_a-\omega_{\eta }-i\gamma_{\eta }}\left(\mathbf{S}^{\rm G-\frac{1}{2}}\right)_{\mu\eta}\right\}\left(\sigma^+\rho_a\sigma^- - \sigma^-\sigma^+\rho_a \right),\nonumber\\
&\equiv \tilde{\Gamma}^{\rm loss}\left(\sigma^-\rho_a\sigma^+ - \sigma^+\sigma^-\rho_a \right)+\tilde{\Gamma}^{\rm loss*}\left( \sigma^-\rho_a\sigma^+ - \rho_a\sigma^+\sigma^-\right) \nonumber\\
&+\tilde{\Gamma}^{\rm gain}\left(\sigma^+\rho_a\sigma^- - \rho_a\sigma^-\sigma^+ \right)+\tilde{\Gamma}^{\rm gain*}\left( \sigma^+\rho_a\sigma^- - \sigma^-\sigma^+\rho_a\right)
.\label{eq: LetaetaFin1}
\end{align}
Using the definition of the symmetrized light-TLS coupling constants (cf. Secs.~\ref{Subsec: SEP_Gain_Loss} and \ref{Subsec: UNI_Gain_Loss}), we obtain 
\begin{align}
    \tilde{\Gamma}^{\rm loss}&=\sum_{\eta\eta'}\tilde{g}_\eta S_{\eta\eta'}^{\rm L}\tilde{g}_{\eta'}^*\frac{-i}{\Delta_{\eta \rm a}-i\gamma_\eta}\equiv  \frac{\Gamma^{\rm loss}}{2}+i\omega_{\rm LS}^{\rm loss},\\
    \tilde{\Gamma}^{\rm gain}&=\sum_{\eta\eta'}\tilde{g}_\eta S_{\eta'\eta}^{\rm G}\tilde{g}_{\eta'}^*\frac{-i}{\Delta_{\eta \rm a}-i\gamma_\eta}\equiv \frac{\Gamma^{\rm gain}}{2}+i\omega_{\rm LS}^{\rm gain},
\end{align}
which leads finally to the bad cavity limit master equation \eqref{eq: BadCavMaster}, when neglecting the photonic Lamb shift.

In contrast to the separated gain-loss operator approach, the bad cavity limit derivation from the lossy mode quantization cannot be directly applied to the unified gain-loss operator approach and must be generalized, since there are also reversed Lindblad dissipator terms $\mathcal{L}[\mathbf{a}^{\prime \dagger}]$, describing incoherent pumping. To be more specific, the action of $\mathcal{L}_{\rm em}$ on the TLS-QNM interaction Liouvillian $\mathcal{L}_I$ changes significantly. For instance, $\mathcal{L}_{\rm em}(\mathbf{a}'\rho)$ will couple to terms  $\mathbf{a}'\mathcal{L}_{\rm em}\rho$ and $i[\boldsymbol{\chi}^{\prime+}-i\boldsymbol{\chi}^{\prime\rm L-}]\cdot\mathbf{a}'\rho$ (similar to the separated gain-loss operator approach), but it will also couple to $-\rho\boldsymbol{\chi}^{\prime\rm G-}\cdot\mathbf{a}'$ in a different operator ordering. Consequently $\mathcal{L}_{\rm em}(\rho\mathbf{a}')$ must be also taken into account, resulting in a linear system of equations for the underlying superoperators that must be solved to obtain the explicit expression of $\exp(-\mathcal{L}_{\rm em}t)\mathcal{L}_I\rho$. However, after the more complicated algebra is performed the remaining procedure of the above derivation is formally identical (cf. Ref.~\onlinecite{carmichael2009statistical} for a single mode case).

Below we show a different method to obtain the bad cavity limit, where it is explicit proven for a single mode case, that both approaches coincide.

\section{Bad cavity limit within a Bloch equation treatment for the quantized QNM models\label{app: BlochEquations}}
To demonstrate the agreement of the bad cavity limit between the unified and separated gain-loss operator approach, we employ a Bloch equation treatment for the special case of a single QNM (the corresponding derivation for the multi-mode case is straightforward). 

\subsection{Unified gain-loss operator approach}
We start with the unified gain-loss operator approach and investigate the equation of motion for the TLS lowering operator expectation value in a frame rotating with $\omega_{\rm a}$,
\begin{equation}
    \frac{{\rm d}}{{\rm d}t}\langle\tilde{\sigma}^-\rangle = -\frac{\Gamma^{\rm B}}{2}\langle\tilde{\sigma}^-\rangle +i\sqrt{S'}\tilde g \langle \sigma_z \tilde{a}'\rangle,\label{eq: SigmaTimeEvoSingleQNM}
\end{equation}
where 
\begin{equation}
   \tilde{\sigma}^-(t)=e^{i\omega_{\rm a}t} \sigma^-(t),~\tilde{a}'(t)=e^{i\omega_{\rm a}t}a'(t),
\end{equation}
are the 
{\it slowly varying} operators and $S'=S^{\rm L}-S^{\rm G}$. Here, we have chosen normal ordering for the photon operator.
Next, we formally solve the Heisenberg equation of motion for the QNM operator $\tilde{a}'$ to get
\begin{align}
    \tilde{a}'(t) =& e^{i(\omega_{\rm a}-\tilde{\omega}_{\rm c})t}\tilde{a}'(0)-i\sqrt{S'}\tilde{g}^* \int_0^{t}{\rm d}t' e^{i(\omega_{\rm a}-\tilde{\omega}_{\rm c})(t-t')}\tilde{\sigma}^-(t') \\
    &+\sqrt{2(S^{\rm L}/S')\gamma_{\rm c}}e^{i\omega_{\rm a}t}\int_0^{t}{\rm d}t' e^{-i\tilde{\omega}_{\rm c}(t-t')}c^{\prime\rm L}_{\rm in}(t')+ \sqrt{2(S^{\rm G}/S')\gamma_{\rm c}}e^{i\omega_{\rm a}t}\int_0^{t}{\rm d}t' e^{-i\tilde{\omega}_{\rm c}(t-t')}c^{\prime\rm G\dagger}_{\rm in}(t').\label{eq: FormalaPrimeSol}
\end{align}

Inserting back into Eq.~\eqref{eq: SigmaTimeEvoSingleQNM} yields 
\begin{align}
    \frac{{\rm d}}{{\rm d}t}\langle\tilde{\sigma}^-\rangle =& -\frac{\Gamma^{\rm B}}{2}\langle\tilde{\sigma}^-\rangle+i\sqrt{S'}\tilde g e^{i(\omega_{\rm a}-\tilde{\omega}_{\rm c})t}\langle \sigma_z(t)a'(0) \rangle + S'|\tilde{g}|^2 \int_0^{t}{\rm d}t' e^{i(\omega_{\rm a}-\tilde\omega_{\rm c})(t-t')} \langle\sigma_z(t) \tilde{\sigma}^-(t')\rangle \nonumber\\
    &+i\tilde g \sqrt{2S^{\rm L}\gamma_{\rm c}}e^{i\omega_{\rm a}t}\int_0^{t}{\rm d}t'e^{-i\tilde\omega_{\rm c}(t-t')} \langle \sigma_z(t)c^{\prime\rm L}_{\rm in}(t') \rangle\nonumber\\
    &+i\tilde g \sqrt{2S^{\rm G}\gamma_{\rm c}}e^{i\omega_{\rm a}t}\int_0^{t}{\rm d}t'e^{-i\tilde\omega_{\rm c}(t-t')} \langle \sigma_z(t) c^{\prime\rm G\dagger}_{\rm in}(t').\rangle \label{eq: SigmaTimeEvoSingleQNM3}
\end{align}
The second term on the right-hand side quickly decays to zero and is thus neglected in the following, while the third term can be rewritten via a coordinate transformation $t-t'\rightarrow \tau$:
\begin{equation}
 \int_0^{t}{\rm d}t' e^{i(\omega_{\rm a}-\tilde\omega_{\rm c})(t-t')} \langle\sigma_z(t) \tilde{\sigma}^-(t')\rangle= \int_0^{t}{\rm d}\tau e^{i(\omega_{\rm a}-\tilde\omega_{\rm c})\tau} \langle\sigma_z(t) \tilde{\sigma}^-(t-\tau)\rangle.
\end{equation}
Within a Markov approximation we approximately replace $\tilde{\sigma}^-(t-\tau)\approx \tilde{\sigma}^-(t)$ under the integral and extend the upper integral boundary to $+\infty$ to obtain
\begin{equation}
    \int_0^{t}{\rm d}\tau e^{i(\omega_{\rm a}-\tilde\omega_{\rm c})\tau} \langle\sigma_z(t) \tilde{\sigma}^-(t-\tau)\rangle\approx \langle\sigma_z(t) \tilde{\sigma}^-(t)\rangle \int_0^{\infty}{\rm d}\tau e^{i(\omega_{\rm a}-\tilde\omega_{\rm c})\tau}=\frac{i}{\omega_{\rm a}-\tilde{\omega}_{\rm c}}\langle\sigma_z(t) \tilde{\sigma}^-(t)\rangle.
\end{equation}
Using properties of the Pauli matrices, we can further simplify this term as
\begin{equation}
    \frac{i}{\omega_{\rm a}-\tilde{\omega}_{\rm c}}\langle\sigma_z(t) \tilde{\sigma}^-(t)\rangle=-\frac{i}{\omega_{\rm a}-\tilde{\omega}_{\rm c}}\langle \tilde{\sigma}^-(t)\rangle.
\end{equation}
Inserting back into Eq.~\eqref{eq: SigmaTimeEvoSingleQNM3} yields
\begin{align}
    \frac{{\rm d}}{{\rm d}t}\langle\tilde{\sigma}^-\rangle \approx& -\frac{\Gamma^{\rm B}}{2}\langle\tilde{\sigma}^-\rangle - i\frac{S'|\tilde{g}|^2}{\omega_{\rm a}-\tilde{\omega}_{\rm c}}  \langle \tilde{\sigma}^-\rangle \nonumber\\
    &+i\tilde g \sqrt{2S^{\rm L}\gamma_{\rm c}}e^{i\omega_{\rm a}t}\int_0^{t}{\rm d}t'e^{-i\tilde\omega_{\rm c}(t-t')} \langle \sigma_z(t)c^{\prime\rm L}_{\rm in}(t') \rangle\nonumber\\
    &+i\tilde g \sqrt{2S^{\rm G}\gamma_{\rm c}}e^{i\omega_{\rm a}t}\int_0^{t}{\rm d}t'e^{-i\tilde\omega_{\rm c}(t-t')} \langle \sigma_z(t) c^{\prime\rm G\dagger}_{\rm in}(t')\rangle. \label{eq: SigmaTimeEvoSingleQNM4}
\end{align}

Thus, to obtain a closed equation of motion, we are left with terms proportional to the photon reservoir operators. Assuming the reservoir is initially in the vacuum state (consistent with the assumptions to derive the QNM master equation), we get $\langle \sigma_z(t)c^{\prime\rm L}_{\rm in}(t') \rangle=0$, so that 
\begin{align}
    \frac{{\rm d}}{{\rm d}t}\langle\tilde{\sigma}^-\rangle =& -\frac{\Gamma^{\rm B}}{2}\langle\tilde{\sigma}^-\rangle - i\frac{S'|\tilde{g}|^2}{\omega_{\rm a}-\tilde{\omega}_{\rm c}}  \langle \tilde{\sigma}^-\rangle +i\tilde g \sqrt{2S^{\rm G}\gamma_{\rm c}}e^{i\omega_{\rm a}t}\int_0^{t}{\rm d}t'e^{-i\tilde\omega_{\rm c}(t-t')} \langle \sigma_z(t) c^{\prime\rm G\dagger}_{\rm in}(t')\rangle .\label{eq: SigmaTimeEvoSingleQNM5}
\end{align}
We see a potential difficulty of the Fock space construction here, namely that the last expectation value does not vanish, although we have chosen normal ordering of the system photon operator. This is because there appears an adjoint reservoir operator on the very right of the operator product, which represents the pumping mechanism. 
To proceed, we first formally write the above equations into the form 
\begin{equation}
     \frac{{\rm d}}{{\rm d}t}\langle\tilde{\sigma}^-\rangle = -\frac{\Gamma^{\rm B}}{2}\langle\tilde{\sigma}^-\rangle - i\frac{(S^{\rm L}-S^{\rm G})|\tilde{g}|^2}{\omega_{\rm a}-\tilde{\omega}_{\rm c}}  \langle \tilde{\sigma}^-\rangle +K^{\rm gain}(\omega_{\rm a}),
\end{equation}
with the gain-induced term
\begin{equation}
    K^{\rm gain}(\omega_{\rm a})=i\tilde g \sqrt{2S^{\rm G}\gamma_{\rm c}}e^{i\omega_{\rm a}t}\int_0^{t}{\rm d}t'e^{-i\tilde\omega_{\rm c}(t-t')} \langle \sigma_z(t) c^{\prime\rm G\dagger}_{\rm in}(t')\rangle.
\end{equation}
Transforming back into the non-rotating frame yields
\begin{align}
     \frac{{\rm d}}{{\rm d}t}\langle\sigma^-\rangle = -i\omega_{\rm a}\langle\sigma^-\rangle-\frac{\Gamma^{\rm B}}{2}\langle\sigma^-\rangle - i\frac{(S^{\rm L}-S^{\rm G})|\tilde{g}|^2}{\omega_{\rm a}-\tilde{\omega}_{\rm c}}  \langle \sigma^-\rangle + e^{-i\omega_{\rm a}t}K^{\rm gain}(\omega_{\rm a}).\label{eq: BadCavityRes1UnifiedPre}
\end{align}

Finally, we split the prefactor of the third term on the right-hand side into a real and imaginary part to obtain
\begin{align}
     \frac{{\rm d}}{{\rm d}t}\langle\sigma^-\rangle = -i\left[\omega_{\rm a}+\Delta_{\rm LS}^{\rm loss}-\Delta_{\rm LS}^{\rm gain}\right]\langle\sigma^-\rangle-\frac{\Gamma^{\rm B}}{2}\langle\sigma^-\rangle - \frac{\Gamma^{\rm loss}-\Gamma^{\rm gain}}{2}  \langle \sigma^-\rangle + e^{-i\omega_{\rm a}t}K^{\rm gain}(\omega_{\rm a}),\label{eq: BadCavityRes1Unified}
\end{align}
where 
\begin{align}
   \Delta_{\rm LS}^{\rm loss}&= \frac{S^{\rm L}|\tilde{g}|^2(\omega_{\rm a}-\omega_{\rm c})}{(\omega_{\rm a}-\omega_{\rm c})^2+\gamma_{\rm c}^2},\\
    \Delta_{\rm LS}^{\rm gain}&= \frac{S^{\rm G}|\tilde{g}|^2(\omega_{\rm a}-\omega_{\rm c})}{(\omega_{\rm a}-\omega_{\rm c})^2+\gamma_{\rm c}^2},\\
   \Gamma^{\rm loss}&=\frac{2S^{\rm L}|\tilde{g}|^2\gamma_{\rm c}}{(\omega_{\rm a}-\omega_{\rm c})^2+\gamma_{\rm c}^2},\\
   \Gamma^{\rm gain}&=\frac{2S^{\rm G}|\tilde{g}|^2\gamma_{\rm c}}{(\omega_{\rm a}-\omega_{\rm c})^2+\gamma_{\rm c}^2}.
\end{align}

\subsection{Separated gain-loss operator approach - Normal operator ordering}
Before we proceed to investigate $K^{\rm gain}(\omega_{\rm a})$ in more detail, we now turn to the separated gain-loss operator approach, where the equation of motion for the TLS lowering operator expectation value, in a frame rotating with $\omega_{\rm a}$, reads
\begin{equation}
    \frac{{\rm d}}{{\rm d}t}\langle\tilde{\sigma}^-\rangle = -\frac{\Gamma^{\rm B}}{2}\langle\tilde{\sigma}^-\rangle +i\sqrt{S^{\rm L}} \tilde g \langle \sigma_z \tilde{a}_{\rm L}\rangle+i\sqrt{S^{\rm G}} \tilde g\langle\tilde{a}_{\rm G}^\dagger\sigma_z \rangle,\label{eq: SigmaTimeEvoSingleQNMSep}
\end{equation}
with
\begin{equation}
   \tilde{\sigma}^-(t)=e^{i\omega_{\rm a}t} \sigma^-(t),~\tilde{a}_{\rm L}(t)=e^{i\omega_{\rm a}t}a_{\rm L}(t),~\tilde{a}_{\rm G}^\dagger(t)=e^{i\omega_{\rm a}t}a_{\rm G}^\dagger(t).
\end{equation}
Here, we have again chosen normal ordering of the system photon operators. Next, we formally solve the Heisenberg equation of motion for $a_{\rm L}$ and $a_{\rm G}^\dagger$:
\begin{align}
    \tilde{a}_{\rm L}(t)=e^{i(\omega_{\rm a}-\tilde{\omega}_{\rm c})t} a_{\rm L}(0)-i\sqrt{S^{\rm L}}\tilde{g}^*\int_0^t  {\rm d}t' e^{i(\omega_{\rm a}-\tilde{\omega}_{\rm c})(t-t')} \tilde{\sigma}^-(t')+\sqrt{2\gamma_{\rm c}}e^{i\omega_{\rm a}t}\int_0^t  {\rm d}t' e^{-i\tilde{\omega}_{\rm c}(t-t')}c_{\rm L}^{\rm in}(t'),\label{eq: a_L_time_evo}\\
     \tilde{a}_{\rm G}^\dagger(t)=e^{i(\omega_{\rm a}-\tilde{\omega}_{\rm c})t}a_{\rm G}^\dagger(0)+i\sqrt{S^{\rm G}}\tilde{g}^*\int_0^t  {\rm d}t' e^{i(\omega_{\rm a}-\tilde{\omega}_{\rm c})(t-t')} \tilde{\sigma}^-(t')+\sqrt{2\gamma_{\rm c}}e^{i\omega_{\rm a}t}\int_0^t  {\rm d}t' e^{-i\tilde{\omega}_{\rm c}(t-t')}c_{\rm G}^{\rm in\dagger}(t')\label{eq: a_G_dagger_time_evo}.
\end{align}
Inserting back into Eq.~\eqref{eq: SigmaTimeEvoSingleQNMSep} yields 
\begin{align}
    \frac{{\rm d}}{{\rm d}t}\langle\tilde{\sigma}^-\rangle =& -\frac{\Gamma^{\rm B}}{2}\langle\tilde{\sigma}^-\rangle +i\sqrt{S^{\rm L}} \tilde g e^{i(\omega_{\rm a}-\tilde{\omega}_{\rm c})t} \langle \sigma_z (t) a_{\rm L}(0)\rangle+i\sqrt{S^{\rm G}} \tilde g e^{i(\omega_{\rm a}-\tilde{\omega}_{\rm c})t}\langle a_{\rm G}^\dagger(0)\sigma_z \rangle\nonumber\\
    & +S^{\rm L}|\tilde g|^2\int_0^t  {\rm d}t' e^{i(\omega_{\rm a}-\tilde{\omega}_{\rm c})(t-t')} \langle\sigma_z(t)\tilde{\sigma}^-(t')\rangle- S^{\rm G}|\tilde g|^2 \int_0^t  {\rm d}t' e^{i(\omega_{\rm a}-\tilde{\omega}_{\rm c})(t-t')} \langle\tilde{\sigma}^-(t')\sigma_z(t)\rangle\nonumber\\
    &+i\sqrt{S^{\rm L}}\tilde g\sqrt{2\gamma_{\rm c}}e^{i\omega_{\rm a}t}\int_0^t  {\rm d}t' e^{-i\tilde{\omega}_{\rm c}(t-t')}\langle \sigma_z(t)c_{\rm L}^{\rm in}(t')\rangle\nonumber\\
    &+i\sqrt{S^{\rm G}}\tilde g\sqrt{2\gamma_{\rm c}}e^{i\omega_{\rm a}t}\int_0^t  {\rm d}t' e^{-i\tilde{\omega}_{\rm c}(t-t')}\langle c_{\rm G}^{\rm in\dagger}(t')\sigma_z(t)\rangle.\label{eq: SigmaTimeEvoSingleQNMSep2}
\end{align}

Similar to the unified gain-loss operator approach, the second and third term on the right-hand side of the above equation rapidly decay to zero, and thus are neglected in the following. Applying then the same Markov approximation as in the last subsection to the terms in the second line of the above equation yields
\begin{align}
    \frac{{\rm d}}{{\rm d}t}\langle\tilde{\sigma}^-\rangle =& -\frac{\Gamma^{\rm B}}{2}\langle\tilde{\sigma}^-\rangle +i\frac{S^{\rm L}|\tilde g|^2}{\omega_{\rm a}-\tilde{\omega}_{\rm c}}\langle\sigma_z\tilde{\sigma}^-\rangle- i\frac{S^{\rm L}|\tilde g|^2}{\omega_{\rm a}-\tilde{\omega}_{\rm c}}\langle \tilde{\sigma}^-\sigma_z\rangle\nonumber\\
    &+i\sqrt{S^{\rm L}}\tilde g\sqrt{2\gamma_{\rm c}}e^{i\omega_{\rm a}t}\int_0^t  {\rm d}t' e^{-i\tilde{\omega}_{\rm c}(t-t')}\langle \sigma_z(t)c_{\rm L}^{\rm in}(t')\rangle\nonumber\\
    &+i\sqrt{S^{\rm G}}\tilde g\sqrt{2\gamma_{\rm c}}e^{i\omega_{\rm a}t}\int_0^t  {\rm d}t' e^{-i\tilde{\omega}_{\rm c}(t-t')}\langle c_{\rm G}^{\rm in\dagger}(t')\sigma_z(t)\rangle.\label{eq: SigmaTimeEvoSingleQNMSep3}
\end{align}
Next, we use again the properties of the Pauli operators, namely $\tilde{\sigma}^-\sigma_z=\tilde\sigma^-$ and $\sigma_z\tilde{\sigma}^-=-\tilde\sigma^-$ to obtain
\begin{align}
    \frac{{\rm d}}{{\rm d}t}\langle\tilde{\sigma}^-\rangle \approx& -\frac{\Gamma^{\rm B}}{2}\langle\tilde{\sigma}^-\rangle - i\frac{(S^{\rm L}+S^{\rm G})|\tilde{g}|^2}{\omega_{\rm a}-\tilde{\omega}_{\rm c}}  \langle \tilde{\sigma}^-\rangle \nonumber\\
    &+i\sqrt{S^{\rm L}}\tilde g \sqrt{2\gamma_{\rm c}}e^{i\omega_{\rm a}t}\int_0^{t}{\rm d}t'e^{-i\tilde\omega_{\rm c}(t-t')} \langle \sigma_z(t)c_{\rm L}^{\rm in}(t') \rangle\nonumber\\
    &+i\sqrt{S^{\rm G}}\tilde g \sqrt{2\gamma_{\rm c}}e^{i\omega_{\rm a}t}\int_0^{t}{\rm d}t'e^{-i\tilde\omega_{\rm c}(t-t')} \langle c_{\rm G}^{\rm in\dagger}(t')\sigma_z(t) \rangle. \label{eq: SigmaTimeEvoSingleQNM4Sep}
\end{align}

In contrast to the unified gain-loss operator approach, the sum $S^{\rm L}+S^{\rm G}$ appears as the prefactor of the second term on the right-hand side (rather then the difference $S'=S^{\rm L}-S^{\rm G}$). This crucial difference is induced by the different ordering of $\sigma_z$ and $\tilde\sigma^+$ for the gain and loss related part, respectively. Moreover, since we have initially chosen normal ordering for the photon operators, 
the reservoir contributions in the above equation immediately vanish, since we assume the vacuum state, so that
\begin{align}
    \frac{{\rm d}}{{\rm d}t}\langle\tilde{\sigma}^-\rangle = -\frac{\Gamma^{\rm B}}{2}\langle\tilde{\sigma}^-\rangle - i\frac{(S^{\rm L}+S^{\rm G})|\tilde{g}|^2}{\omega_{\rm a}-\tilde{\omega}_{\rm c}}  \langle \tilde{\sigma}^-\rangle.  \label{eq: SigmaTimeEvoSingleQNM5Sep}
\end{align}
Transforming back into the non-rotating frame yields
\begin{align}
     \frac{{\rm d}}{{\rm d}t}\langle\sigma^-\rangle = -i\omega_{\rm a}\langle\sigma^-\rangle-\frac{\Gamma^{\rm B}}{2}\langle\sigma^-\rangle - i\frac{(S^{\rm L}+S^{\rm G})|\tilde{g}|^2}{\omega_{\rm a}-\tilde{\omega}_{\rm c}}  \langle \sigma^-\rangle.
\end{align}

Finally, we again split the prefactor of the third term on the right-hand side into a real and imaginary part to obtain
\begin{align}
     \frac{{\rm d}}{{\rm d}t}\langle\sigma^-\rangle = -i\left[\omega_{\rm a}+\Delta_{\rm LS}^{\rm loss}+\Delta_{\rm LS}^{\rm gain}\right]\langle\sigma^-\rangle-\frac{\Gamma^{\rm B}}{2}\langle\sigma^-\rangle - \frac{\Gamma^{\rm loss}+\Gamma^{\rm gain}}{2}  \langle \sigma^-\rangle,
\end{align}
where the rates are defined at the end of the last subsection.
This is formally identical to the results from Ref.~\onlinecite{franke2021fermi}, namely the dephasing is exponentially damped with $(\Gamma^{\rm loss}+\Gamma^{\rm gain}+\Gamma^{\rm B})/2$, not related to the LDOS, which would be proportional to $(\Gamma^{\rm loss}-\Gamma^{\rm gain}+\Gamma^{\rm B})/2$ (as would be the case for the unified gain-loss operator approach if one neglects $K^{\rm gain}$).

\subsection{Separated gain-loss operator approach - Mixed operator ordering}
To connect the separated and unified gain-loss operator approach, here, we leave the gain photon part in anti-normal ordering and the loss photon part in normal ordering, where the equation of motion for the TLS lowering operator expectation value, in a frame rotating with $\omega_{\rm a}$, reads
\begin{equation}
    \frac{{\rm d}}{{\rm d}t}\langle\tilde{\sigma}^-\rangle = -\frac{\Gamma^{\rm B}}{2}\langle\tilde{\sigma}^-\rangle +i\sqrt{S^{\rm L}} \tilde g \langle \sigma_z\tilde{a}_{\rm L} \rangle+i\sqrt{S^{\rm G}} \tilde g\langle\sigma_z\tilde{a}_{\rm G}^\dagger \rangle.\label{eq: SigmaTimeEvoSingleQNMSepAnti}
\end{equation}
Inserting Eqs.~\eqref{eq: a_L_time_evo} and \eqref{eq: a_G_dagger_time_evo} into Eq.~\eqref{eq: SigmaTimeEvoSingleQNMSepAnti} and applying the same manipulations as in Eq.~\eqref{eq: SigmaTimeEvoSingleQNMSep2}-\eqref{eq: SigmaTimeEvoSingleQNM4Sep} yields 
\begin{align}
    \frac{{\rm d}}{{\rm d}t}\langle\tilde{\sigma}^-\rangle \approx& -\frac{\Gamma^{\rm B}}{2}\langle\tilde{\sigma}^-\rangle - i\frac{(S^{\rm L}-S^{\rm G})|\tilde{g}|^2}{\omega_{\rm a}-\tilde{\omega}_{\rm c}}  \langle \tilde{\sigma}^-\rangle \nonumber\\
    &+i\sqrt{S^{\rm L}}\tilde g \sqrt{2\gamma_{\rm c}}e^{i\omega_{\rm a}t}\int_0^{t}{\rm d}t'e^{-i\tilde\omega_{\rm c}(t-t')} \langle  \sigma_z(t)c_{\rm L}^{\rm in}(t')\rangle\nonumber\\
    &+i\sqrt{S^{\rm G}}\tilde g \sqrt{2\gamma_{\rm c}}e^{i\omega_{\rm a}t}\int_0^{t}{\rm d}t'e^{-i\tilde\omega_{\rm c}(t-t')} \langle \sigma_z(t)c_{\rm G}^{\rm in\dagger}(t') \rangle. \label{eq: SigmaTimeEvoSingleQNM4SepAnti}
\end{align}
In contrast to the normal operator ordering, 
we recognize the same prefactor (proportional to $S^{\rm L}-S^{\rm G}$) in the second term on the right-hand side as in the unified gain-loss operator approach. Moreover, due to the mixed operator ordering of the photon operators, the reservoir contributions in the above equation only vanish for the loss-related part, which is again similar to the unified gain-loss operator description. Reformulating the above equation leads to
\begin{align}
    \frac{{\rm d}}{{\rm d}t}\langle\tilde{\sigma}^-\rangle = -\frac{\Gamma^{\rm B}}{2}\langle\tilde{\sigma}^-\rangle - i\frac{(S^{\rm L}-S^{\rm G})|\tilde{g}|^2}{\omega_{\rm a}-\tilde{\omega}_{\rm c}}  \langle \tilde{\sigma}^-\rangle + M^{\rm gain}(\omega_{\rm a}),  \label{eq: SigmaTimeEvoSingleQNM5SepAnti}
\end{align}
where 
\begin{equation}
    M^{\rm gain}(\omega_{\rm a})=i\sqrt{S^{\rm G}}\tilde g \sqrt{2\gamma_{\rm c}}e^{i\omega_{\rm a}t}\int_0^{t}{\rm d}t'e^{-i\tilde\omega_{\rm c}(t-t')} \langle \sigma_z(t)c_{\rm G}^{\rm in\dagger}(t') \rangle.
\end{equation}
Transforming back into the non-rotating frame yields
\begin{align}
     \frac{{\rm d}}{{\rm d}t}\langle\sigma^-\rangle = -i\omega_{\rm a}\langle\sigma^-\rangle-\frac{\Gamma^{\rm B}}{2}\langle\sigma^-\rangle - i\frac{(S^{\rm L}+S^{\rm G})|\tilde{g}|^2}{\omega_{\rm a}-\tilde{\omega}_{\rm c}}  \langle \sigma^-\rangle+e^{-i\omega_{\rm a}t}M^{\rm gain}(\omega_{\rm a}).
\end{align}

Finally, we again split the prefactor of the third term on the right-hand side into a real and imaginary part to obtain
\begin{align}
     \frac{{\rm d}}{{\rm d}t}\langle\sigma^-\rangle = -i\left[\omega_{\rm a}+\Delta_{\rm LS}^{\rm loss}-\Delta_{\rm LS}^{\rm gain}\right]\langle\sigma^-\rangle-\frac{\Gamma^{\rm B}}{2}\langle\sigma^-\rangle - \frac{\Gamma^{\rm loss}-\Gamma^{\rm gain}}{2}  \langle \sigma^-\rangle+e^{-i\omega_{\rm a}t}M^{\rm gain}(\omega_{\rm a}).
\end{align}
This is formally identical to the results from the unified gain-loss operator approach, with the only formal difference being the definition of the reservoir input operators. However the algebraic properties of $c_{\rm G}^{\rm in}(t')$ and $c_{\rm G}^{\prime\rm in}(t')$ are the same. This fact on its own partially confirms that both approaches yield the same result, and it will be explicitly shown in the next subsection.

\subsection{Derivation of $K^{\rm gain}(\omega_{\rm a})$}
Comparing the results from the separated and unified gain-loss operator approach implies, that $K^{\rm gain}(\omega_{\rm a})$ must be identical to $M^{\rm gain}(\omega_{\rm a})$ and both should take the form
\begin{equation}
    K^{\rm gain}(\omega_{\rm a})=i\tilde g \sqrt{2S^{\rm G}\gamma_{\rm c}}e^{i\omega_{\rm a}t}\int_0^{t}{\rm d}t'e^{-i\tilde\omega_{\rm c}(t-t')} \langle \sigma_z(t) c^{\prime\rm G\dagger}_{\rm in}(t')\rangle=2\left[-i\Delta_{\rm LS}^{\rm gain}-\frac{\Gamma^{\rm gain}}{2}\right]\langle\tilde{\sigma}^-\rangle,
\end{equation}
within a Markov approximation (twice the gain-related part of the complex cavity-induced pump rate) in order to ensure that the unified gain-loss operator approach predicts the correct bad cavity limit.

To further investigate $K^{\rm gain}(\omega_{\rm a})$, we start by solving the equation of motion for $\sigma_z$:
\begin{align}
   \sigma_z(t)=&e^{-\Gamma^{\rm B}t}\sigma_z(0) +2i\sqrt{S'} \int_0^t {\rm d}\tau e^{-\Gamma^{\rm B}(t-\tau)}[-\tilde{g}\tilde{\sigma}^+(\tau)\tilde{a}'(\tau)+\tilde{g}^*\tilde{\sigma}^-(\tau)\tilde{a}^{\prime\dagger}(\tau)]\nonumber\\
   &-\Gamma^{\rm B}\int_0^t{\rm d}\tau e^{-\Gamma^{\rm B}(t-\tau)}-2\sqrt{\Gamma^{\rm B}} \int_0^t {\rm d}\tau e^{-\Gamma^{\rm B}(t-\tau)}[-\sigma^+(\tau)c^{\rm a}_{\rm in}(\tau)+c^{\rm a\dagger}_{\rm in}(\tau)\sigma^-(\tau)].
\end{align}
When inserting back into Eq.~\eqref{eq: SigmaTimeEvoSingleQNM5}, the first and third term on the right-hand side will not contribute to the expectation value because of $\langle c^{\prime\rm G\dagger}_{\rm in}(t'')\rangle=0$. We are left with 
\begin{align}
\langle \sigma_z(t)c^{\prime\rm G\dagger}_{\rm in}(t')\rangle =  &2i\sqrt{S'} \int_0^t {\rm d}\tau e^{-\Gamma^{\rm B}(t-\tau)}[-\tilde{g}\langle \tilde{\sigma}^+(\tau)\tilde{a}'(\tau)c^{\prime\rm G\dagger}_{\rm  in}(t')\rangle+\tilde{g}^*\langle \tilde{\sigma}^-(\tau)\tilde{a}^{\prime\dagger}(\tau)c^{\prime\rm G\dagger}_{\rm in}(t')\rangle]\nonumber\\
   &+2i\sqrt{\Gamma^{\rm B}} \int_0^t {\rm d}\tau e^{-\Gamma^{\rm B}(t-\tau)}[-\langle \sigma^+(\tau)c^{\rm a}_{\rm in}(\tau)c^{\prime\rm G\dagger}_{\rm  in}(t')\rangle+\langle c^{\rm a\dagger}_{\rm in}(\tau)\sigma^-(\tau)c^{\prime\rm G\dagger}_{\rm  in}(t')\rangle].
\end{align}

Since the atomic and QNM input operators are assumed to be statistically independent, the second line of the right-hand side also vanishes, because of the vacuum state assumption for the atomic reservoir. Furthermore, we approximate $\langle\tilde{\sigma}^+(t') \tilde{a}'(t')c^{\prime\rm G\dagger}_{\rm  in}(t'')\rangle\approx \langle \tilde{a}'(t')c^{\prime\rm G\dagger}_{\rm  in}(t'')\rangle\langle\tilde{\sigma}^+(t')\rangle$ and only take the free parts of $a'(t)$ into account (meaning the first, third and fourth term of Eq.~\eqref{eq: FormalaPrimeSol}) to be consistent with second-order pertubation theory: 
\begin{align}
\langle \sigma_z(t)c^{\prime\rm G\dagger}_{\rm in}(t')\rangle \approx  &-2i\sqrt{S'} \int_0^t {\rm d}\tau e^{-\Gamma^{\rm B}(t-\tau)}\tilde{g}e^{i(\omega_{\rm a}-\tilde\omega_{\rm c})\tau}\langle \tilde{a}'(0)c^{\prime\rm G\dagger}_{\rm  in}(t')\rangle\langle\tilde{\sigma}^+(\tau)\rangle\nonumber\\
&+2i\sqrt{S'} \int_0^t {\rm d}\tau e^{-\Gamma^{\rm B}(t-\tau)}\tilde{g}^*e^{-i(\omega_{\rm a}-\tilde\omega_{\rm c}^*)\tau}\langle\tilde{a}^{\prime\dagger}(0) c^{\prime\rm G\dagger}_{\rm in}(t')\rangle\langle\tilde{\sigma}^-(\tau)\rangle]\nonumber\\
&-2i\sqrt{2S^{\rm L}\gamma_c} \int_0^t {\rm d}\tau e^{-\Gamma^{\rm B}(t-\tau)}e^{i\omega_{\rm a}\tau}\int_0^\tau {\rm d}\tau'e^{-i\tilde{\omega}_{\rm c}(\tau-\tau')}\tilde{g}\langle c^{\prime\rm L}_{\rm  in}(\tau')c^{\prime\rm G\dagger}_{\rm  in}(t')\rangle\langle\tilde{\sigma}^+(\tau)\rangle\nonumber\\
&-2i\sqrt{2S^{\rm G}\gamma_c} \int_0^t {\rm d}\tau e^{-\Gamma^{\rm B}(t-\tau)}e^{i\omega_{\rm a}\tau}\int_0^\tau {\rm d}\tau'e^{-i\tilde{\omega}_{\rm c}(\tau-\tau')}\tilde{g}\langle c^{\prime\rm G\dagger}_{\rm  in}(\tau')c^{\prime\rm G\dagger}_{\rm  in}(t')\rangle\langle\tilde{\sigma}^+(\tau)\rangle\nonumber\\
&+2i\sqrt{2S^{\rm L}\gamma_c} \int_0^t {\rm d}\tau e^{-\Gamma^{\rm B}(t-\tau)}e^{-i\omega_{\rm a}\tau}\int_0^\tau {\rm d}\tau'e^{i\tilde{\omega}_{\rm c}^*(\tau-\tau')}\tilde{g}^*\langle c^{\prime\rm L\dagger}_{\rm  in}(\tau')c^{\prime\rm G\dagger}_{\rm  in}(t')\rangle\langle\tilde{\sigma}^-(\tau)\rangle\nonumber\\
&+2i\sqrt{2S^{\rm G}\gamma_c} \int_0^t {\rm d}\tau e^{-\Gamma^{\rm B}(t-\tau)}e^{-i\omega_{\rm a}\tau}\int_0^\tau {\rm d}\tau'e^{i\tilde{\omega}_{\rm c}^*(\tau-\tau')}\tilde{g}^*\langle c^{\prime\rm G}_{\rm  in}(\tau')c^{\prime\rm G\dagger}_{\rm  in}(t')\rangle\langle\tilde{\sigma}^-(\tau)\rangle.
\end{align}

We see that only the last term will contribute to the correlation function $\langle \sigma_z(t)c^{\prime\rm G\dagger}_{\rm in}(t')\rangle$, since $[a'(0),c_{\rm in}^{\prime\rm L(G)}(t)]=0$ (the same applies for any other combination) and  since the loss and gain-related reservoir operators are statistically independent. Using $\langle c^{\prime\rm G}_{\rm  in}(\tau')c^{\prime\rm G\dagger}_{\rm  in}(t')\rangle=\delta(\tau'-t')$, we obtain 
\begin{equation}
    \langle \sigma_z(t)c^{\prime\rm G\dagger}_{\rm in}(t')\rangle = 2i\sqrt{2S^{\rm G}\gamma_c} \int_0^t {\rm d}\tau e^{-\Gamma^{\rm B}(t-\tau)}e^{-i\omega_{\rm a}\tau}\int_0^\tau {\rm d}\tau'e^{i\tilde{\omega}_{\rm c}^*(\tau-\tau')}\tilde{g}^*\delta(\tau'-t')\langle\tilde{\sigma}^-(\tau)\rangle.
\end{equation}
Inserting back into $K^{\rm gain}(\omega_{\rm a})$ yields
\begin{align}
    K^{\rm gain}(\omega_{\rm a})&=-4|\tilde g|^2 S^{\rm G}\gamma_{\rm c}e^{i\omega_{\rm a}t}\int_0^{t}{\rm d}t'e^{-i\tilde\omega_{\rm c}(t-t')}   \int_0^t {\rm d}\tau e^{-\Gamma^{\rm B}(t-\tau)}e^{-i\omega_{\rm a}\tau}\int_0^\tau {\rm d}\tau'e^{i\tilde{\omega}_{\rm c}^*(\tau-\tau')}\tilde{g}^*\delta(\tau'-t')\langle\tilde{\sigma}^-(\tau)\rangle\\
    &=-4|\tilde g|^2 S^{\rm G}\gamma_{\rm c}e^{i\omega_{\rm a}t}\int_0^t {\rm d}\tau\int_0^\tau {\rm d}\tau'e^{-i\tilde\omega_{\rm c}(t-\tau')}    e^{-\Gamma^{\rm B}(t-\tau)}e^{-i\omega_{\rm a}\tau}e^{i\tilde{\omega}_{\rm c}^*(\tau-\tau')}\tilde{g}^*\langle\tilde{\sigma}^-(\tau)\rangle\nonumber\\
    &=-4|\tilde g|^2 S^{\rm G}\gamma_{\rm c}e^{i\omega_{\rm a}t}\int_0^t {\rm d}\tau\left[\int_0^\tau {\rm d}\tau'e^{i(\tilde\omega_{\rm c}-\tilde\omega_{\rm c}^*)\tau'}  \right]e^{-i\tilde\omega_{\rm c}t}    e^{-\Gamma^{\rm B}(t-\tau)}e^{-i\omega_{\rm a}\tau}e^{i\tilde{\omega}_{\rm c}^*\tau}\tilde{g}^*\langle\tilde{\sigma}^-(\tau)\rangle.
\end{align}
The temporal integral with respect to $\tau'$ is performed via 
\begin{equation}
    \int_0^{\tau}{\rm d}\tau'e^{i[\tilde\omega_{\rm c}-\tilde{\omega}_{\rm c}^*]\tau'}=\frac{1}{2\gamma_{\rm c}}\left(e^{2\gamma_{\rm c}\tau}-1\right),
\end{equation}
and we are left with 
\begin{equation}
     K^{\rm gain}(\omega_{\rm a})=-2|\tilde g|^2 S^{\rm G} \int_0^t {\rm d}\tau e^{i(\omega_{\rm a}-\tilde{\omega}_{\rm c}+i\Gamma^{\rm B})(t-\tau)} \langle\tilde{\sigma}^-(\tau)\rangle+2|\tilde g|^2 S^{\rm G} e^{-\gamma_{\rm c}t}\int_0^t  {\rm d}\tau ^{-\gamma_{\rm c}\tau} e^{i(\omega_{\rm a}-\omega_{\rm c}+i\Gamma^{\rm B})(t-\tau)} \langle\tilde{\sigma}^-(\tau)\rangle.
\end{equation}
Within the bad cavity limit regime, we can neglect the second term (which rapidly decays to zero) and applying a Markov approximation to the first term would approximately give
\begin{equation}
     K^{\rm gain}(\omega_{\rm a})\approx -2|\tilde g|^2 S^{\rm G} \langle\tilde{\sigma}^-(t)\rangle\int_0^\infty {\rm d}\tau e^{i(\omega_{\rm a}-\tilde{\omega}_{\rm c}+i\Gamma^{\rm B})(t-\tau)}. 
\end{equation}

To be consistent with the assumption that the atomic reservoir is independent from the photonic reservoir and taking into account the bad cavity limit parameter regime ($\Gamma^{\rm B}\ll\gamma_{\rm c}$), we neglect $\Gamma^{\rm B}$ 
and arrive at
\begin{equation}
     K^{\rm gain}(\omega_{\rm a})\approx -2|\tilde g|^2 S^{\rm G} \langle\tilde{\sigma}^-(t)\rangle\frac{i}{\omega_{\rm a}-\tilde{\omega}_{\rm c}}.
\end{equation}
This is precisely twice the gain-related part of the TLS rate. Indeed, inserting  $K^{\rm gain}(\omega_{\rm a})$ into Eq.~\eqref{eq: BadCavityRes1UnifiedPre} leads to
\begin{align}
     \frac{{\rm d}}{{\rm d}t}\langle\sigma^-\rangle &= -i\omega_{\rm a}\langle\sigma^-\rangle-\frac{\Gamma^{\rm B}}{2}\langle\sigma^-\rangle - i\frac{(S^{\rm L}-S^{\rm G})|\tilde{g}|^2}{\omega_{\rm a}-\tilde{\omega}_{\rm c}}  \langle \sigma^-\rangle -2|\tilde g|^2 S^{\rm G} \langle\sigma^-\rangle\frac{i}{\omega_{\rm a}-\tilde{\omega}_{\rm c}}, \\
     &=-i\omega_{\rm a}\langle\sigma^-\rangle-\frac{\Gamma^{\rm B}}{2}\langle\sigma^-\rangle - i\frac{(S^{\rm L}+S^{\rm G})|\tilde{g}|^2}{\omega_{\rm a}-\tilde{\omega}_{\rm c}}  \langle \sigma^-\rangle \label{eq: BadCavityRes1UnifiedFinal},
\end{align}
which is identical to the expression derived from the separated gain-loss operator approach, and which completes this proof.

\section{Bad cavity limit within a Bloch equation treatment for the improved phenomenological quantum gain model\label{app: BlochEquationsPheno}}
In this section, we show a derivation of the cavity-enhanced spontaneous emission rate of the TLS within the improved phenomenological quantum gain model by explicitly taking the bad cavity limit of the corresponding full Bloch equations. For the derivation of the a bad cavity limit master equation, one could again apply a similar derivation as in Ref.~\onlinecite{Cirac} and Appendix~\ref{app: MarkovBad} (not explicitly shown here).

We start again with the Heisenberg equation of motion for the expectation value of the TLS lowering operator $\langle\sigma^-\rangle$, in a rotating frame with $\omega_{\rm a}$:
\begin{equation}
    \partial_t \langle\tilde{\sigma}^-\rangle = -\frac{\Gamma^{\rm B}}{2}\langle\tilde\sigma^-\rangle-\sum_{i}g_i\langle \sigma_z \tilde{a}_i(t)\rangle\label{eq: Blocheq1},
\end{equation}
where 
\begin{equation}
    \tilde{a}_{i}=e^{i\omega_{\rm a}t}a_i,~\tilde{\sigma}^- = e^{i\omega_{\rm a} t}\sigma^-,
\end{equation}
and $i={\rm L,G}$. Note again, that $\Gamma^{\rm B}$ is included ad-hoc into the phenomenological quantum gain model.
Next, we look at the Heisenberg equation of motion for the photon operator $a_i$:
\begin{equation}
    \partial_t a_i=-i\sum_{j=\rm L,R}\tilde{\Omega}_{ij}a_j - g_i\sigma^- +  \xi_i,
\end{equation}
where $\xi_{\rm L}=F_{\rm L}$ is a phenomenological (quantum) noise force, which counteracts the loss, and $\xi_{\rm G}=F_{\rm G}^\dagger$ is the adjoint operator of a phenomenological (quantum) noise force, which counteracts the amplification. 

The noise forces are assumed to fulfil
\begin{align}
     \langle \xi_{i}(t)\xi_j^\dagger(t')\rangle=2\delta_{ij}\delta_{i\rm L}\gamma_{\rm L}\delta(t-t'),~
     \langle \xi_{i}^\dagger(t)\xi_j(t')\rangle=2\delta_{ij}\delta_{i\rm G}\gamma_{\rm G}\delta(t-t'),
\end{align}
and thus represents white noise with a reservoir state being the vacuum state. Next, we recall the photon-photon matrix from the improved phenomenological quantum gain approach,
\begin{equation}\label{eq: photonmatrix}
    \tilde{\boldsymbol{\Omega}}=\begin{pmatrix}
    \tilde{\omega}_{\rm L} & -\kappa\\
    -\kappa & \tilde{\omega}_{\rm G}
    \end{pmatrix}.
\end{equation}

Formally solving the Heisenberg equations of motion in vector form yields (for $\mathbf{a}=[a_{\rm L},a_{\rm G}]$):
\begin{equation}
    \tilde{\mathbf{a}}(t)=e^{i\omega_{\rm a}t}e^{-i\tilde{\boldsymbol\Omega}t}\cdot\mathbf{a}(0) -\int_0^{t}{\rm d}t'e^{i\omega_{\rm a}(t-t')}e^{-i\tilde{\boldsymbol\Omega}(t-t')}\cdot \mathbf{g}\tilde{\sigma}^-(t') +  \int_0^{t}{\rm d}t'e^{i\omega_{\rm a} t}e^{-i\tilde{\boldsymbol\Omega}(t-t')}\cdot \boldsymbol{\xi}(t').
\end{equation}
Next, we write $\tilde{\boldsymbol\Omega}$ in its diagonal form via $\tilde{\boldsymbol\Omega}=\mathbf{V}\cdot\tilde{\mathbf{D}}\cdot\mathbf{V}^{-1}$ and use 
\begin{equation}
    e^{-i\mathbf{V}\cdot\tilde{\mathbf{D}}\cdot\mathbf{V}^{-1}t}=\mathbf{V}\cdot e^{-i\tilde{\mathbf{D}}t}\cdot\mathbf{V}^{-1},
\end{equation}
to get 
\begin{equation}
    \tilde{\mathbf{a}}(t)=e^{i\omega_{\rm a}t}\mathbf{V}\cdot e^{-i\tilde{\mathbf {D}}t}\cdot\mathbf{V}^{-1}\cdot\mathbf{a}(0) -\int_0^{t}{\rm d}t'e^{i\omega_{\rm a}(t-t')}\mathbf{V}\cdot e^{-i\tilde{\mathbf{D}}(t-t')}\cdot\mathbf{V}^{-1}\cdot\mathbf{g}\tilde{\sigma}^-(t') +  \int_0^{t}{\rm d}t'e^{i\omega_{\rm a} t}\mathbf{V}\cdot e^{-i\tilde{\mathbf{D}}(t-t')}\cdot\mathbf{V}^{-1}\cdot \boldsymbol{\xi}(t').
\end{equation}
Here, $\mathbf{V}$ is the right eigenmatrix of $\tilde{\boldsymbol\Omega}$ and $\tilde{\mathbf{D}}$ contains the eigenvalues of $\tilde{\boldsymbol\Omega}$.
In component form, the above equation reads
\begin{align}
    \tilde{a}_i(t)=&\sum_{j,k}[\mathbf{V}]_{ij} e^{i(\omega_{\rm a}-\tilde{\Omega}^{\rm eig}_j) t}[\mathbf{V}^{-1}]_{jk}a_k(0)\nonumber\\
    &-\sum_{j,k}\int_0^{t}{\rm d}t'[\mathbf{V}]_{ij} e^{i(\omega_{\rm a}-\tilde{\Omega}^{\rm eig}_j)(t-t')}[\mathbf{V}^{-1}]_{jk}g_k\tilde{\sigma}^-(t') \nonumber\\
    &+\sum_{j,k}  e^{i\omega_{\rm a} t}\int_0^{t}{\rm d}t'[\mathbf{V}]_{ij} e^{-i\tilde{\Omega}^{\rm eig}_j(t-t')}[\mathbf{V}^{-1}]_{jk} \xi_k(t').\label{eq: PhotonOpFormalSolPheno}
\end{align}

When inserting back into Eq.~\eqref{eq: Blocheq1}, we can neglect the first part of the above formal solution (which rapidly decays to zero), and obtain 
\begin{align}
    \partial_t \langle\tilde{\sigma}^-\rangle =&-\frac{\Gamma^{\rm B}}{2}\langle\tilde{\sigma}^-\rangle+ \sum_{i,j,k}g_i[\mathbf{V}]_{ij}\left[\int_0^{t}{\rm d}t' e^{i(\omega_{\rm a}-\tilde{\Omega}^{\rm eig}_j)(t-t')}\langle \sigma_z \tilde{\sigma}^-(t') \rangle\right][\mathbf{V}^{-1}]_{jk}g_k\nonumber\\
    &-e^{i\omega_{\rm a}t}\sum_{i,j,k}g_i[\mathbf{V}]_{ij}\left[\int_0^{t}{\rm d}t' e^{-i\tilde{\Omega}^{\rm eig}_j(t-t')}\langle \sigma_z \xi_k(t') \rangle\right][\mathbf{V}^{-1}]_{jk}g_k\label{eq: Blocheq2}.
\end{align}

We first concentrate on the first term and the appearing temporal integral. We substitute the integral variable $t'\rightarrow \tau=t-t'$, so that the term in brackets can be rewritten as
\begin{equation}
    \int_0^{t}{\rm d}t' e^{i(\omega_0-\tilde{\Omega}^{\rm eig}_j)(t-t')}\langle \sigma_z \tilde{\sigma}^-(t') \rangle= \int_0^{t}{\rm d}\tau e^{i(\omega_0-\tilde{\Omega}^{\rm eig}_j)\tau}\langle \sigma_z \tilde{\sigma}^-(t-\tau) \rangle.
\end{equation}
Within a Markov approximation, we replace $\langle \sigma_z \tilde{\sigma}^-(t-\tau) \rangle$ with $\langle \sigma_z \tilde{\sigma}^-(t) \rangle$ and extend the upper integral limit to $t\rightarrow \infty$, so that
\begin{equation}
   \int_0^{t}{\rm d}\tau e^{i(\omega_0-\tilde{\Omega}^{\rm eig}_j)\tau}\langle \sigma_z \tilde{\sigma}^-(t-\tau) \rangle=\langle \sigma_z \tilde{\sigma}^-\rangle \int_0^{\infty}{\rm d}\tau e^{i(\omega_0-\tilde{\Omega}^{\rm eig}_j)\tau}=\frac{i}{\omega_0-\tilde{\Omega}^{\rm eig}_j}\langle \sigma_z \tilde{\sigma}^-\rangle.
\end{equation}

The above approximation is valid for weak light-matter coupling, which is consistent with the bad cavity limit assumptions $|{\rm Im}[\tilde{\Omega}^{\rm eig}_i]|\gg g_i$.
Inserting back into Eq.~\eqref{eq: Blocheq2} yields 
\begin{align}
    \partial_t \langle\tilde{\sigma}^-\rangle =&-\frac{\Gamma^{\rm B}}{2}\langle\tilde{\sigma}^-\rangle+ \sum_{i,j,k}g_i[\mathbf{V}]_{ij}\frac{i}{\omega_0-\tilde{\Omega}^{\rm eig}_j}[\mathbf{V}^{-1}]_{jk}g_k\langle \sigma_z \tilde{\sigma}^- \rangle  \nonumber\\
    &-e^{i\omega_{\rm a}t}\sum_{i,j,k}g_i[\mathbf{V}]_{ij}\left[\int_0^{t}{\rm d}t' e^{-i\tilde{\Omega}^{\rm eig}_j(t-t')}\langle \sigma_z \xi_k(t') \rangle\right][\mathbf{V}^{-1}]_{jk}g_k\label{eq: Blocheq3}.
\end{align}
Next, $\langle \sigma_z \tilde{\sigma}^- \rangle=-\langle \tilde{\sigma}^- \rangle$ is used to get
\begin{align}
    \partial_t \langle\tilde{\sigma}^-\rangle =&-\frac{\Gamma^{\rm B}}{2}\langle\tilde{\sigma}^-\rangle-\frac{1}{2} \left[2\sum_{i,j,k}g_ig_k[\mathbf{V}]_{ij}\frac{i}{\omega_{\rm a}-\tilde{\Omega}^{\rm eig}_j}[\mathbf{V}^{-1}]_{jk}\right]\langle  \tilde{\sigma}^- \rangle-K_{\rm phen}.\label{eq: Blocheq4}
\end{align}
After some algebra, one can check that the real part of the term in brackets is identical to the LDOS part of the cavity-enhanced spontaneous emission rate in a non-diagonal GF form (Eq.~\eqref{eq: GFNonDiagNM}),
\begin{equation}
    2\sum_{i,j,k}g_ig_k[\mathbf{V}]_{ij}\frac{i}{\omega_{\rm a}-\tilde{\Omega}^{\rm eig}_j}[\mathbf{V}^{-1}]_{jk}=\frac{2}{\hbar\epsilon_0}\mathbf{d}\cdot{\rm Im}\left[\mathbf{G}_{\rm phen}(\mathbf{r}_{\rm a},\mathbf{r}_{\rm a},\omega_{\rm a})\right]\cdot\mathbf{d}\equiv \Gamma^{\rm LDOS}_{\rm phen},
\end{equation}
and 
\begin{equation}
    K_{\rm phen}=e^{i\omega_{\rm a}t}\sum_{i,j,k}g_i[\mathbf{V}]_{ij}\left[\int_0^{t}{\rm d}t' e^{-i\tilde{\Omega}^{\rm eig}_j(t-t')}\langle \sigma_z \xi_k(t') \rangle\right][\mathbf{V}^{-1}]_{jk}.
\end{equation}

Next, we investigate $K_{\rm phen}$ by formally solving the equation of motion for $\sigma_z(t)$: Similar to the unified operator approach, we only take into account the terms that couple to $a_l,a_l^\dagger$ to obtain for $\langle \sigma_z \xi_k(t') \rangle$:
\begin{equation}
    \langle \sigma_z \xi_k(t') \rangle=2\sum_l \int_0^t {\rm d}\tau e^{-\Gamma^{\rm B}(t-\tau)}g_l[\langle \tilde{\sigma}^+(\tau)\tilde{a}_l(\tau)\xi_k(t')\rangle+\langle\tilde{\sigma}^-(\tau)\tilde{a}^{\dagger}_l(\tau)\xi_k(t')\rangle].
\end{equation}
Once again, we  approximate $\langle\tilde{\sigma}^+(\tau) \tilde{a}_l(\tau)\xi_k(t')\rangle\approx \langle \tilde{a}_l(\tau)\xi_k(t')\rangle\langle\tilde{\sigma}^+(t')\rangle$ and only take the free parts of $a_l(\tau)$ into account (meaning the first, third and fourth term of Eq.~\eqref{eq: PhotonOpFormalSolPheno}) to be consistent with second-order perturbation theory:
\begin{align}
\langle \sigma_z(t)\xi_k(t')\rangle \approx  &2 \sum_l\int_0^t {\rm d}\tau e^{-\Gamma^{\rm B}(t-\tau)}g_l \sum_{n,m}[\mathbf{V}]_{ln} e^{i(\omega_{\rm a}-\tilde{\Omega}^{\rm eig}_n) t}[\mathbf{V}^{-1}]_{nm}\langle \tilde{a}_m(0)\xi_k(t')\rangle\langle\tilde{\sigma}^+(\tau)\rangle\nonumber\\
&+2\sum_l \int_0^t {\rm d}\tau e^{-\Gamma^{\rm B}(t-\tau)}g_l \sum_{j,m}[\mathbf{V}]_{ln}^* e^{-i(\omega_{\rm a}-\tilde{\Omega}^{\rm eig*}_n) t}[\mathbf{V}^{-1}]_{nm}^*\langle\tilde{a}^{\dagger}_m(0) \xi_k(t')\rangle\langle\tilde{\sigma}^-(\tau)\rangle]\nonumber\\
&+2 \sum_l\int_0^t {\rm d}\tau e^{-\Gamma^{\rm B}(t-\tau)}e^{i\omega_{\rm a}\tau}g_l\sum_{n,m}  \int_0^{\tau}{\rm d}\tau'[\mathbf{V}]_{ln} e^{-i\tilde{\Omega}^{\rm eig}_n(\tau-\tau')}[\mathbf{V}^{-1}]_{nm} \langle \xi_m(\tau') \xi_k(t')\rangle\langle\tilde{\sigma}^+(\tau)\rangle\nonumber\\
&+2\sum_l \int_0^t {\rm d}\tau e^{-\Gamma^{\rm B}(t-\tau)}e^{-i\omega_{\rm a}\tau}g_l\sum_{n,m}  \int_0^{\tau}{\rm d}\tau'[\mathbf{V}]_{ln}^* e^{i\tilde{\Omega}^{\rm eig*}_n(\tau-\tau')}[\mathbf{V}^{-1}]_{nm}^* \langle \xi_m^\dagger(\tau')\xi_k(t')\rangle\langle\tilde{\sigma}^-(\tau)\rangle.
\end{align}

As before, only the last term will survive leading to 
\begin{align}
    K_{\rm phen}=e^{i\omega_{\rm a}t}\sum_{i,j,k}\sum_{l,n,m}g_ig_l[\mathbf{V}]_{ij}[\mathbf{V}]_{ln}^*&\int_0^{t}{\rm d}t' e^{-i\tilde{\Omega}^{\rm eig}_j(t-t')}2 \int_0^t {\rm d}\tau e^{-\Gamma^{\rm B}(t-\tau)}e^{-i\omega_{\rm a}\tau}  \nonumber\\
    &\times\int_0^{\tau}{\rm d}\tau' e^{i\tilde{\Omega}^{\rm eig*}_n(\tau-\tau')}\langle \xi_m^\dagger(\tau')\xi_k(t')\rangle\langle\tilde{\sigma}^-(\tau)\rangle[\mathbf{V}^{-1}]_{nm}^* [\mathbf{V}^{-1}]_{jk}.
\end{align}
This can be simplified to
\begin{align}
    K_{\rm phen}=4e^{i\omega_{\rm a}t}\sum_{i,j,k}\sum_{l,n,m}g_ig_l[\mathbf{V}]_{ij}[\mathbf{V}]_{ln}^*& \int_0^t {\rm d}\tau \left(\int_0^{\tau}{\rm d}\tau'e^{i(\tilde{\Omega}^{\rm eig}_j-\tilde{\Omega}^{\rm eig*}_n)\tau'}\right)e^{-\Gamma^{\rm B}(t-\tau)}e^{-i\omega_{\rm a}\tau}e^{-i\tilde{\Omega}^{\rm eig}_j t}  \nonumber\\
    &\times e^{i\tilde{\Omega}^{\rm eig*}_n \tau}\delta_{km}\delta_{k\rm G}\gamma_{\rm G}\langle\tilde{\sigma}^-(\tau)\rangle[\mathbf{V}^{-1}]_{nm}^* [\mathbf{V}^{-1}]_{jk}.
\end{align}
The integral in brackets can be solved as 
\begin{equation}
    \int_0^{\tau}{\rm d}\tau'e^{i(\tilde{\Omega}^{\rm eig}_j-\tilde{\Omega}^{\rm eig*}_n)\tau'}=\frac{1}{i(\tilde{\Omega}^{\rm eig}_j-\tilde{\Omega}^{\rm eig*}_n)}\left[e^{i(\tilde{\Omega}^{\rm eig}_j-\tilde{\Omega}^{\rm eig*}_n)\tau}-1\right].
\end{equation}
We again neglect the second term and arrive at 
\begin{align}
    K_{\rm phen}=4\sum_{i,j,k}\sum_{l,n,m}g_ig_l[\mathbf{V}]_{ij}[\mathbf{V}]_{ln}^*& \frac{\delta_{km}\delta_{k\rm G}\gamma_{\rm G}}{i(\tilde{\Omega}^{\rm eig}_j-\tilde{\Omega}^{\rm eig*}_n)}\int_0^t {\rm d}\tau e^{i(\omega_{\rm a}-\tilde{\Omega}^{\rm eig}_j+i\Gamma^{\rm B}) (t-\tau)}  \langle\tilde{\sigma}^-(\tau)\rangle[\mathbf{V}^{-1}]_{nm}^* [\mathbf{V}^{-1}]_{jk}.
\end{align}
Applying a Markov approximation, we then get 
\begin{align}
    K_{\rm phen}=4\sum_{i,j}\sum_{l,n}g_ig_l[\mathbf{V}]_{ij}[\mathbf{V}]_{ln}^* \frac{\gamma_{\rm G}}{i(\tilde{\Omega}^{\rm eig}_j-\tilde{\Omega}^{\rm eig*}_n)}\frac{i}{\omega_{\rm a}-\tilde{\Omega}^{\rm eig}_j}  [\mathbf{V}^{-1}]_{j\rm G}[\mathbf{V}^{-1}]_{n\rm G}^* \langle\tilde{\sigma}^-\rangle\equiv 2\left[i\Delta^{\rm gain}_{\rm LS,phen}+\frac{\Gamma^{\rm gain}_{\rm phen}}{2}\right]\langle\tilde\sigma^-\rangle.
\end{align}
After inserting back into Eq.~\eqref{eq: Blocheq4}, and transforming into a non-rotating picture, we finally arrive at 
\begin{align}
    \partial_t \langle\sigma^-\rangle =&-i\left[\Delta_{\rm LS,phen}^{\rm LDOS}+2\Delta_{\rm LS,phen}^{\rm gain}\right]\langle\sigma^-\rangle-\frac{\Gamma^{\rm B}+\Gamma^{\rm LDOS}_{\rm phen}+2\Gamma^{\rm gain}_{\rm phen}}{2}\langle\sigma^-\rangle\label{eq: Blocheq5},
\end{align}
which completes this derivation.

\twocolumngrid


\begin{thebibliography}{79}%
\makeatletter
\providecommand \@ifxundefined [1]{%
 \@ifx{#1\undefined}
}%
\providecommand \@ifnum [1]{%
 \ifnum #1\expandafter \@firstoftwo
 \else \expandafter \@secondoftwo
 \fi
}%
\providecommand \@ifx [1]{%
 \ifx #1\expandafter \@firstoftwo
 \else \expandafter \@secondoftwo
 \fi
}%
\providecommand \natexlab [1]{#1}%
\providecommand \enquote  [1]{``#1''}%
\providecommand \bibnamefont  [1]{#1}%
\providecommand \bibfnamefont [1]{#1}%
\providecommand \citenamefont [1]{#1}%
\providecommand \href@noop [0]{\@secondoftwo}%
\providecommand \href [0]{\begingroup \@sanitize@url \@href}%
\providecommand \@href[1]{\@@startlink{#1}\@@href}%
\providecommand \@@href[1]{\endgroup#1\@@endlink}%
\providecommand \@sanitize@url [0]{\catcode `\\12\catcode `\$12\catcode
  `\&12\catcode `\#12\catcode `\^12\catcode `\_12\catcode `\%12\relax}%
\providecommand \@@startlink[1]{}%
\providecommand \@@endlink[0]{}%
\providecommand \url  [0]{\begingroup\@sanitize@url \@url }%
\providecommand \@url [1]{\endgroup\@href {#1}{\urlprefix }}%
\providecommand \urlprefix  [0]{URL }%
\providecommand \Eprint [0]{\href }%
\providecommand \doibase [0]{https://doi.org/}%
\providecommand \selectlanguage [0]{\@gobble}%
\providecommand \bibinfo  [0]{\@secondoftwo}%
\providecommand \bibfield  [0]{\@secondoftwo}%
\providecommand \translation [1]{[#1]}%
\providecommand \BibitemOpen [0]{}%
\providecommand \bibitemStop [0]{}%
\providecommand \bibitemNoStop [0]{.\EOS\space}%
\providecommand \EOS [0]{\spacefactor3000\relax}%
\providecommand \BibitemShut  [1]{\csname bibitem#1\endcsname}%
\let\auto@bib@innerbib\@empty
%</preamble>
\bibitem [{\citenamefont {Yoshie}\ \emph {et~al.}(2004)\citenamefont {Yoshie},
  \citenamefont {Scherer}, \citenamefont {Hendrickson}, \citenamefont
  {Khitrova}, \citenamefont {Gibbs}, \citenamefont {Rupper}, \citenamefont
  {Ell}, \citenamefont {Shchekin},\ and\ \citenamefont
  {Deppe}}]{Yoshie_Nature_432_200_2004}%
  \BibitemOpen
  \bibfield  {author} {\bibinfo {author} {\bibfnamefont {T.}~\bibnamefont
  {Yoshie}}, \bibinfo {author} {\bibfnamefont {A.}~\bibnamefont {Scherer}},
  \bibinfo {author} {\bibfnamefont {J.}~\bibnamefont {Hendrickson}}, \bibinfo
  {author} {\bibfnamefont {G.}~\bibnamefont {Khitrova}}, \bibinfo {author}
  {\bibfnamefont {H.~M.}\ \bibnamefont {Gibbs}}, \bibinfo {author}
  {\bibfnamefont {G.}~\bibnamefont {Rupper}}, \bibinfo {author} {\bibfnamefont
  {C.}~\bibnamefont {Ell}}, \bibinfo {author} {\bibfnamefont {O.~B.}\
  \bibnamefont {Shchekin}},\ and\ \bibinfo {author} {\bibfnamefont {D.~G.}\
  \bibnamefont {Deppe}},\ }\bibfield  {title} {\bibinfo {title} {Vacuum rabi
  splitting with a single quantum dot in a photonic crystal nanocavity},\
  }\href@noop {} {\bibfield  {journal} {\bibinfo  {journal} {Nature}\ }\textbf
  {\bibinfo {volume} {432}},\ \bibinfo {pages} {200} (\bibinfo {year}
  {2004})}\BibitemShut {NoStop}%
\bibitem [{\citenamefont {Manga~Rao}\ and\ \citenamefont
  {Hughes}(2007)}]{sh2007}%
  \BibitemOpen
  \bibfield  {author} {\bibinfo {author} {\bibfnamefont {V.~S.~C.}\
  \bibnamefont {Manga~Rao}}\ and\ \bibinfo {author} {\bibfnamefont
  {S.}~\bibnamefont {Hughes}},\ }\bibfield  {title} {\bibinfo {title} {Single
  quantum dot spontaneous emission in a finite-size photonic crystal waveguide:
  Proposal for an efficient ``on chip'' single photon gun},\ }\href
  {https://doi.org/10.1103/PhysRevLett.99.193901} {\bibfield  {journal}
  {\bibinfo  {journal} {Phys. Rev. Lett.}\ }\textbf {\bibinfo {volume} {99}},\
  \bibinfo {pages} {193901} (\bibinfo {year} {2007})}\BibitemShut {NoStop}%
\bibitem [{\citenamefont {{Kamandar Dezfouli}}\ \emph
  {et~al.}(2017)\citenamefont {{Kamandar Dezfouli}}, \citenamefont {Gordon},\
  and\ \citenamefont {Hughes}}]{KamandarDezfouli2017}%
  \BibitemOpen
  \bibfield  {author} {\bibinfo {author} {\bibfnamefont {M.}~\bibnamefont
  {{Kamandar Dezfouli}}}, \bibinfo {author} {\bibfnamefont {R.}~\bibnamefont
  {Gordon}},\ and\ \bibinfo {author} {\bibfnamefont {S.}~\bibnamefont
  {Hughes}},\ }\bibfield  {title} {\bibinfo {title} {{Modal theory of modified
  spontaneous emission for a hybrid plasmonic photonic-crystal cavity
  system}},\ }\href {https://doi.org/10.1103/PhysRevA.95.013846} {\bibfield
  {journal} {\bibinfo  {journal} {Phys. Rev. A}\ }\textbf {\bibinfo {volume}
  {95}},\ \bibinfo {pages} {013846} (\bibinfo {year} {2017})}\BibitemShut
  {NoStop}%
\bibitem [{\citenamefont {Reitzenstein}\ \emph {et~al.}(2007)\citenamefont
  {Reitzenstein}, \citenamefont {Hofmann}, \citenamefont {Gorbunov},
  \citenamefont {Strau\ss}, \citenamefont {Kwon}, \citenamefont {Schneider},
  \citenamefont {L\"offler}, \citenamefont {H\"ofling}, \citenamefont {Kamp},\
  and\ \citenamefont {Forchel}}]{micropillars}%
  \BibitemOpen
  \bibfield  {author} {\bibinfo {author} {\bibfnamefont {S.}~\bibnamefont
  {Reitzenstein}}, \bibinfo {author} {\bibfnamefont {C.}~\bibnamefont
  {Hofmann}}, \bibinfo {author} {\bibfnamefont {A.}~\bibnamefont {Gorbunov}},
  \bibinfo {author} {\bibfnamefont {M.}~\bibnamefont {Strau\ss}}, \bibinfo
  {author} {\bibfnamefont {S.~H.}\ \bibnamefont {Kwon}}, \bibinfo {author}
  {\bibfnamefont {C.}~\bibnamefont {Schneider}}, \bibinfo {author}
  {\bibfnamefont {A.}~\bibnamefont {L\"offler}}, \bibinfo {author}
  {\bibfnamefont {S.}~\bibnamefont {H\"ofling}}, \bibinfo {author}
  {\bibfnamefont {M.}~\bibnamefont {Kamp}},\ and\ \bibinfo {author}
  {\bibfnamefont {A.}~\bibnamefont {Forchel}},\ }\bibfield  {title} {\bibinfo
  {title} {{AlAs/GaAs} micropillar cavities with quality factors exceeding
  150,000},\ }\href@noop {} {\bibfield  {journal} {\bibinfo  {journal} {Appl.
  Phys. Lett.}\ }\textbf {\bibinfo {volume} {90}},\ \bibinfo {pages} {251109}
  (\bibinfo {year} {2007})}\BibitemShut {NoStop}%
\bibitem [{\citenamefont {Bajoni}\ \emph {et~al.}(2008)\citenamefont {Bajoni},
  \citenamefont {Senellart}, \citenamefont {Wertz}, \citenamefont {Sagnes},
  \citenamefont {Miard}, \citenamefont {Lema\^{\i}tre},\ and\ \citenamefont
  {Bloch}}]{micropillars2}%
  \BibitemOpen
  \bibfield  {author} {\bibinfo {author} {\bibfnamefont {D.}~\bibnamefont
  {Bajoni}}, \bibinfo {author} {\bibfnamefont {P.}~\bibnamefont {Senellart}},
  \bibinfo {author} {\bibfnamefont {E.}~\bibnamefont {Wertz}}, \bibinfo
  {author} {\bibfnamefont {I.}~\bibnamefont {Sagnes}}, \bibinfo {author}
  {\bibfnamefont {A.}~\bibnamefont {Miard}}, \bibinfo {author} {\bibfnamefont
  {A.}~\bibnamefont {Lema\^{\i}tre}},\ and\ \bibinfo {author} {\bibfnamefont
  {J.}~\bibnamefont {Bloch}},\ }\bibfield  {title} {\bibinfo {title} {Polariton
  laser using single micropillar {GaAs - GaAlAs} semiconductor cavities},\
  }\href {https://doi.org/10.1103/PhysRevLett.100.047401} {\bibfield  {journal}
  {\bibinfo  {journal} {Phys. Rev. Lett.}\ }\textbf {\bibinfo {volume} {100}},\
  \bibinfo {pages} {047401} (\bibinfo {year} {2008})}\BibitemShut {NoStop}%
\bibitem [{\citenamefont {Reithmaier}\ \emph {et~al.}(2004)\citenamefont
  {Reithmaier}, \citenamefont {Sek}, \citenamefont {L\"offler}, \citenamefont
  {Hofmann}, \citenamefont {Kuhn}, \citenamefont {Reitzenstein}, \citenamefont
  {Keldysh}, \citenamefont {Kulakovskii}, \citenamefont {Reinecke},\ and\
  \citenamefont {Forchel}}]{Reithmaier_Nature_432_197_2004}%
  \BibitemOpen
  \bibfield  {author} {\bibinfo {author} {\bibfnamefont {J.~P.}\ \bibnamefont
  {Reithmaier}}, \bibinfo {author} {\bibfnamefont {G.}~\bibnamefont {Sek}},
  \bibinfo {author} {\bibfnamefont {A.}~\bibnamefont {L\"offler}}, \bibinfo
  {author} {\bibfnamefont {C.}~\bibnamefont {Hofmann}}, \bibinfo {author}
  {\bibfnamefont {S.}~\bibnamefont {Kuhn}}, \bibinfo {author} {\bibfnamefont
  {S.}~\bibnamefont {Reitzenstein}}, \bibinfo {author} {\bibfnamefont {L.~V.}\
  \bibnamefont {Keldysh}}, \bibinfo {author} {\bibfnamefont {V.~D.}\
  \bibnamefont {Kulakovskii}}, \bibinfo {author} {\bibfnamefont {T.~L.}\
  \bibnamefont {Reinecke}},\ and\ \bibinfo {author} {\bibfnamefont
  {A.}~\bibnamefont {Forchel}},\ }\bibfield  {title} {\bibinfo {title} {Strong
  coupling in a single quantum dot-semiconductor microcavity system},\
  }\href@noop {} {\bibfield  {journal} {\bibinfo  {journal} {Nature}\ }\textbf
  {\bibinfo {volume} {432}},\ \bibinfo {pages} {197} (\bibinfo {year}
  {2004})}\BibitemShut {NoStop}%
\bibitem [{\citenamefont {Faraon}\ \emph {et~al.}(2008)\citenamefont {Faraon},
  \citenamefont {Fushman}, \citenamefont {Englund}, \citenamefont {Stoltz},
  \citenamefont {Petroff},\ and\ \citenamefont
  {Vu{\v{c}}kovi{\'c}}}]{faraon2008coherent}%
  \BibitemOpen
  \bibfield  {author} {\bibinfo {author} {\bibfnamefont {A.}~\bibnamefont
  {Faraon}}, \bibinfo {author} {\bibfnamefont {I.}~\bibnamefont {Fushman}},
  \bibinfo {author} {\bibfnamefont {D.}~\bibnamefont {Englund}}, \bibinfo
  {author} {\bibfnamefont {N.}~\bibnamefont {Stoltz}}, \bibinfo {author}
  {\bibfnamefont {P.}~\bibnamefont {Petroff}},\ and\ \bibinfo {author}
  {\bibfnamefont {J.}~\bibnamefont {Vu{\v{c}}kovi{\'c}}},\ }\bibfield  {title}
  {\bibinfo {title} {Coherent generation of non-classical light on a chip via
  photon-induced tunnelling and blockade},\ }\href@noop {} {\bibfield
  {journal} {\bibinfo  {journal} {Nature Physics}\ }\textbf {\bibinfo {volume}
  {4}},\ \bibinfo {pages} {859} (\bibinfo {year} {2008})}\BibitemShut {NoStop}%
\bibitem [{\citenamefont {Brooks}\ \emph {et~al.}(2012)\citenamefont {Brooks},
  \citenamefont {Botter}, \citenamefont {Schreppler}, \citenamefont {Purdy},
  \citenamefont {Brahms},\ and\ \citenamefont {Stamper-Kurn}}]{brooks2012non}%
  \BibitemOpen
  \bibfield  {author} {\bibinfo {author} {\bibfnamefont {D.~W.}\ \bibnamefont
  {Brooks}}, \bibinfo {author} {\bibfnamefont {T.}~\bibnamefont {Botter}},
  \bibinfo {author} {\bibfnamefont {S.}~\bibnamefont {Schreppler}}, \bibinfo
  {author} {\bibfnamefont {T.~P.}\ \bibnamefont {Purdy}}, \bibinfo {author}
  {\bibfnamefont {N.}~\bibnamefont {Brahms}},\ and\ \bibinfo {author}
  {\bibfnamefont {D.~M.}\ \bibnamefont {Stamper-Kurn}},\ }\bibfield  {title}
  {\bibinfo {title} {Non-classical light generated by quantum-noise-driven
  cavity optomechanics},\ }\href@noop {} {\bibfield  {journal} {\bibinfo
  {journal} {Nature}\ }\textbf {\bibinfo {volume} {488}},\ \bibinfo {pages}
  {476} (\bibinfo {year} {2012})}\BibitemShut {NoStop}%
\bibitem [{\citenamefont {Imamo\u{g}lu}\ \emph {et~al.}(1999)\citenamefont
  {Imamo\u{g}lu}, \citenamefont {Awschalom}, \citenamefont {Burkard},
  \citenamefont {DiVincenzo}, \citenamefont {Loss}, \citenamefont {Sherwin},\
  and\ \citenamefont {Small}}]{quantinfo}%
  \BibitemOpen
  \bibfield  {author} {\bibinfo {author} {\bibfnamefont {A.}~\bibnamefont
  {Imamo\u{g}lu}}, \bibinfo {author} {\bibfnamefont {D.~D.}\ \bibnamefont
  {Awschalom}}, \bibinfo {author} {\bibfnamefont {G.}~\bibnamefont {Burkard}},
  \bibinfo {author} {\bibfnamefont {D.~P.}\ \bibnamefont {DiVincenzo}},
  \bibinfo {author} {\bibfnamefont {D.}~\bibnamefont {Loss}}, \bibinfo {author}
  {\bibfnamefont {M.}~\bibnamefont {Sherwin}},\ and\ \bibinfo {author}
  {\bibfnamefont {A.}~\bibnamefont {Small}},\ }\bibfield  {title} {\bibinfo
  {title} {Quantum information processing using quantum dot spins and cavity
  {QED}},\ }\href {https://doi.org/10.1103/PhysRevLett.83.4204} {\bibfield
  {journal} {\bibinfo  {journal} {Phys. Rev. Lett.}\ }\textbf {\bibinfo
  {volume} {83}},\ \bibinfo {pages} {4204} (\bibinfo {year}
  {1999})}\BibitemShut {NoStop}%
\bibitem [{\citenamefont {Loss}\ and\ \citenamefont
  {DiVincenzo}(1998)}]{loss1998quantum}%
  \BibitemOpen
  \bibfield  {author} {\bibinfo {author} {\bibfnamefont {D.}~\bibnamefont
  {Loss}}\ and\ \bibinfo {author} {\bibfnamefont {D.~P.}\ \bibnamefont
  {DiVincenzo}},\ }\bibfield  {title} {\bibinfo {title} {Quantum computation
  with quantum dots},\ }\href@noop {} {\bibfield  {journal} {\bibinfo
  {journal} {Physical Review A}\ }\textbf {\bibinfo {volume} {57}},\ \bibinfo
  {pages} {120} (\bibinfo {year} {1998})}\BibitemShut {NoStop}%
\bibitem [{\citenamefont {{Peng, B. and Özdemir, . K. and Rotter, S. and
  Yilmaz, H. and Liertzer, M. and Monifi, F. and Bender, C. M. and Nori, F. and
  Yang, L.}}(2014)}]{peng_loss-induced_2014}%
  \BibitemOpen
  \bibfield  {author} {\bibinfo {author} {\bibnamefont {{Peng, B. and Özdemir,
  . K. and Rotter, S. and Yilmaz, H. and Liertzer, M. and Monifi, F. and
  Bender, C. M. and Nori, F. and Yang, L.}}},\ }\bibfield  {title} {\bibinfo
  {title} {Loss-induced suppression and revival of lasing},\ }\href
  {https://doi.org/10.1126/science.1258004} {\bibfield  {journal} {\bibinfo
  {journal} {Science}\ }\textbf {\bibinfo {volume} {346}},\ \bibinfo {pages}
  {328} (\bibinfo {year} {2014})}\BibitemShut {NoStop}%
\bibitem [{\citenamefont {{Peng, Bo and Özdemir, Şahin Kaya and Lei, Fuchuan
  and Monifi, Faraz and Gianfreda, Mariagiovanna and Long, Gui Lu and Fan,
  Shanhui and Nori, Franco and Bender, Carl M. and Yang,
  Lan}}(2014)}]{peng_paritytime-symmetric_2014}%
  \BibitemOpen
  \bibfield  {author} {\bibinfo {author} {\bibnamefont {{Peng, Bo and Özdemir,
  Şahin Kaya and Lei, Fuchuan and Monifi, Faraz and Gianfreda, Mariagiovanna
  and Long, Gui Lu and Fan, Shanhui and Nori, Franco and Bender, Carl M. and
  Yang, Lan}}},\ }\bibfield  {title} {\bibinfo {title} {Parity–time-symmetric
  whispering-gallery microcavities},\ }\href
  {https://doi.org/10.1038/nphys2927} {\bibfield  {journal} {\bibinfo
  {journal} {Nature Phys.}\ }\textbf {\bibinfo {volume} {10}},\ \bibinfo
  {pages} {394} (\bibinfo {year} {2014})}\BibitemShut {NoStop}%
\bibitem [{\citenamefont {Chang}\ \emph {et~al.}(2014)\citenamefont {Chang},
  \citenamefont {Jiang}, \citenamefont {Hua}, \citenamefont {Yang},
  \citenamefont {Wen}, \citenamefont {Jiang}, \citenamefont {Li}, \citenamefont
  {Wang},\ and\ \citenamefont {Xiao}}]{chang_paritytime_2014}%
  \BibitemOpen
  \bibfield  {author} {\bibinfo {author} {\bibfnamefont {L.}~\bibnamefont
  {Chang}}, \bibinfo {author} {\bibfnamefont {X.}~\bibnamefont {Jiang}},
  \bibinfo {author} {\bibfnamefont {S.}~\bibnamefont {Hua}}, \bibinfo {author}
  {\bibfnamefont {C.}~\bibnamefont {Yang}}, \bibinfo {author} {\bibfnamefont
  {J.}~\bibnamefont {Wen}}, \bibinfo {author} {\bibfnamefont {L.}~\bibnamefont
  {Jiang}}, \bibinfo {author} {\bibfnamefont {G.}~\bibnamefont {Li}}, \bibinfo
  {author} {\bibfnamefont {G.}~\bibnamefont {Wang}},\ and\ \bibinfo {author}
  {\bibfnamefont {M.}~\bibnamefont {Xiao}},\ }\bibfield  {title} {\bibinfo
  {title} {Parity–time symmetry and variable optical isolation in
  active–passive-coupled microresonators},\ }\href
  {https://doi.org/10.1038/nphoton.2014.133} {\bibfield  {journal} {\bibinfo
  {journal} {Nature Photon.}\ }\textbf {\bibinfo {volume} {8}},\ \bibinfo
  {pages} {524} (\bibinfo {year} {2014})}\BibitemShut {NoStop}%
\bibitem [{\citenamefont {{Chen, Weijian and Kaya Özdemir, Şahin and Zhao,
  Guangming and Wiersig, Jan and Yang, Lan}}(2017)}]{chen_exceptional_2017}%
  \BibitemOpen
  \bibfield  {author} {\bibinfo {author} {\bibnamefont {{Chen, Weijian and Kaya
  Özdemir, Şahin and Zhao, Guangming and Wiersig, Jan and Yang, Lan}}},\
  }\bibfield  {title} {\bibinfo {title} {Exceptional points enhance sensing in
  an optical microcavity},\ }\href {https://doi.org/10.1038/nature23281}
  {\bibfield  {journal} {\bibinfo  {journal} {Nature}\ }\textbf {\bibinfo
  {volume} {548}},\ \bibinfo {pages} {192} (\bibinfo {year}
  {2017})}\BibitemShut {NoStop}%
\bibitem [{\citenamefont {{Chen, Weijian and Zhang, Jing and Peng, Bo and
  Özdemir, Şahin Kaya and Fan, Xudong and Yang,
  Lan}}(2018)}]{chen_parity-time-symmetric_2018}%
  \BibitemOpen
  \bibfield  {author} {\bibinfo {author} {\bibnamefont {{Chen, Weijian and
  Zhang, Jing and Peng, Bo and Özdemir, Şahin Kaya and Fan, Xudong and Yang,
  Lan}}},\ }\bibfield  {title} {\bibinfo {title} {Parity-time-symmetric
  whispering-gallery mode nanoparticle sensor [invited]},\ }\href
  {https://doi.org/10.1364/PRJ.6.000A23} {\bibfield  {journal} {\bibinfo
  {journal} {Photonics Research}\ }\textbf {\bibinfo {volume} {6}},\ \bibinfo
  {pages} {A23} (\bibinfo {year} {2018})}\BibitemShut {NoStop}%
\bibitem [{\citenamefont {Miri}\ and\ \citenamefont
  {Alù}(2019)}]{miri_exceptional_2019}%
  \BibitemOpen
  \bibfield  {author} {\bibinfo {author} {\bibfnamefont {M.-A.}\ \bibnamefont
  {Miri}}\ and\ \bibinfo {author} {\bibfnamefont {A.}~\bibnamefont {Alù}},\
  }\bibfield  {title} {\bibinfo {title} {Exceptional points in optics and
  photonics},\ }\href {https://doi.org/10.1126/science.aar7709} {\bibfield
  {journal} {\bibinfo  {journal} {Science}\ }\textbf {\bibinfo {volume}
  {363}},\ \bibinfo {pages} {eaar7709} (\bibinfo {year} {2019})}\BibitemShut
  {NoStop}%
\bibitem [{\citenamefont {Carmichael}(2009)}]{carmichael2009statistical}%
  \BibitemOpen
  \bibfield  {author} {\bibinfo {author} {\bibfnamefont {H.~J.}\ \bibnamefont
  {Carmichael}},\ }\href@noop {} {\emph {\bibinfo {title} {Statistical methods
  in quantum optics 2: Non-classical fields}}}\ (\bibinfo  {publisher}
  {Springer Science \& Business Media},\ \bibinfo {year} {2009})\BibitemShut
  {NoStop}%
\bibitem [{\citenamefont {Agarwal}(2013)}]{GirishBook1}%
  \BibitemOpen
  \bibfield  {author} {\bibinfo {author} {\bibfnamefont {G.~S.}\ \bibnamefont
  {Agarwal}},\ }\href@noop {} {\emph {\bibinfo {title} {Quantum Optics}}}\
  (\bibinfo  {publisher} {Cambridge University Press},\ \bibinfo {year}
  {2013})\BibitemShut {NoStop}%
\bibitem [{\citenamefont {Ho}\ \emph {et~al.}(1998)\citenamefont {Ho},
  \citenamefont {Leung}, \citenamefont {Maassen van~den Brink},\ and\
  \citenamefont {Young}}]{2ndquanho}%
  \BibitemOpen
  \bibfield  {author} {\bibinfo {author} {\bibfnamefont {K.~C.}\ \bibnamefont
  {Ho}}, \bibinfo {author} {\bibfnamefont {P.~T.}\ \bibnamefont {Leung}},
  \bibinfo {author} {\bibfnamefont {A.}~\bibnamefont {Maassen van~den Brink}},\
  and\ \bibinfo {author} {\bibfnamefont {K.}~\bibnamefont {Young}},\ }\bibfield
   {title} {\bibinfo {title} {Second quantization of open systems using
  quasinormal modes},\ }\href {https://doi.org/10.1103/PhysRevE.58.2965}
  {\bibfield  {journal} {\bibinfo  {journal} {Phys. Rev. E}\ }\textbf {\bibinfo
  {volume} {58}},\ \bibinfo {pages} {2965} (\bibinfo {year}
  {1998})}\BibitemShut {NoStop}%
\bibitem [{\citenamefont {Franke}\ \emph {et~al.}(2019)\citenamefont {Franke},
  \citenamefont {Hughes}, \citenamefont {Dezfouli}, \citenamefont {Kristensen},
  \citenamefont {Busch}, \citenamefont {Knorr},\ and\ \citenamefont
  {Richter}}]{PhysRevLett.122.213901}%
  \BibitemOpen
  \bibfield  {author} {\bibinfo {author} {\bibfnamefont {S.}~\bibnamefont
  {Franke}}, \bibinfo {author} {\bibfnamefont {S.}~\bibnamefont {Hughes}},
  \bibinfo {author} {\bibfnamefont {M.~K.}\ \bibnamefont {Dezfouli}}, \bibinfo
  {author} {\bibfnamefont {P.~T.}\ \bibnamefont {Kristensen}}, \bibinfo
  {author} {\bibfnamefont {K.}~\bibnamefont {Busch}}, \bibinfo {author}
  {\bibfnamefont {A.}~\bibnamefont {Knorr}},\ and\ \bibinfo {author}
  {\bibfnamefont {M.}~\bibnamefont {Richter}},\ }\bibfield  {title} {\bibinfo
  {title} {Quantization of quasinormal modes for open cavities and plasmonic
  cavity quantum electrodynamics},\ }\href
  {https://doi.org/10.1103/PhysRevLett.122.213901} {\bibfield  {journal}
  {\bibinfo  {journal} {Phys. Rev. Lett.}\ }\textbf {\bibinfo {volume} {122}},\
  \bibinfo {pages} {213901} (\bibinfo {year} {2019})}\BibitemShut {NoStop}%
\bibitem [{\citenamefont {Kepesidis}\ \emph {et~al.}(2016)\citenamefont
  {Kepesidis}, \citenamefont {Milburn}, \citenamefont {Huber}, \citenamefont
  {Makris}, \citenamefont {Rotter},\ and\ \citenamefont
  {Rabl}}]{Kepesidis_2016}%
  \BibitemOpen
  \bibfield  {author} {\bibinfo {author} {\bibfnamefont {K.~V.}\ \bibnamefont
  {Kepesidis}}, \bibinfo {author} {\bibfnamefont {T.~J.}\ \bibnamefont
  {Milburn}}, \bibinfo {author} {\bibfnamefont {J.}~\bibnamefont {Huber}},
  \bibinfo {author} {\bibfnamefont {K.~G.}\ \bibnamefont {Makris}}, \bibinfo
  {author} {\bibfnamefont {S.}~\bibnamefont {Rotter}},\ and\ \bibinfo {author}
  {\bibfnamefont {P.}~\bibnamefont {Rabl}},\ }\bibfield  {title} {\bibinfo
  {title}
  {{\textdollar}$\lbrace${\textbackslash}mathscr$\lbrace$p$\rbrace$$\rbrace$$\lbrace${\textbackslash}mathscr$\lbrace$t$\rbrace$$\rbrace${\textdollar}-symmetry
  breaking in the steady state of microscopic gain{\textendash}loss systems},\
  }\href {https://doi.org/10.1088/1367-2630/18/9/095003} {\bibfield  {journal}
  {\bibinfo  {journal} {New Journal of Physics}\ }\textbf {\bibinfo {volume}
  {18}},\ \bibinfo {pages} {095003} (\bibinfo {year} {2016})}\BibitemShut
  {NoStop}%
\bibitem [{\citenamefont {Vashahri-Ghamsari}\ \emph {et~al.}(2017)\citenamefont
  {Vashahri-Ghamsari}, \citenamefont {He},\ and\ \citenamefont
  {Xiao}}]{PhysRevA.96.033806}%
  \BibitemOpen
  \bibfield  {author} {\bibinfo {author} {\bibfnamefont {S.}~\bibnamefont
  {Vashahri-Ghamsari}}, \bibinfo {author} {\bibfnamefont {B.}~\bibnamefont
  {He}},\ and\ \bibinfo {author} {\bibfnamefont {M.}~\bibnamefont {Xiao}},\
  }\bibfield  {title} {\bibinfo {title} {Continuous-variable entanglement
  generation using a hybrid $\mathcal{PT}$-symmetric system},\ }\href
  {https://doi.org/10.1103/PhysRevA.96.033806} {\bibfield  {journal} {\bibinfo
  {journal} {Phys. Rev. A}\ }\textbf {\bibinfo {volume} {96}},\ \bibinfo
  {pages} {033806} (\bibinfo {year} {2017})}\BibitemShut {NoStop}%
\bibitem [{\citenamefont {Arkhipov}\ \emph {et~al.}(2019)\citenamefont
  {Arkhipov}, \citenamefont {Miranowicz}, \citenamefont {Di~Stefano},
  \citenamefont {Stassi}, \citenamefont {Savasta}, \citenamefont {Nori},\ and\
  \citenamefont {\"Ozdemir}}]{PhysRevA.99.053806}%
  \BibitemOpen
  \bibfield  {author} {\bibinfo {author} {\bibfnamefont {I.~I.}\ \bibnamefont
  {Arkhipov}}, \bibinfo {author} {\bibfnamefont {A.}~\bibnamefont
  {Miranowicz}}, \bibinfo {author} {\bibfnamefont {O.}~\bibnamefont
  {Di~Stefano}}, \bibinfo {author} {\bibfnamefont {R.}~\bibnamefont {Stassi}},
  \bibinfo {author} {\bibfnamefont {S.}~\bibnamefont {Savasta}}, \bibinfo
  {author} {\bibfnamefont {F.}~\bibnamefont {Nori}},\ and\ \bibinfo {author}
  {\bibfnamefont {i.~m. c.~K.}\ \bibnamefont {\"Ozdemir}},\ }\bibfield  {title}
  {\bibinfo {title} {Scully-lamb quantum laser model for parity-time-symmetric
  whispering-gallery microcavities: Gain saturation effects and
  nonreciprocity},\ }\href {https://doi.org/10.1103/PhysRevA.99.053806}
  {\bibfield  {journal} {\bibinfo  {journal} {Phys. Rev. A}\ }\textbf {\bibinfo
  {volume} {99}},\ \bibinfo {pages} {053806} (\bibinfo {year}
  {2019})}\BibitemShut {NoStop}%
\bibitem [{\citenamefont {Arkhipov}\ \emph {et~al.}(2020)\citenamefont
  {Arkhipov}, \citenamefont {Miranowicz}, \citenamefont {Minganti},\ and\
  \citenamefont {Nori}}]{PhysRevA.101.013812}%
  \BibitemOpen
  \bibfield  {author} {\bibinfo {author} {\bibfnamefont {I.~I.}\ \bibnamefont
  {Arkhipov}}, \bibinfo {author} {\bibfnamefont {A.}~\bibnamefont
  {Miranowicz}}, \bibinfo {author} {\bibfnamefont {F.}~\bibnamefont
  {Minganti}},\ and\ \bibinfo {author} {\bibfnamefont {F.}~\bibnamefont
  {Nori}},\ }\bibfield  {title} {\bibinfo {title} {Quantum and semiclassical
  exceptional points of a linear system of coupled cavities with losses and
  gain within the {Scully-Lamb} laser theory},\ }\href
  {https://doi.org/10.1103/PhysRevA.101.013812} {\bibfield  {journal} {\bibinfo
   {journal} {Phys. Rev. A}\ }\textbf {\bibinfo {volume} {101}},\ \bibinfo
  {pages} {013812} (\bibinfo {year} {2020})}\BibitemShut {NoStop}%
\bibitem [{\citenamefont {Dezfouli}\ \emph {et~al.}(2019)\citenamefont
  {Dezfouli}, \citenamefont {Gordon},\ and\ \citenamefont
  {Hughes}}]{dezfouli2019molecular}%
  \BibitemOpen
  \bibfield  {author} {\bibinfo {author} {\bibfnamefont {M.~K.}\ \bibnamefont
  {Dezfouli}}, \bibinfo {author} {\bibfnamefont {R.}~\bibnamefont {Gordon}},\
  and\ \bibinfo {author} {\bibfnamefont {S.}~\bibnamefont {Hughes}},\
  }\bibfield  {title} {\bibinfo {title} {Molecular optomechanics in the
  anharmonic cavity-qed regime using hybrid metal--dielectric cavity modes},\
  }\href@noop {} {\bibfield  {journal} {\bibinfo  {journal} {ACS Photonics}\
  }\textbf {\bibinfo {volume} {6}},\ \bibinfo {pages} {1400} (\bibinfo {year}
  {2019})}\BibitemShut {NoStop}%
\bibitem [{\citenamefont {Dung}\ \emph {et~al.}(1998)\citenamefont {Dung},
  \citenamefont {Kn\"oll},\ and\ \citenamefont {Welsch}}]{Dung}%
  \BibitemOpen
  \bibfield  {author} {\bibinfo {author} {\bibfnamefont {H.~T.}\ \bibnamefont
  {Dung}}, \bibinfo {author} {\bibfnamefont {L.}~\bibnamefont {Kn\"oll}},\ and\
  \bibinfo {author} {\bibfnamefont {D.-G.}\ \bibnamefont {Welsch}},\ }\bibfield
   {title} {\bibinfo {title} {Three-dimensional quantization of the
  electromagnetic field in dispersive and absorbing inhomogeneous
  dielectrics},\ }\href {https://doi.org/10.1103/PhysRevA.57.3931} {\bibfield
  {journal} {\bibinfo  {journal} {Phys. Rev. A}\ }\textbf {\bibinfo {volume}
  {57}},\ \bibinfo {pages} {3931} (\bibinfo {year} {1998})}\BibitemShut
  {NoStop}%
\bibitem [{\citenamefont {Scheel}\ \emph {et~al.}(1998)\citenamefont {Scheel},
  \citenamefont {Kn{\"o}ll},\ and\ \citenamefont {Welsch}}]{scheel1998qed}%
  \BibitemOpen
  \bibfield  {author} {\bibinfo {author} {\bibfnamefont {S.}~\bibnamefont
  {Scheel}}, \bibinfo {author} {\bibfnamefont {L.}~\bibnamefont {Kn{\"o}ll}},\
  and\ \bibinfo {author} {\bibfnamefont {D.-G.}\ \bibnamefont {Welsch}},\
  }\bibfield  {title} {\bibinfo {title} {{QED} commutation relations for
  inhomogeneous kramers-kronig dielectrics},\ }\href@noop {} {\bibfield
  {journal} {\bibinfo  {journal} {Physical Review A}\ }\textbf {\bibinfo
  {volume} {58}},\ \bibinfo {pages} {700} (\bibinfo {year} {1998})}\BibitemShut
  {NoStop}%
\bibitem [{\citenamefont {Vogel}\ and\ \citenamefont
  {Welsch}(2006)}]{vogel2006}%
  \BibitemOpen
  \bibfield  {author} {\bibinfo {author} {\bibfnamefont {W.}~\bibnamefont
  {Vogel}}\ and\ \bibinfo {author} {\bibfnamefont {D.-G.}\ \bibnamefont
  {Welsch}},\ }\href@noop {} {\emph {\bibinfo {title} {Quantum optics}}}\
  (\bibinfo  {publisher} {John Wiley \& Sons},\ \bibinfo {year}
  {2006})\BibitemShut {NoStop}%
\bibitem [{\citenamefont {Huttner}\ and\ \citenamefont
  {Barnett}(1992)}]{Huttner}%
  \BibitemOpen
  \bibfield  {author} {\bibinfo {author} {\bibfnamefont {B.}~\bibnamefont
  {Huttner}}\ and\ \bibinfo {author} {\bibfnamefont {S.~M.}\ \bibnamefont
  {Barnett}},\ }\bibfield  {title} {\bibinfo {title} {Quantization of the
  electromagnetic field in dielectrics},\ }\href
  {https://doi.org/10.1103/PhysRevA.46.4306} {\bibfield  {journal} {\bibinfo
  {journal} {Phys. Rev. A}\ }\textbf {\bibinfo {volume} {46}},\ \bibinfo
  {pages} {4306} (\bibinfo {year} {1992})}\BibitemShut {NoStop}%
\bibitem [{\citenamefont {Barnett}\ \emph {et~al.}(1992)\citenamefont
  {Barnett}, \citenamefont {Huttner},\ and\ \citenamefont
  {Loudon}}]{barnett1992spontaneous}%
  \BibitemOpen
  \bibfield  {author} {\bibinfo {author} {\bibfnamefont {S.~M.}\ \bibnamefont
  {Barnett}}, \bibinfo {author} {\bibfnamefont {B.}~\bibnamefont {Huttner}},\
  and\ \bibinfo {author} {\bibfnamefont {R.}~\bibnamefont {Loudon}},\
  }\bibfield  {title} {\bibinfo {title} {Spontaneous emission in absorbing
  dielectric media},\ }\href@noop {} {\bibfield  {journal} {\bibinfo  {journal}
  {Physical review letters}\ }\textbf {\bibinfo {volume} {68}},\ \bibinfo
  {pages} {3698} (\bibinfo {year} {1992})}\BibitemShut {NoStop}%
\bibitem [{\citenamefont {Suttorp}\ and\ \citenamefont {van
  Wonderen}(2004)}]{suttorp}%
  \BibitemOpen
  \bibfield  {author} {\bibinfo {author} {\bibfnamefont {L.~G.}\ \bibnamefont
  {Suttorp}}\ and\ \bibinfo {author} {\bibfnamefont {A.~J.}\ \bibnamefont {van
  Wonderen}},\ }\bibfield  {title} {\bibinfo {title} {Fano diagonalization of a
  polariton model for an inhomogeneous absorptive dielectric},\ }\href
  {http://stacks.iop.org/0295-5075/67/i=5/a=766} {\bibfield  {journal}
  {\bibinfo  {journal} {EPL (Europhysics Letters)}\ }\textbf {\bibinfo {volume}
  {67}},\ \bibinfo {pages} {766} (\bibinfo {year} {2004})}\BibitemShut
  {NoStop}%
\bibitem [{\citenamefont {Philbin}(2010)}]{philbin2010canonical}%
  \BibitemOpen
  \bibfield  {author} {\bibinfo {author} {\bibfnamefont {T.~G.}\ \bibnamefont
  {Philbin}},\ }\bibfield  {title} {\bibinfo {title} {Canonical quantization of
  macroscopic electromagnetism},\ }\href@noop {} {\bibfield  {journal}
  {\bibinfo  {journal} {New Journal of Physics}\ }\textbf {\bibinfo {volume}
  {12}},\ \bibinfo {pages} {123008} (\bibinfo {year} {2010})}\BibitemShut
  {NoStop}%
\bibitem [{\citenamefont {Hopfield}(1958)}]{hopfield1958theory}%
  \BibitemOpen
  \bibfield  {author} {\bibinfo {author} {\bibfnamefont {J.}~\bibnamefont
  {Hopfield}},\ }\bibfield  {title} {\bibinfo {title} {Theory of the
  contribution of excitons to the complex dielectric constant of crystals},\
  }\href@noop {} {\bibfield  {journal} {\bibinfo  {journal} {Physical Review}\
  }\textbf {\bibinfo {volume} {112}},\ \bibinfo {pages} {1555} (\bibinfo {year}
  {1958})}\BibitemShut {NoStop}%
\bibitem [{\citenamefont {Fano}(1947)}]{PhysRev.72.26}%
  \BibitemOpen
  \bibfield  {author} {\bibinfo {author} {\bibfnamefont {U.}~\bibnamefont
  {Fano}},\ }\bibfield  {title} {\bibinfo {title} {Ionization yield of
  radiations. ii. the fluctuations of the number of ions},\ }\href
  {https://doi.org/10.1103/PhysRev.72.26} {\bibfield  {journal} {\bibinfo
  {journal} {Phys. Rev.}\ }\textbf {\bibinfo {volume} {72}},\ \bibinfo {pages}
  {26} (\bibinfo {year} {1947})}\BibitemShut {NoStop}%
\bibitem [{\citenamefont {Raabe}\ and\ \citenamefont
  {Welsch}(2008)}]{raabe2008qed}%
  \BibitemOpen
  \bibfield  {author} {\bibinfo {author} {\bibfnamefont {C.}~\bibnamefont
  {Raabe}}\ and\ \bibinfo {author} {\bibfnamefont {D.-G.}\ \bibnamefont
  {Welsch}},\ }\bibfield  {title} {\bibinfo {title} {{QED} in arbitrary linear
  media: Amplifying media},\ }\href@noop {} {\bibfield  {journal} {\bibinfo
  {journal} {The European Physical Journal Special Topics}\ }\textbf {\bibinfo
  {volume} {160}},\ \bibinfo {pages} {371} (\bibinfo {year}
  {2008})}\BibitemShut {NoStop}%
\bibitem [{\citenamefont {Amooghorban}\ \emph {et~al.}(2011)\citenamefont
  {Amooghorban}, \citenamefont {Wubs}, \citenamefont {Mortensen},\ and\
  \citenamefont {Kheirandish}}]{PhysRevA.84.013806}%
  \BibitemOpen
  \bibfield  {author} {\bibinfo {author} {\bibfnamefont {E.}~\bibnamefont
  {Amooghorban}}, \bibinfo {author} {\bibfnamefont {M.}~\bibnamefont {Wubs}},
  \bibinfo {author} {\bibfnamefont {N.~A.}\ \bibnamefont {Mortensen}},\ and\
  \bibinfo {author} {\bibfnamefont {F.}~\bibnamefont {Kheirandish}},\
  }\bibfield  {title} {\bibinfo {title} {Casimir forces in multilayer
  magnetodielectrics with both gain and loss},\ }\href
  {https://doi.org/10.1103/PhysRevA.84.013806} {\bibfield  {journal} {\bibinfo
  {journal} {Phys. Rev. A}\ }\textbf {\bibinfo {volume} {84}},\ \bibinfo
  {pages} {013806} (\bibinfo {year} {2011})}\BibitemShut {NoStop}%
\bibitem [{\citenamefont {Franke}\ \emph {et~al.}(2021)\citenamefont {Franke},
  \citenamefont {Ren}, \citenamefont {Richter}, \citenamefont {Knorr},\ and\
  \citenamefont {Hughes}}]{franke2021fermi}%
  \BibitemOpen
  \bibfield  {author} {\bibinfo {author} {\bibfnamefont {S.}~\bibnamefont
  {Franke}}, \bibinfo {author} {\bibfnamefont {J.}~\bibnamefont {Ren}},
  \bibinfo {author} {\bibfnamefont {M.}~\bibnamefont {Richter}}, \bibinfo
  {author} {\bibfnamefont {A.}~\bibnamefont {Knorr}},\ and\ \bibinfo {author}
  {\bibfnamefont {S.}~\bibnamefont {Hughes}},\ }\bibfield  {title} {\bibinfo
  {title} {Fermi's golden rule for spontaneous emission in absorptive and
  amplifying media},\ }\href@noop {} {\bibfield  {journal} {\bibinfo  {journal}
  {arXiv preprint arXiv:2102.13015}\ } (\bibinfo {year} {2021})}\BibitemShut
  {NoStop}%
\bibitem [{\citenamefont {Scheel}\ \emph {et~al.}(1999)\citenamefont {Scheel},
  \citenamefont {Kn\"oll},\ and\ \citenamefont {Welsch}}]{Scheel}%
  \BibitemOpen
  \bibfield  {author} {\bibinfo {author} {\bibfnamefont {S.}~\bibnamefont
  {Scheel}}, \bibinfo {author} {\bibfnamefont {L.}~\bibnamefont {Kn\"oll}},\
  and\ \bibinfo {author} {\bibfnamefont {D.-G.}\ \bibnamefont {Welsch}},\
  }\bibfield  {title} {\bibinfo {title} {Spontaneous decay of an excited atom
  in an absorbing dielectric},\ }\href
  {https://doi.org/10.1103/PhysRevA.60.4094} {\bibfield  {journal} {\bibinfo
  {journal} {Phys. Rev. A}\ }\textbf {\bibinfo {volume} {60}},\ \bibinfo
  {pages} {4094} (\bibinfo {year} {1999})}\BibitemShut {NoStop}%
\bibitem [{\citenamefont {Dung}\ \emph {et~al.}(2000)\citenamefont {Dung},
  \citenamefont {Kn{\"o}ll},\ and\ \citenamefont
  {Welsch}}]{dung2000spontaneous}%
  \BibitemOpen
  \bibfield  {author} {\bibinfo {author} {\bibfnamefont {H.~T.}\ \bibnamefont
  {Dung}}, \bibinfo {author} {\bibfnamefont {L.}~\bibnamefont {Kn{\"o}ll}},\
  and\ \bibinfo {author} {\bibfnamefont {D.-G.}\ \bibnamefont {Welsch}},\
  }\bibfield  {title} {\bibinfo {title} {Spontaneous decay in the presence of
  dispersing and absorbing bodies: General theory and application to a
  spherical cavity},\ }\href@noop {} {\bibfield  {journal} {\bibinfo  {journal}
  {Physical Review A}\ }\textbf {\bibinfo {volume} {62}},\ \bibinfo {pages}
  {053804} (\bibinfo {year} {2000})}\BibitemShut {NoStop}%
\bibitem [{\citenamefont {Philbin}(2011)}]{philbin2011casimir}%
  \BibitemOpen
  \bibfield  {author} {\bibinfo {author} {\bibfnamefont {T.~G.}\ \bibnamefont
  {Philbin}},\ }\bibfield  {title} {\bibinfo {title} {Casimir effect from
  macroscopic quantum electrodynamics},\ }\href@noop {} {\bibfield  {journal}
  {\bibinfo  {journal} {New Journal of Physics}\ }\textbf {\bibinfo {volume}
  {13}},\ \bibinfo {pages} {063026} (\bibinfo {year} {2011})}\BibitemShut
  {NoStop}%
\bibitem [{\citenamefont {Ge}\ and\ \citenamefont
  {Hughes}(2015)}]{PhysRevB.92.205420}%
  \BibitemOpen
  \bibfield  {author} {\bibinfo {author} {\bibfnamefont {R.-C.}\ \bibnamefont
  {Ge}}\ and\ \bibinfo {author} {\bibfnamefont {S.}~\bibnamefont {Hughes}},\
  }\bibfield  {title} {\bibinfo {title} {Quantum dynamics of two quantum dots
  coupled through localized plasmons: An intuitive and accurate quantum optics
  approach using quasinormal modes},\ }\href
  {https://doi.org/10.1103/PhysRevB.92.205420} {\bibfield  {journal} {\bibinfo
  {journal} {Phys. Rev. B}\ }\textbf {\bibinfo {volume} {92}},\ \bibinfo
  {pages} {205420} (\bibinfo {year} {2015})}\BibitemShut {NoStop}%
\bibitem [{\citenamefont {Franke}\ \emph
  {et~al.}(2020{\natexlab{a}})\citenamefont {Franke}, \citenamefont {Ren},
  \citenamefont {Hughes},\ and\ \citenamefont
  {Richter}}]{franke2020fluctuation}%
  \BibitemOpen
  \bibfield  {author} {\bibinfo {author} {\bibfnamefont {S.}~\bibnamefont
  {Franke}}, \bibinfo {author} {\bibfnamefont {J.}~\bibnamefont {Ren}},
  \bibinfo {author} {\bibfnamefont {S.}~\bibnamefont {Hughes}},\ and\ \bibinfo
  {author} {\bibfnamefont {M.}~\bibnamefont {Richter}},\ }\bibfield  {title}
  {\bibinfo {title} {Fluctuation-dissipation theorem and fundamental photon
  commutation relations in lossy nanostructures using quasinormal modes},\
  }\href {https://doi.org/10.1103/PhysRevResearch.2.033332} {\bibfield
  {journal} {\bibinfo  {journal} {Phys. Rev. Research}\ }\textbf {\bibinfo
  {volume} {2}},\ \bibinfo {pages} {033332} (\bibinfo {year}
  {2020}{\natexlab{a}})}\BibitemShut {NoStop}%
\bibitem [{\citenamefont {Franke}\ \emph
  {et~al.}(2020{\natexlab{b}})\citenamefont {Franke}, \citenamefont {Richter},
  \citenamefont {Ren}, \citenamefont {Knorr},\ and\ \citenamefont
  {Hughes}}]{franke2020quantized}%
  \BibitemOpen
  \bibfield  {author} {\bibinfo {author} {\bibfnamefont {S.}~\bibnamefont
  {Franke}}, \bibinfo {author} {\bibfnamefont {M.}~\bibnamefont {Richter}},
  \bibinfo {author} {\bibfnamefont {J.}~\bibnamefont {Ren}}, \bibinfo {author}
  {\bibfnamefont {A.}~\bibnamefont {Knorr}},\ and\ \bibinfo {author}
  {\bibfnamefont {S.}~\bibnamefont {Hughes}},\ }\bibfield  {title} {\bibinfo
  {title} {Quantized quasinormal mode description of non-linear cavity qed
  effects from coupled resonators with a fano-like resonance},\ }\href@noop {}
  {\bibfield  {journal} {\bibinfo  {journal} {arXiv preprint arXiv:2006.04506}\
  } (\bibinfo {year} {2020}{\natexlab{b}})}\BibitemShut {NoStop}%
\bibitem [{\citenamefont {Lai}\ \emph {et~al.}(1990)\citenamefont {Lai},
  \citenamefont {Leung}, \citenamefont {Young}, \citenamefont {Barber},\ and\
  \citenamefont {Hill}}]{Lai}%
  \BibitemOpen
  \bibfield  {author} {\bibinfo {author} {\bibfnamefont {H.~M.}\ \bibnamefont
  {Lai}}, \bibinfo {author} {\bibfnamefont {P.~T.}\ \bibnamefont {Leung}},
  \bibinfo {author} {\bibfnamefont {K.}~\bibnamefont {Young}}, \bibinfo
  {author} {\bibfnamefont {P.~W.}\ \bibnamefont {Barber}},\ and\ \bibinfo
  {author} {\bibfnamefont {S.~C.}\ \bibnamefont {Hill}},\ }\bibfield  {title}
  {\bibinfo {title} {Time-independent perturbation for leaking electromagnetic
  modes in open systems with application to resonances in microdroplets},\
  }\href {https://doi.org/10.1103/PhysRevA.41.5187} {\bibfield  {journal}
  {\bibinfo  {journal} {Phys. Rev. A}\ }\textbf {\bibinfo {volume} {41}},\
  \bibinfo {pages} {5187} (\bibinfo {year} {1990})}\BibitemShut {NoStop}%
\bibitem [{\citenamefont {Leung}\ \emph
  {et~al.}(1994{\natexlab{a}})\citenamefont {Leung}, \citenamefont {Liu},\ and\
  \citenamefont {Young}}]{LeungSP1D}%
  \BibitemOpen
  \bibfield  {author} {\bibinfo {author} {\bibfnamefont {P.~T.}\ \bibnamefont
  {Leung}}, \bibinfo {author} {\bibfnamefont {S.~Y.}\ \bibnamefont {Liu}},\
  and\ \bibinfo {author} {\bibfnamefont {K.}~\bibnamefont {Young}},\ }\bibfield
   {title} {\bibinfo {title} {Completeness and orthogonality of quasinormal
  modes in leaky optical cavities},\ }\href
  {https://doi.org/10.1103/PhysRevA.49.3057} {\bibfield  {journal} {\bibinfo
  {journal} {Phys. Rev. A}\ }\textbf {\bibinfo {volume} {49}},\ \bibinfo
  {pages} {3057} (\bibinfo {year} {1994}{\natexlab{a}})}\BibitemShut {NoStop}%
\bibitem [{\citenamefont {Leung}\ \emph
  {et~al.}(1994{\natexlab{b}})\citenamefont {Leung}, \citenamefont {Liu},\ and\
  \citenamefont {Young}}]{Leung3}%
  \BibitemOpen
  \bibfield  {author} {\bibinfo {author} {\bibfnamefont {P.~T.}\ \bibnamefont
  {Leung}}, \bibinfo {author} {\bibfnamefont {S.~Y.}\ \bibnamefont {Liu}},\
  and\ \bibinfo {author} {\bibfnamefont {K.}~\bibnamefont {Young}},\ }\bibfield
   {title} {\bibinfo {title} {Completeness and time-independent perturbation of
  the quasinormal modes of an absorptive and leaky cavity},\ }\href
  {https://doi.org/10.1103/PhysRevA.49.3982} {\bibfield  {journal} {\bibinfo
  {journal} {Phys. Rev. A}\ }\textbf {\bibinfo {volume} {49}},\ \bibinfo
  {pages} {3982} (\bibinfo {year} {1994}{\natexlab{b}})}\BibitemShut {NoStop}%
\bibitem [{\citenamefont {Ching}\ \emph {et~al.}(1998)\citenamefont {Ching},
  \citenamefont {Leung}, \citenamefont {Maassen van~den Brink}, \citenamefont
  {Suen}, \citenamefont {Tong},\ and\ \citenamefont {Young}}]{2ndquant2}%
  \BibitemOpen
  \bibfield  {author} {\bibinfo {author} {\bibfnamefont {E.~S.~C.}\
  \bibnamefont {Ching}}, \bibinfo {author} {\bibfnamefont {P.~T.}\ \bibnamefont
  {Leung}}, \bibinfo {author} {\bibfnamefont {A.}~\bibnamefont {Maassen van~den
  Brink}}, \bibinfo {author} {\bibfnamefont {W.~M.}\ \bibnamefont {Suen}},
  \bibinfo {author} {\bibfnamefont {S.~S.}\ \bibnamefont {Tong}},\ and\
  \bibinfo {author} {\bibfnamefont {K.}~\bibnamefont {Young}},\ }\bibfield
  {title} {\bibinfo {title} {Quasinormal-mode expansion for waves in open
  systems},\ }\href {https://doi.org/10.1103/RevModPhys.70.1545} {\bibfield
  {journal} {\bibinfo  {journal} {Rev. Mod. Phys.}\ }\textbf {\bibinfo {volume}
  {70}},\ \bibinfo {pages} {1545} (\bibinfo {year} {1998})}\BibitemShut
  {NoStop}%
\bibitem [{\citenamefont {Kristensen}\ \emph {et~al.}(2020)\citenamefont
  {Kristensen}, \citenamefont {Herrmann}, \citenamefont {Intravaia},\ and\
  \citenamefont {Busch}}]{Kristensen:20}%
  \BibitemOpen
  \bibfield  {author} {\bibinfo {author} {\bibfnamefont {P.~T.}\ \bibnamefont
  {Kristensen}}, \bibinfo {author} {\bibfnamefont {K.}~\bibnamefont
  {Herrmann}}, \bibinfo {author} {\bibfnamefont {F.}~\bibnamefont
  {Intravaia}},\ and\ \bibinfo {author} {\bibfnamefont {K.}~\bibnamefont
  {Busch}},\ }\bibfield  {title} {\bibinfo {title} {Modeling electromagnetic
  resonators using quasinormal modes},\ }\href
  {https://doi.org/10.1364/AOP.377940} {\bibfield  {journal} {\bibinfo
  {journal} {Adv. Opt. Photon.}\ }\textbf {\bibinfo {volume} {12}},\ \bibinfo
  {pages} {612} (\bibinfo {year} {2020})}\BibitemShut {NoStop}%
\bibitem [{\citenamefont {Muljarov}\ \emph {et~al.}(2010)\citenamefont
  {Muljarov}, \citenamefont {Langbein},\ and\ \citenamefont
  {Zimmermann}}]{muljarovPert}%
  \BibitemOpen
  \bibfield  {author} {\bibinfo {author} {\bibfnamefont {E.~A.}\ \bibnamefont
  {Muljarov}}, \bibinfo {author} {\bibfnamefont {W.}~\bibnamefont {Langbein}},\
  and\ \bibinfo {author} {\bibfnamefont {R.}~\bibnamefont {Zimmermann}},\
  }\bibfield  {title} {\bibinfo {title} {Brillouin-wigner perturbation theory
  in open electromagnetic systems},\ }\href@noop {} {\bibfield  {journal}
  {\bibinfo  {journal} {EPL (Europhysics Letters)}\ }\textbf {\bibinfo {volume}
  {92}},\ \bibinfo {pages} {50010} (\bibinfo {year} {2010})}\BibitemShut
  {NoStop}%
\bibitem [{\citenamefont {Kristensen}\ \emph
  {et~al.}(2012{\natexlab{a}})\citenamefont {Kristensen}, \citenamefont
  {Van~Vlack},\ and\ \citenamefont {Hughes}}]{KristensenHughes}%
  \BibitemOpen
  \bibfield  {author} {\bibinfo {author} {\bibfnamefont {P.~T.}\ \bibnamefont
  {Kristensen}}, \bibinfo {author} {\bibfnamefont {C.}~\bibnamefont
  {Van~Vlack}},\ and\ \bibinfo {author} {\bibfnamefont {S.}~\bibnamefont
  {Hughes}},\ }\bibfield  {title} {\bibinfo {title} {Generalized effective mode
  volume for leaky optical cavities},\ }\href
  {https://doi.org/10.1364/OL.37.001649} {\bibfield  {journal} {\bibinfo
  {journal} {Opt. Lett.}\ }\textbf {\bibinfo {volume} {37}},\ \bibinfo {pages}
  {1649} (\bibinfo {year} {2012}{\natexlab{a}})}\BibitemShut {NoStop}%
\bibitem [{\citenamefont {Sauvan}\ \emph {et~al.}(2013)\citenamefont {Sauvan},
  \citenamefont {Hugonin}, \citenamefont {Maksymov},\ and\ \citenamefont
  {Lalanne}}]{SauvanNorm}%
  \BibitemOpen
  \bibfield  {author} {\bibinfo {author} {\bibfnamefont {C.}~\bibnamefont
  {Sauvan}}, \bibinfo {author} {\bibfnamefont {J.~P.}\ \bibnamefont {Hugonin}},
  \bibinfo {author} {\bibfnamefont {I.~S.}\ \bibnamefont {Maksymov}},\ and\
  \bibinfo {author} {\bibfnamefont {P.}~\bibnamefont {Lalanne}},\ }\bibfield
  {title} {\bibinfo {title} {Theory of the spontaneous optical emission of
  nanosize photonic and plasmon resonators},\ }\href
  {https://doi.org/10.1103/PhysRevLett.110.237401} {\bibfield  {journal}
  {\bibinfo  {journal} {Phys. Rev. Lett.}\ }\textbf {\bibinfo {volume} {110}},\
  \bibinfo {pages} {237401} (\bibinfo {year} {2013})}\BibitemShut {NoStop}%
\bibitem [{\citenamefont {Kristensen}\ and\ \citenamefont
  {Hughes}(2014)}]{NormKristHughes}%
  \BibitemOpen
  \bibfield  {author} {\bibinfo {author} {\bibfnamefont {P.~T.}\ \bibnamefont
  {Kristensen}}\ and\ \bibinfo {author} {\bibfnamefont {S.}~\bibnamefont
  {Hughes}},\ }\bibfield  {title} {\bibinfo {title} {Modes and mode volumes of
  leaky optical cavities and plasmonic nanoresonators},\ }\href@noop {}
  {\bibfield  {journal} {\bibinfo  {journal} {ACS Photonics}\ }\textbf
  {\bibinfo {volume} {1}},\ \bibinfo {pages} {2} (\bibinfo {year}
  {2014})}\BibitemShut {NoStop}%
\bibitem [{\citenamefont {Zschiedrich}\ \emph {et~al.}(2018)\citenamefont
  {Zschiedrich}, \citenamefont {Binkowski}, \citenamefont {Nikolay},
  \citenamefont {Benson}, \citenamefont {Kewes},\ and\ \citenamefont
  {Burger}}]{PhysRevA.98.043806}%
  \BibitemOpen
  \bibfield  {author} {\bibinfo {author} {\bibfnamefont {L.}~\bibnamefont
  {Zschiedrich}}, \bibinfo {author} {\bibfnamefont {F.}~\bibnamefont
  {Binkowski}}, \bibinfo {author} {\bibfnamefont {N.}~\bibnamefont {Nikolay}},
  \bibinfo {author} {\bibfnamefont {O.}~\bibnamefont {Benson}}, \bibinfo
  {author} {\bibfnamefont {G.}~\bibnamefont {Kewes}},\ and\ \bibinfo {author}
  {\bibfnamefont {S.}~\bibnamefont {Burger}},\ }\bibfield  {title} {\bibinfo
  {title} {Riesz-projection-based theory of light-matter interaction in
  dispersive nanoresonators},\ }\href
  {https://doi.org/10.1103/PhysRevA.98.043806} {\bibfield  {journal} {\bibinfo
  {journal} {Phys. Rev. A}\ }\textbf {\bibinfo {volume} {98}},\ \bibinfo
  {pages} {043806} (\bibinfo {year} {2018})}\BibitemShut {NoStop}%
\bibitem [{\citenamefont {Lalanne}\ \emph {et~al.}(2018)\citenamefont
  {Lalanne}, \citenamefont {Yan}, \citenamefont {Vynck}, \citenamefont
  {Sauvan},\ and\ \citenamefont {Hugonin}}]{Lalanne_review}%
  \BibitemOpen
  \bibfield  {author} {\bibinfo {author} {\bibfnamefont {P.}~\bibnamefont
  {Lalanne}}, \bibinfo {author} {\bibfnamefont {W.}~\bibnamefont {Yan}},
  \bibinfo {author} {\bibfnamefont {K.}~\bibnamefont {Vynck}}, \bibinfo
  {author} {\bibfnamefont {C.}~\bibnamefont {Sauvan}},\ and\ \bibinfo {author}
  {\bibfnamefont {J.-P.}\ \bibnamefont {Hugonin}},\ }\bibfield  {title}
  {\bibinfo {title} {Light interaction with photonic and plasmonic
  resonances},\ }\href@noop {} {\bibfield  {journal} {\bibinfo  {journal}
  {Laser \& Photonics Reviews}\ }\textbf {\bibinfo {volume} {12}},\ \bibinfo
  {pages} {1700113} (\bibinfo {year} {2018})}\BibitemShut {NoStop}%
\bibitem [{\citenamefont {Carlson}\ and\ \citenamefont
  {Hughes}(2019)}]{carlson2019dissipative}%
  \BibitemOpen
  \bibfield  {author} {\bibinfo {author} {\bibfnamefont {C.}~\bibnamefont
  {Carlson}}\ and\ \bibinfo {author} {\bibfnamefont {S.}~\bibnamefont
  {Hughes}},\ }\bibfield  {title} {\bibinfo {title} {Dissipative modes, purcell
  factors and directional beta factors in gold bowtie nanoantenna structures},\
  }\href@noop {} {\bibfield  {journal} {\bibinfo  {journal} {arXiv preprint
  arXiv:1910.10110}\ } (\bibinfo {year} {2019})}\BibitemShut {NoStop}%
\bibitem [{\citenamefont {Ren}\ \emph {et~al.}(2021{\natexlab{a}})\citenamefont
  {Ren}, \citenamefont {Franke},\ and\ \citenamefont
  {Hughes}}]{EPClassicalPaper}%
  \BibitemOpen
  \bibfield  {author} {\bibinfo {author} {\bibfnamefont {J.}~\bibnamefont
  {Ren}}, \bibinfo {author} {\bibfnamefont {S.}~\bibnamefont {Franke}},\ and\
  \bibinfo {author} {\bibfnamefont {S.}~\bibnamefont {Hughes}},\ }\href@noop {}
  {\bibinfo {title} {Quasinormal modes and {Purcell factors} of coupled loss
  and gain resonators near an exceptional point}} (\bibinfo {year}
  {2021}{\natexlab{a}}),\ \Eprint {https://arxiv.org/abs/arXiv:2101.07633}
  {arXiv:2101.07633} \BibitemShut {NoStop}%
\bibitem [{\citenamefont {Minganti}\ \emph {et~al.}(2019)\citenamefont
  {Minganti}, \citenamefont {Miranowicz}, \citenamefont {Chhajlany},\ and\
  \citenamefont {Nori}}]{PhysRevA.100.062131}%
  \BibitemOpen
  \bibfield  {author} {\bibinfo {author} {\bibfnamefont {F.}~\bibnamefont
  {Minganti}}, \bibinfo {author} {\bibfnamefont {A.}~\bibnamefont
  {Miranowicz}}, \bibinfo {author} {\bibfnamefont {R.~W.}\ \bibnamefont
  {Chhajlany}},\ and\ \bibinfo {author} {\bibfnamefont {F.}~\bibnamefont
  {Nori}},\ }\bibfield  {title} {\bibinfo {title} {Quantum exceptional points
  of {non-Hermitian} hamiltonians and liouvillians: The effects of quantum
  jumps},\ }\href {https://doi.org/10.1103/PhysRevA.100.062131} {\bibfield
  {journal} {\bibinfo  {journal} {Phys. Rev. A}\ }\textbf {\bibinfo {volume}
  {100}},\ \bibinfo {pages} {062131} (\bibinfo {year} {2019})}\BibitemShut
  {NoStop}%
\bibitem [{\citenamefont {Zhang}\ and\ \citenamefont
  {Li}(2010)}]{PhysRevA.81.033843}%
  \BibitemOpen
  \bibfield  {author} {\bibinfo {author} {\bibfnamefont {K.}~\bibnamefont
  {Zhang}}\ and\ \bibinfo {author} {\bibfnamefont {Z.-Y.}\ \bibnamefont {Li}},\
  }\bibfield  {title} {\bibinfo {title} {Transfer behavior of quantum states
  between atoms in photonic crystal coupled cavities},\ }\href
  {https://doi.org/10.1103/PhysRevA.81.033843} {\bibfield  {journal} {\bibinfo
  {journal} {Phys. Rev. A}\ }\textbf {\bibinfo {volume} {81}},\ \bibinfo
  {pages} {033843} (\bibinfo {year} {2010})}\BibitemShut {NoStop}%
\bibitem [{\citenamefont {Barlow}\ \emph {et~al.}(2015)\citenamefont {Barlow},
  \citenamefont {Bennett},\ and\ \citenamefont
  {Beige}}]{doi:10.1080/09500340.2014.992992}%
  \BibitemOpen
  \bibfield  {author} {\bibinfo {author} {\bibfnamefont {T.~M.}\ \bibnamefont
  {Barlow}}, \bibinfo {author} {\bibfnamefont {R.}~\bibnamefont {Bennett}},\
  and\ \bibinfo {author} {\bibfnamefont {A.}~\bibnamefont {Beige}},\ }\bibfield
   {title} {\bibinfo {title} {A master equation for a two-sided optical
  cavity},\ }\href {https://doi.org/10.1080/09500340.2014.992992} {\bibfield
  {journal} {\bibinfo  {journal} {Journal of Modern Optics}\ }\textbf {\bibinfo
  {volume} {62}},\ \bibinfo {pages} {S11} (\bibinfo {year} {2015})},\ \bibinfo
  {note} {pMID: 25892851},\ \Eprint
  {https://arxiv.org/abs/https://doi.org/10.1080/09500340.2014.992992}
  {https://doi.org/10.1080/09500340.2014.992992} \BibitemShut {NoStop}%
\bibitem [{\citenamefont {Kristensen}\ \emph {et~al.}(2015)\citenamefont
  {Kristensen}, \citenamefont {Ge},\ and\ \citenamefont
  {Hughes}}]{Kristensen2015}%
  \BibitemOpen
  \bibfield  {author} {\bibinfo {author} {\bibfnamefont {P.~T.}\ \bibnamefont
  {Kristensen}}, \bibinfo {author} {\bibfnamefont {R.-C.}\ \bibnamefont {Ge}},\
  and\ \bibinfo {author} {\bibfnamefont {S.}~\bibnamefont {Hughes}},\
  }\bibfield  {title} {\bibinfo {title} {Normalization of quasinormal modes in
  leaky optical cavities and plasmonic resonators},\ }\href@noop {} {\bibfield
  {journal} {\bibinfo  {journal} {Phys. Rev. A}\ }\textbf {\bibinfo {volume}
  {92}},\ \bibinfo {pages} {053810} (\bibinfo {year} {2015})}\BibitemShut
  {NoStop}%
\bibitem [{\citenamefont {Bai}\ \emph {et~al.}(2013)\citenamefont {Bai},
  \citenamefont {Perrin}, \citenamefont {Sauvan}, \citenamefont {Hugonin},\
  and\ \citenamefont {Lalanne}}]{Bai}%
  \BibitemOpen
  \bibfield  {author} {\bibinfo {author} {\bibfnamefont {Q.}~\bibnamefont
  {Bai}}, \bibinfo {author} {\bibfnamefont {M.}~\bibnamefont {Perrin}},
  \bibinfo {author} {\bibfnamefont {C.}~\bibnamefont {Sauvan}}, \bibinfo
  {author} {\bibfnamefont {J.-P.}\ \bibnamefont {Hugonin}},\ and\ \bibinfo
  {author} {\bibfnamefont {P.}~\bibnamefont {Lalanne}},\ }\bibfield  {title}
  {\bibinfo {title} {Efficient and intuitive method for the analysis of light
  scattering by a resonant nanostructure},\ }\href
  {https://doi.org/10.1364/OE.21.027371} {\bibfield  {journal} {\bibinfo
  {journal} {Opt. Express}\ }\textbf {\bibinfo {volume} {21}},\ \bibinfo
  {pages} {27371} (\bibinfo {year} {2013})}\BibitemShut {NoStop}%
\bibitem [{\citenamefont {Lee}\ \emph {et~al.}(1999)\citenamefont {Lee},
  \citenamefont {Leung},\ and\ \citenamefont {Pang}}]{MDR1}%
  \BibitemOpen
  \bibfield  {author} {\bibinfo {author} {\bibfnamefont {K.~M.}\ \bibnamefont
  {Lee}}, \bibinfo {author} {\bibfnamefont {P.~T.}\ \bibnamefont {Leung}},\
  and\ \bibinfo {author} {\bibfnamefont {K.~M.}\ \bibnamefont {Pang}},\
  }\bibfield  {title} {\bibinfo {title} {Dyadic formulation of
  morphology-dependent resonances. i. completeness relation},\ }\href
  {https://doi.org/10.1364/JOSAB.16.001409} {\bibfield  {journal} {\bibinfo
  {journal} {J. Opt. Soc. Am. B}\ }\textbf {\bibinfo {volume} {16}},\ \bibinfo
  {pages} {1409} (\bibinfo {year} {1999})}\BibitemShut {NoStop}%
\bibitem [{\citenamefont {Ge}\ \emph {et~al.}(2014)\citenamefont {Ge},
  \citenamefont {Kristensen}, \citenamefont {Young},\ and\ \citenamefont
  {Hughes}}]{GeNJP2014}%
  \BibitemOpen
  \bibfield  {author} {\bibinfo {author} {\bibfnamefont {R.-C.}\ \bibnamefont
  {Ge}}, \bibinfo {author} {\bibfnamefont {P.~T.}\ \bibnamefont {Kristensen}},
  \bibinfo {author} {\bibfnamefont {J.~F.}\ \bibnamefont {Young}},\ and\
  \bibinfo {author} {\bibfnamefont {S.}~\bibnamefont {Hughes}},\ }\bibfield
  {title} {\bibinfo {title} {Quasinormal mode approach to modelling
  light-emission and propagation in nanoplasmonics},\ }\href
  {http://stacks.iop.org/1367-2630/16/i=11/a=113048} {\bibfield  {journal}
  {\bibinfo  {journal} {New Journal of Physics}\ }\textbf {\bibinfo {volume}
  {16}},\ \bibinfo {pages} {113048} (\bibinfo {year} {2014})}\BibitemShut
  {NoStop}%
\bibitem [{\citenamefont {Ren}\ \emph {et~al.}(2020{\natexlab{a}})\citenamefont
  {Ren}, \citenamefont {Franke}, \citenamefont {Knorr}, \citenamefont
  {Richter},\ and\ \citenamefont {Hughes}}]{ren_near-field_2020}%
  \BibitemOpen
  \bibfield  {author} {\bibinfo {author} {\bibfnamefont {J.}~\bibnamefont
  {Ren}}, \bibinfo {author} {\bibfnamefont {S.}~\bibnamefont {Franke}},
  \bibinfo {author} {\bibfnamefont {A.}~\bibnamefont {Knorr}}, \bibinfo
  {author} {\bibfnamefont {M.}~\bibnamefont {Richter}},\ and\ \bibinfo {author}
  {\bibfnamefont {S.}~\bibnamefont {Hughes}},\ }\bibfield  {title} {\bibinfo
  {title} {Near-field to far-field transformations of optical quasinormal modes
  and efficient calculation of quantized quasinormal modes for open cavities
  and plasmonic resonators},\ }\href
  {https://doi.org/10.1103/PhysRevB.101.205402} {\bibfield  {journal} {\bibinfo
   {journal} {Physical Review B}\ }\textbf {\bibinfo {volume} {101}},\ \bibinfo
  {pages} {205402} (\bibinfo {year} {2020}{\natexlab{a}})}\BibitemShut
  {NoStop}%
\bibitem [{\citenamefont {Ren}\ \emph {et~al.}(2020{\natexlab{b}})\citenamefont
  {Ren}, \citenamefont {Franke}, \citenamefont {Knorr}, \citenamefont
  {Richter},\ and\ \citenamefont {Hughes}}]{PhysRevB.101.205402}%
  \BibitemOpen
  \bibfield  {author} {\bibinfo {author} {\bibfnamefont {J.}~\bibnamefont
  {Ren}}, \bibinfo {author} {\bibfnamefont {S.}~\bibnamefont {Franke}},
  \bibinfo {author} {\bibfnamefont {A.}~\bibnamefont {Knorr}}, \bibinfo
  {author} {\bibfnamefont {M.}~\bibnamefont {Richter}},\ and\ \bibinfo {author}
  {\bibfnamefont {S.}~\bibnamefont {Hughes}},\ }\bibfield  {title} {\bibinfo
  {title} {Near-field to far-field transformations of optical quasinormal modes
  and efficient calculation of quantized quasinormal modes for open cavities
  and plasmonic resonators},\ }\href
  {https://doi.org/10.1103/PhysRevB.101.205402} {\bibfield  {journal} {\bibinfo
   {journal} {Phys. Rev. B}\ }\textbf {\bibinfo {volume} {101}},\ \bibinfo
  {pages} {205402} (\bibinfo {year} {2020}{\natexlab{b}})}\BibitemShut
  {NoStop}%
\bibitem [{\citenamefont {Matloob}\ \emph {et~al.}(1997)\citenamefont
  {Matloob}, \citenamefont {Loudon}, \citenamefont {Artoni}, \citenamefont
  {Barnett},\ and\ \citenamefont {Jeffers}}]{PhysRevA.55.1623}%
  \BibitemOpen
  \bibfield  {author} {\bibinfo {author} {\bibfnamefont {R.}~\bibnamefont
  {Matloob}}, \bibinfo {author} {\bibfnamefont {R.}~\bibnamefont {Loudon}},
  \bibinfo {author} {\bibfnamefont {M.}~\bibnamefont {Artoni}}, \bibinfo
  {author} {\bibfnamefont {S.~M.}\ \bibnamefont {Barnett}},\ and\ \bibinfo
  {author} {\bibfnamefont {J.}~\bibnamefont {Jeffers}},\ }\bibfield  {title}
  {\bibinfo {title} {Electromagnetic field quantization in amplifying
  dielectrics},\ }\href {https://doi.org/10.1103/PhysRevA.55.1623} {\bibfield
  {journal} {\bibinfo  {journal} {Phys. Rev. A}\ }\textbf {\bibinfo {volume}
  {55}},\ \bibinfo {pages} {1623} (\bibinfo {year} {1997})}\BibitemShut
  {NoStop}%
\bibitem [{\citenamefont {Kristensen}\ \emph
  {et~al.}(2012{\natexlab{b}})\citenamefont {Kristensen}, \citenamefont
  {Van~Vlack},\ and\ \citenamefont {Hughes}}]{kristensen2012generalized}%
  \BibitemOpen
  \bibfield  {author} {\bibinfo {author} {\bibfnamefont {P.~T.}\ \bibnamefont
  {Kristensen}}, \bibinfo {author} {\bibfnamefont {C.}~\bibnamefont
  {Van~Vlack}},\ and\ \bibinfo {author} {\bibfnamefont {S.}~\bibnamefont
  {Hughes}},\ }\bibfield  {title} {\bibinfo {title} {Generalized effective mode
  volume for leaky optical cavities},\ }\href@noop {} {\bibfield  {journal}
  {\bibinfo  {journal} {Optics letters}\ }\textbf {\bibinfo {volume} {37}},\
  \bibinfo {pages} {1649} (\bibinfo {year} {2012}{\natexlab{b}})}\BibitemShut
  {NoStop}%
\bibitem [{\citenamefont {Tao}\ \emph {et~al.}(2020)\citenamefont {Tao},
  \citenamefont {Zhu}, \citenamefont {Zhong},\ and\ \citenamefont
  {Liu}}]{tao_coupling_2020}%
  \BibitemOpen
  \bibfield  {author} {\bibinfo {author} {\bibfnamefont {C.}~\bibnamefont
  {Tao}}, \bibinfo {author} {\bibfnamefont {J.}~\bibnamefont {Zhu}}, \bibinfo
  {author} {\bibfnamefont {Y.}~\bibnamefont {Zhong}},\ and\ \bibinfo {author}
  {\bibfnamefont {H.}~\bibnamefont {Liu}},\ }\bibfield  {title} {\bibinfo
  {title} {Coupling theory of quasinormal modes for lossy and dispersive
  plasmonic nanoresonators},\ }\href
  {https://doi.org/10.1103/PhysRevB.102.045430} {\bibfield  {journal} {\bibinfo
   {journal} {Physical Review B}\ }\textbf {\bibinfo {volume} {102}},\ \bibinfo
  {pages} {045430} (\bibinfo {year} {2020})}\BibitemShut {NoStop}%
\bibitem [{\citenamefont {{COMSOL Inc.}}()}]{comsol}%
  \BibitemOpen
  \bibfield  {author} {\bibinfo {author} {\bibnamefont {{COMSOL Inc.}}},\
  }\href {www.comsol.com} {\bibinfo {title} {Comsol multiphysics v
  5.4}}\BibitemShut {NoStop}%
\bibitem [{\citenamefont {Ren}\ \emph {et~al.}(2021{\natexlab{b}})\citenamefont
  {Ren}, \citenamefont {Franke},\ and\ \citenamefont {Hughes}}]{2108.10194}%
  \BibitemOpen
  \bibfield  {author} {\bibinfo {author} {\bibfnamefont {J.}~\bibnamefont
  {Ren}}, \bibinfo {author} {\bibfnamefont {S.}~\bibnamefont {Franke}},\ and\
  \bibinfo {author} {\bibfnamefont {S.}~\bibnamefont {Hughes}},\ }\href@noop {}
  {\bibinfo {title} {Connecting classical and quantum mode theories for coupled
  lossy cavity resonators using quasinormal modes}} (\bibinfo {year}
  {2021}{\natexlab{b}}),\ \Eprint {https://arxiv.org/abs/arXiv:2108.10194}
  {arXiv:2108.10194} \BibitemShut {NoStop}%
\bibitem [{\citenamefont {Cirac}(1992)}]{Cirac}%
  \BibitemOpen
  \bibfield  {author} {\bibinfo {author} {\bibfnamefont {J.~I.}\ \bibnamefont
  {Cirac}},\ }\bibfield  {title} {\bibinfo {title} {Interaction of a two-level
  atom with a cavity mode in the bad-cavity limit},\ }\href
  {https://doi.org/10.1103/PhysRevA.46.4354} {\bibfield  {journal} {\bibinfo
  {journal} {Phys. Rev. A}\ }\textbf {\bibinfo {volume} {46}},\ \bibinfo
  {pages} {4354} (\bibinfo {year} {1992})}\BibitemShut {NoStop}%
\bibitem [{\citenamefont {Bardroff}\ and\ \citenamefont
  {Stenholm}(1999)}]{PhysRevA.60.2529}%
  \BibitemOpen
  \bibfield  {author} {\bibinfo {author} {\bibfnamefont {P.~J.}\ \bibnamefont
  {Bardroff}}\ and\ \bibinfo {author} {\bibfnamefont {S.}~\bibnamefont
  {Stenholm}},\ }\bibfield  {title} {\bibinfo {title} {Quantum theory of excess
  noise},\ }\href {https://doi.org/10.1103/PhysRevA.60.2529} {\bibfield
  {journal} {\bibinfo  {journal} {Phys. Rev. A}\ }\textbf {\bibinfo {volume}
  {60}},\ \bibinfo {pages} {2529} (\bibinfo {year} {1999})}\BibitemShut
  {NoStop}%
\bibitem [{\citenamefont {Bardroff}\ and\ \citenamefont
  {Stenholm}(2000)}]{PhysRevA.61.023806}%
  \BibitemOpen
  \bibfield  {author} {\bibinfo {author} {\bibfnamefont {P.~J.}\ \bibnamefont
  {Bardroff}}\ and\ \bibinfo {author} {\bibfnamefont {S.}~\bibnamefont
  {Stenholm}},\ }\bibfield  {title} {\bibinfo {title} {Quantum langevin theory
  of excess noise in lasers},\ }\href
  {https://doi.org/10.1103/PhysRevA.61.023806} {\bibfield  {journal} {\bibinfo
  {journal} {Phys. Rev. A}\ }\textbf {\bibinfo {volume} {61}},\ \bibinfo
  {pages} {023806} (\bibinfo {year} {2000})}\BibitemShut {NoStop}%
\bibitem [{\citenamefont {Buhmann}(2013)}]{buhmann2013dispersionII}%
  \BibitemOpen
  \bibfield  {author} {\bibinfo {author} {\bibfnamefont {S.}~\bibnamefont
  {Buhmann}},\ }\href@noop {} {\emph {\bibinfo {title} {Dispersion Forces II:
  Many-Body Effects, Excited Atoms, Finite Temperature and Quantum
  Friction}}},\ Vol.\ \bibinfo {volume} {248}\ (\bibinfo  {publisher}
  {Springer},\ \bibinfo {year} {2013})\BibitemShut {NoStop}%
\bibitem [{\citenamefont {Gardiner}\ and\ \citenamefont
  {Collett}(1985)}]{Gardiner1}%
  \BibitemOpen
  \bibfield  {author} {\bibinfo {author} {\bibfnamefont {C.~W.}\ \bibnamefont
  {Gardiner}}\ and\ \bibinfo {author} {\bibfnamefont {M.~J.}\ \bibnamefont
  {Collett}},\ }\bibfield  {title} {\bibinfo {title} {Input and output in
  damped quantum systems: Quantum stochastic differential equations and the
  master equation},\ }\href {https://doi.org/10.1103/PhysRevA.31.3761}
  {\bibfield  {journal} {\bibinfo  {journal} {Phys. Rev. A}\ }\textbf {\bibinfo
  {volume} {31}},\ \bibinfo {pages} {3761} (\bibinfo {year}
  {1985})}\BibitemShut {NoStop}%
\bibitem [{\citenamefont {Amooghorban}\ \emph {et~al.}(2013)\citenamefont
  {Amooghorban}, \citenamefont {Mortensen},\ and\ \citenamefont
  {Wubs}}]{PhysRevLett.110.153602}%
  \BibitemOpen
  \bibfield  {author} {\bibinfo {author} {\bibfnamefont {E.}~\bibnamefont
  {Amooghorban}}, \bibinfo {author} {\bibfnamefont {N.~A.}\ \bibnamefont
  {Mortensen}},\ and\ \bibinfo {author} {\bibfnamefont {M.}~\bibnamefont
  {Wubs}},\ }\bibfield  {title} {\bibinfo {title} {Quantum optical
  effective-medium theory for loss-compensated metamaterials},\ }\href
  {https://doi.org/10.1103/PhysRevLett.110.153602} {\bibfield  {journal}
  {\bibinfo  {journal} {Phys. Rev. Lett.}\ }\textbf {\bibinfo {volume} {110}},\
  \bibinfo {pages} {153602} (\bibinfo {year} {2013})}\BibitemShut {NoStop}%
\bibitem [{\citenamefont {Abraham}\ and\ \citenamefont
  {Penrose}(2017)}]{PhysRevE.95.012125}%
  \BibitemOpen
  \bibfield  {author} {\bibinfo {author} {\bibfnamefont {E.}~\bibnamefont
  {Abraham}}\ and\ \bibinfo {author} {\bibfnamefont {O.}~\bibnamefont
  {Penrose}},\ }\bibfield  {title} {\bibinfo {title} {Physics of negative
  absolute temperatures},\ }\href {https://doi.org/10.1103/PhysRevE.95.012125}
  {\bibfield  {journal} {\bibinfo  {journal} {Phys. Rev. E}\ }\textbf {\bibinfo
  {volume} {95}},\ \bibinfo {pages} {012125} (\bibinfo {year}
  {2017})}\BibitemShut {NoStop}%
\bibitem [{\citenamefont {Hilbert}\ \emph {et~al.}(2014)\citenamefont
  {Hilbert}, \citenamefont {H\"anggi},\ and\ \citenamefont
  {Dunkel}}]{PhysRevE.90.062116}%
  \BibitemOpen
  \bibfield  {author} {\bibinfo {author} {\bibfnamefont {S.}~\bibnamefont
  {Hilbert}}, \bibinfo {author} {\bibfnamefont {P.}~\bibnamefont {H\"anggi}},\
  and\ \bibinfo {author} {\bibfnamefont {J.}~\bibnamefont {Dunkel}},\
  }\bibfield  {title} {\bibinfo {title} {Thermodynamic laws in isolated
  systems},\ }\href {https://doi.org/10.1103/PhysRevE.90.062116} {\bibfield
  {journal} {\bibinfo  {journal} {Phys. Rev. E}\ }\textbf {\bibinfo {volume}
  {90}},\ \bibinfo {pages} {062116} (\bibinfo {year} {2014})}\BibitemShut
  {NoStop}%
\bibitem [{\citenamefont {Hall}(2003)}]{hall2015lie}%
  \BibitemOpen
  \bibfield  {author} {\bibinfo {author} {\bibfnamefont {B.}~\bibnamefont
  {Hall}},\ }\href@noop {} {\emph {\bibinfo {title} {Lie groups, Lie algebras,
  and representations: an elementary introduction}}},\ Vol.\ \bibinfo {volume}
  {222}\ (\bibinfo  {publisher} {Springer},\ \bibinfo {year}
  {2003})\BibitemShut {NoStop}%
\end{thebibliography}
\end{document}